\documentclass{article}
\usepackage{comment}
\usepackage{wrapfig}
\usepackage{enumerate}
\usepackage{paralist}
\usepackage{color}
\usepackage{amsmath}
\usepackage{amssymb}
\usepackage{booktabs}
\usepackage{curves}
\usepackage{epubtk}
\usepackage{epic}
\usepackage{graphicx}
\usepackage{longtable}
\usepackage{textcomp}

\numberwithin{equation}{section}
\numberwithin{figure}{section}
\numberwithin{table}{section}

\renewcommand{\vec}[1]{\boldsymbol{\mathrm{#1}}}

\newcommand{\theappendix}{}

\begin{document}

\title{The Pioneer Anomaly}

\author{%
\epubtkAuthorData{Slava G.\ Turyshev}{%
Jet Propulsion Laboratory,
California Institute of Technology,\\
4800 Oak Grove Drive, Pasadena, CA 91109, USA}{%
turyshev@jpl.nasa.gov}{%
http://science.jpl.nasa.gov/people/Turyshev/}%
\\
~\\
\epubtkAuthorData{Viktor T.\ Toth}{%
Ottawa, ON  K1N 9H5, Canada%
}{%
vttoth@vttoth.com}{%
http://www.vttoth.com/}%
}

\date{}
\maketitle

\begin{abstract}
Radio-metric Doppler tracking data received from the Pioneer~10 and 11
spacecraft from heliocentric distances of 20\,--\,70~AU has
consistently indicated the presence of a small, anomalous,
blue-shifted frequency drift uniformly changing with a rate of
$\sim6\times 10^{-9}$~Hz/s. Ultimately, the drift was interpreted as a
constant sunward deceleration of each particular spacecraft at the
level of $a_P  = (8.74 \pm 1.33)\times 10^{-10}$~m/s$^2$. This
apparent violation of the Newton's gravitational inverse-square law
has become known as the Pioneer anomaly; the nature of this anomaly
remains unexplained. In this review, we summarize the current
knowledge of the physical properties of the anomaly and the
conditions that led to its detection and characterization. We  review
various mechanisms proposed to explain the anomaly and discuss the
current state of efforts to determine its nature. A comprehensive new
investigation of the anomalous behavior of the two Pioneers has begun
recently. The new efforts rely on the much-extended set of
radio-metric Doppler data for both spacecraft in conjunction with the
newly available complete record of their telemetry files and a large
archive of original project documentation. As the new study is yet to
report its findings, this review provides the necessary background for
the new results to appear in the near future. In particular, we
provide a significant amount of information on the design, operations
and behavior of the two Pioneers during their entire missions,
including descriptions of various data formats and techniques used for
their navigation and radio-science data analysis. As most of this
information was recovered relatively recently, it was not used in the
previous studies of the Pioneer anomaly, but it is critical for the
new investigation.
\end{abstract}


\epubtkKeywords{Pioneer Anomaly}


\newpage


\section{Introduction}
\label{sec:intro}

For generations of researchers our solar system provided opportunities
to establish and test fundamental laws of gravity. By studying the
motion of planets, their moons, and comets, astronomers learned the
basic rules that govern the dynamics of a system of gravitating
bodies. Today we apply this knowledge to study the universe around us,
expecting that the same laws of gravity govern the behavior of the
universe on large scales, from planetary systems similar to our own to
galaxies and to the entire cosmos as a whole.

Astronomers, however, do not normally discover new laws of nature. We
are not yet able to manipulate the objects of our scrutiny. The
telescopes and detectors we operate are simply passive probes that
cannot order the cosmos what to do. Yet they \textit{can} tell us when
something isn't following established rules. For example, take the
planet Uranus, whose discovery is credited to the English astronomer
William Herschel and dated to 1781 (others had already noted its
presence in the sky but misidentified it as a star).  As observational
data about its orbit accumulated over the following decades, people
began to notice that Uranus's orbit deviated slightly from the
dictates of Newton's gravity, which by then had withstood a century's
worth of testing on the other planets and their moons. Some prominent
astronomers suggested that perhaps Newton's laws begin to break down
at such great distances from the Sun.

This led immediately to the question: What is there to do? Abandon or
modify Newton's laws and come up with new rules of gravity? Or
postulate a yet-to-be-discovered planet in the outer solar system,
whose gravity was absent from the calculations for Uranus's orbit? The
answer came in 1846, when astronomers discovered the planet Neptune
just where a planet had to be for its gravity to perturb Uranus in
just the ways measured. Newton's laws were safe\ldots for the time being.

Then there is Mercury, the planet closest to the Sun. Its orbit, too,
habitually disobeyed Newton's laws of gravity resulting in an
anomalous precession of its perihelion. This anomaly was known for a
long time; it amounts to 43~seconds of arc (") per century and cannot
be explained within Newton's gravity, thereby presenting a challenge
for physicists and astronomers. In 1855, the French astronomer Urbain
Jean Joseph Le Verrier, who in 1846 predicted Neptune's position in
the sky within one degree, wrote that the anomalous residue of the
Mercurial precession would be accounted for if yet another as-yet
undiscovered planet -- call it Vulcan -- revolves inside the Mercurial
orbit so close to the Sun that it would be practically impossible to
discover in the solar glare, or perhaps it was an entire uncatalogued
belt of asteroids orbiting between Mercury and the Sun.

It turns out that Le Verrier was wrong on both counts. This time he
really did need a new understanding of gravity. Within the limits of
precision that our measuring tools impose, Newton's laws behave well
in the outer solar system. However, they break down in the inner solar
system, where the Sun's gravitational field is so powerful that it
warps space. And that is where we cannot ignore the effects of general
relativity. It took another 60 years to solve this puzzle. In 1915,
before publishing the historical paper with the field equations of the
general theory of relativity (e.g., \cite{Einstein-1915,
  Einstein-1916}), Albert Einstein computed the expected perihelion
precession of Mercury's orbit. This was not the first time Einstein
tackled this problem: indeed, earlier versions of his gravity theory
were rejected, in part, because they predicted the wrong value (often
with the wrong sign) for Mercury's perihelion
advance~\cite{Earmann1993}. However, when he obtained the famous
43"/century needed to account for the anomaly, he realized that a new
era in gravitational physics had just begun.

The stories of these two planets, Mercury and Uranus, involve two
similar-looking anomalies, yet two completely different solutions.

Ever since its original publication on November~25,
1915~\cite{Einstein-1915, Einstein-1916}, Einstein's general theory of
relativity continues to be an active area of both theoretical and
experimental research~\cite{Turyshev:2008dr, Turyshev-UFN-2008}. Even
after nearly a century since its discovery, the theory successfully
accounts for all solar system observations gathered to date; it is
remarkable that Einstein's theory has survived every
test~\cite{Will-lrr-2006-3}.  In fact, both in the weak field limit
evident in our solar system and with the stronger fields present in
systems of binary pulsars the predictions of general relativity have
been extremely well tested. Such longevity and success make general
relativity the \textit{de facto} ``standard'' theory of gravitation
for all practical purposes involving spacecraft navigation, astronomy,
astrophysics, cosmology and fundamental
physics~\cite{Turyshev-etal-2007}.

Remarkably, even after more than 300 years since the publication of
Newton's ``Principia'' and nearly 100 years after the discovery of
Einstein's general theory of relativity, our knowledge of gravitation
is still incomplete. Many challenges remain, leading us to explore the
physics beyond Einstein's theory~\cite{Turyshev:2008dr,
  Turyshev-UFN-2008}. In fact, growing observational evidence points
to the need for new physics. Multiple dedicated efforts to discover
new fundamental symmetries, investigations of the limits of
established symmetries, tests of the general theory of relativity,
searches for gravitational waves, and attempts to understand the
nature of dark matter were among the topics that had been the focus of
scientific research at the end of the last century. These efforts have
further intensified with the unexpected discovery in the late 1990s of
a small acceleration rate of our expanding Universe, which triggered
many new activities aimed at answering important questions related to
the most fundamental laws of Nature~\cite{Turyshev-UFN-2008,
  Turyshev-etal-2007}.

Many modern theories of gravity that were proposed to address the
challenges above, including string theory, supersymmetry, and
brane-world theories, suggest that new physical interactions will
appear at different ranges.  For instance, this may happen because at
sub-millimeter distances new dimensions can exist, thereby changing
the gravitational inverse-square law~\cite{ADD-1998,
  ADD-1999}. Similar forces that act at short distances are predicted
in supersymmetric theories with weak scale
compactifications~\cite{AnDD-1998}, in some theories with very low
energy supersymmetry breaking~\cite{Dimopoulos-Giudice-1998}, and also
in theories of very low quantum gravity scale~\cite{Dvali-etal-1998,
  Randall-Sundrum-1999, Sundrum-1999}.

Although much of the research effort was devoted to the study of the
behavior of gravity at very short distances, notably on
millimeter-to-micrometer ranges, it is possible that tiny deviations
from the inverse-square law occur at much larger distances. In fact,
there is a possibility that noncompact extra dimensions could produce
such deviations at astronomical
distances~\cite{Dvali-Gruzinov-Zaldarriaga-2003}. By far the most
stringent constraints to date on deviations from the inverse-square
law come from very precise measurements of the Moon's orbit about the
Earth. Analysis of lunar laser ranging data tests the gravitational
inverse-square law  on scales of the Earth-Moon
distance~\cite{Williams-Turyshev-Murphy-2004}, so far reporting no
anomaly at the level of accuracy of 3~\texttimes~10\super{-11} of the
gravitational field strength.

While most of the modern experiments in the solar system do not show
disagreements with general relativity, there are puzzles that require
further investigation. One such puzzle was presented by the Pioneer~10
and 11 spacecraft. The radiometric tracking data received from these
spacecraft while they were at heliocentric distances of 20\,--\,70
astronomical units (AU) have consistently indicated the presence of a
small, anomalous, Doppler frequency drift. The drift was interpreted
as a constant sunward acceleration of $a_P=(8.74 \pm 1.33)\times
10^{-10}\mathrm{\ m/s}^2$ experienced by both
spacecraft~\cite{pioprl,pioprd,moriond}. This apparent violation of
the inverse-square law has become known as the Pioneer anomaly; the
nature of this anomaly remains unexplained.

Before Pioneer~10 and 11, Newtonian gravity was not measured with
great precision over great distances and was therefore never
confirmed. The unique ``built-in'' navigation capabilities of the two
Pioneers allowed them to reach the levels of $\sim10^{-10}\mathrm{\ m/s}^2$ in
acceleration sensitivity. Such an exceptional sensitivity allowed
researchers to use Pioneer~10 and 11 to test the gravitational inverse
square law in the largest-scale gravity experiment ever conducted.
However, the experiment failed to confirm the validity of this
fundamental law of Newtonian gravity in the outer regions of the solar
system. Thus, the nagging question remains: Just how well do we know
gravity?

One can demonstrate that beyond 15~AU the difference between the
predictions of Newton and Einstein are negligible. So, at the moment,
two forces seem to be at play in deep space: Newton's law of gravity
and the Pioneer anomaly. Until the anomaly is thoroughly accounted for
by conventional causes, and can therefore be eliminated from
consideration, the validity of Newton's laws in the outer solar system
will remain in doubt. This fact justifies the importance of the
investigation of the nature of the Pioneer anomaly.

However, the Pioneer anomaly is not the only unresolved puzzle. Take
the dark matter and dark energy problem. While extensive efforts to
detect the dark matter that is believed to be responsible for the
puzzling observations of galaxy rotation curves have not met with
success so far, modifications of gravitational laws have also been
proposed as a solution to this puzzle. We still do not know for sure
whether or not the ultimate solution for the dark matter problem will
require a modification of the Standard Model of cosmology, but some
suggested that new gravitational laws are at play in the arms of
spiral galaxies.

A similar solution was proposed to explain the cosmological
observations that indicate that the expansion of the universe is
accelerating. There is now a great deal of evidence indicating that
over 70\% of the critical density of the universe is in the form of a
``negative-pressure'' dark energy component; we have no understanding
of its origin or nature. Given the profound challenge presented by the
dark energy problem, a number of authors have considered the
possibility that cosmic acceleration is not due to a particular
substance, but rather that it arises from new gravitational physics
(see discussion in~\cite{Turyshev-UFN-2008}).

Many of the models that were proposed to explain the observed
acceleration of the universe without dark energy or the observed
deviation from Newtonian laws of gravity in the arms of spiral
galaxies without dark matter may also produce measurable gravitational
effects on the scale of the solar system. These effects could manifest
themselves as a ``dark force'', similar to the one detected by the
Pioneer~10 and 11 spacecraft. Some believe that the Pioneer anomaly
may be a critical piece of evidence as it may indicate a deviation
from Einstein's gravity theory on the scales of the solar system. But
is it? Or can the Pioneer anomaly be explained by the mundane physics
of a previously unaccounted-for on-board systematic effect? In this
review we summarize the current knowledge of the anomaly and explore
possible ways to answer this question.

The review is organized as follows. We begin with descriptions of the
Pioneer~10 and 11 spacecraft and the strategies for obtaining and
analyzing their data. In Section~\ref{sec:pio-project} we  describe
the Pioneer spacecraft. We provide a significant amount of information
on the design, operations and behavior of Pioneer~10 and 11 during
their entire missions, including information from original project
documentation, descriptions of various data formats, and techniques
used for their navigation. This information is critical to the
ongoing investigation of the Pioneer anomaly.

In Section~\ref{sec:pio-data} we describe the techniques used for
acquisition of the Pioneer data. In particular, we discuss the Deep
Space Network (DSN), its history and current status, describe the DSN
tracking stations and details of their operations in support of deep
space missions. We present the available radiometric Doppler data and
describe techniques for data preparation and analysis.  We also
discuss the Pioneer telemetry data and its value for the anomaly
investigation.

In Section~\ref{sec:navigation} we address the basic elements of the
theoretical foundation for precision spacecraft navigation. In
particular, we discuss the observational techniques and physical
models that were used for precision tracking of the Pioneer spacecraft
and analysis of their data. We describe models of gravitational forces
and those that are of nongravitational nature.

In Section~\ref{sec:anomaly} we focus on the detection and initial
characterization of the Pioneer anomaly. We describe how the anomalous
acceleration was originally identified in the data. We continue by
summarizing the current knowledge of the physical properties of the
Pioneer anomaly. We briefly review the original efforts to understand
the signal.

In Section~\ref{sec:theory-explain} we review various mechanisms
proposed to explain the anomaly that use unmodeled forces with origin
either external to the spacecraft or those generated on-board. We
discuss theoretical proposals that include modifications of general
relativity, modified Newtonian gravity, cosmological theories,
theories of dark matter and other similar activities. We review the
efforts at independent confirmation with other spacecraft, planets,
and other bodies in the solar system.

In Section~\ref{sec:current-status}, we describe the results of
various independent studies of the Pioneer anomaly. We discuss new
Pioneer~10 and 11 radiometric Doppler data that recently became
available. This much extended set of Pioneer Doppler data is the
primary source for the new investigation of the anomaly. A near
complete record of the flight telemetry that was received from the two
Pioneers is also available. Together with original project
documentation and newly developed software tools, this additional
information is now used to reconstruct the engineering history of both
spacecraft, with special emphasis on the possible contribution to the
anomalous acceleration by on-board systematic effects. We review the
current status of these efforts to investigate the anomaly.

In Section~\ref{sec:conclusions}, we present our summary and
conclusions.

In Appendices~\ref{app:geometry}\,--\,\ref{app:formatCE}, we provide additional
information on the geometry and design of Pioneer~10 and 11 spacecraft
and describe various data formats used for mission operations.

\newpage
\section{The Pioneer~10 and 11 Project}
\label{sec:pio-project}

NASA's Pioneer program began in 1958, in the earliest days of the
space age, with experimental spacecraft that were designed to reach
Earth escape velocity and perform explorations of the interplanetary
space beyond the Earth's orbit. Several of these launch attempts ended
in failure; the five spacecraft that reached space later became known
as Pioneers~1\,--\,5. These were followed in the second half of the 1960s
by Pioneers~6\,--\,9, a series of significantly more sophisticated
spacecraft that were designed to be launched into solar orbit and make
solar observations. These spacecraft proved extremely
robust\epubtkFootnote{See details on the Pioneer missions at
  \url{http://www.nasa.gov/centers/ames/missions/archive/pioneer.html}.
Note that Pioneer~6 remained operational for more than 35 years after
launch.} and paved the way for the most ambitious projects yet in the
unmanned space program: Pioneers~10 and 11.

The Pioneer~10 and 11 spacecraft were the first two man-made objects
designed to explore the outer solar system (see details
in~\cite{PC223, PC261, PC220, PC224, PC202, PC260, JUP1974, SAT1979,
  TM81233, EP264, P11SAT, SP446, SP349, HM1967, SP268, SNAP19,
  HISTORY,11339, TCSDR3, PFG100}). Their objectives were to conduct,
during the 1972\,--\,73 Jovian opportunities, exploratory
investigation beyond the orbit of Mars of the interplanetary medium,
the nature of the asteroid belt, the environmental and atmospheric
characteristics of Jupiter and Saturn (for Pioneer~11), and
investigate the solar system beyond Jupiter's orbit.

In this section we review the Pioneer~10 and 11 missions. We present
information about the spacecraft design. Our discussion focuses on
subsystems that played important roles in the continued functioning of
the vehicles and on subsystems that may have affected their dynamical
behavior: specifically, we review the propulsion, attitude control,
power, communication, and thermal subsystems. We also provide
information about the history of these systems throughout the two
spacecrafts' exceptionally lengthy missions.

\subsection{The Pioneer~10 and 11 missions}

The Pioneer missions were the first to cross the asteroid belt,
perform in situ observations of the interplanetary medium in the outer
solar system, and close-up observations of the gas giant
Jupiter. Their mission design was characterized by simplicity: a
powerful launch vehicle placed the spacecraft on an hyperbolic
trajectory aimed directly at Jupiter, which the spacecraft were
expected to reach approximately 21 months after launch.

At the distance of Jupiter, operating a spacecraft using solar panels
is no longer practical (certainly not at the level of technology that
was available to the designers of the Pioneer missions in the late
1960s.) For this reason, nuclear power was chosen as the means to
provide electrical power to the spacecraft, in the form of \super{238}Pu
powered radioisotope thermoelectric generators (RTGs). As even this
was relatively new technology at the time the missions were designed,
the power subsystem was suitably over-engineered, the design
requirement being a completely functional spacecraft capable of
performing all planned science observations with only three (out of
four) RTGs operating.

Such conservative engineering characterized the entire design of these
spacecraft and their missions, and it was likely responsible for the
two spacecrafts' exceptional longevity, and their ability to deliver
science results that far exceeded the expectations of their
designers. The original plan envisioned a primary mission of
600\,--\,900 days in duration. Nevertheless, following its encounter
with Jupiter, Pioneer~10 remained functional for over 30 years;
meanwhile Pioneer~11, though not as long lived as its sister craft,
successfully navigated a path across the solar system for another
encounter with Saturn, offering the first close-up observations of the
ringed planet.

\epubtkImage{}{%
  \begin{figure}[t]
    \centerline{\includegraphics[width=0.80\textwidth]{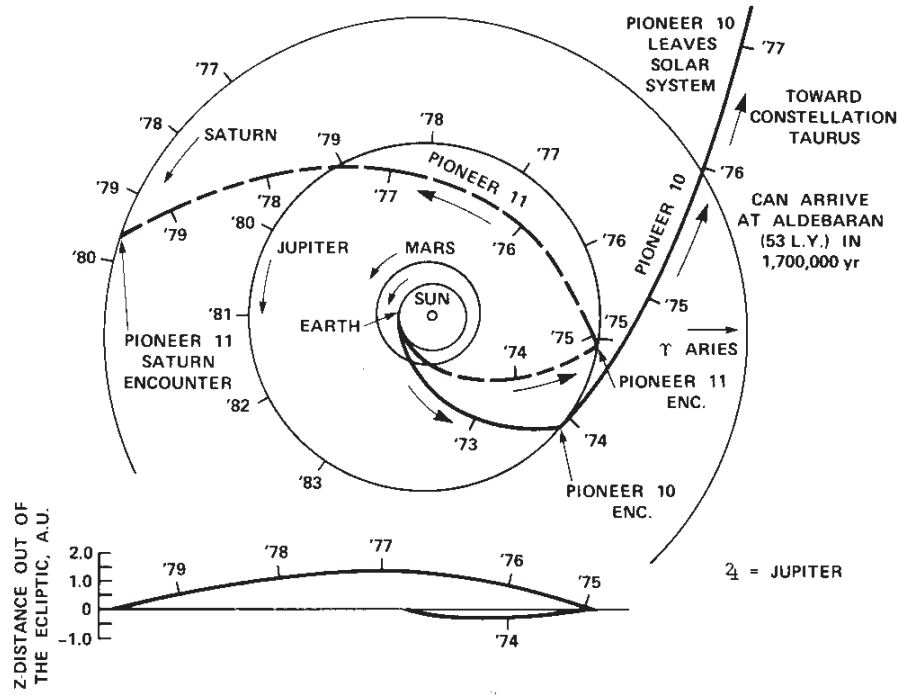}}
    \caption{Trajectories of Pioneer~10 and 11 during their primary
    missions in the solar system (from~\cite{SP349}). The time ticks
    shown along the trajectories and planetary orbits represent the
    distance traveled during each year.}
    \label{fig:pioneer_inner_path}
\end{figure}}

\epubtkImage{}{%
  \begin{figure}[t]
    \centerline{\includegraphics[width=0.90\linewidth]{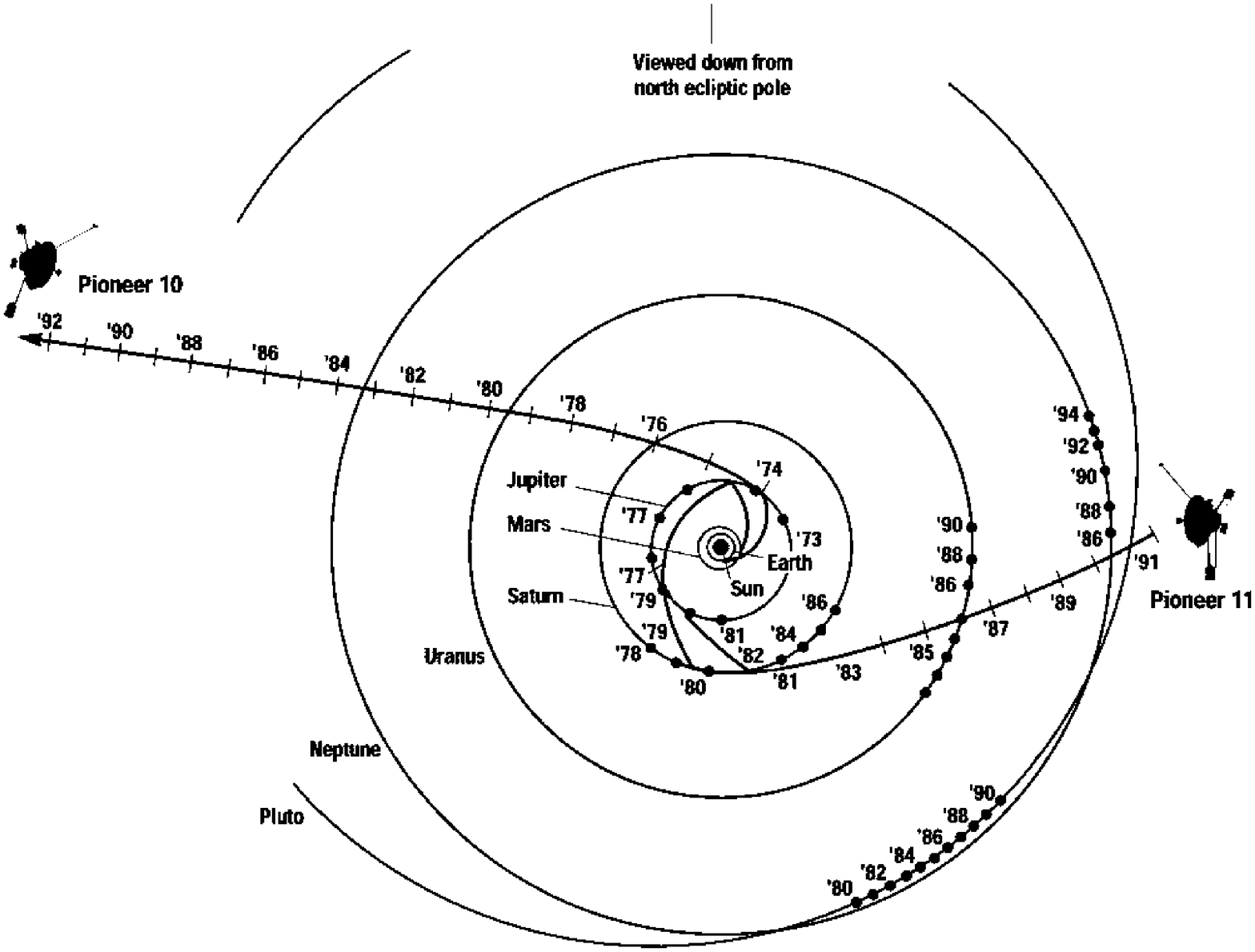}}
    \caption{Ecliptic pole view of the Pioneer~10 and Pioneer~11
    trajectories during major parts of their extended
    missions. Pioneer~10 is traveling in a direction almost opposite
    to the galactic center, while Pioneer~11 is heading approximately
    in the shortest direction to the heliopause. The direction of the
    solar system's motion in the galaxy is approximately towards the
    top. (From~\cite{pioprd}.) }
    \label{fig:pioneer_path}
\end{figure}}

After the Jupiter and Saturn (for Pioneer~11) encounters (see
Figure~\ref{fig:pioneer_inner_path}), the craft followed escape
hyperbolic orbits near the plane of the ecliptic on opposite sides of
the solar system, continuing their extended
missions~\cite{SP349}. (See Figure~\ref{fig:pioneer_path}.) The
spacecraft explored the outer regions of the solar system, studying
energetic particles from the Sun (solar wind), and cosmic rays
entering our portion of the Milky Way. Major milestones of the two
Pioneer projects are shown in Table~\ref{tb:milestones}.

\begin{table}[h]
  \caption{Major milestones of the Pioneer~10 and 11 projects.}
  \label{tb:milestones}
  \centering
  \begin{tabular}{lcc}
    \toprule
    Event & Pioneer~10 & Pioneer~11\\
    \midrule
    Launch & March 3, 1972 & March 6, 1973\\
    Jupiter encounter & December 3, 1973 & December 4, 1974\\
    Saturn encounter & N/A & September 1, 1979\\
    Last telemetry received & April 27, 2002 & September 30, 1995\\
    \bottomrule
  \end{tabular}
\end{table}

The Pioneers were excellent vehicles for the purposes of precision
celestial mechanics experiments~\cite{pioprl,
pioprd,
2002IJMPD..11.1545A,
2002gr.qc.....8046M,   
2005CQGra..22.5343N,   
2007arXiv0709.3866N,   
2007arXiv0709.1917N,   
2001hep.ph...10373N,   
2004CQGra..21.4005N,   
2005PhLB..613...11N,   
2005AIPC..758..113N,   
2007AA...463..393O,  
moriond,
Turyshev:2005zm,   
Turyshev:2005zk,   
Turyshev:2005vj}.
This was due to a combination of many factors, including the presence
of a coherent mode transceiver on board, their attitude control
(spin-stabilized, with a minimum number of attitude correction
maneuvers using thrusters), power design (the RTGs being on extended
booms aided the stability of craft and also reduced thermal effects on
the craft; see Figure~\ref{fig:pio-craft}), and precise Doppler
tracking (with the accuracy of post-fit Doppler residuals at the level
of mHz).  The exceptional ``built-in'' acceleration sensitivity of the
Pioneer~10 and 11 spacecraft naturally allowed them to reach a level
of accuracy of $\sim\,10^{-10}\mathrm{\ m/s}^2$.  The result was one of the
most precise spacecraft navigations in deep space to
date~\cite{1976AJ.....81.1153N}.

\subsubsection{Pioneer~10 mission details}

Pioneer~10 was launched on 2 March 1972 (3 March 1972 at 01:49
Universal Coordinated Time) from Cape Canaveral on top of an
Atlas/Centaur/TE364-4 launch vehicle~\cite{NLVS1972}. The launch
marked the first use of the Atlas-Centaur as a three-stage launch
vehicle.

The launch vehicle configuration was an Atlas launch vehicle equipped
with a Centaur~D upper stage and a TE364-4 solid-fuel third stage that
provided additional thrust and also supplied the initial spin of the
spacecraft. The third stage was required to accelerate Pioneer~10 to
the speed of 14.39~km/s, needed for the flight to Jupiter.

After a powered flight of approximately 14 minutes, the spacecraft was
separated from its launch vehicle; its initial spin
$\sim$~60 revolutions per minute (rpm) was reduced by thrusters, and
then reduced further when the magnetometer and RTG booms were
extended~\cite{SEQ, HESPRICH}. The spacecraft was then oriented to
ensure that its high-gain antenna pointed towards the Earth. Thus, the
initial cruise phase from the Earth to Jupiter began.

The first interplanetary cruise phase of Pioneer~10 took approximately
21 months. During this time, Pioneer~10 successfully crossed the
asteroid belt, demonstrating for the first time that this region of
the solar system is safe for spacecraft to travel through.

Pioneer~10 arrived at Jupiter in late November,
1973~\cite{JUP1974}. Its closest approach to the red giant occurred on
4~December 1973, at 02:25 UTC. It performed the first ever close-up
observations of the gas giant, before continuing its journey out of
the solar system on an hyperbolic escape trajectory
(Figure~\ref{fig:pioneer_path}). During the planetary encounter,
Pioneer~10 took several photographs of the planet and its moons,
measured Jupiter's magnetic fields, and observed the planet's
radiation belts. Radiation in the Jovian environment, potentially
damaging to the spacecraft's electronics, was a concern to the mission
designers. However, Pioneer~10 survived the planetary encounter
without significant damage, although its star sensor became
inoperative shortly afterwards~\cite{AIAA87-0502}, a likely result of
excessive radiation exposure near Jupiter.

The encounter with Jupiter changed Pioneer~10's trajectory as was
planned by JPL navigators~\cite{VanAllen-2003}. As a result,
Pioneer~10 was now on an hyperbolic escape trajectory that took it to
ever more distant parts of the solar system. Originally, signal loss
was expected before Pioneer~10 reached twice the heliocentric distance
of Jupiter (the downlink telecommunication power margin was 6~dB at
the time of Jupiter encounter); however, continuing upgrades to the
facilities of the Deep Space Network (DSN) permitted tracking of
Pioneer~10 until the official termination of Pioneer~10's science
mission in 1997 and even beyond.

Pioneer~10 continued to make valuable scientific investigations until
its science mission ended on March~31, 1997. After this date,
Pioneer~10's weak signal was tracked by the NASA's DSN as part of an
advanced concept study of communication technology in support of
NASA's future interstellar probe mission. Pioneer~10 eventually became
the first man-made object to leave the solar system.

During one of the last attempts to contact Pioneer~10, in April 2001,
at first no signal was detected; however, the spacecraft's signal did
appear once it detected a signal from the Earth and its radio system
switched to coherent mode. From this, it was concluded that the
on-board transmitter frequency reference (temperature controlled
crystal oscillator) failed, possibly due to the combined effects of
aging, the extreme cold environment of deep space, and a drop in the
main bus voltage due to the depletion of the spacecraft's RTG power
source. This failure had no impact on the ability to obtain precision
Doppler measurements from Pioneer~10.

On March~2, 2002 NASA's DSN made another contact with Pioneer~10 and
confirmed that the spacecraft was still operational thirty years after
its launch on March~3, 1972 (UT). The uplink signal was transmitted on
March~1 from the DSN's Goldstone, California facility and a downlink
response was received twenty-two hours later by the 70-meter antenna
at Madrid, Spain. At this time the spacecraft was 11.9~billion
kilometers from Earth at about 79.9~AU from the Sun and heading
outward into interstellar space in the general direction of Aldebaran
at a distance of about 68 light years from the Earth, and a travel
time of two million years. The last telemetry data point was obtained
from Pioneer~10 on 27~April 2002 when the craft was 80~AU from the
Sun.

The last signal from Pioneer~10 was received on Earth on 23~January
2003, when NASA's DSN received a very weak signal from the venerable
spacecraft from the distance of $\sim$~82.1~AU from the Sun. The
previous three contacts had very faint signals with no telemetry
received.  At that time, NASA engineers reported that Pioneer~10's RTG
has decayed to the point where it may not have enough power to send
additional transmissions to Earth. Consequently, the DSN did not
detect a signal during a contact attempt on 7 February 2003. Thus,
after more than 30 years in space, the Pioneer~10 spacecraft sent its
last signal to Earth.

The final attempt to contact Pioneer~10 took place on the
34th anniversary of its launch, on 3\,--\,5 March
2006~\cite{MDR2005}. At that time, the spacecraft was 90.08~AU from
the Sun, moving at 12.08~km/s. The round-trip light time (i.e., time
needed for a DSN radio signal to reach Pioneer~10 and return back to
the Earth) was approximately 24~h~56~m, so the same antenna, DSS-14 at
Goldstone, CA, was used for the track. Unfortunately, no signal was
received. Given the age of the spacecraft's power source, it was clear
that there was no longer sufficient electrical power on board to
operate the transmitter~\cite{2006CaJPh..84.1063T}.

\subsubsection{Pioneer~11 mission details}

Pioneer~11 followed its older sister approximately one year later. It
was launched on 5 April 1973 (on April 6, 1973 at 02:11~UTC), also on
top of an Atlas/Centaur/TE364-4 launch vehicle. The second stage used
for Pioneer~11 was a Centaur D-1A, while the third stage was a TE364-4
solid fuel vehicle.

After safe passage through the asteroid belt on 19 April 1974,
Pioneer~11's thrusters were fired to add another $\sim$~65~m/s to the
spacecraft's velocity. This adjusted the aiming point at Jupiter to
43,000~km  above the cloud tops. The close approach also allowed the
spacecraft to be accelerated by Jupiter to a velocity of 48.06~km/s,
so that it would be carried across the solar system some
2.4~billion~km to Saturn.

Early in its mission, Pioneer~11 suffered a propulsion system anomaly
that caused the spin rate of the spacecraft to increase significantly
(see Figure~\ref{fig:spin}). Fortunately, the spin rate was not high
enough to endanger the spacecraft or compromise its mission
objectives.

Pioneer~11's first interplanetary cruise phase lasted approximately 20
months. During this time, a major trajectory correction maneuver was
performed, aiming Pioneer~11 for a precision encounter with
Jupiter. Pioneer~11's closest approach to Jupiter occurred on 2
December 1974 at 17:22 UTC. This encounter provided the necessary
gravity assist to alter Pioneer~11's trajectory for a planned
encounter with Saturn (see Figure~\ref{fig:pioneer_inner_path}).

The second interplanetary cruise phase of Pioneer~11's mission took it
across the solar system. Initially, Pioneer~11's heliocentric distance
was actually decreasing as it followed an hyperbolic trajectory taking
the spacecraft more than 1~AU above the plane of the ecliptic. This
phase of the mission culminated in a successful encounter with
Saturn. Pioneer~11's closest approach to the ringed planet occurred on
1 September 1979, at 16:31~UTC.  Still fully operational, Pioneer~11
was able to make close-up observations of the ringed planet.

After this second planetary encounter, Pioneer~11 continued to escape
the solar system on an hyperbolic escape trajectory, and remained
operational for many years. Pioneer~11 explored the outer regions of
our solar system, studying the solar wind and cosmic rays.

The spacecraft sent its last coherent Doppler data on October 1, 1990
while at 31.7~AU from the Sun\epubtkFootnote{To generate ephemerides
  for solar-system bodies, including many spacecraft, one can use
  JPL's HORIZONS system, which is located at:
  \url{http://ssd.jpl.nasa.gov/?horizons}.}. In October 1990 a
microwave relay switch failed on board Pioneer~11, in its
communications subsystem. The most notable consequence of this failure
is that it was no longer possible to operate this spacecraft's radio
system in coherent mode, which is required for precision Doppler
observations. Therefore, after this event, precision Doppler data was
no longer produced by the Pioneer~11 spacecraft.

The spacecraft continued to provide science observations until the end
of its mission in 1995. In September 1995, Pioneer~11 was at a
distance of 6.5~billion~km from Earth. At that distance, it takes over
6~hours for the radio signal to reach Earth. However, by September
1995, Pioneer~11 could no longer make any scientific observations as
its power supply was nearly depleted. On 30 September 1995, routine
daily mission operations were stopped. Intermittent contact continued
until November 1995, at which time the last communication with
Pioneer~11 took place. There has been no communication with Pioneer~11
since. The Earth's motion has carried our planet out of the view of
the spacecraft antenna.

\subsubsection{Pioneer~10 and 11 project documentation}
\label{sec:pio-docs}

Up until 2005, very little documentation on the Pioneer spacecraft was
available to researchers. Indeed, around this time much of the Pioneer
archival material stored at NASA's Ames Research Center was scheduled
for destruction due to budget constraints.

The growing interest in the Pioneer anomaly helped to initiate an
effort at the NASA Ames Research Center to recover the entire archive
of the Pioneer Project documents for the period from 1966 to 2003 (see
details in~\cite{2007arXiv0710.2656T, MDR2005}).  This massive archive
contains all Pioneer~10 and 11 project documents discussing the
spacecraft and mission design, fabrication of various components,
results of various tests performed during fabrication, assembly,
pre-launch, as well as calibrations performed on the vehicles; and
also administrative documents including quarterly reports, memoranda,
etc. Most of the maneuver records, spin rate data, significant events
of the craft, etc., have also been identified.

A complete set of Pioneer-related documentation is listed in the
Bibliography. Here, we mention some of the more significant pieces of
documentation that are essential to understanding the Pioneer~10 and
11 spacecraft and their anomalous accelerations:

\begin{itemize}
\item The first document to be mentioned is entitled ``Pioneer F/G:
  Spacecraft Operational Characteristics''~\cite{PC202} (colloquially
  referred to by its identifier as ``PC-202''), and contains a
  complete description of the Pioneer~10 and 11 spacecraft and their
  subsystems. The document was last revised in mid-1971, just months
  before the launch of Pioneer~10, indicating that it reflects
  accurately the configuration of the Pioneer~10 spacecraft as it
  flew.

\item Valuable information was found in the TRW Systems Group's
  document entitled ``Pioneer Project Flights F and G Final
  Report''~\cite{PFG100}, which contained post-launch information
  about both spacecraft, including, among other things, detailed
  information about their exact launch configuration.

\item Details about the SNAP-19 radioisotope thermoelectric generators
  can be found in Teledyne Isotopes Energy Systems Division's ``SNAP
  19 Pioneer F \& G Final Report''~\cite{SNAP19}, which has been
  released for public distribution on 9 February 2006.

\item Much additional detail about the thermal design of the
  spacecraft was obtained from TRW Systems Group's ``Pioneer F/G
  Thermal Control Subsystem Design Review Number 3''~\cite{TCSDR3}.

\item The Master Data Record (MDR) file format is described in
  Alliedsignal Technical Services Corporation's
  document~\cite{ARC221}. Sensor calibration data for Pioneer~10 is
  provided by BENDIX Field Engineering Corporation~\cite{ARC037}; for
  Pioneer~11, the same information was provided privately by
  L.\ Kellogg~\cite{LK2003}. The data format used by scientific
  instruments in the MDRs is described in ``Pioneer F/G: EGSE Computer
  Programming Specifications for Scientific
  Instruments''~\cite{PC260}; further information about both
  scientific and engineering data words is present in ``Pioneer F/G:
  On-line Ground Data System Software
  Specification''~\cite{PC261}. Together, these resources make it
  possible to read and interpret the entire preserved telemetry record
  of Pioneer~10 and 11 using modern software~\cite{MDR2005}.

\item Operational details about the Pioneer~10 and 11 missions were
  recorded in meeting presentations, many of which have been
  preserved~\cite{PSG87B, PSG88B, PSG90B, PSG93B, PSG94A, PSG94D,
    PSG95D, PSG96B}. Additional details are provided in~\cite{%
    TM108108,               
    ASTRO,                  
    PNL,                    
    PSB,                    
    NASA102269,             
    JAVA1984}.              

\item Details about tracking and data acquisition were published in
  the \textit{JPL Deep Space Network's Technical Reports}
  series~\cite{
JPL42-35-D,
JPL42-37-E,
JPL42-39-C,
JPL42-41-F,
JPL42-44-E,
JPL32-1526-IX-FF,
JPL42-20-CC,
JPL42-27-Y,
JPL42-36-R,
JPL42-35-P,
JPL32-1526-XV-E,
JPL32-1526-XVI-BB,
JPL42-31-J,
JPL42-45-G,
JPL42-100-K,
JPL32-1526-V-W,
JPL42-93-F,
JPL42-46-E,
JPL32-1526-XIX-I,
JPL42-21-G,
JPL32-1526-XV-D,
JPL32-1526-XVI-E,
JPL32-1526-XVII-D,
JPL32-1526-XVIII-C,
JPL32-1526-XIX-D,
JPL42-21-D,
JPL42-22-C,
JPL42-24-D,
JPL42-26-D,
JPL42-30-E,
JPL42-33-D,
JPL42-47-D,
JPL42-52-B,
JPL42-54-D,
JPL42-58-I,
JPL42-61-C,
JPL42-63-D,
JPL42-65-F,
JPL42-22-H,
JPL42-106-W,
JPL32-1526-XIX-O,
JPL42-53-C,
JPL42-55-C,
JPL32-1526-V-B,
JPL32-1526-III-C,
JPL32-1526-IV-D,
JPL32-1526-II-C,
JPL32-1526-VI-D,
JPL32-1526-XI-D,
JPL32-1526-X-G,
JPL32-1526-X-C,
JPL32-1526-XI-C,
JPL32-1526-VII-B,
JPL32-1526-VIII-C,
JPL32-1526-IX-E,
JPL32-1526-IX-AA,
JPL32-1526-X-KK,
JPL42-91-O,
JPL42-111-O}. Further, relevant details can be found in~\cite{%
CM1969,                 
DSN810-005,             
JPL72-012A-09A,         
LP1974B,                
LP1974A,                
TDA42120,               
NASA4263,               
WB1995}.                 
Details about orbit estimation procedures are provided in~\cite{%
JPL7726,                
EPS2005,                
RBF1980,                
FB1977,                 
FF1974,                 
FF1975,                 
TMX65361,               
TMX2024,                
WL1974}.                

\end{itemize}

The on-going study of the Pioneer anomaly would not be possible
without these resources, much of which was preserved only because of
the labors of dedicated individuals.

\subsection{The Pioneer spacecraft}

In this section we discuss the details of the Pioneer~10 and 11
spacecraft, focusing only on those most relevant to the study of the
Pioneer anomaly.

\subsubsection{General characteristics}
\label{sec:pio-gen-character}

Externally, the shape of the Pioneer~10 and 11 spacecraft was
dominated by the large (2.74~m diameter) high-gain antenna
(HGA)\epubtkFootnote{Customarily, the direction in which the high-gain
  antenna points is referred to as the ``fore'' (also, $+z$)
  direction, whereas the direction towards the ``bottom'' of the
  spacecraft is referred to as the ``aft'' (also $-z$) direction. Note
  that most of the time, the spacecraft is, in fact, traveling in the
  aft direction, as the HGA is pointing in the general direction
  towards the Earth, in a direction opposite to the direction of
  travel (especially in deep space).}, behind which most of the
spacecrafts' instrumentation was housed in two adjoining hexagonal
compartments (Figure~\ref{fig:pio-craft}). The main compartment, in
the shape of a regular hexagonal block, contained the fuel tank,
electrical power supplies, and most control and navigation
electronics. The adjoining compartment, shaped as an irregular
hexagonal block, contained science instruments. Several openings were
provided for science instrument sensors. The internal arrangement of
spacecraft components is shown in Figure~\ref{fig:internal}.

The main and science compartments collectively formed the spacecraft
body, which was covered by multilayer thermal insulation on all sides
except part of the aft side, where a passive thermal control louver
system was situated.

\epubtkImage{}{%
  \begin{figure}[t!]
    \centering{\includegraphics[width=1.0\textwidth]{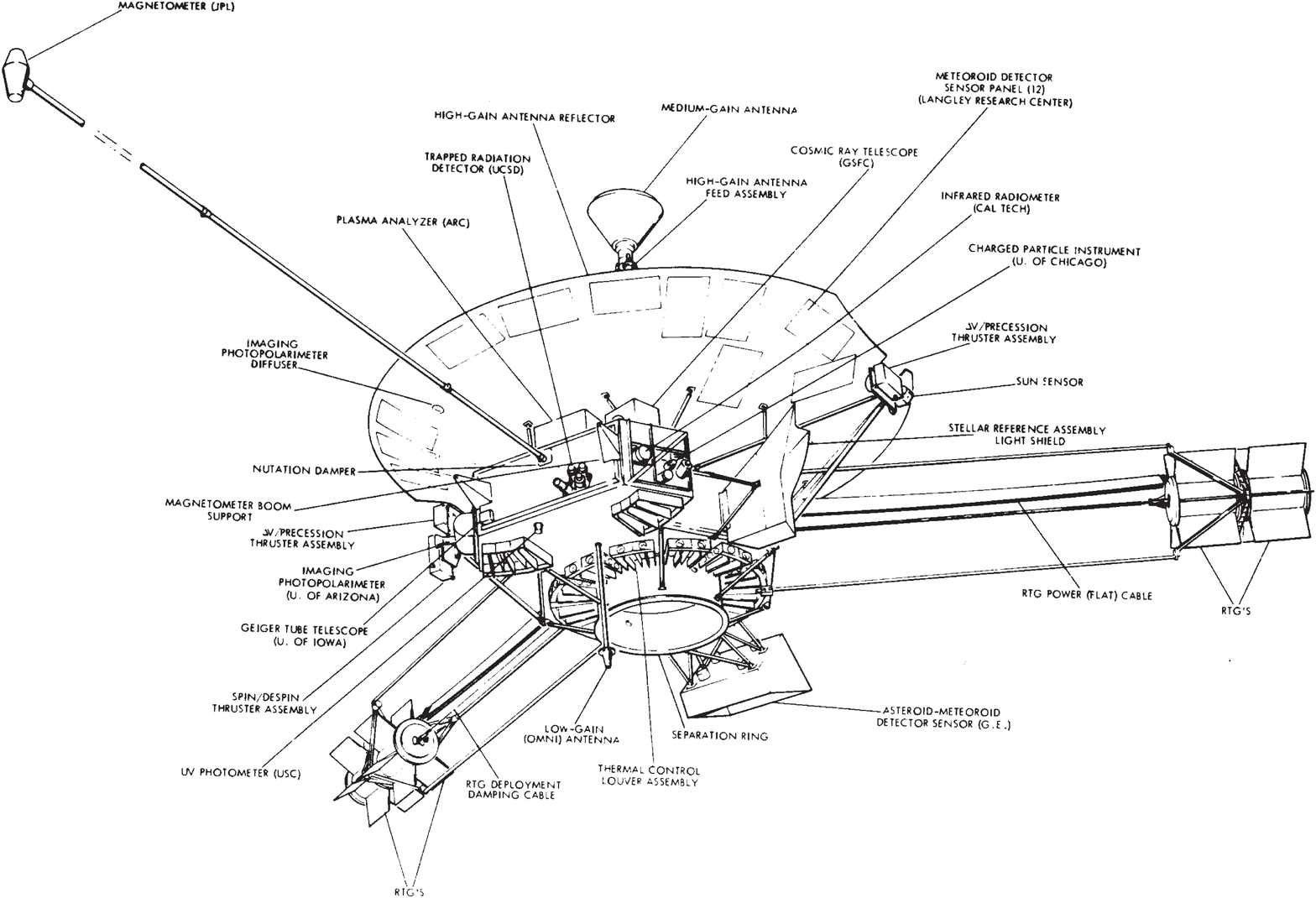}}
    \caption{A drawing of the Pioneer
    spacecraft. (From~\cite{PC202}.)}
    \label{fig:pio-craft}
\end{figure}}

For the purposes of attitude control, the entire spacecraft was
designed to spin in the plane of the HGA.

Three extensible booms were attached to the main compartment, spaced
at 120\textdegree. Two of these booms, both approximately 3~m long, each
held two radioisotope thermoelectric generators (RTGs). This design
was dictated mainly by concerns about the effects of radiation from
the RTGs on the spacecrafts' instruments, but it also had the
beneficial side effect of minimizing radiative heat exchange between
the RTGs and the spacecraft body. The third boom, approximately 6~m in
length, held the magnetometer sensor. The length of this boom ensured
that the sensor was not responding to magnetic fields originating in
the spacecraft itself.

The total mass of Pioneer~10 and 11 was approximately 260~kg at the
time of launch, of which approximately 30~kg was propellant and
pressurant. The masses of the spacecraft slowly varied throughout
their missions primarily due to propellant usage (for details, see
Section~\ref{sec:sc-masses}).

The propulsion system was designed to perform three types of
maneuvers: spin/despin (setting the initial spin rate shortly after
launch), precession (to keep the HGA pointing towards the Earth, and
also to orient the spacecraft during orbit correction maneuvers) and
velocity changes. Of the two spacecraft, Pioneer~11 used more of its
propellant, in the course of velocity correction maneuvers that were
used to adjust the spacecraft's trajectory for its eventual encounter
with Saturn.

\epubtkImage{}{%
\begin{figure}[t!]
\begin{minipage}{0.59\linewidth}
\includegraphics[width=\linewidth]{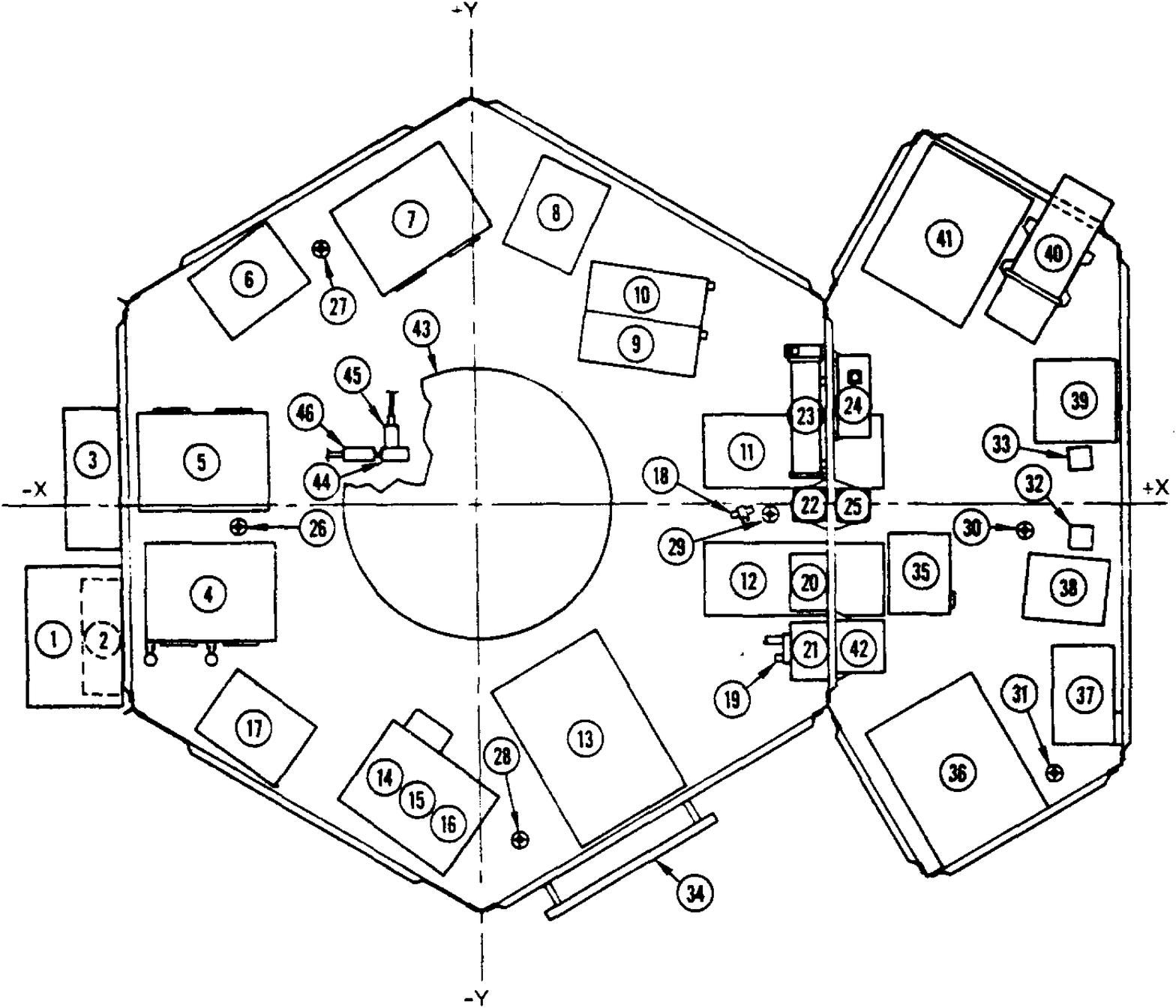}
\end{minipage}
\hskip 0.05\linewidth
\begin{minipage}{0.39\linewidth}
\begin{tiny}
\begin{tabular}{|r|l|}\hline
1 & Data Storage Unit (DSU)\\
2 & Asteroid/Meteoroid Detector Electronics\\
3 & Battery\\
4 & Power Control Units (PCU)\\
5 & Central TRF Unit\\
6 & Inverter Assembly No. 2\\
7 & Command Distribution Unit (CDU)\\
8 & Stellar Reference Assembly (SRA)\\
9 & Receiver No. 1\\
10 & Receiver No. 2\\
11 & TWTA No. 1\\
12 & TWTA No. 2\\
13 & Digital Telemetry Unit (DTU)\\
14 & Control Electronics Assembly (CEA)\\
15 & Conscan Signal Processor\\
16 & Digital Decoder Unit\\
17 & Inverter Assembly No. 1\\
18 & Attenuator TWT No. 1\\
19 & Attenuator TWT No. 2\\
20 & Transmitter Driver No. 1\\
21 & Transmitter Driver No. 2\\
22 & Transfer Switch -- Receive\\
23 & Diplexer No. 2/Coupler\\
24 & Diplexer No. 1\\
25 & Transfer Switch -- Transmit\\
26 & Thermistor No. 1\\
27 & Thermistor No. 2\\
28 & Thermistor No. 3\\
29 & Thermistor No. 4\\
30 & Thermistor No. 5\\
31 & Thermistor No. 6\\
32 & Despin Sensor No. 1\\
33 & Despin Sensor No. 2\\
34 & Shunt Radiator Assembly\\
35 & Magnetometer Electronics\\
36 & Imaging Photo-Polarimeter\\
37 & Geiger Tube Telescope\\
38 & Ultraviolet Photometer\\
39 & Trapped Radiation Detector\\
40 & Infrared Radiometer\\
41 & Charged Particle Instrument\\
42 & Meteoroid Detector Electronics\\
43 & Propellant Tank\\
44 & Temperature Transducer\\
45 & Filter -- Propellant\\
46 & Pressure Transducer\\
\hline
\end{tabular}
\end{tiny}
\end{minipage}
\caption{Pioneer~10 and 11 internal equipment
  arrangement. (From~\cite{PC202}.)}
\label{fig:internal}
\end{figure}}

\subsubsection{Science instruments}

The Pioneer spacecraft carried an identical set of 11 science
instruments, with a 12th instrument present only on Pioneer~11,
namely:

\begin{enumerate}
\item JPL Helium Vector Magnetometer
\item \vskip -6pt ARC Plasma Analyzer
\item \vskip -6pt U/Chicago Charged Particle Experiment
\item \vskip -6pt U/Iowa Geiger Tube Telescope
\item \vskip -6pt GSFC Cosmic Ray Telescope
\item \vskip -6pt UCSD Trapped Radiation Detector
\item \vskip -6pt UCS Ultraviolet Photometer
\item \vskip -6pt U/Arizona Imaging Photopolarimeter
\item \vskip -6pt CIT Jovian Infrared Radiometer
\item \vskip -6pt GE Asteroid/Meteoroid Detector
\item \vskip -6pt LaRC Meteoroid Detector
\item \vskip -6pt Flux-Gate Magnetometer (Pioneer~11 only)
\end{enumerate}

The power system of the Pioneer spacecraft was designed to ensure that
a successful encounter with Jupiter (with all science instruments
operating) can be carried out with only three (out of four)
functioning radioisotope thermoelectric generators. Instruments could
be commanded on or off by ground control; late in the extended
mission, when sufficient power to operate all instruments
simultaneously was no longer available, a power sharing plan was
implemented to ensure that the power demand on board would not exceed
available power levels.

At the end of its mission, only one instrument on board Pioneer~10
remained in operation; the University of Iowa Geiger Tube
Telescope. An attempt was made to power down this instrument, in order
to improve the available power margin. It is not known if this command
was received or executed by the spacecraft.

\subsubsection{Radioisotope thermoelectric generators (RTGs)}
\label{sec:RTG}

\epubtkImage{}{%
  \begin{figure}[t!]
    \centerline{\includegraphics[width=0.8\linewidth]{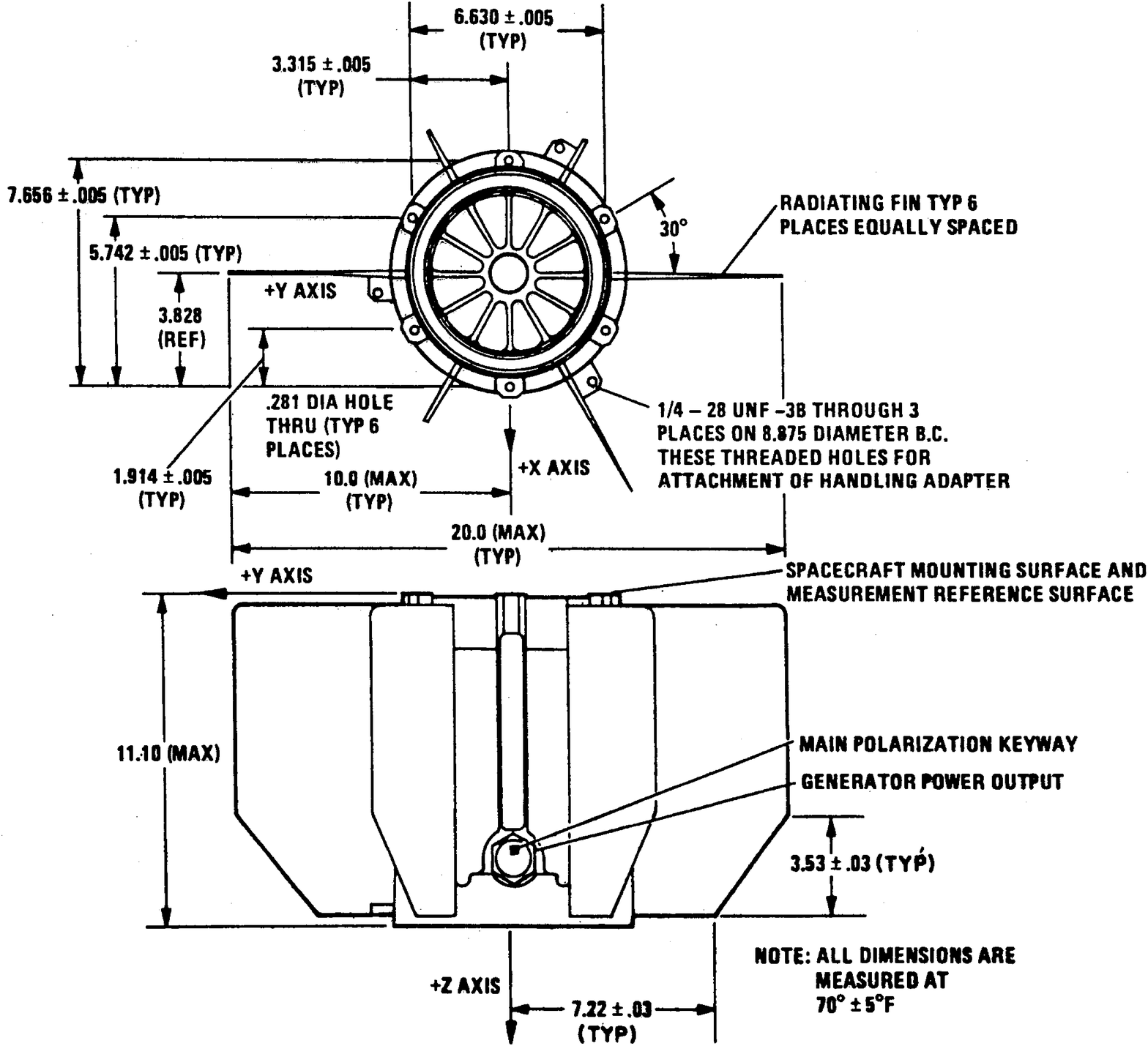}}
    \caption{The SNAP-19 RTGs used on Pioneer~10 and 11
    (from~\cite{SNAP19}). Note the enlarged fin structure. Dimensions
    are in inches (1"~=~2.54~cm).}
    \label{fig:RTGDIM}
\end{figure}}

The primary electrical source on board Pioneer~10 and 11 was a set of
four radioisotope thermoelectric generators~\cite{JPL0410, FW1999,
  SNAP19, CR162891, DOE32117}. Each of these RTGs contains 18
\super{238}Pu capsules, approximately two inches (5.08~cm) in diameter
and 0.2~inches (0.51~cm) thick. The total thermal power of each RTG,
when freshly fueled, was $\sim$~650~W. Each RTG contains two sets of 45
bimetallic thermocouples, connected in series. The thermocouples
initially operated at $\sim$~6\% efficiency; the nominal RTG output is
$\sim$~4~V, 10~A. The total available power at launch on board each
spacecraft was $\sim$~160~W (Figure~\ref{fig:elec}).

Excess heat from each RTG is radiated into space by a set of six heat
radiating fins. The fins provide the necessary radiating area to
ensure that a sufficient temperature differential is present on the
thermocouples, for efficient operation.

The thermal power of the RTGs is a function of the total power
produced by the \super{238}Pu fuel, and the amount of power removed in
the form of electrical energy by the thermocouples. The half-life of
\super{238}Pu is 87.74~years. The efficiency of the thermocouples
decreased over the years as a result of the decreasing temperature
differential between the hot and cold ends, and also as a result of
aging. At the time of last transmission, each RTG on board Pioneer~10
produced less than 15~W of electrical power.

The actual shape of the RTGs is shown in Figure~\ref{fig:RTGDIM}. It
is important to note that the RTGs on board Pioneer~10 and 11 were
built with the large fins that are depicted in this figure. These
enlarged fins are not shown in many drawings and photographs
(including Figure~\ref{fig:pio-craft}).

\subsubsection{The electrical subsystem}

\epubtkImage{}{%
  \begin{figure}[t!]
    \centerline{\includegraphics[width=\linewidth]{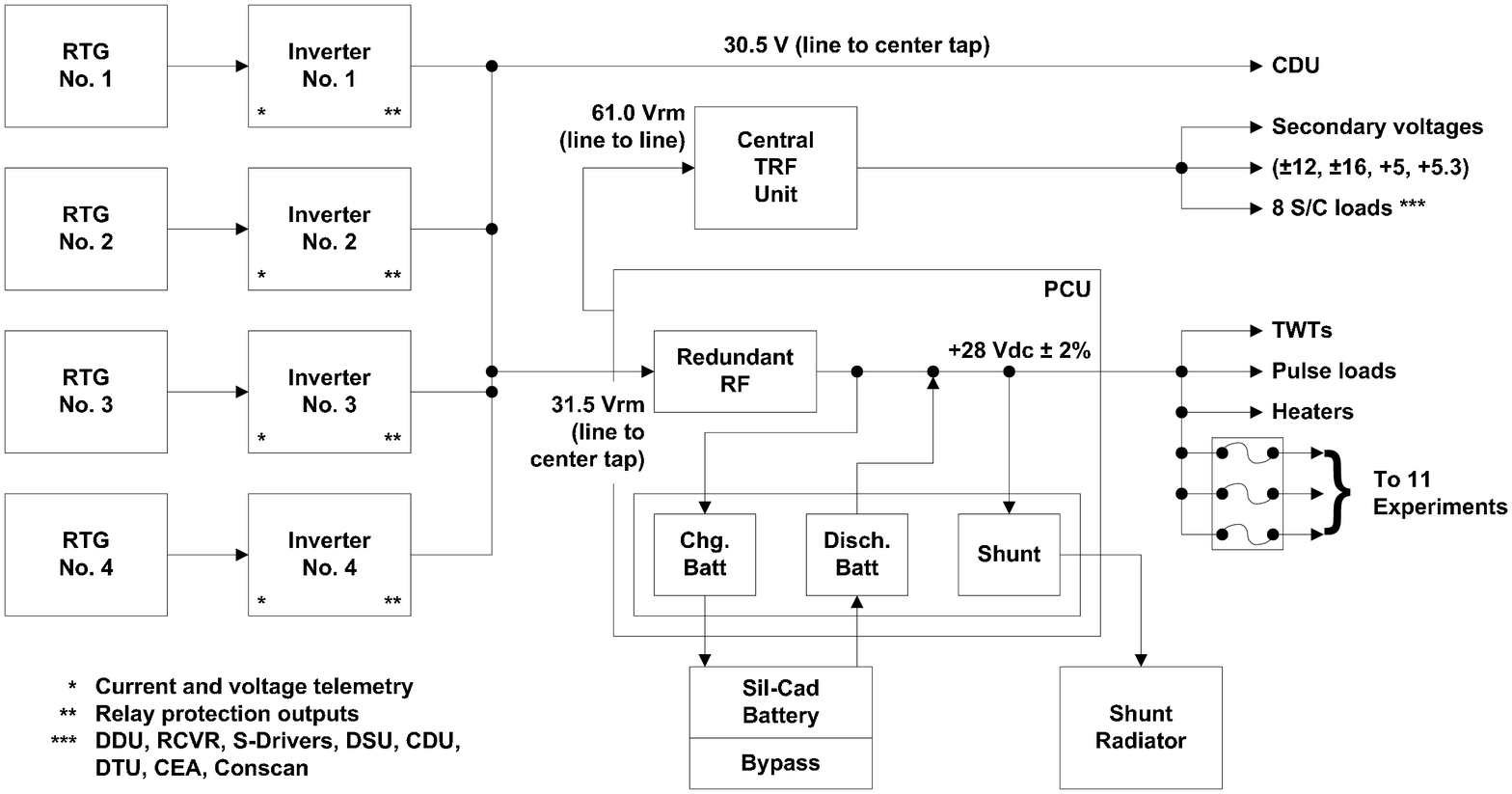}}
    \caption{Overview of the Pioneer~10 and 11 electrical subsystem
    (from~\cite{PC202}).}
    \label{fig:electrical}
\end{figure}}

Electrical power from the RTGs reached the spacecraft body via a set
of ribbon cables. There, raw power from the RTGs was fed into a series
of electric power supplies that produced 28~VDC for the spacecraft's
main bus, and also other voltages on secondary power buses.

Electrical power consumption on board the spacecraft was regulated to
ensure a constant voltage on the various power buses on the one hand,
and an optimal current draw from the RTGs on the other hand. The
electrical power subsystem was designed such that the spacecraft could
perform its primary mission, namely close-up observations of the
planet Jupiter approximately 21 months after launch, using only three
RTGs, while operating a full compliment of science
instruments. Consequently, in the early years of their mission,
significant amounts of excess electrical power were available on both
spacecraft. The power regulation circuitry diverted this excess power
to a shunt circuit, which dissipated some excess power internally,
while routing the remaining excess power to an externally mounted
shunt radiator.

\epubtkImage{}{%
  \begin{figure}
    \centerline{\includegraphics[width=\linewidth]{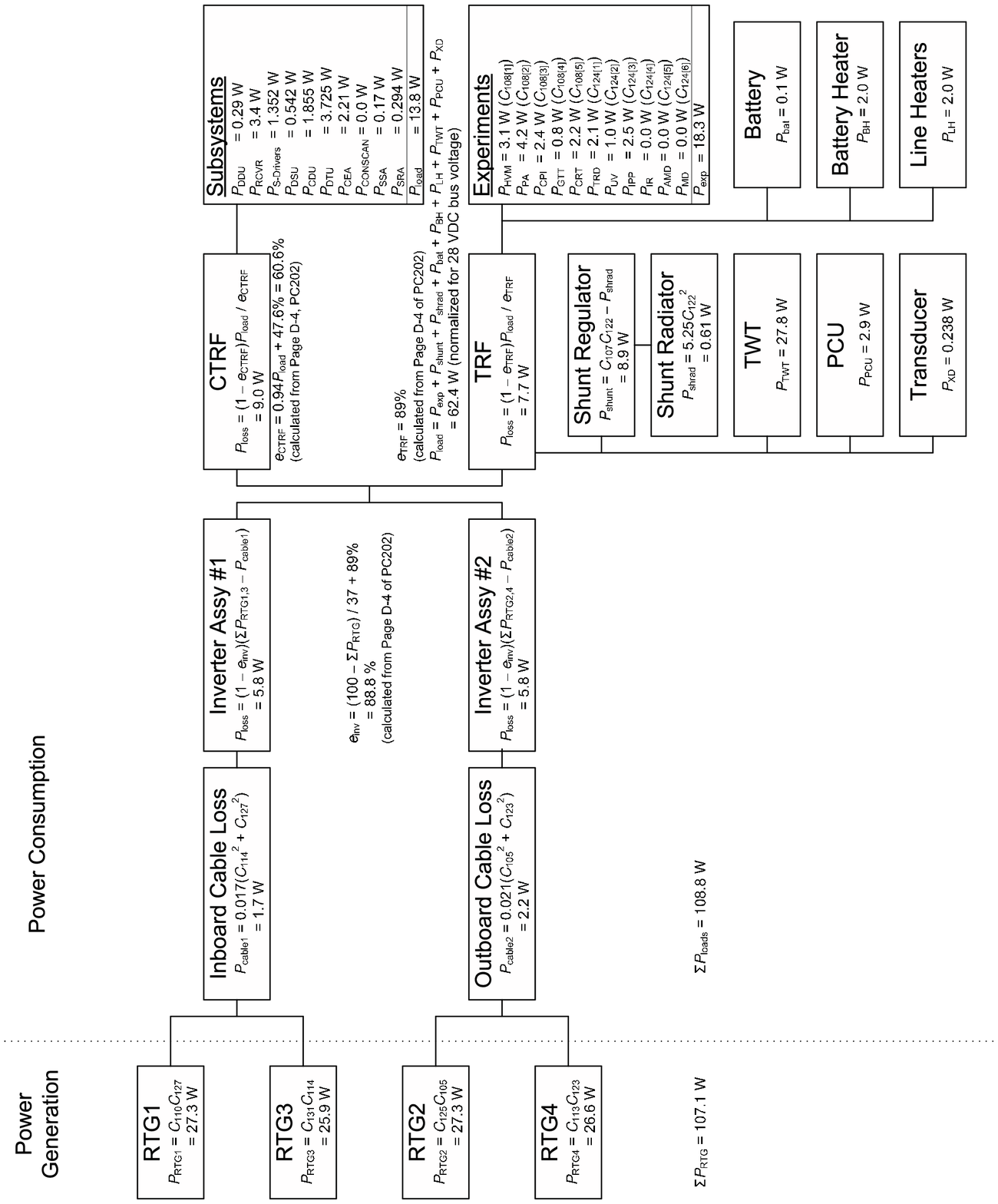}}
    \caption{Pioneer~10 power budget on July~25, 1981, taken as an
    example. Power readings that were obtained from spacecraft
    telemetry are indicated by the telemetry word in the form
    $C_{nnn}$. The discrepancy between generated power and power
    consumption is due to rounding errors and uncertainties in the
    nominal vs. actual power consumption of various subsystems.}
    \label{fig:PIOPOWER}
\end{figure}}

The total power available on board at the time of launch was in excess
of $\sim$~160~W. This figure decreased steadily throughout the
missions, obeying an approximate negative exponential law. The actual
amount of power available was a function of the decay of the
radioisotope fuel, the decreasing temperature differential between the
hot and cold ends of thermoelectric elements, and degradation of the
elements themselves. At the end of its mission, the power available on
board Pioneer~10 was less than 60~W. (Indeed, a drop in the main bus
voltage is the most likely reason that Pioneer~10 eventually fell
silent, as the reduced voltage was no longer sufficient to operate the
spacecraft's transmitter.)

The RTGs generated electrical power at $\sim$~4~VDC
(Figure~\ref{fig:electrical}). Power output from each RTG was fed to a
separate inverter circuit, producing 61~VAC (peak-to-peak) at
$\sim$~2.5~kHz. Output from the four inverters was combined and fed to
the Power Control Unit (PCU), which generated the 28~VDC main bus
voltage, managed the on-board battery, controlled the dissipation of
excess power via a shunt circuit, and also provided power to the
Central Transformer Rectifier (CTRF) component, which, in turn,
supplied power at various voltages (e.g., \textpm~16~VDC, \textpm~12~VDC,
+5~VDC) to other subsystems and instruments.

The power budget at any given time was a function of available power
vs. spacecraft load. For instance, Figure~\ref{fig:PIOPOWER} shows
Pioneer~10's power budget on July~25, 1981.

The on-board battery was designed to help with transient peak loads
that temporarily exceeded the capabilities of the RTGs. The battery
was composed of eight silver-cadmium cells, each of which had a
capacity of 5~Ah and was equipped with an individual charge/discharge
bypass circuitry.

\subsubsection{Propulsion and attitude control}
\label{sec:propsubsys}

\epubtkImage{}{%
  \begin{figure}[t!]
    \centerline{\includegraphics[width=0.8\linewidth]{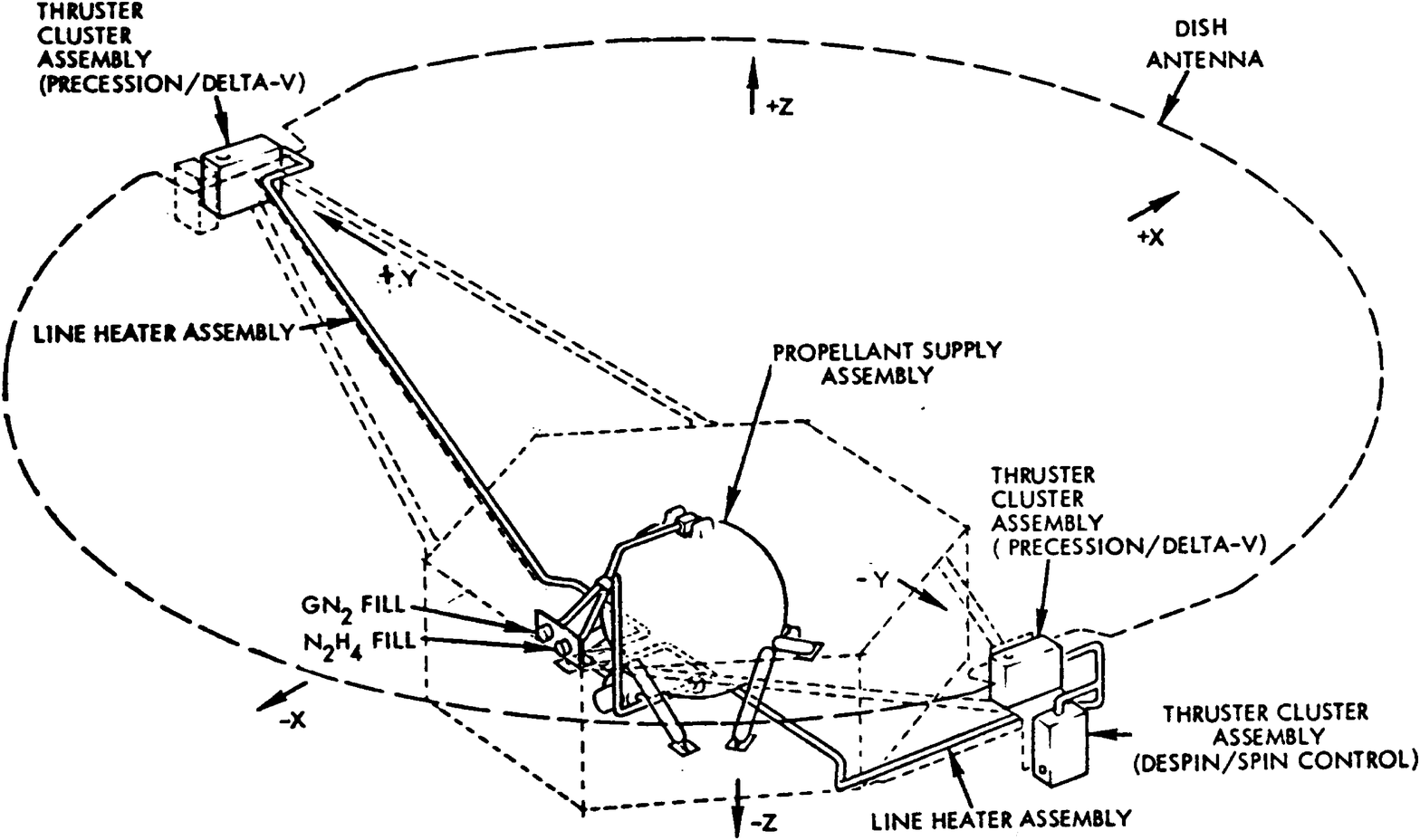}}
    \caption{An overview of the Pioneer~10 and 11 propulsion
    subsystem (from~\cite{PC202}).}
    \label{fig:propulsion}
\end{figure}}

After separation from their respective launch vehicles, the orbits of
Pioneer~10 and 11 were determined by the laws of celestial
mechanics. The spacecraft had only a small amount of fuel on board,
used by their propulsion system designed to control the spacecraft's
spin and orientation, and execute minor course correction maneuvers.

The propulsion system consisted of three thruster cluster assemblies
(see Figure~\ref{fig:propulsion}), each comprising two 1~lb
($\sim$~4.5~N) thrusters. All thruster cluster assemblies were mounted
along the rim of the HGA. One pair of clusters was oriented
tangentially along the antenna perimeter, and its two thrusters were
intended to be used to increase or decrease the spin rate of the
spacecraft. The remaining two pairs were oriented perpendicular to the
antenna plane, on opposite sides of the antenna. These two thruster
cluster assemblies were used in pairs. If two thrusters pointing in
the same direction were fired simultaneously, this resulted in a net
change in the spacecraft's velocity in a direction perpendicular to
the antenna plane. If two clusters were fired in the opposite
direction, this caused the spacecraft's spin axis to precess. This
latter type of maneuver was used, in particular, to maintain an
Earth-pointing orientation of the HGA to ensure good reception of
radio signals.

The thrusters were labeled VPT (velocity and precession thruster) and
SCT (spin control thruster.) VPT~1 and VPT~3 were oriented in the same
direction as the HGA (the {\it +z} direction), while VPT~2 and VPT~4 were
oriented in the opposite direction.

The propulsion system utilized hydrazine (N\sub{2}H\sub{4}) monopropellant
fuel, of which $\sim$~27~kg was available on board, in a 38~liter tank
that was pressurized with N\sub{2}. The propellant and pressurant were
separated by a flexible membrane, which prevented the mixing of the
liquid propellant and gaseous pressurant in the weightless environment
of space. The fuel tank was located at the center of the spacecraft,
and was heated by the spacecraft's electrical equipment. Fuel lines
leading to the thruster cluster assemblies were heated electrically,
while the thruster cluster assemblies were equipped with small (1~W)
radioisotope heating units (RHUs) containing \super{238}Pu fuel.

The capabilities of the propulsion system are summarized in
Table~\ref{tb:propcap}.

\begin{table}[h]
  \caption{Capabilities of the Pioneer~10 and 11 propulsion system.}
  \label{tb:propcap}
  \centering
  \begin{tabular}{llc}
    \toprule
    Maneuver     & Thrusters & Max.\ capability\\
    \midrule
    Despin       & SCT~1 or 2                 & 58~rpm\\
    Spin control & SCT~1 or 2                 & 14~rpm\\
    Precession   & VPT~1 and 4 or VPT~2 and 3 & 1250\textdegree\\
    Delta-\it{v} & VPT~1 and 3 or VPT~2 and 4 & 250~m/s\\
    \bottomrule
\end{tabular}
\end{table}

The spacecraft's Earth-pointing attitude was maintained as the
spacecraft were spinning in the plane of the HGA, at a nominal rate of
4.8 revolutions per minute (rpm). The propulsion system had the
capability to adjust the spin rate of the spacecraft, and to precess
the spin axis, in order to correct for orientation errors, and to
ensure that the spacecraft followed the Earth's position in the sky as
seen from on board.

The spin axis perpendicular to the plane of the HGA is one of the
spacecraft's principal axis of inertia. A wobble damper
mechanism~\cite{PC202} dampened rotations around any axis other than
this principal axis of inertia, ensuring a stable attitude even after
a precession maneuver.

\subsubsection{Navigation}

The Pioneer~10 and 11 spacecraft relied on standard methods of deep
space navigation~\cite{MG2005, Moyer-2003} (see
Section~\ref{sec:navigation}). The spacecraft's position was
determined using the spacecraft's radio signal and the laws of
celestial mechanics. The radio signal offered a precision Doppler
observable, from which the spacecraft's velocity relative to an Earth
station along the line-of-sight could be computed. Repeated observations
and knowledge of the spacecraft's prior trajectory were sufficient to
obtain highly accurate solutions of the spacecraft's orbit.

The orientation of the spacecraft was estimated from the quality of
the radio communication link (i.e., the spacecraft had to be
approximately Earth-pointing in order for the Earth to fall within the
HGA radiation pattern.) The rate and phase of the spacecraft's
rotation was established by a redundant pair of sun sensors and a star
sensor on board. These sensors (selectable by ground command) provided
a roll reference pulse that was used for navigation purposes (as
explained in the next paragraph) as well as by on-board science
instruments. The time between two subsequent roll reference pulses was
measured and telemetered to the ground.

The spacecraft also had minimal autonomous (closed loop) navigation
capability, designed to make it possible for the spacecraft to restore
its orientation by ``homing in'' on an Earth-based signal. The
maneuver, called a conical scan (CONSCAN) maneuver, utilized a piston
mechanism~\cite{Acker1972} with electrically heated freon gas that
displaced the feed horn located at the focal point of the high-gain
antenna. Unless the Earth was exactly on the HGA centerline, this
introduced a sinusoidal modulation in the amplitude of the signal
received from the Earth. A simple integrator circuit, utilizing this
sinusoidal modulation and the roll reference pulse, triggered firings
of the precession thrusters, adjusting the spacecraft's axis of
rotation until it coincided with the direction of the Earth.

Frequently, instead of CONSCAN maneuvers, ``open loop'' attitude
correction maneuvers were used, calculated to ensure that after the
maneuver, the spacecraft was oriented to ``lead'' the Earth, allowing
the Earth to move through the antenna pattern subsequently. This
reduced the frequency of attitude correction maneuvers; further, open
loop maneuvers generally consumed less propellant than autonomous
CONSCAN maneuvers.

Early in the mission, attitude correction maneuvers had to be executed
regularly, due to the combined motion of the Earth and the
spacecraft. Late in the extended mission, only two attitude correction
maneuvers were needed annually, to compensate for the Earth's motion
around the Sun, and for the spacecraft's ``sideways'' motion along its
hyperbolic escape trajectory.

\subsubsection{Communication system}
\label{sec:comm}

The spacecraft maintained its communication link with the Earth using
a set of S-band transmitters and receivers on board, in combination
with three antennae.

The main communication antenna of the spacecraft was the 2.74~m
diameter high-gain antenna. The antenna's narrow beamwidth
(3.3\textdegree\ downlink, 3.5\textdegree\ uplink) ensured an effective
radiated power of 70~dBm, allowing communication with the spacecraft
over interplanetary distances. (The original mission design
anticipated signal loss some time after Jupiter encounter, but still
within the orbit of Saturn; increases in the sensitivity of Earth
stations allowed communication with Pioneer~10 up until 2003, when the
spacecraft was over 70~AU from the Earth.)

Mounted along the centerline of the HGA was the horn of a medium-gain
antenna (MGA). On the opposite ({\it --z}) side of the spacecraft, at the
bottom of the main compartment was mounted a third, low-gain
omnidirectional antenna (LGA). This antenna was used during the
initial mission phases, before the HGA was oriented towards the Earth.

The spacecraft had two receivers and two transmitters on board,
switchable by ground command. While one receiver was connected to the
HGA, the other was connected to the MGA/LGA. The sensitivity of the
receiver was --149~dBm; at the time of the last transmission, the
spacecraft detected the Earth station's signal at a strength of
--131.7~dBm.

The spacecraft utilized two traveling wave tube (TWT) transmitters for
microwave transmission. The TWTs were selectable by ground command; it
was possible to power off both TWTs to conserve power (such as when
CONSCAN maneuvers were performed late in the mission, when the
available electrical power on board was no longer sufficient to
operate a TWT transmitter and the feed movement mechanism
simultaneously.)

The radio systems operated in the S-band, utilizing a frequency of
$\sim$~2.1~GHz for uplink, and $\sim$~2.3~GHz for downlink. The
transmitter frequency was synthesized on board by an independent
oscillator. However, the spacecraft's radio system could also operate
in a  {\it coherent mode}: in this mode, the downlink signal's
carrier frequency was phase-coherently synchronized to the uplink
frequency, at the exact frequency ratio of 240/221. In this mode, the
precision and stability of the downlink signal's carrier frequency was
not limited by the equipment on board. This mode allowed precision
Doppler frequency measurements with millihertz accuracy.

The main function of the spacecraft's communication system was to
provide two-way data communication between the ground and the
spacecraft. Data communication was performed at a rate of
16\,--\,2048~bits per second (bps). Communication from the ground
consisted of commands that were decoded by the spacecraft's radio
communication subsystem. Communication to the ground consisted of
measurement results from the spacecraft's suite of science
instruments, and engineering telemetry.

\subsubsection{Telemetry data and its interpretation}

Communication between the spacecraft and a DSN antenna took place
using a variety of data formats; the format selected depended on the
type of experiment that the spacecraft was ordered to perform, but
typically the format gave precedence to science results over
telemetry. However, telemetry information was continuously transmitted
to the Earth at all times, using a small portion of the available data
bandwidth.

The telemetry data stream was assembled on board by the digital
telemetry unit (DTU), which comprised some $\sim$~800
TTL\epubtkFootnote{Transistor-Transistor-Logic, or TTL integrated
  circuits are standardized logic circuits built using bipolar
  transistors that have been in widespread use since the 1960s in the
  form of small-scale integration (SSI) integrated circuits containing
  a small number of logic gates per chip.} integrated circuits, and
was also equipped with 49,152 bits of ferrite core data memory. A
total of 10 science data formats (of which 5 were utilized) and 4
engineering data formats, yielding a total number of 18 different
valid format combinations, was selectable by ground command. The DTU
could operate in three modes: realtime (passing through science
measurements as received from instruments), store (storing
measurements in the on-board memory) and readout (transmitting
measurements previously stored in on-board memory.)

When one of the engineering data formats was selected, the spacecraft
transmitted only engineering telemetry. These formats were utilized,
for instance, during maneuvers or spacecraft troubleshooting.

Most of the time, a science data format was used, in which case most
of the bits in the telemetry data stream contained science data. A
small portion, called the subcommutator, was reserved for engineering
data; this part of the telemetry record cycled through all telemetry
data words in sequence.

The telemetry record size was 192 bits, divided into 36 6-bit
words. Depending on the telemetry format chosen, either all 36 6-bit
words contained engineering telemetry, or a single engineering
telemetry word was transmitted in every (or every second) telemetry
record.

There were 128 distinct engineering telemetry words. Depending on the
science format used, the subcommutator was present either in every
transmitted record or every second record. Therefore, it may have
taken as many as 256 telemetry records before a particular engineering
word was transmitted. At the lowest data rate of 16~bits per second,
this meant that any given parameter was telemetered to the Earth once
every 51.2~minutes.

An additional 64 telemetry words were used to transmit engineering
information from science instruments. These parameters were only
transmitted in the subcommutator, and at the lowest available data
rate, a particular science instrument telemetry word was repeated
every 25.6~minutes.

The 128 engineering telemetry words were organized into 4 groups of 32
words each; it is customary to denote them using the notation
$C_{mnn}$, where $m=1...4$ is the group number, and $nn=01...32$ is
the telemetry word. Similarly, science instrument telemetry words were
labeled $E_{mnn}$, with $m=1...2$ and $nn=01...32$.

When a science or engineering telemetry word was used to convey the
reading from an analog (temperature, voltage, current, pressure, etc.)
sensor, the reading was digitized with 6-bit resolution~\cite{ARC223,
  ARC221, LK2003}, and the resulting 6-bit data word was
transmitted. Associated with each telemetry word representing an
analog measurement was a set of calibration coefficients that formed a
5th-order calibration polynomial. For each calibration polynomial, a
range was also defined that established valid readings that could be
decoded by that polynomial.

Appendix~\ref{app:formatCE} lists selected engineering and science
telemetry data words that may be relevant to the analysis of the
Pioneer anomaly.

\subsubsection{Thermal subsystem}

The Pioneer spacecraft were equipped with a thermal control system
comprising a variety of active and passive thermal control
devices. The purpose of these devices was to maintain the required
operating temperatures for all vital subsystems of the spacecraft.

The main spacecraft body was covered by multilayer insulating
blankets~\cite{AIAA72-285, STOCHL1974}. These blankets were designed
to retain heat within the spacecraft when it was situated in deep
space, far from the Sun.

\epubtkImage{}{%
\begin{figure}[t]
\begin{center}
\includegraphics[width=0.65\linewidth]{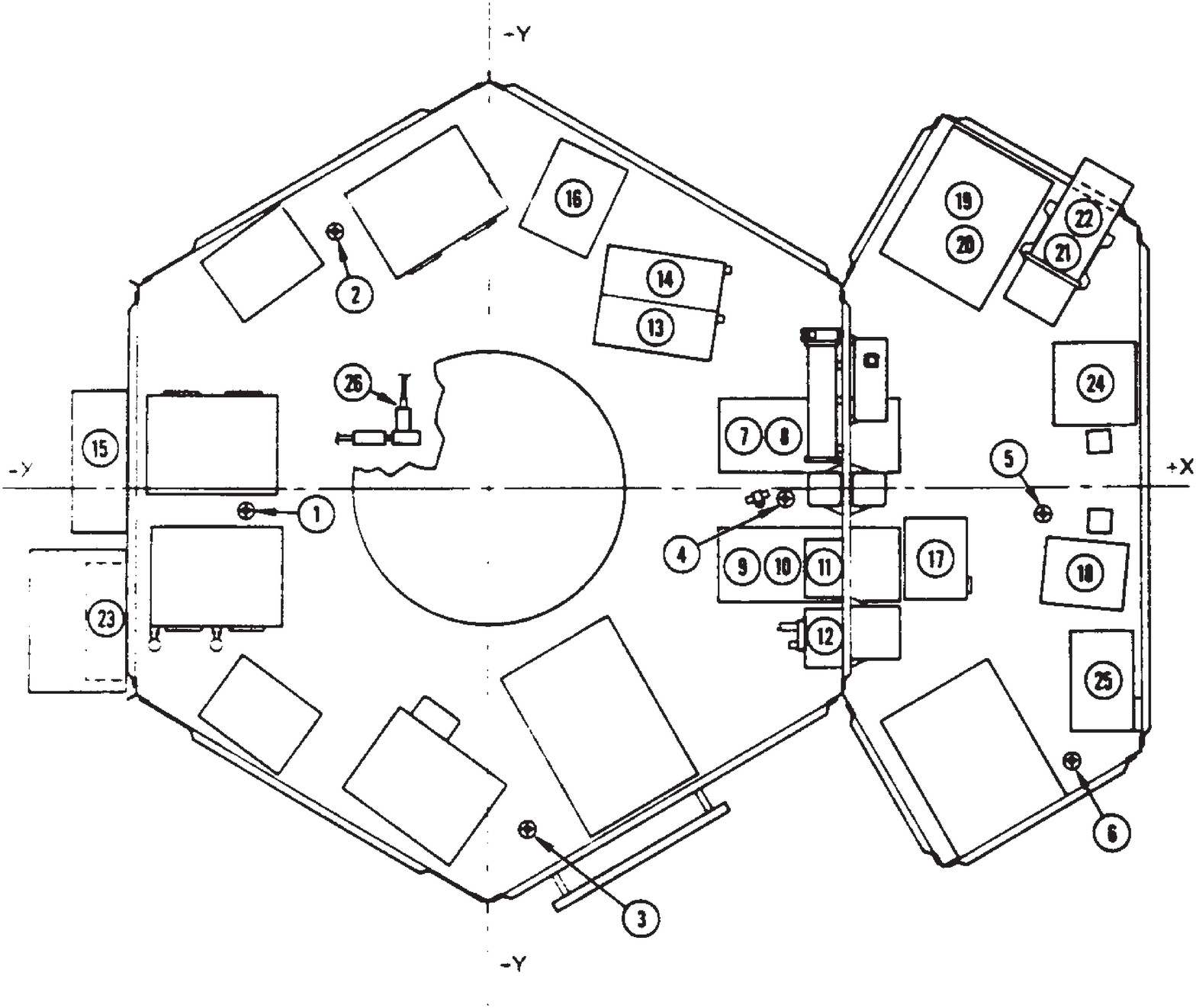}
\end{center}
\caption{Location of thermal sensors in the instrument compartment of
  the Pioneer~10 and 11 spacecraft (from~\cite{PC202}).  Platform
  temperature sensors are mounted at locations 1 to 6. Some locations
  (i.e., end of RTG booms, propellant tank interior, etc.) not shown.}
\label{fig:sensors}
\end{figure}}

To prevent overheating of the spacecraft interior near the Sun, a
thermal louver system~\cite{PC202, AIAA69-697, JPL32-555, TCSDR3} was
utilized. The louvers were located at the bottom of the spacecraft,
organized in a circular pattern, with additional louvers on the
science compartment (Figure~\ref{fig:louvers}). These louvers were
actuated by bimetallic springs that were thermally (radiatively)
coupled to the main electronics platform behind the louvers. The
louvers were designed to be fully open when the platform temperature
exceeded 90\textdegree\,F, and fully close when the temperature fell below
40\textdegree\,F (Figure~\ref{fig:langle}).

\epubtkImage{}{%
\begin{figure}[t!]
\centering
\includegraphics[width=0.6\linewidth]{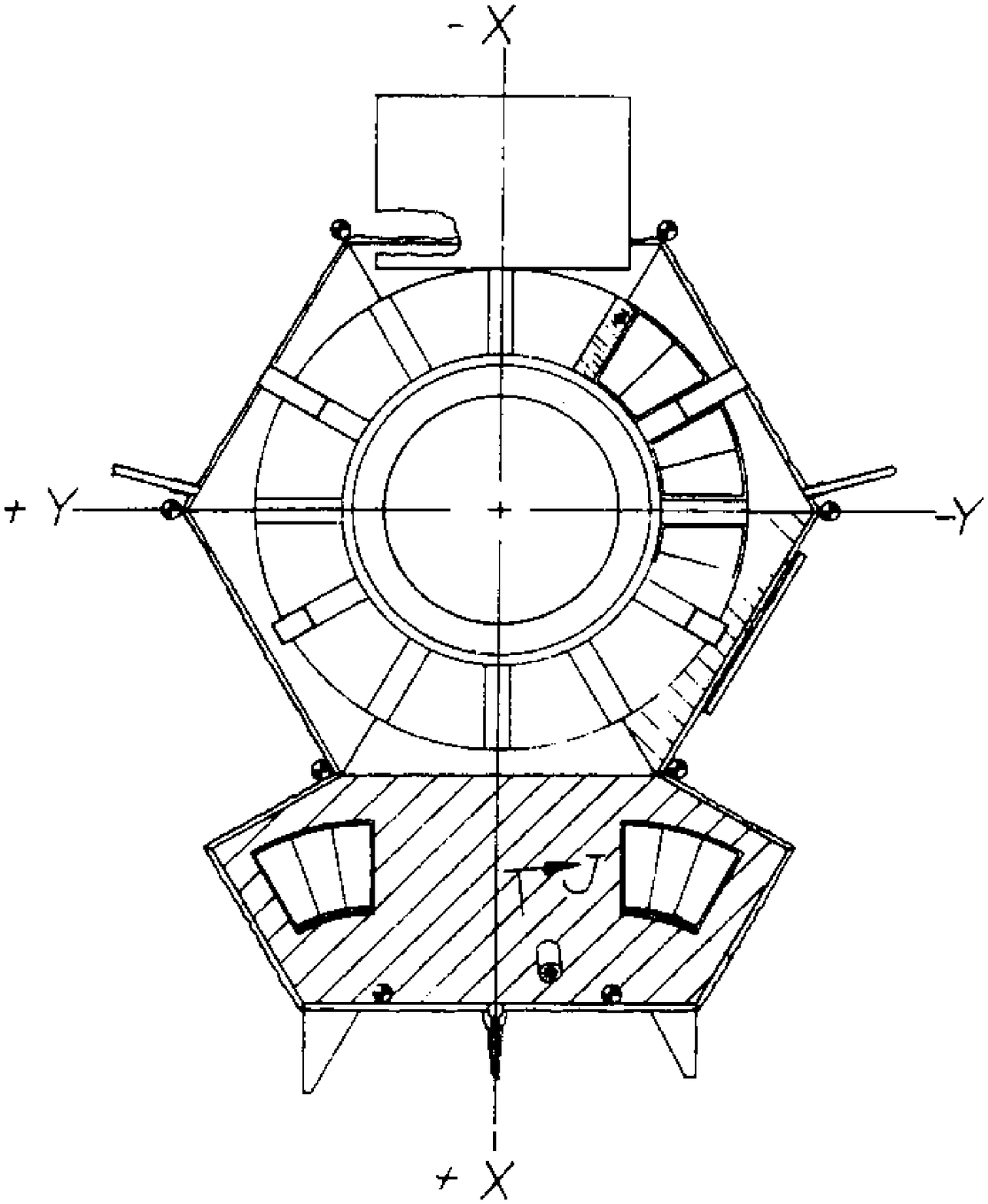}
\caption{The Pioneer~10 and 11 thermal control louver system, as seen
  from the aft ({\it --z}) direction (from~\cite{PC202}).}
\label{fig:louvers}
\end{figure}}

\epubtkImage{}{%
\begin{figure}[t!]
\centering
\includegraphics[width=0.6\linewidth]{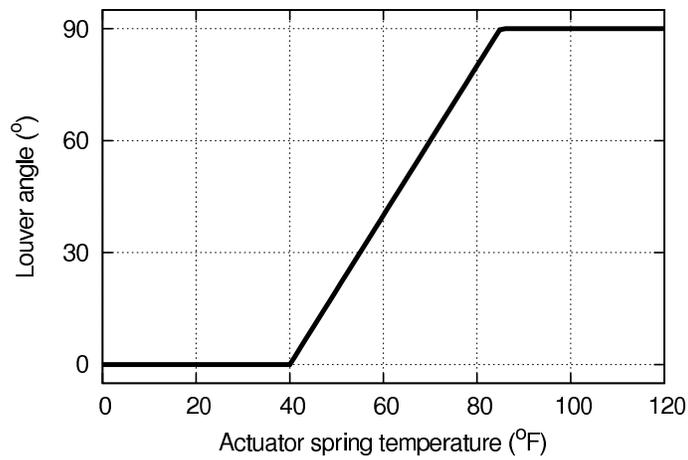}
\caption{Louver blade angle as a function of platform temperature
  (from~\cite{TCSDR3}). Temperatures in \textdegree\,F
  ([\textdegree\,C]~=~([\textdegree\,F]--32)~\texttimes~5/9).}
\label{fig:langle}
\end{figure}}

\epubtkImage{}{%
\begin{figure}[t!]
\centering
\includegraphics[width=0.6\linewidth]{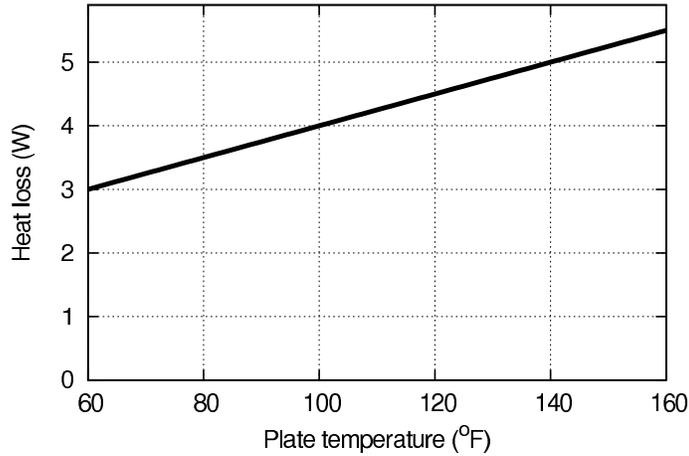}
\caption{Louver structure heat loss as a function of platform
  temperature (from~\cite{TCSDR3}). Temperatures in \textdegree\,F
  ([\textdegree\,C]~=~([\textdegree\,F]--32)~\texttimes~5/9).}
\label{fig:lloss}
\end{figure}}

\epubtkImage{}{%
\begin{figure}[t!]
\centering
\includegraphics[width=0.6\linewidth]{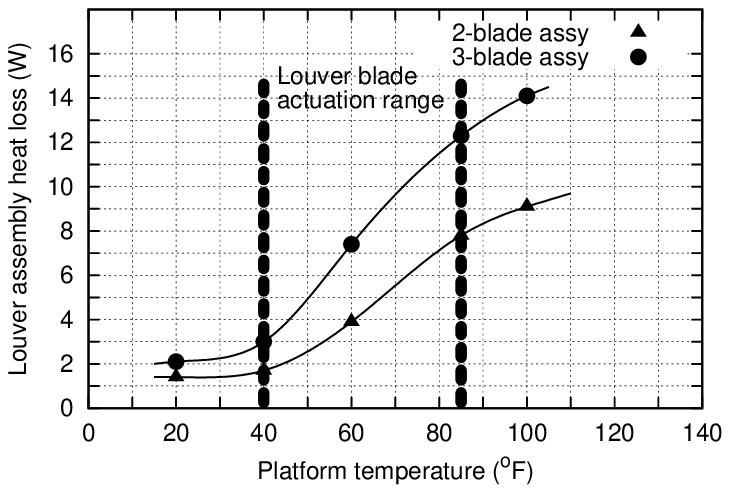}
\caption{Louver assembly performance
  (from~\cite{TCSDR3}). Temperatures in \textdegree\,F
  ([\textdegree\,C]~=~([\textdegree\,F]--32)~\texttimes~5/9).}
\label{fig:lperf}
\end{figure}}

The louver system emits heat in two ways: through structural
components (Figure~\ref{fig:lloss}), and through the louver blade
assemblies (Figure~\ref{fig:lperf}). The total heat emitted by the
louver system is the sum of the heat emitted via these two mechanisms.

The fuel lines extending from the spacecraft body to the thruster
cluster assemblies located along the rim of the HGA were insulated by
multilayer thermal blankets and were kept warm by electric heater
lines. The thruster cluster assemblies contained radioisotope heater
units (RHUs), designed to prevent the propellant from
freezing. Additional RHUs heated the star sensor and the magnetometer
assembly at the end of the magnetometer boom.

The spacecraft's battery was mounted on the outside of the main
spacecraft body, and heated by an electrical heater.

Most heat produced by electrical equipment on board was released
within the insulated interior of the spacecraft. The requirement for
science instruments was to leak no more than the instrument's own
power consumption plus $0.5$~W to space.

Some excess electrical power was radiated away as heat by an
externally mounted shunt radiator plate, which formed part of the
spacecraft's electrical power subsystem.

Most heat on board was produced by the four radioisotope
thermoelectric generators. Electrical power in these generators was
produced by bimetallic thermocouples that relied on the temperature
difference between their hot and cold ends for power
generation. Therefore, it was essential that the cold ends of the
thermocouples were connected to RTG radiator fins that radiated heat
into space with high efficiency.

Temperature readings from many locations throughout the spacecraft,
including temperatures at six key locations on the main electronics
platform, were telemetered to the ground (Figure~\ref{fig:sensors}).

With the exception of the louver system, the two adjacent hexagonal
parts of the spacecraft body are covered by multilayer
insulation. There is no insulation between the main and science
compartments of the spacecraft body.

The surface materials and paints used to cover most major exterior
surfaces are documented~\cite{PC202, CR1786, NASA32873, NASA1121,
  NASAD1116, TCSDR3}. In particular, the surfaces of the spacecraft
body, HGA, and RTGs are well described, along with the thermal control
louver system in terms of solar absorptance and infrared emittance
(Table~\ref{tb:radiometric}). Solar absorptance is characterized by a
dimensionless number (usually denoted by $\alpha$) between 0 and 1
representing the efficiency with which a particular material absorbs
the radiant energy of the Sun when compared to an ideal black
body. Infrared emittance is similarly characterized by a dimensionless
number (usually dented by $\epsilon$) between 0 and 1 that represents
the efficiency with which a material radiates heat at lower
(typically, room) temperatures as compared to an ideal black body.

\begin{table}[htbp]
  \caption{Radiometric properties of Pioneer~10 and 11 major exterior
  surfaces (at launch): solar absorptance ($\alpha$) and infrared
  emittance ($\epsilon$).}
  \label{tb:radiometric}
  \centering
  \begin{tabular}{lccc}
    \toprule
    Surface & Area (m\super{2}) & $\alpha$ & $\epsilon$\\
    \midrule
    HGA front (white paint)                               & 5.91 & 0.21 & 0.85\\
    HGA rear (bare aluminum)                              & 5.91 & 0.17 & 0.04\\
    Spacecraft body, front (2-mil aluminized Mylar)       & 1.55 & 0.17 & 0.70\\
    Spacecraft body, rear (2-mil aluminized Kapton)       & 1.19 & 0.40 & 0.70\\
    Spacecraft body, RTG sides (2-mil aluminized Kapton)  & 0.65 & 0.40 & 0.70\\
    Spacecraft body, other sides (2-mil aluminized Mylar) & 1.21 & 0.17 & 0.70\\
    RTG fin surfaces (white paint)                        & ~    & 0.20 & 0.83\\
    Louver blades, closed (bare aluminum)                 & 0.36 & 0.17 & 0.04\\
    \bottomrule
  \end{tabular}
\end{table}

The exterior surfaces of the Pioneer~10 and 11 spacecraft are covered
by a variety of materials and paints. Changes in the spacecraft
properties can affect/induce forces that are both of on-board and of
external origin. Therefore, it is of great importance to establish the
extent to which any of the spacecraft's properties might have been
changing over time.

\subsection{Spacecraft operating history}

The Pioneer~10 and 11 spacecraft spent about three decades in deep
space. Most of what we know about the anomalous acceleration is based
on data that was collected when the spacecraft were well into their
second and third decades of operation. This raises an obvious
question: what are the effects of aging and to what extent may aging
be responsible for the anomalous acceleration?

In this section, we summarize our knowledge about the effects of aging
on the spacecrafts' subsystems.

\subsubsection{Spacecraft physical configuration}

The overall shape of the Pioneer~10 and 11 spacecraft is not expected
to change significantly with age. The spacecraft is fundamentally a
rigid body; other than the constant centrifugal force that arises as a
result of the spacecraft's rotation, there are no forces stretching,
bending, or otherwise acting on the spacecraft structurally.

The Pioneer spacecraft had few moving parts. After initial boom
deployment, the spacecrafts' physical configurations remained largely
unchanging, with a few notable exceptions.

Consumption of the fuel load on board resulted in small changes in the
spacecrafts' mass distribution during the large course correction
maneuvers early in the Pioneer~10 and 11 missions. As the amount of
fuel on board was small, and the fuel tank was situated near the
spacecraft's center-of-gravity, the effects of later attitude
correction maneuvers, which consumed minuscule amounts of fuel, were
likely negligible.

Some instruments had moving parts: notably, the Imaging
Photopolarimeter (IPP) instrument had a telescope that was mounted on
a scan platform, allowing it to be used for Jupiter imaging. Operating
this instrument's small moving parts, however, would have introduced
only minute changes in the spacecraft's mass distribution and thermal
properties.

More notable was the spacecraft's passive louver system. As discussed
in Section~\ref{sec:heat}, the louver system was designed to vent
excess heat radiatively from the interior of the spacecraft. The state
of the louver system can be determined as a function of the
electronics platform temperatures. The position of the louver blades
can significantly alter the thermal behavior of the spacecraft, by
allowing a higher proportion of interior heat to escape through the
louver system.

At large heliocentric distances (beyond $\sim$~25~AU), the louver
system is always closed, and the spacecraft's physical configuration
remains constant.

\subsubsection{Changes in spacecraft mass}
\label{sec:sc-masses}

As we discussed in Section~\ref{sec:pio-gen-character}, the nominal
launch mass of the Pioneer~10 and 11 spacecraft was $\sim$~260~kg, of
which $\sim$~30~kg was propellant and pressurant.\epubtkFootnote{Note
  that the acceleration due to nongravitational forces, regardless of
  their external or internal origin, is dependent on the actual mass
  of the spacecraft.} The spacecraft mass slowly decreased, primarily
as a result of propellant usage. Additionally, small mass losses may
occur due to fuel leaks, pressurant outgassing, He outgassing
($\alpha$ particles) from the RTGs and RHUs, and possibly, outgassing
from the spacecraft batteries.

Pioneer~10 used only a moderate amount of propellant, as it performed
no major trajectory correction maneuvers. If the propellant used
amounted to one quarter of the propellant on board, this means that
Pioneer~10's mass would have decreased to $\sim$~250~kg late in its
mission (the 2002 JPL study used the nominal value of
251.883~kg). Further, it should be noted that most propellant usage
occurred prior to Jupiter encounter; afterwards, Pioneer~10 only used
minimal amounts of propellant for precession maneuvers, needed to keep
its antenna aimed at the Earth.

Pioneer~11 performed major trajectory correction maneuvers en route to
Jupiter and Saturn. The maneuvers were in order to allow it to follow a
precisely calculated orbit that utilized a gravitational assist from
Jupiter that was needed to set up the spacecraft for its encounter
with Saturn. As a result, Pioneer~11 is believed to have used
significantly more propellant than its twin, perhaps three quarters of
the total available on board. The spacecraft's mass, therefore, may
have decreased to $\sim$~232~kg following its encounter with Saturn
(the mass used in the 2002 JPL study was somewhat higher,
239.73~kg~\cite{pioprd}).

We note that these figures are crude estimates, as the actual amount
of propellant on board is not telemetered. Sensors inside the
propellant tank did offer temperature and pressure telemetry, but
these sensors were not sufficiently reliable for a precise estimate of
the remaining fuel on board.

The spacecraft can also lose mass due to outgassing. Two possible
sources of outgassing are helium outgassing from the radioactive fuel
on board in the radioisotope thermoelectric generators and
radioisotope heater units, and chemical outgassing from the
spacecraft's battery. An upper limit of 18~g on helium outgassing can
be established using the known physical properties of the \super{238}Pu
fuel (see Section~\ref{sec:navigation}), which is not
significant. Similarly, the amount of gas that can escape from the
spacecraft's batteries is small, especially in view of the fact that
the batteries performed nominally far longer than anticipated: under
no circumstances can it exceed the battery mass, but in all
likelihood, and especially in view of the fact that the batteries
performed nominally throughout the mission, any outgassing is
necessarily limited to a very small fraction of the $\sim 2.35$~kg
battery mass (see Section~\ref{sec:navigation}). Therefore, outgassing
cannot have played a major role in the evolution of the spacecraft
mass; furthermore, any mass loss due to outgassing is dwarfed by
uncertainties in the mass of the remaining fuel inventory.

Could the spacecraft have gained mass, for instance, by collecting
dust particles from the interplanetary medium? In situ measurements by
the Pioneer spacecraft themselves provide an upper limit on the amount
of dust encountered by the spacecraft. After passing Jupiter's orbit,
the dust flux measured by Pioneer~10 remained approximately constant,
at $3\times 10^{-6}$~m$^{-2}$s$^{-1}$
particles~\cite{2002AJ....123.2857L}. The upper limit on particle
sizes is $10^{-4}$~kg. Even assuming that all particles had masses
near this upper limit, the total amount of mass that could have
accumulated on the spacecraft over the course of 20 years would be no
more than $\sim$~1~kg. However, given that the spacecraft is moving
through interplanetary space at a velocity of 10~km/s or higher, these
assumptions on particle size would imply a dust density of $\sim
10^{-14}$~kg/m$^3$, which is many orders of magnitude higher than more
realistic estimates (see Section~\ref{sec:force_drag}). Therefore, the
actual dust mass accumulated on the spacecraft cannot be more than a
few grams at the most; consequently, this mechanism for mass increase
can also be safely ignored.

\subsubsection{Instrumentation}

It is not known how on-board instrumentation (i.e., telemetry sensors)
respond to aging. What we know is that many sensors stopped providing
usable readings when measured values (e.g., temperatures) dropped
outside calibrated
ranges~\cite{2006CaJPh..84.1063T,2007arXiv0710.2656T,2007arXiv0710.0191T,
  MDR2005}. Other sensors continued to provide consistent readings,
with no indication of sensor failure.

However, there were a few sensor anomalies that may be due to
age-related sensor defects. Most notable among these are the anomalous
readings from the propellant tank of Pioneer~10 (described in
Section~\ref{sec:propulsion_syst}).

It is also unknown how on-board instrumentation responds when their
supply voltage drops below the nominal level. Late in its mission, the
electrical power subsystem on board Pioneer~10 no longer had
sufficient power to maintain the nominal main bus voltage of
28~VDC. As this coincides with changes in physical sensor readings
(e.g., drops in temperature), the extent to which those readings are
affected by the drop in voltage is not readily evident.

\subsubsection{Electrical system}

Insofar as we can determine from telemetry, the electrical subsystems
on board Pioneer~10 and 11 performed nominally throughout the
missions, so long as sufficient electrical power was available from
the RTGs. The on-board chemical batteries remained functional for many
years; eventually, due to irreversible chemical changes and decreasing
temperatures, the batteries ceased functioning.

\begin{figure*}
\hskip -6pt
\begin{minipage}[b]{.5\linewidth}
\centering \includegraphics[width=\linewidth]{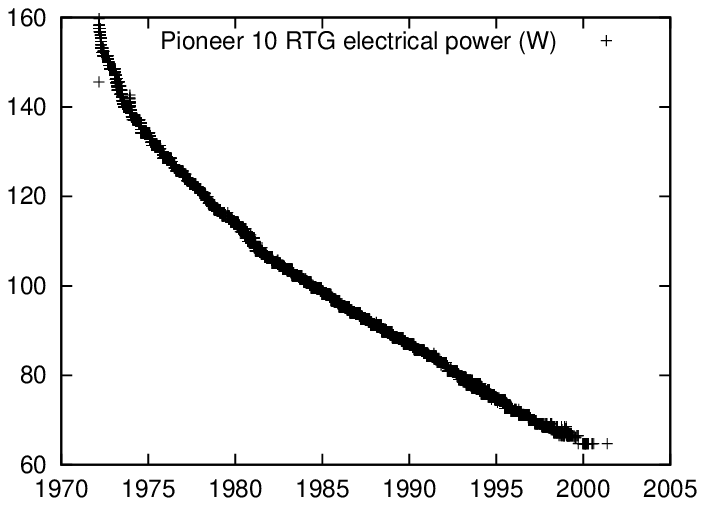}
\end{minipage}
\hskip 0.001\linewidth
\begin{minipage}[b]{.5\linewidth}
\centering \includegraphics[width=\linewidth]{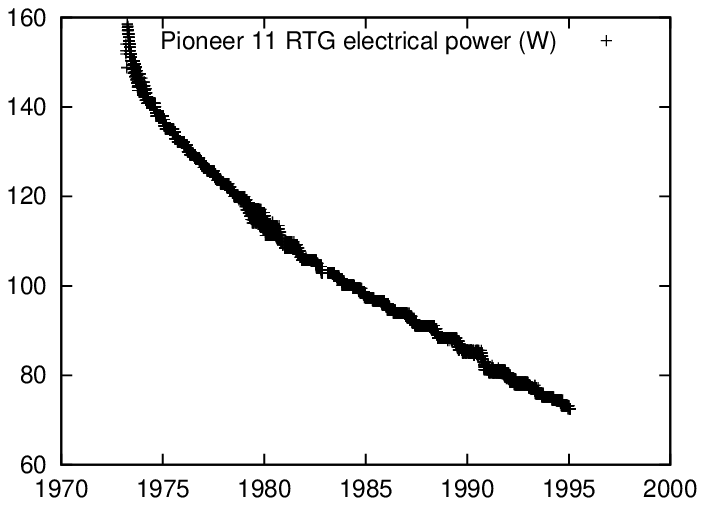}
\end{minipage}
\caption{Changes in total RTG electrical output (in W) on board Pioneer~10 (left) and 11 (right), as computed using the missions' on-board telemetry.}
\label{fig:elec}
\vskip -5pt
\end{figure*}

\subsubsection{Radioisotope thermoelectrical generators}

The effects of aging on the RTGs are complex. Internally, aging causes
a degradation of the bimetallic thermocouples, contributing to their
loss of efficiency and the decrease in RTG electrical power
output. Externally, it has been conjectured~\cite{2003PhRvD..67h4021S}
that the RTG exterior surfaces may have aged due to solar bleaching
and impact by dust particles. The extent to which such degradation may
have occurred (if at all) is unknown. The resulting fore-aft asymmetry
may be a significant source of unaccounted-for acceleration in the
approximately sunward direction.

\subsubsection{Propulsion system}
\label{sec:propulsion_syst}

\epubtkImage{}{%
\begin{figure}[t!]
\centering
\includegraphics[width=\linewidth]{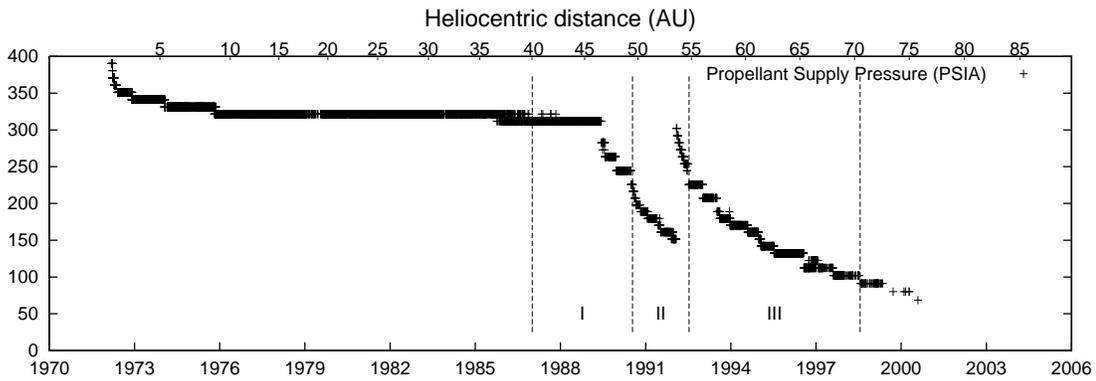}
\caption{Propulsion tank pressure (in pounds per square inch absolute; 1~psia${}=6.895$~kPa) on board Pioneer~10. The three intervals studied in~\cite{pioprd} are marked by roman numerals and separated by vertical lines.}
\label{fig:c210}
\end{figure}}

The propellant pressure sensor on board Pioneer~10 began to show
anomalous behavior in June 1989. Propellant pressure, which remained
steady up to this point, only decreasing slowly as a result of cooling
and occasional propellant usage, suddenly began to show a sharp decay,
dropping from over 300~psia down to about 150~psia by January 1992
(Figure~\ref{fig:c210}). At this time, the propellant pressure
instantaneously increased to its pre-1989 value of $\sim$~310~psia,
after which it began dropping again. There were no corresponding
changes or other anomalies in the observed propellant temperature and
expellant temperature. Therefore, the most likely explanation for this
anomaly is a sensor malfunction, not a real loss of fuel or
propellant.

On December 18, 1975, subsequent to an attempted maneuver, and as a
result of a stuck thruster valve, the spin of Pioneer~11 increased
dramatically, from a spin rate of $\sim$~5.5 revolutions per minute
(rpm) to $\sim$~7.7~rpm (Figure~\ref{fig:spin}, right
panel). Fortunately, the thruster ceased firing before the spin rate
increased to a value that would have threatened the spacecraft's
structural integrity or compromised its ability to carry out its
mission.

\subsubsection{Attitude control and spin}
\label{sec:spin}

Encounter with the intense radiation environment in the vicinity of
Jupiter damaged the star sensor on board Pioneer~10. As the sun
sensors operate only up to a distance of $\sim$~30~AU, Pioneer~10 had
operated without a primary roll reference for many years before the
end of its mission. (The rate of Pioneer~10's spin was determined from
measurements taken by its Imaging Photo-Polarimeter (IPP) instrument
and other methods~\cite{AIAA87-0502}.)

\begin{figure*}[t!]
\hskip -6pt
\begin{minipage}[b]{.5\linewidth}
\centering \includegraphics[width=\linewidth]{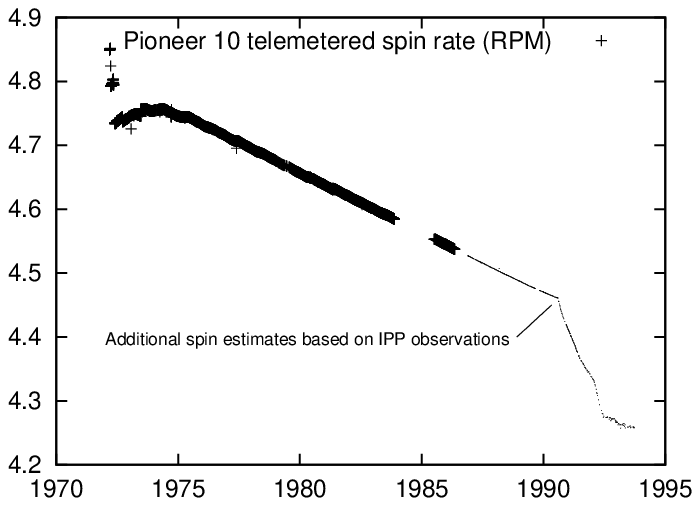}
\end{minipage}
\hskip 0.001\linewidth
\begin{minipage}[b]{.5\linewidth}
\centering \includegraphics[width=\linewidth]{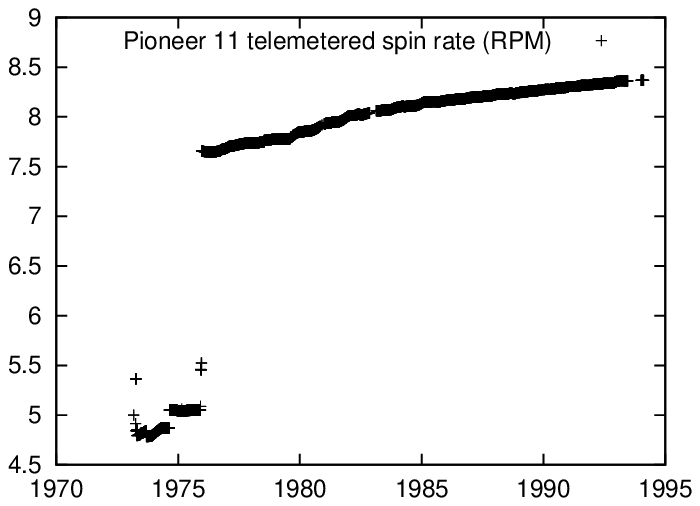}
\end{minipage}
\caption{On-board spin rate measurements (in rpm) for Pioneer~10 (left) and Pioneer~11 (right). The sun sensor used on Pioneer~10 for spin determination was temporarily disabled between November 1983 and July 1985, and was turned off in May 1986, resulting in a `frozen' value being telemetered that no longer reflected the actual spin rate of the spacecraft. Continuing spot measurements of the spin rate were made using the Imaging Photo-Polarimeter (IPP) until 1993. The anomalous increase in Pioneer~11's spin rate early in the mission was due to a failed spin thruster. Continuing increases in the spin rate were due to maneuvers; when the spacecraft was undisturbed, its spin rate slowly decreased,
 as seen in Figure~\ref{fig:spinzoom}.}
\label{fig:spin}
\vskip -5pt
\end{figure*}

\begin{figure*}[t!]
\hskip -6pt
\begin{minipage}[b]{.5\linewidth}
\centering \includegraphics[width=\linewidth]{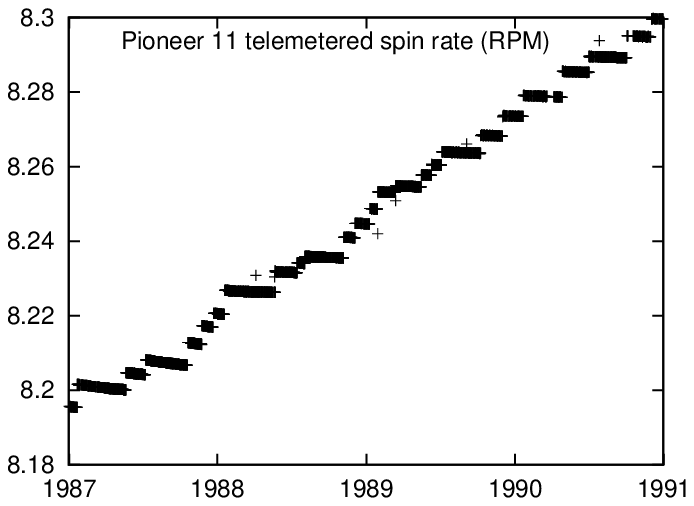}
\end{minipage}
\hskip 0.001\linewidth
\begin{minipage}[b]{.5\linewidth}
\centering \includegraphics[width=\linewidth]{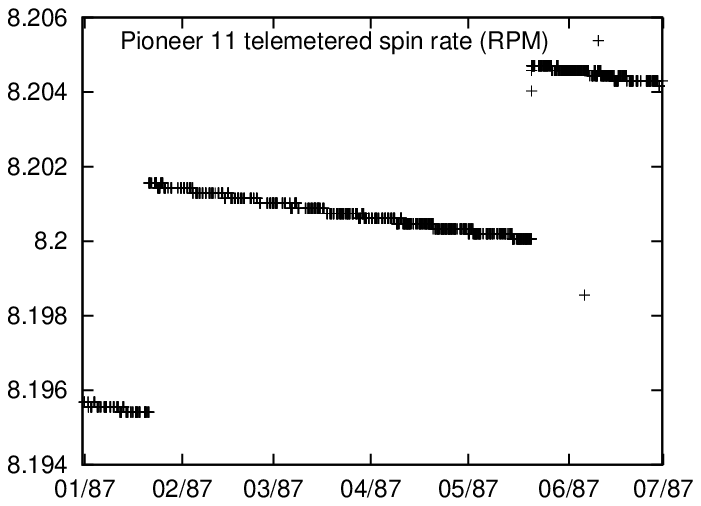}
\end{minipage}
\caption{Zoomed plots of the spin rate of Pioneer~11. On the left, the interval examined in~\cite{pioprd} is shown; maneuvers are clearly visible, resulting in discrete jumps in the spin rate. The figure on the right focuses on the first half of 1987; the decrease in the spin rate when the spacecraft was undisturbed is clearly evident.}
\label{fig:spinzoom}
\vskip -5pt
\end{figure*}

The nominal spin rate of the Pioneer~10 and 11 spacecraft was
4.8~rpm. This spin rate was achieved by reducing the spacecraft's
initial rate of spin, provided by the launch vehicle, in successive
stages, first by firing spin/despin thrusters, and then by extending
the RTG and magnetometer booms. Later, spin could be precisely
adjusted and corrected by the spin/despin thrusters.

On Pioneer~11, due to the spin thruster anomaly described in the
previous section, the spacecraft's spin remained at an abnormally high
value. In fact, it further increased, presumably as a result of fuel
leaks, all the way up to $\sim$~8.4~rpm at the time of the last
available telemetry data point, on February 11, 1994.

Meanwhile, Pioneer~10's spin slowly decreased over time, probably due
to a combination of effects that may include fuel leaks as well as the
thermal recoil force and associated change in angular momentum.

The spin rates of Pioneer~10 and 11 are shown in
Figure~\ref{fig:spin}. Spin was measured on board using one of several
sensors, namely a star sensor and two sun sensors. The purpose of
these sensors was to provide a roll reference pulse that could then be
used to synchronize other equipment, including the IPP instrument and
the navigational system.

The sun sensors required a minimum angle between the spacecraft's spin
axis and the spacecraft-Sun line. Further, they required that the
spacecraft be within a certain distance of the Sun, in order for a
reliable roll reference pulse to be generated. For these reasons, the
sun sensors could not be used to provide a roll reference pulse once
the spacecraft were more than $\sim$~30~AU from the Sun.

There were no such limitations on the star sensor; however, the star
sensor on board Pioneer~10 ceased functioning shortly after Jupiter
encounter, probably due to radiation damage suffered while the
spacecraft traversed the intense radiation environment in the gas
giant's vicinity.

As a result, Pioneer~10 lost its roll reference source when its
distance from the Sun increased beyond $\sim$~30~AU. Although the roll
reference assembly continued to provide roll reference pulses at the
last ``frozen in'' rate, this rate no longer matched the actual rate
of revolution of the spacecraft.

Nevertheless, it was important to know the spin rate of the spacecraft
with reasonable precision, in order to be able to carry out precession
maneuvers reliably, and also because the spacecraft's spin affected
the spacecraft's radio signal and the Doppler observable. For this
reason, the IPP was reused as a surrogate star sensor, its images of
the star field providing a reference that could then be used by the
navigation team to compute the actual spin rate of the spacecraft on
the Earth~\cite{AIAA87-0502}. It was during the time when the IPP
instrument was used to compute the rate of spin that a spin anomaly
was detected. The spin-down rate of Pioneer~10 suddenly grew in 1990,
and then eventually returned to its approximate previous value
(Figure~\ref{fig:spin}, left panel.)

Very late in Pioneer~10's mission, when the power on board was no
longer sufficient to operate the IPP instrument, crude estimates of
the spin rate were made using navigational data.

In contrast to the spin behavior of Pioneer~10, the spin rate of
Pioneer~11 continued to increase after the initial jump following the
thruster anomaly. However, a close look at detailed plots of the spin
rate reveal a more intricate picture. It seems that Pioneer~11's spin
rate was actually {\em decreasing} between maneuvers, when the
spacecraft was undisturbed; however, each precession maneuver
increased the spacecraft's spin rate by a notable amount. The rate of
decrease between successive maneuvers was not constant, suggesting
that fuel leaks played a more significant role in Pioneer~11's spin
behavior than in Pioneer~10's.

\subsection{Thermal control subsystem}
\label{sec:heat}

A quick look at the Pioneer spacecraft (Figure~\ref{fig:pio-craft}) is
sufficient to see that the spacecraft's thermal radiation pattern may
be anisotropic: whereas one side is dominated by the high-gain
antenna, the other side contains a thermal louver system designed to
vent excess heat into space. The RTGs are also positioned slightly
behind the HGA, making it likely that at least some of their heat is
reflected in the $-z$ direction.

This leads to the conclusion that anisotropically rejected thermal
radiation cannot be ignored when we evaluate the evolution of the
Pioneer~10 and 11 trajectories, and must be accounted for with as much
precision as possible.

Far from the Sun and planets, the only notable heat sources on board
Pioneer~10 and 11 are internal to the spacecraft. We enumerate the
following heat sources:

\begin{itemize}
\item Waste heat from the radioisotope thermoelectric generators;
\item \vskip -6pt Heat produced by electrical equipment on board;
\item \vskip -6pt Heat from small radioisotope heater units;
\item \vskip -6pt Transient heating from the propulsion system.
\end{itemize}

While strictly speaking it is not a heat source, one must also
consider microwave radiation from the spacecraft's radio transmitter
and HGA, as this radiation also removes what would otherwise appear as
thermal energy from within the spacecraft.

\subsubsection{Waste heat from the RTGs}

The RTGs are the most substantial sources of heat on board. Each of
the four generators on board the spacecraft produced $\sim$~650~W of
power at the time of launch, of which $\sim$~40~W was converted into
electrical energy; the rest was radiated into space as waste heat. The
radiation pattern of the RTGs is determined by the shape and
composition of their radiating fins (see Figure~\ref{fig:RTGDIM}),
but, notwithstanding the possible effects of aging, it is believed to
be fore-aft symmetrical.

\begin{table}[htbp]
  \caption[RTG total power measurements prior to
  launch.]{RTG total power measurements prior to
  launch~\cite{SNAP19}. The reported accuracy of these measurements is
  0.1~W, but values are rounded to the nearest W. RTG numbering
  corresponds to the actual number of units built.}
  \label{tb:rtgheat}
  \centering
  \begin{tabular}{ccllc}
    \toprule
    RTG\# & Spacecraft & Location & Test date & Thermal power (W)\\
    \midrule
    44 & Pioneer~10 & Outboard & Oct 1971 & 649\\
    45 & Pioneer~10 & Inboard  & Nov 1971 & 646\\
    46 & Pioneer~10 & Outboard & Nov 1971 & 647\\
    48 & Pioneer~10 & Inboard  & Dec 1971 & 649\\
    \midrule
    49 & Pioneer~11 & Outboard & Sep 1972 & 649\\
    51 & Pioneer~11 & Inboard  & Oct 1972 & 650\\
    52 & Pioneer~11 & Outboard & Oct 1972 & 649\\
    53 & Pioneer~11 & Inboard  & Oct 1972 & 649\\
    \bottomrule
  \end{tabular}
\end{table}

The amount of power generated by the RTGs is determined by the
radioactive decay of the \super{238}Pu fuel on board. The radioactive
half-life of the fuel is precisely known ($\sim$~87.74 years), and the
total power output of each RTG was measured before launch
(Table~\ref{tb:rtgheat}).

The amount of power removed from each RTG in the form of electrical
energy is telemetered to the ground. (Specifically, the RTG output
voltage and current for each individual RTG is telemetered.) The
remainder of the RTG power is radiated in the form of waste heat.

For each RTG, two temperature measurements are also available. One
sensor, internal to the RTG, measures the temperature at the hot end
of the bimetallic thermocouples. The other sensor measures the
temperature near the root of one of the RTG radiating fins.

\subsubsection{Electrical heat}

Next to the RTGs, the second most significant source of thermal
radiation on board is the set of electrical equipment operating on the
spacecraft. (Figure~\ref{fig:internal} shows the internal arrangement
of components.) From a thermal perspective, nearly all electrical
systems on board perform only one function: they convert electrical
energy into waste heat. The one exception is the spacecraft's radio
transmitter, converting some electrical energy into a directed beam of
radio frequency energy.

Early in the missions, the RTGs produced more electrical power than
what was needed on board. The main bus voltage on board was maintained
at a constant value by a power supply circuit, while excess electrical
power was converted into heat. The shunt regulator controlled a
variable current, which flowed through the regulator itself and an
external radiator plate. The external radiator plate acted as an ohmic
resistor, and the power radiated by it can be easily computed from the
shunt current telemetry (Figure~\ref{fig:PIOPOWER}). The shunt
radiator is mounted such that its thermal radiation is primarily in a
direction that is perpendicular to the spacecraft spin axis.

Some electrical heat is generated outside the spacecraft body: notable
items include electrical heaters for the fuel lines that deliver fuel
to the thruster cluster assemblies along the antenna rim, an
electrical heater for the spacecraft battery, and some scientific
instruments.

Most of the electrical power not radiated away by the spacecraft
antenna or the shunt radiator is converted into heat inside the
spacecraft body. As the spacecraft is very close to a thermal steady
state, all the electrical heat produced internally must be radiated
into space. This thermal radiation is likely to cause a measurable
sunward acceleration, for several reasons. First, the spacecraft body
is heavily insulated by multilayer thermal insulation; however, at the
bottom of the spacecraft body is the thermal louver system designed to
vent excess heat. Even in its closed state, the effective emissivity
of this louver system is much higher than that of the multilayer
insulation, so this is the preferred direction in which heat escapes
the spacecraft. Second, the spacecraft body is situated behind the
high-gain antenna; even if the thermal radiation pattern of the
spacecraft body were isotropic, the back of the HGA would
preferentially reflect a significant portion of this heat in the aft
direction.

Figure~\ref{fig:PIOPOWER} also indicates how the power consumption by
various pieces of equipment on board can be computed from telemetry.

The thermal state of the interior of the spacecraft is characterized
in a redundant manner. In addition to the thermal power of various
components that can be computed from telemetry, there exist numerous
temperature sensors on board, most notably among them six platform
temperature sensors, which are the most likely to measure ambient
temperatures (as opposed to sensors that, say, are designed to measure
the temperature of a specific electrical component, such as a
transistor amplifier.)

Much less is known about heat escaping the interior of the spacecraft
through other routes. The spacecraft's science instruments utilize
various holes and openings in the spacecraft body in order to collect
information from the environment. Design documentation prescribes that
no instrument can lose more heat through the opening than its own
power consumption plus 0.5~W; however, the actually heat loss per
instrument is not known.

\subsubsection{Radioisotope heater units}
\label{sec:rhu}

\epubtkImage{}{%
\begin{figure}[t!]
\centering
\includegraphics[width=0.45\linewidth]{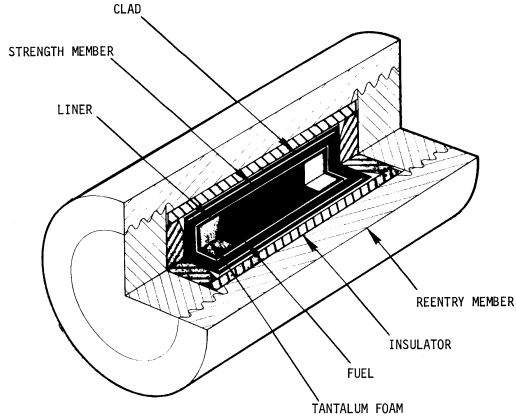}
\caption{1~W radioisotope heater unit (RHU). From~\cite{PC202}.}
\label{fig:rhu}
\end{figure}}

A further source of continuous heat on board is the set of
radioisotope heater units (RHUs) placed at strategic locations to
maintain operating temperatures. These RHUs are capsules containing a
small amount of \super{238}Pu fuel, generating $\sim$~1~W of heat
(Figure~\ref{fig:rhu}).

Each thruster cluster assembly housed three RHUs that prevented the
freezing of thruster valves. One RHU was located at the sun sensor,
while an additional RHU was placed at the magnetometer. (Some reports
suggest that a 12th RHU may also have been on board one or both
spacecraft.)

The thruster cluster assemblies were designed to radiate heat in a
direction perpendicular to the spin axis\epubtkFootnote{Private
  communication in 2008 with Jim Moses, a TRW retiree.}. There are no
known asymmetries of any thermal radiation from the magnetometer
assembly at the end of the magnetometer boom. Therefore, it is
unlikely that thermal radiation from the RHUs in general contributed
much thermal recoil force in the fore-aft direction.

\subsubsection{Waste heat from the propulsion system}

The spacecraft's propulsion system, when operating, produced
significant amounts of heat; as the hydrazine monopropellant underwent
a chemical reaction in the presence of a catalyst in the spacecraft's
thrusters, the thrusters and thruster cluster assemblies warmed up to
several hundred degrees Centigrade. This heat was radiated into space
as the assemblies cooled down to their pre-maneuver temperatures after
a thruster firing event over the course of several hours.

However, the uncertainty in the magnitude of velocity change caused by
the thruster event itself dwarfs any acceleration produced by the
radiation of this residual heat. For this reason, heat from the
propulsion system does not need to be considered when accounting for
thermal recoil forces.

\subsubsection{The energy radiated in the radio beam}
\label{sec:radiobeam}

\epubtkImage{}{%
  \begin{figure}[t!]
    \centerline{
      \includegraphics[width=.5\linewidth]{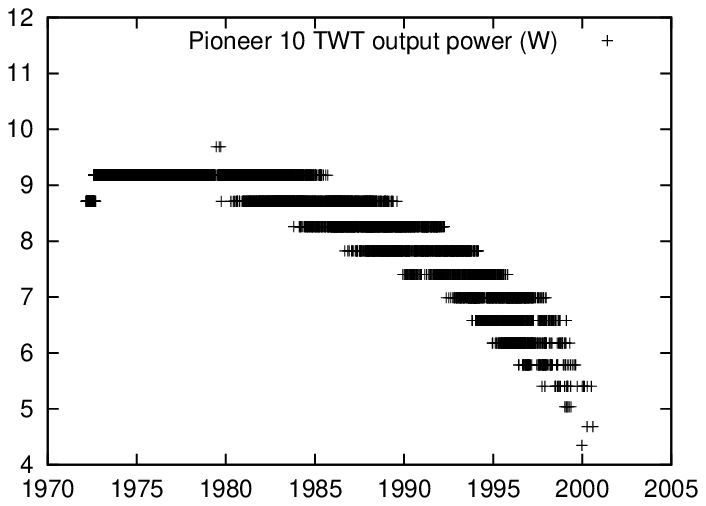}
      \includegraphics[width=.5\linewidth]{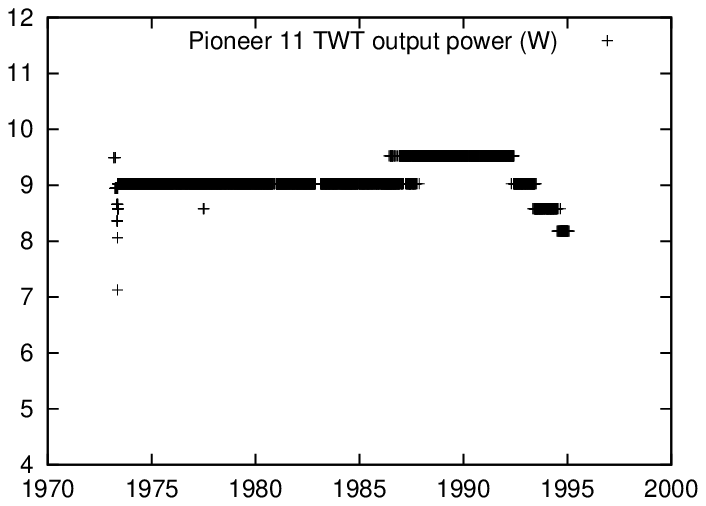}
    }
    \caption{The emitted power (measured in dBm, converted to W) of
    the traveling wave tube transmitter throughout the mission, as
    measured by on-board telemetry. \emph{Left:} Pioneer~10, which
    used TWT~A (telemetry word C\sub{231}). \emph{Right:} Pioneer~11,
    initially using TWT~A but switching to TWT~B (telemetry word
    C\sub{214}) early in its mission.}
    \label{fig:TWT}
\end{figure}}

To complete our discussion of thermal radiation emitted by the
spacecraft, we must also consider the spacecraft's radio beam, for
two reasons. First, any energy radiated by the radio beam is energy
that is not converted into heat inside the spacecraft body. Second,
the radio beam itself produces a recoil force that is similar in
nature to the thermal recoil force.

The nominal power of the radio transmitter is 8~W, and the radio beam
is highly collimated by the high gain antenna. Nevertheless, some loss
occurs around the antenna fringes, and the beam itself also has a
spread. Assuming that $\sim$~10\% of the radio beam emitted by the
feedhorn misses the antenna dish, at an approximate angle of
45$^\circ$, we can calculate an efficiency of $\sim$~0.83 at which the
antenna converts the emitted radio energy into
momentum~\cite{2003PhRvD..67h4021S}.

The actual power of the radio beam may not have been exactly 8~W. The
output of the traveling wave tube oscillator that generated this
microwave energy was measured on board and telemetered to the ground
(Figure~\ref{fig:TWT}). Especially in the case of Pioneer~10, we note
the variability of the TWT power near the end of mission. This
corresponds to the drop in main bus voltage when the power available
on board was no longer sufficient to maintain nominal voltages. It is
unclear, therefore, if the measured drop in output power is an actual
drop or a sensor artifact.

\subsubsection{Thermal measurements in the telemetry}

The thermal history of Pioneer~10 and 11 can be characterized
accurately, and in detail, with the help of the recently recovered
telemetry files and project documentation.

The telemetry data offer a redundant picture of the spacecrafts'
thermal state. On the one hand, the amount of power available on board
can be computed from electrical readings. On the other hand, a number
of temperature sensors on board offer a coarse temperature map of the
spacecraft.

The electrical state of Pioneer~10, when the spacecraft was at 25~AU
from the Sun, is shown in Figure~\ref{fig:PIOPOWER}. While this figure
represents a snapshot of the spacecraft's state at a particular moment
in time, this information is available for the entire duration of both
Pioneer missions.

The thermal state of the spacecraft body can be verified using the
readings of six temperature sensors that were located at various
points on the electronics platform: four in the main compartment, two
in the adjacent compartment that housed science instruments (see
locations 1\,--\,6 in Figure~\ref{fig:sensors}). The temporal
evolution of these temperature readings agrees with expectations:
outlying sensors show consistently lower temperatures, and all
temperatures are dropping steadily, as the distance between the
spacecraft and the Sun increases while the amount of power available
on board decreases (Figures~\ref{fig:23temp} and \ref{fig:24temp}).

\epubtkImage{}{%
\begin{figure}[t!]
\hskip -6pt
\begin{minipage}[b]{.5\linewidth}
\centering \includegraphics[width=\linewidth]{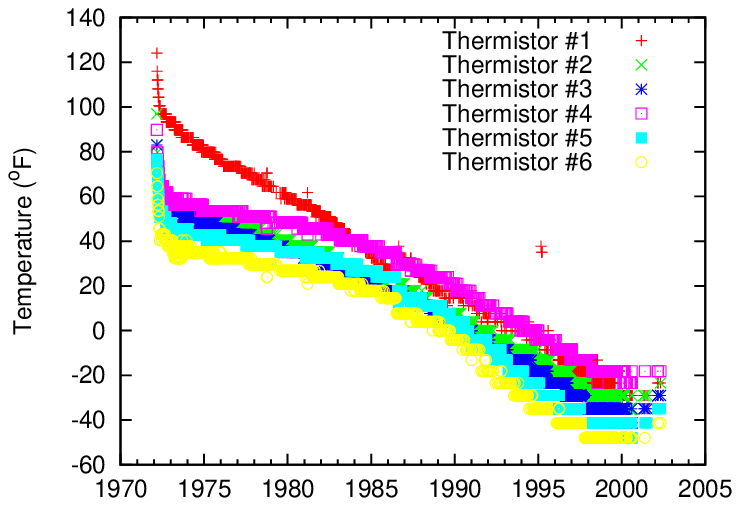}
\end{minipage}
\hskip 0.001\linewidth
\begin{minipage}[b]{.5\linewidth}
\centering \includegraphics[width=\linewidth]{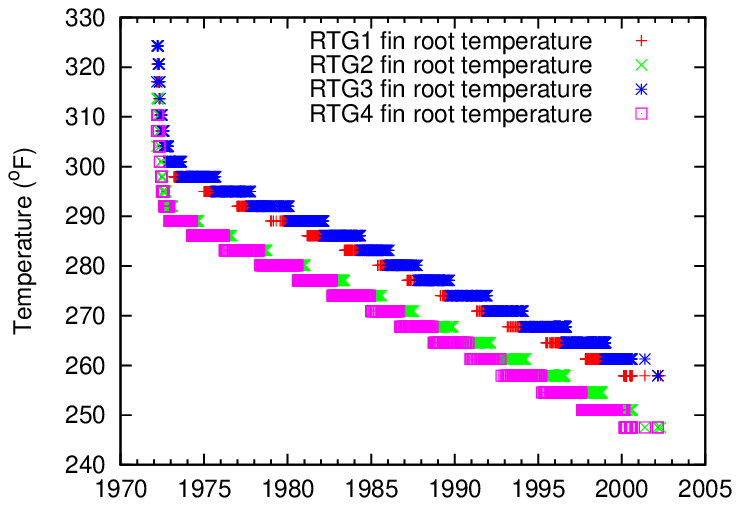}
\end{minipage}
\caption{Platform temperatures (left) and RTG fin root temperatures (right) on board Pioneer~10. Temperatures in \textdegree\,F ([\textdegree\,C]~=~([\textdegree\,F]--32)~\texttimes~5/9).}
\label{fig:23temp}
\end{figure}}

\epubtkImage{}{%
\begin{figure}[t!]
\hskip -6pt
\begin{minipage}[b]{.5\linewidth}
\centering \includegraphics[width=\linewidth]{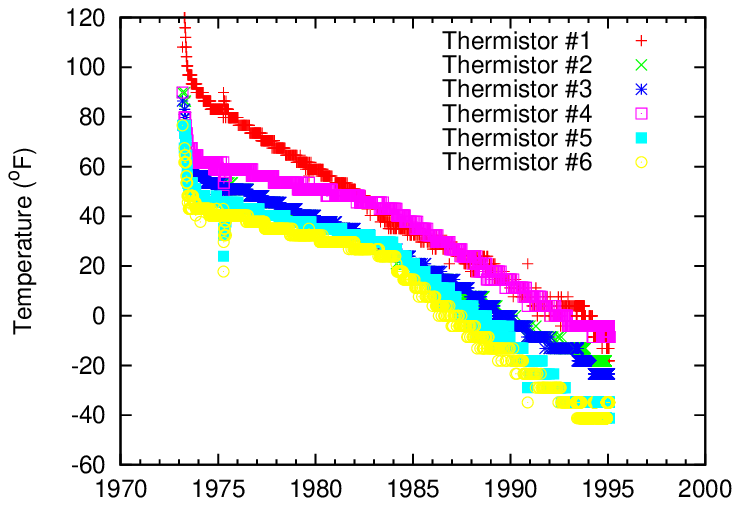}
\end{minipage}
\hskip 0.001\linewidth
\begin{minipage}[b]{.5\linewidth}
\centering \includegraphics[width=\linewidth]{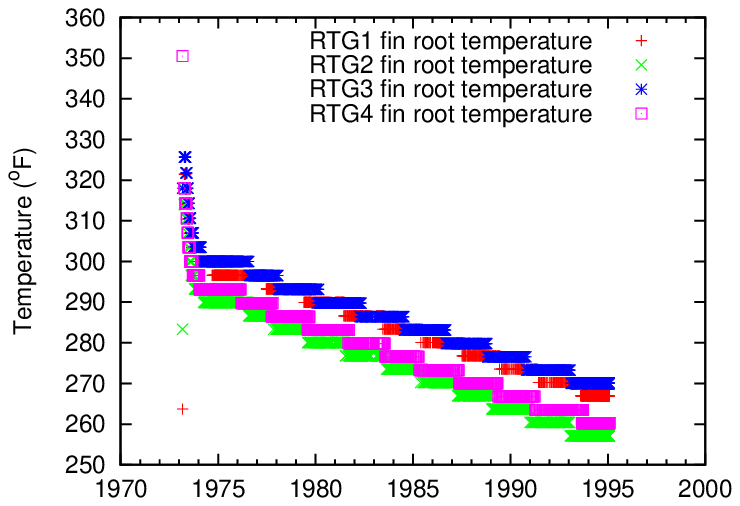}
\end{minipage}
\caption{Platform temperatures (left) and RTG fin root temperatures (right) on board Pioneer~11. Temperatures in \textdegree\,F ([\textdegree\,C]~=~([\textdegree\,F]--32)~\texttimes~5/9).}
\label{fig:24temp}
\end{figure}}

There was a large number of additional temperature sensors on board
(see Appendix~\ref{app:formatCE}). However, these temperature sensors
measured the internal temperatures of on-board equipment and science
instruments. One example is the hot junction temperature sensor inside
the RTGs, measuring the thermocouple temperature at its hot end. These
readings may be of limited use when assessing the overall thermal
state of the spacecraft. Nevertheless, they are also available in the
telemetry.

The thermal behavior of the spacecraft was an important concern when
Pioneer~10 and 11 were designed. Going beyond a general
description~\cite{PC202}, much of the work that was done to ensure
proper thermal behavior is documented in a review document of the
Pioneer~10 and 11 thermal control
subsystem~\cite{TCSDR3}. Specifically, this document provides detailed
information about the thermal properties of components internal to the
spacecraft, and about the anticipated maximum heat losses through
various spacecraft structural and other components.

Detailed information about the SNAP-19 RTGs used on board Pioneer~10
and 11 is also available~\cite{SNAP19}.

\subsubsection{Surface radiometric properties and thermal behavior}

It has been suggested (see, e.g.,~\cite{2003PhRvD..67h4021S}) that the
material properties of these surfaces may have changed over time,
introducing a time-dependent anisotropy in the spacecraft's thermal
properties. For instance, solar bleaching may affect spacecraft
surfaces facing the Sun~\cite{CR1786}, while surfaces facing the
direction of motion may be affected by the impact of interplanetary
dust particles.

Note that through most of their operating lives, the high-gain
antennas of the Pioneer~10 and 11 spacecraft were always pointing in
the approximate direction of the Sun. Therefore, the same side of the
spacecraft: notably, the interior surfaces of their HGAs and one side
of each RTG, were exposed to the effects of solar light, including
ultraviolet radiation, and charged particles. It is not unreasonable
to assume that this continuous exposure may have altered the visible
light and infrared optical properties of these surfaces, introducing
in particular, a fore-aft asymmetry in the thermal radiation pattern
of the RTGs. On the other hand, any such effects would be mitigated by
the fact that the spacecraft were receding from the Sun very rapidly,
and spent most of their operating lives at several AUs or more from
the Sun, receiving only a fraction of the solar radiation that is
received, for instance, by Earth-orbiting satellites of comparable
age.

Another possible effect may have altered the infrared radiometric
properties of spacecraft surfaces facing in the direction of motion
(which, most of the time, would be surfaces facing away from the Sun.)
As the spacecraft travels through the interplanetary medium at speeds
in excess of 12~km/s, impact by charged particles and dust may have
corroded these surfaces. Once again, this may have introduced a
fore-aft anisotropy in the infrared radiometric properties of the
spacecraft, most notably their RTGs.

NASA conducted several experiments to investigate the effect of space
exposure on the thermal and optical properties of various
materials. One investigation, called the Thermal Control Surfaces
(TCS) experiment~\cite{CR-1999-209008}, examined the long-term effects
of exposure of different materials placed on an external palette on
the International Space Station. Although results of this test are
quite important, the near-Earth environment is quite different from
that of deep space. Concerning solar bleaching, although the
spacecraft spent decades in space, most of the time was spent far away
from the Sun, resulting in a low number of ``equivalent Sun hours''
(ESH). The cumulative exposure to solar radiation of the Sun-facing
surfaces of Pioneer~10 and 11 is less than the amount of solar
radiation test surfaces were exposed to during the TCS experiment, in
which test surfaces were also exposed to the atomic oxygen of the
upper atmosphere, which is not a consideration in the case of the
Pioneer spacecraft.

The most pronounced effect on coatings in the deep space environment
is due to exposure to solar ultraviolet radiation~\cite{CR1786}. The
ESH for the Pioneer~10 and 11 spacecraft is approximately 3000 hours,
most of which ($>$~95\%) were accumulated when the spacecraft were
relatively close ($<$~15~AU) to the Sun. The exterior surfaces of the
RTGs were covered by a zirconium coating in a sodium silicate
binder~\cite{SNAP19}. For similar inorganic coatings, the most
pronounced effect of prolonged solar exposure is an increase
($\Delta\alpha\sim 0.05$) in solar absorptance. No data suggest a
noticeable change in infrared emittance.

The effects of exposure to dust and micrometeoroid impacts in the
interplanetary environment do not result in significant optical
damage, defined in terms of changes in solar absorptance or infrared
emittance~\cite{CR1786}. This is confirmed by the Voyager~1 and 2
spacecraft that had unprotected camera lenses that were facing the
direction of motion, yet suffered no observable optical
degradation. This fact suggests that the optical effects of exposure
to the interplanetary medium on the spacecraft are negligible (see
discussion in~\cite{pioprd, Turyshev:2005zk}).

\newpage
\section{Pioneer Data Acquisition and Preparation}
\label{sec:pio-data}

Discussions of radio science experiments with spacecraft in the solar
system require a general knowledge of the sophisticated experimental
techniques used by NASA's Deep Space Network\epubtkFootnote{A
  technical description, with a history and photographs, of NASA's
  Deep Space Network can be found at
  \url{http://deepspace.jpl.nasa.gov/dsn/}. The document describing
  the radio science system can be found at
  \url{http://deepspace.jpl.nasa.gov/dsndocs/810-5/810-5.html}.}
(DSN). Specifically, for the purposes of the Pioneer Doppler data
analysis one needs a general knowledge of the methods and techniques
implemented in the radio science subsystem of the DSN. One also needs
an understanding of how spacecraft telemetry is collected at the DSN,
distributed and used to assess the state of the spacecraft systems at
a particular epoch.

Since its beginnings in 1958, the DSN underwent a number of major
upgrades and additions. This was necessitated by the needs of
particular space missions\epubtkFootnote{The last such upgrade was
  conducted for the Cassini mission when the DSN capabilities were
  extended to cover the Ka radio frequency bandwidth. For more
  information on DSN methods, techniques, and present capabilities,
  see~\cite{Asmar2005, THORNTON2000}.}. The history of the Pioneer~10
and 11 projects is inextricably connected to that of DSN. Due to the
continuing improvements of the entire network, Pioneer~10 was able to
communicate with the project team for over 30 years -- far beyond the
originally planned operational life of 3 years or less.

In this section, we discuss the history of the DSN, its current
status, and describe the DSN antennas and operations in support of
deep space missions. We also review the methods and techniques
implemented in the radio science subsystem of the DSN that is used to
obtain the radio tracking data, from which, after analysis, results
are generated.

\subsection{The Deep Space Network}
\label{sec:ground}

Conceived at the dawn of the space era in the late 1950s, the DSN is a
collection of radio tracking stations positioned around the
globe~\cite{TDA42128, IMBRIALE2002, THORNTON2000}, along with ground
data transfer and data processing systems, designed to maintain
continuous two-way communication, including commands and telemetry,
with a variety of deep space vehicles located throughout the solar
system.

The first DSN complex was constructed near the Jet Propulsion
Laboratory (JPL), in Goldstone, California. It was here that the first
large (26~m) DSN antenna, DSS-11 ``Pioneer'' (named after,
unsurprisingly, the early Pioneer program) was constructed.

The first overseas DSN location was in Woomera, Australia, about
350~km north of Adelaide. DSS-41, another 26~m antenna, was
constructed there in 1960. The second overseas DSN complex was built
near Johannesburg, South Africa. The 26~m antenna of DSS-26 was
constructed there in 1961. Together, Goldstone, Woomera and
Johannesburg made it possible to initiate and maintain continuous
communication with distant spacecraft at any time of the
day\epubtkFootnote{The station location in Johannesburg was a
  compromise; Spain would have been preferred for technical
  reasons. As a result of historical changes in South Africa and
  Spain, construction of a new DSN complex near Madrid began, and
  eventually, the Johannesburg facility was closed. In the meantime,
  there was a strong rationale to construct additional antennas in
  Australia at a more accessible location. The Woomera complex
  remained operational until Pioneer~10 was well on its way towards
  Jupiter, but eventually, it was superseded by a new DSN complex on
  Australian soil, in Tidbinbilla near Canberra.}, marking the
beginning of the DSN.

Today, three locations -- Goldstone, California; Madrid, Spain; and
Canberra, Australia -- form the backbone of the DSN. Presently, each of
the three DSN complexes hosts several tracking stations with different
capabilities and antennas of different sizes~\cite{IMBRIALE2002,
  SP00014227}.

\subsubsection{DSN tracking stations}
\label{sec:tracking-stations}

The primary purpose of the DSN is to maintain two-way communication
with distant spacecraft. The DSN can send, or uplink, command
instructions and data, and it can receive, or downlink, engineering
telemetry and scientific observations from instruments on board these
spacecraft.

The DSN can also be used for precision radio science measurements,
including measurements of a signal's frequency and timing
observations. These capabilities allow the DSN to perform, for
instance, Doppler measurements of line-of-sight velocity, ranging
measurements to suitably equipped spacecraft, occultation experiments
when a spacecraft flies behind a planetary body, and planetary radar
observations~\cite{Asmar2005}.

\epubtkImage{}{%
  \begin{figure}[t!]
    \centerline{\includegraphics[width=\linewidth]{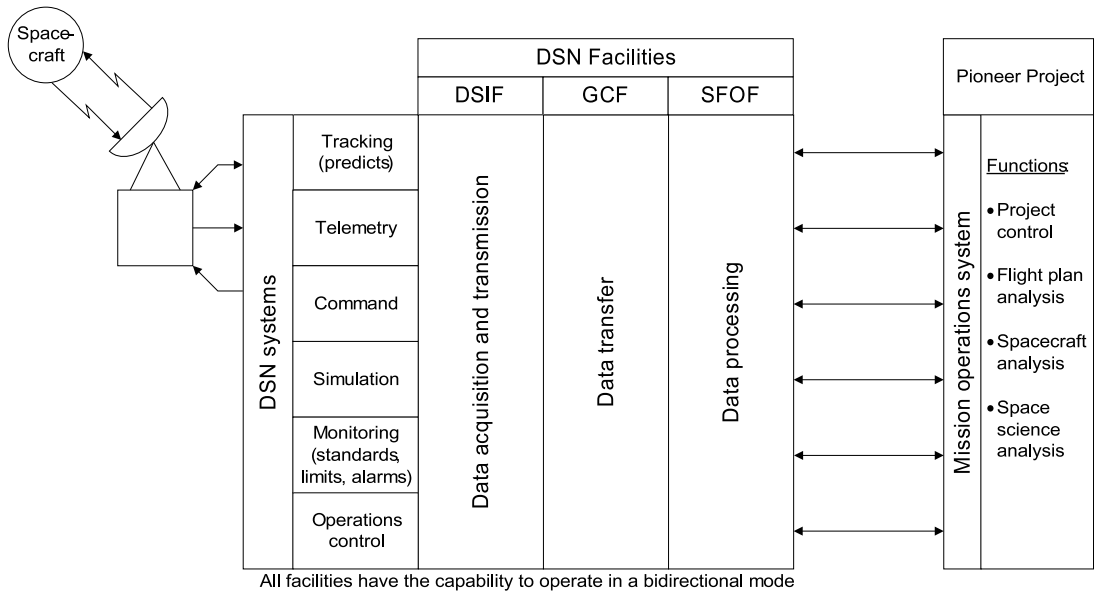}}
    \caption{DSN facilities planned to be used by the Pioneer project
    in 1972~\cite{JPL32-1526-V-B}. DSIF stands for Deep Space
    Instrumentation Facility, GCF is the abbreviation for Ground
    Communications Facility, while SFOF stands for the Space Flight Operations
    Facility.}
    \label{fig:dsnfac}
\end{figure}}

The antennas of the DSN are capable of bidirectional communication
using a variety of frequencies in the L-band (1\,--\,2~GHz), S-band
(2\,--\,4~GHz), X-band (8\,--\,12~GHz) and, more recently, K-band
(12\,--\,40~GHz). Early spacecraft used the L-band for
communication~\cite{SP00014227} but within a few years, S-band
replaced L-band as the preferred frequency band. The Pioneer~10 and 11
spacecraft used S-band transmitters and receivers. X-band and, more
recently, K-band is used on more modern spacecraft.

In addition to the permanent DSN complexes that presently exist at
Goldstone, California, Madrid, Spain, and Canberra, Australia, and the
now defunct complexes in Woomera, Australia and Johannesburg, South
Africa, occasionally, non-DSN facilities (e.g., the Parkes radio
observatory in Australia) were also utilized for communication and
navigation. During the long lifetime of the Pioneer project, nearly
all the large antennas of DSN tracking stations and also some non-DSN
facilities participated in the tracking of the Pioneer~10 and 11
spacecraft at one time or another.

The capabilities of the DSN evolved over the years. By the time of the
launch of Pioneer~10, the DSN was a mature network comprising a number
of 26~m tracking stations at four locations around the globe, a new
64~m tracking station in operation at Goldstone, California, and two
more 64~m tracking stations under construction in Australia and
Spain. The Goldstone facility was connected to the then new Space
Flight Operations Facility (SFOF) built at the JPL via a pair of
16.2~kbps communication links (Figure~\ref{fig:dsnfac}), allowing for
the real-time monitoring of spacecraft telemetry~\cite{SP00014227}.

The most important characteristics of a DSN tracking station can be
described using parameters such as antenna size, antenna (mechanical)
stability, receiver sensitivity, and oscillator stability. These
characteristics determine the accuracy with which the DSN can perform
radio science investigations and, in particular with respect to the
Pioneer~10 and 11 Doppler frequency measurements.

\subsubsection{Antennas of the DSN}
\label{sec:ant}

Large, precision-steerable parabolic dish antennas are the most
recognizable feature of a DSN tracking station. In 1972, several 26~m
antennas were in existence at the Goldstone, Madrid, Johannesburg and
Woomera facilities. Additionally, the 64~m antenna at Goldstone was
already operational, while two 64~m antennas at Madrid and Woomera
were under construction (Figure~\ref{tb:inidsn}).

\begin{table}
  \caption[Pre-launch planned use of Deep Space Instrumentation
  Facilities in support of the Pioneer~10 and 11 projects.]{Pre-launch
  planned use of Deep Space Instrumentation Facilities in support of
  the Pioneer~10 and 11 projects (from~\cite{JPL32-1526-V-B}). Some
  stations only became operational in 1973. Several stations were
  decommissioned during the lifetime of the Pioneer project (DSS-11
  and DSS-62 in 1981, DSS-12 in 1996, and DSS-42 and DSS-61 in 1999).}
  \label{tb:inidsn}
  \centering
  \begin{tabular}{lccl}
    \toprule
    Station & Location & Size & Pioneer support function\\
    \midrule
    DSS-11 ``Pioneer'' & Goldstone & 26~m & Cruise\\
    DSS-12 ``Echo'' & Goldstone & 26~m & Cruise\\
    DSS-14 ``Mars'' & Goldstone & 64~m & Mission enhancement and\\
    ~ & ~ & ~ & Jupiter encounter\\
    \midrule
    DSS-41 & Woomera & 26~m & Cruise\\
    DSS-42 & ``Weemala'' & 26~m & Cruise\\
    DSS-43\super{*} & ``Ballima'' & 64~m & Mission enhancement and\\
    ~ & ~ & ~ & Jupiter encounter\\
    \midrule
    DSS-51 & Johannesburg & 26~m & Launch and cruise\\
    DSS-61 & Robledo & 26~m & Cruise\\
    DSS-62 & Cebreros & 26~m & Cruise\\
    DSS-63\super{*} & Robledo & 64~m & Mission enhancement and\\
    ~ & ~ & ~ & Jupiter encounter\\
    \bottomrule
    \multicolumn{4}{l}{\super{*}After July 1973.}\\
  \end{tabular}
\end{table}

\epubtkImage{}{%
  \begin{figure}[t]
    \centerline{\includegraphics[width=1.01\linewidth]{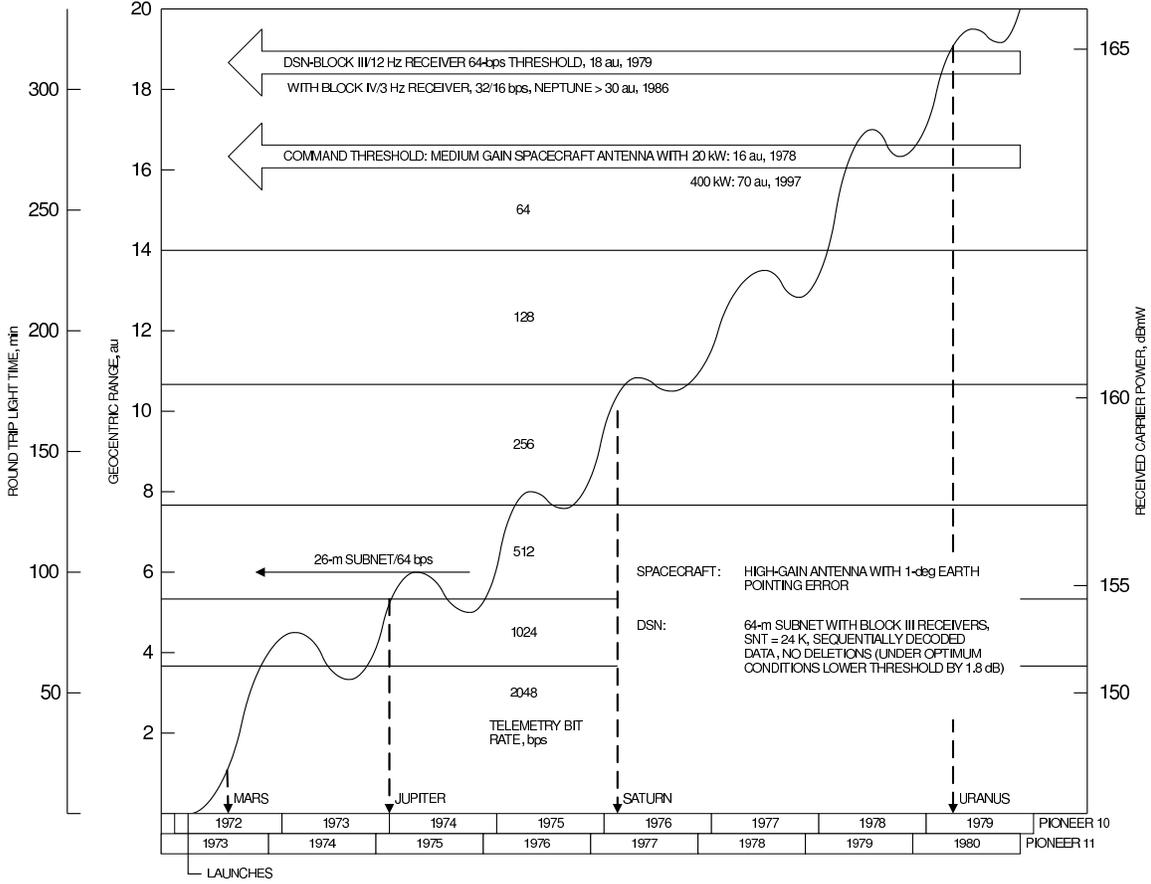}}
    \caption{DSN performance estimate throughout the primary missions
    of Pioneer~10 and 11. Adapted from~\cite{JPL32-1526-XI-D}.}
    \label{fig:dsnperf}
\end{figure}}

To appreciate the impressive performance of the DSN in support of the
Pioneer~10 and 11 missions in deep space, one needs to be able to
evaluate the factors that contribute to the sensitivity of an
antenna. The maximum strength of a signal received by a tracking
station or spacecraft is a function of antenna area and distance
between the transmitting and receiving stations. The gain $G$ of a
parabolic antenna of diameter $d$ at wavelength $\lambda$ is
calculated as
\begin{equation}
G=\pi^2\frac{d^2}{\lambda^2}=4\pi\frac{A}{\lambda^2},
\end{equation}
where $A=\pi d^2/4$ is the antenna area. Space loss due to the
distance $R$ between a transmitting and a receiving antenna is
calculated, in turn, as
\begin{equation}
L_\mathrm{spc}=\frac{\lambda^2}{(4\pi R)^2}.
\end{equation}
The thermal noise power $N$ (also known as the Johnson--Nyquist noise)
of a receiver is calculated from the receiver's system
noise temperature $T_s$ using
\begin{equation}
N=kBT_s,
\label{eq:noise}
\end{equation}
where $B$ is the bandwidth (Hz) and $k=1.381\times
10^{-23}\mathrm{\ JK}^{-1}$ is Boltzmann's constant. The S-band system
noise temperature of a 64~m antenna (DSS-14) was 28~K in
1971~\cite{JPL32-1526-III-C}. Later, with the installation of new
receivers, for an antenna at 60\textdegree\ elevation, the system noise
was reduced to 12.9~K~\cite{JPL32-1526-XIX-O}. The receiver bandwidth
in 1971 was 12~Hz, reduced to 3~Hz for the Block~IV receivers and then
eventually to as low as 0.1~Hz for the Block V receivers (details are
in Section~\ref{sec:dsn-receivers}).

Antenna sensitivity is measured by its signal-to-noise ratio, relating
the power $S$ of the received signal to the noise power $N$ of the
receiver. The strength of the received signal can be calculated
as~\cite{IMBRIALE2002}:
\begin{equation}
\frac{S}{N}=G_TG_RL_\mathrm{spc}\frac{P_T}{N}=\frac{P_TA_TA_R}{\lambda^2R^2N},
\label{eq:snr}
\end{equation}
where $P_T$ is the transmitter power, $A_T$ is the effective area of
the transmitting antenna, and $A_R$ is the effective area of the
receiving antenna.

To calculate the actual signal-to-noise ratio, one must also take into
account additional losses. The effective area of an antenna may be
less than the area of the dish proper, for instance due to
obstructions in front of the antenna surface (e.g., struts,
assemblies, other structural elements). For a 64~m DSN antenna, these
losses amount to 2.7~dB~\cite{JPL32-1526-III-C}, whereas for the
2.74~m parabolic dish antenna of Pioneer~10 and 11, these losses are
about 3.7~dB.

Further (circuit, modulation, pointing) losses must also be
considered. For the downlink from Pioneer~10 and 11 to the ground, the
sum total of these losses is about 10.4~dB.

The signal-to-noise ratio also determines the bit error rate at
various bit rates through the equation
\begin{equation}
\frac{S}{N}=\frac{E_b}{N_0}\frac{f_b}{B},
\label{eq:CN}
\end{equation}
where $E_b$ is energy per bit, $N_0$ is the noise energy per Hz, and
$f_b$ is the bit rate. From this and Equation~(\ref{eq:noise}), the bit
error rate $p_e$ can be computed as~\cite{THORNTON2000}:
\begin{equation}
p_e=0.5~\mathrm{erfc}\bigg(\sqrt{\frac{E_b}{N_0}}\bigg)=0.5~\mathrm{erfc}\bigg(\sqrt{\frac{S}{kf_bT_s}}\bigg).
\end{equation}

The discussion above allows one to present a typical downlink
communications power budget for Pioneer~10, which is given in
Table~\ref{tb:commbudget} (adapted from~\cite{JPL32-1526-IV-D}).

\begin{table}[h]
  \caption[Pioneer~10 Jupiter downlink carrier power budget for
  tracking system.]{Pioneer~10 Jupiter downlink carrier power budget for
  tracking system~\cite{JPL32-1526-IV-D}.}
  \label{tb:commbudget}
  \centering
  \begin{tabular}{lrlrl}
    \toprule
    Link & \multicolumn{2}{c}{Carrier} & \multicolumn{2}{c}{Subcarrier}\\
    \midrule
    Transmitter             &    38.9 & dBm &    38.9 & dBm\\
    Modulation loss         &   --7.3 & dB  &  --27.8 & dB\\
    Circuit loss            &   --2.0 & dB  &   --2.0 & dB\\
    2.75~m antenna gain     &   +32.7 & dB  &   +32.7 & dB\\
    Space loss              & --278.5 & dB  & --278.5 & dB\\
    Pointing loss           &   --1.0 & dB  &    -1.0 & dB\\
    64~m atenna gain        &   +61.0 & dB  &   +61.0 & dB\\
    \midrule
    Received signal level   & --156.2 & dBm & --176.7 & dBm\\
    Less: S/N Ratio         &  --17.1 & dB  &   --6.9 & dB\\
    Detection efficiency    &     N/A &   ~ &   --0.5 & dB\\
    Noise level at receiver & --173.3 & dBm & --184.1 & dBm\\
    Bit rate                &     N/A &   ~ &   512   & bits\\
    Error rate              &     N/A &   ~ & 10\super{-3} & ~\\
    \bottomrule
  \end{tabular}
\end{table}

Using the facilities in existence in 1973, the DSN would have been
able to track Pioneer~10 and 11 up to a geocentric distance of
$\sim$~22~AU, but not beyond. However, due to numerous improvements of
the DSN it was in fact possible to track Pioneer~10 all the way to
over $\sim$~83~AU from the Earth. Not the least of these improvements
was an increase in antenna size, when the DSN's 64~m antennas were
enlarged to 70~m, and many of the 26~m antennas (notably, DSS-12,
DSS-42, and DSS-61 from Table~\ref{tb:inidsn}) were enlarged to 34~m.

In addition to the stations used initially (see Table~\ref{tb:inidsn})
for communication with Pioneer~10 and 11, over the years many other
stations were utilized, which are listed in Table~\ref{tb:moredsn}.

\begin{table}[h]
  \caption{Additional DSN stations in operation during the Pioneer
  mission lifetime that may have been used for tracking Pioneer~10 and
  11.}
  \label{tb:moredsn}
  \centering
  \begin{tabular}{lcc}
    \toprule
    Station & Location & Size\\
    \midrule
    DSS-13 ``Venus'' & Goldstone & 26/34~m\\
    DSS-15 ``Uranus'' & Goldstone & 34~m\\
    DSS-16 ``Apollo'' & Goldstone & 26~m\\
    DSS-17 & Goldstone & 9~m\\
    DSS-23 & Goldstone & 34~m\\
    DSS-24 & Goldstone & 34~m\\
    DSS-25 & Goldstone & 34~m\\
    DSS-26 & Goldstone & 34~m\\
    DSS-27 & Goldstone & 34~m\\
    DSS-28 & Goldstone & 34~m\\
    \midrule
    DSS-33 & Tidbinbilla & 11~m\\
    DSS-34 & Tidbinbilla & 34~m\\
    DSS-44 & Honeysuckle Creek & 26~m\\
    DSS-45 & Tidbinbilla & 34~m\\
    DSS-46 & Tidbinbilla & 26~m\\
    DSS-49 & Parkes & 64~m\\
    \midrule
    DSS-53 & Madrid & 11~m\\
    DSS-54 & Madrid & 26~m\\
    DSS-55 & Madrid & 34~m\\
    DSS-65 & Madrid & 34~m\\
    DSS-66 & Madrid & 26~m\\
    \bottomrule
  \end{tabular}
\end{table}

In order to perform precision tracking, the location of these radio
stations must be known to high accuracy in the same coordinate frame
(e.g., a solar system barycentric frame) in which spacecraft orbits
are calculated.

This task is accomplished in two stages. First, station locations are
given relative to a a geocentric coordinate system, such as the
International Terrestrial Reference Frame (ITRF). Second, a conversion
from the geocentric coordinate system to the appropriate solar system
barycentric reference frame is performed.

For operating DSN stations, station location information is readily
available, e.g., from NASA's Navigation and Ancillary Information
Facility (NAIF\epubtkFootnote{See
  \url{http://naif.jpl.nasa.gov/naif/data\_generic.html}.}). Such
information usually consists of station coordinates at a given epoch,
and station drift (e.g., as a result of continental drift.) Higher
precision station data (e.g., taking into account the effects of tide)
is also available, but such precision is not required for the tracking
of Pioneer~10 and 11.

Station data is harder to come by for stations that have been
decommissioned or moved. Some decommissioned stations are listed in
Table~\ref{tb:DSN}\epubtkFootnote{Information on the location of DSN
  stations that are no longer in service was provided by
  W.M.~Folkner of JPL via private communication. Relocation date for
  DSS-12 is from~\cite{SP00014227}.}.

\begin{table}[h]
  \caption{Station location information for a set of decommissioned
  DSN stations. ITRF93 Cartesian coordinates (in km) and velocities
  (north, east, vertical, in cm/year) are shown. For DSS-12 and
  DSS-61, station information is for dates prior to July 1, 1978 and
  August 9, 1979, respectively.}
  \label{tb:DSN}
  \centering
  \begin{tabular}{lrrrrrr}
    \toprule
    ID & \it{X} & \it{Y} & \it{Z} & $v_N$ & $v_E$ & $v_\mathrm{vert}$\\
    \midrule
    DSS-11 & --2351.429165 &  --4645.078818~ &   3673.764161 & --0.45 & --1.90 & --0.03\\
    DSS-12 & --2350.442882 &  --4651.978543~ &   3665.629171 & --0.45 & --1.90 & --0.03\\
    DSS-41 & --3978.720526 &    3724.850894~ & --3302.171112 &   4.74 &   2.08 & --0.12\\
    DSS-44 & --4451.074302 &    2676.823855~ & --3691.346779 &   4.74 &   2.08 & --0.12\\
    DSS-51 &   5085.442790 &    2668.263555~ & --2768.696907 &   2.02 &   1.74 &   0.04\\
    DSS-61 &   4849.242766 &   --360.277773~ &   4114.882600 &   1.95 &   2.34 &   0.12\\
    DSS-62 &   4846.700295 &   --370.1962103 &   4116.905713 &   1.95 &   2.34 &   0.12\\
    \bottomrule
  \end{tabular}
\end{table}

\subsubsection{DSN receivers}
\label{sec:dsn-receivers}

Antenna size may have been the visually most apparent change at a DSN
complex, but it was upgrades of DSN receiver hardware that resulted in
a really significant improvement in a tracking station's capabilities.

Receivers of the DSN have been improved on a continuous basis during
the lifetime of Pioneer~10 and 11. These improvements were a result of
other mission requirements, but the Pioneer project benefited: far
beyond the predicted range of 22~AU, it was possible to maintain
communication with Pioneer~10 when it was at an incredible 83~AU from
the Sun. (The same DSN complex is now used to communicate with
Voyager~I spacecraft from heliocentric distances of over 110~AU.)

The receiver can contribute to a tracking station's sensitivity in
three different ways. First, the bandwidth of the receiver (loop
bandwidth) can be reduced, increasing the signal-to-noise ratio. Two
additional improvements, lowering the system noise temperature and
eliminating other sources of signal attenuation, not only increase the
signal-to-noise ratio but also decrease the bit error rate.

The ``Block~III'' DSN receivers in use at the time of the launch of
Pioneer~10 and 11 had an S-band system noise temperature of 28~K, and
a receiver loop bandwidth of 12~Hz. These receivers were replaced in
1983\,--\,1985 by ``Block~IV'' receivers in which the S-band system noise
was reduced to 14.5~K in receive only mode, and the receiver loop
bandwidth was reduced to 3~Hz. Further improvements came with the
all-digital ``Block~V'' receivers (also known as the Advanced
Receivers~\cite{JPL42-100-K}) installed in the early 1990s that had,
under ideal circumstances, an S-band system noise of only 12.9~K, and
a receiver loop bandwidth of 0.1~Hz.

Together with the enlarged 70~m antenna, these improved receivers made
it possible to receive the signal of Pioneer~10 at 83~AU with a bit
error rate of $\sim$~1\%. This error rate was reduced further by the
convolutional code in use by Pioneer~10 and 11, which amounted to a
3.8~dB improvement in the signal-to-noise ratio, resulting in an error
rate of about one in 10\super{4} bits.

During planetary encounters, the rapidly changing velocity of the
spacecraft can result in a Doppler shift of its received
frequency. During Pioneer~11's close encounter with Jupiter, this rate
could be as high as $df/dt=0.52\mathrm{\ Hz/s}$~\cite{JPL42-27-Y}. Such a rapid
change in received frequency can exceed the capabilities of the DSN
closed loop receivers to remain ``in lock''. To maintain continuous
communication with such spacecraft, a ``ramping'' technique was
implemented that allowed the tuned frequency of the DSN receiver to
follow closely the predicted frequency of the spacecraft's
transmission~\cite{JPL32-1526-XIX-I}. This ramping technique was used
successfully during the Pioneer~10 and 11 planetary encounters. Later
it became routine operating procedure for the DSN. (As late as
Pioneer~11's 1979 encounter with Saturn, ramp frequencies at DSS-62
were tuned manually by relays of operators, as an automatic tuning
capability was not yet available~\cite{SP00014227}).

\subsubsection{DSN transmitters}

To maintain continuous communication with distant spacecraft, the DSN
must be able to transmit a signal to the spacecraft. The 70~m stations
of the DSN are equipped with transmitters with a maximum transmitting
power of 400~kW in the S-band\epubtkFootnote{The full 400-kW power of
  the DSN was used on March 3\,--\,5 2006, when the DSN attempted to
  contact Pioneer~10 for the last time. Unfortunately, the lack of
  sufficient power resources on-board the spacecraft prevented a
  successful two-way communication~\cite{2006CaJPh..84.1063T}.}. This
is sufficient power to maintain continuous communication with distant
spacecraft even when the spacecraft are not oriented favorably
relative to the Earth, and must use a low-gain omnidirectional antenna
to receive ground commands.

Just like its receivers, the DSN's transmitters are also capable of
ramping. Ramped transmissions are necessary during planetary
encounters in order to ensure that the spacecraft's closed loop
receiver remains in lock even as its line-of-sight velocity relative
to the Earth is changing rapidly\epubtkFootnote{Ramped transmissions
  were also used to compensate for a partial failure of the on-board
  receiver of Voyager~2.}.

\subsubsection{Data communication}

Results of Pioneer radio observations were packaged in data files (see
Figure~\ref{fig:pio-data-format-flow}). Initially, these data files
were transcribed to magnetic tape and delivered physically to the
project site where they were processed. The present-day DSN uses
electronic ground communication networks for this purpose.

\subsection{Acquisition of radiometric Doppler data}
\label{sec:data-acquisition}

Radiometric observations are performed by the DSN stations located at
several DSN sites around the world~\cite{MG2005, Moyer-2003}. Routine
radiometric tracking and navigation of the Pioneer~10 and 11
spacecraft was performed using Doppler observations. The Doppler
extraction process is schematically depicted in
Figure~\ref{fig:doppler}. A reference signal of a known frequency
(usually a frequency close to that of the original transmission) is
mixed with the received signal. The resulting beat frequency is than
measured by a frequency counter. The count over a given time interval
is recorded as the Doppler observation.

\epubtkImage{}{%
  \begin{figure}
    \centerline{\includegraphics[width=0.95\linewidth]{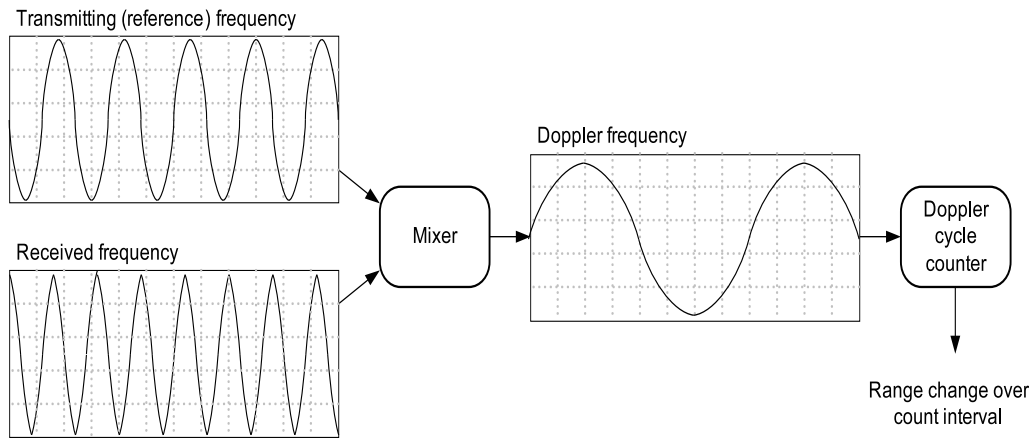}}
    \caption{The Doppler extraction process. Adapted
    from~\cite{THORNTON2000}.}
    \label{fig:doppler}
\end{figure}}

Receiving stations of the DSN are equipped with ultra-stable
oscillators, allowing very precise measurements of the frequency of a
received signal. One way to accomplish this measurement is by
comparing the frequency of the received signal against a reference
signal of known frequency, and count the number of cycles of the
resulting beat frequency for a set period of time. This Doppler count
is then stored in a file for later analysis.

The operating principles of the DSN evolved over time, and modern
receivers may not utilize a beat frequency. However, all Pioneer data
was stored using a format that incorporates an actual or simulated
reference frequency.

Below we discuss the way Doppler observables are formed at the
radio-science subsystem of the DSN. This description applies primarily
to Block~III and Block~IV receivers, which were used for Pioneer
radiometric observations throughout most of the Pioneer~10 and 11
missions.

\epubtkImage{}{%
  \begin{figure}[t!]
    \centerline{\includegraphics[width=\linewidth]{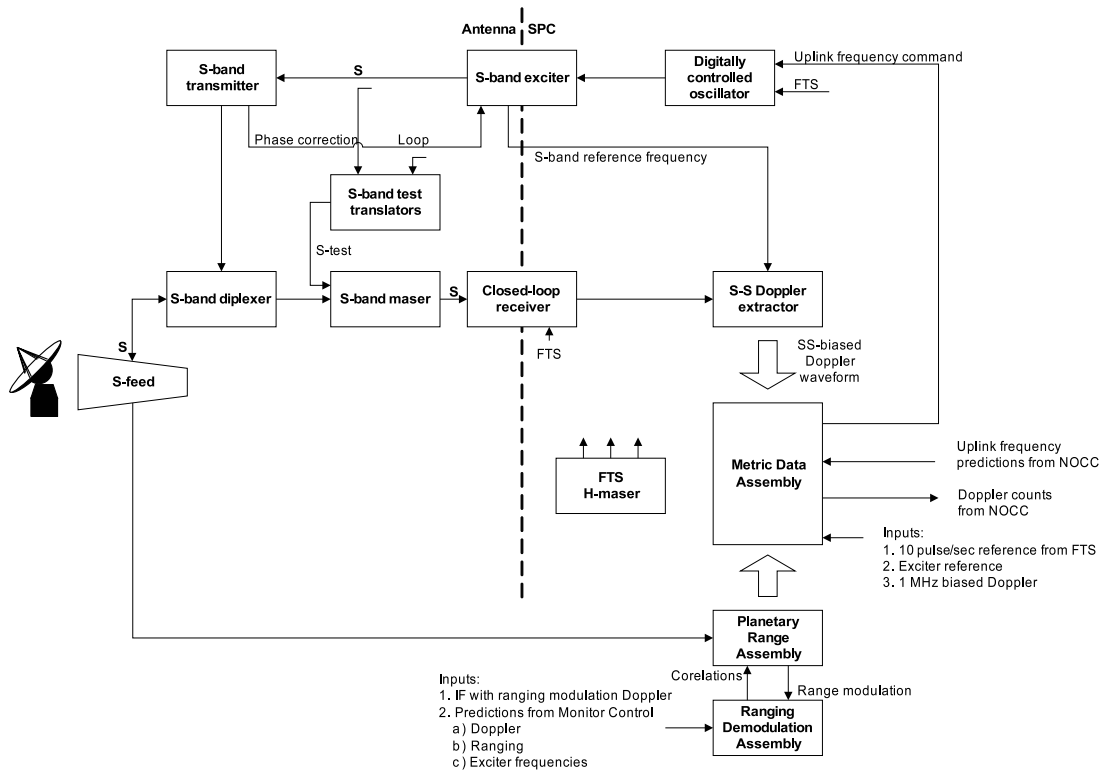}}
    \caption{Block diagram of the DSN baseline configuration as used
    for radio Doppler tracking of the Pioneer~10 and 11
    spacecraft. Adapted from~\cite{pioprd,616-55}. (IF stands for
    Intermediate Frequency.)}
    \label{fig:dsn_block}
\end{figure}}

\subsubsection[DSN frequency and timing system]{DSN frequency and
    timing system\epubtkFootnote{We thank Craig G.\ Markwardt for
    helping us to improve this section significantly.}}

Present day radiometric tracking of spacecraft requires highly
accurate timing and frequency standards at tracking stations. For a
two-way Doppler experiment, for instance, that involves transmitting a
signal to a distant spacecraft and then receiving a response perhaps
several hours later, long-term stability of tracking station
oscillators is essential. Furthermore, if more than one tracking
station is involved in an observation (e.g., ``three-way'' Doppler in
which case one station is used to uplink a signal to the spacecraft,
and another station receives the spacecraft's response), frequency and
clock synchronization between the participating tracking stations must
be very precise~\cite{Asmar2005}.

Originally, the DSN used crystal oscillators. In the 1960s, these were
replaced by rubidium or cesium oscillators that offered much improved
frequency stability, and made precision radio-tracking of distant
spacecraft possible. Rubidium and cesium oscillators offered a
frequency stability of one part in $\sim$~10\super{12} and one part in
$\sim$~10\super{13}, respectively~\cite{SP00014227}, over typical Doppler
counting intervals of 10\super{2}\,--\,10\super{3} seconds. (This is
in agreement with the expected Allan deviation for the S-band signals.)

A further improvement occurred in the 1970s, when the rubidium and
cesium oscillators were in turn replaced by hydrogen masers, which
offered another order of magnitude improvement in frequency
stability~\cite{SP00014227}. Today, the DSN's frequency and timing
system (FTS) is the source for the high accuracy mentioned above (see
Figure~\ref{fig:dsn_block}). At its center is a hydrogen maser that
produces a precise and stable reference frequency. These devices have
Allan deviations (see Section~\ref{sec:clocks}) of approximately
3~\texttimes~10\super{-15} to 1~\texttimes~10\super{-15} for
integration times of 10\super{2} to 10\super{3} seconds,
respectively~\cite{Asmar2005, THORNTON2000}. These masers are good
enough that the quality of Doppler-measurement data is limited by
thermal or plasma noise, and not by the inherent instability of the
frequency references.

However, for two-way Doppler analysis, the relevant quantity is the
stability of the station's frequency standard during the round-trip
light travel time. During the 1980s, hydrogen maser frequency
standards were required to have Allan deviations of less than
10\super{-14} over time scales of
3\,--\,12~hours~\cite{1981ptti.nasa..591C, 1985tdar.rept..113F}. This
corresponds to a Doppler frequency noise of 2~\texttimes~10\super{-5}~Hz.

For three-way Doppler analysis, where the transmitting and receiving
stations are different, the frequency offset between stations is the
relevant quantity.  During the 1980s, station frequencies were
controlled to be the same, to a fractional error of
10\super{-12}~\cite{1983STIN...8336302R, 1980DSNPR..56....7S}. This
corresponds to a Doppler frequency bias of up to 2~\texttimes~10\super{-3}~Hz
between station pairs.  Station keepers did maintain frequency offset
knowledge to a tighter level, but for the most part this knowledge is
not available to analysts today.

Three-way Doppler analysis is weakly sensitive to clock offsets
between stations. By 1968, the operational technique for time
synchronization was the ``Moon Bounce Time Synch'' technique, in
which a precision-timed X-band signal from DSS-13 (which served as
master timekeeper) was transmitted to overseas stations by way of a
lunar reflection, achieving a clock accuracy of 5~$\mu$s between
stations~\cite{SP00014227}. Such timing offsets would produce Doppler
errors of less than 10\super{-6}~Hz. Later, DSN stations were
synchronized utilizing the Global Positioning Satellite (GPS) network,
achieving a synchronization accuracy of 1~$\mu$s or
better~\cite{THORNTON2000}.

\subsubsection{The digitally controlled oscillator and exciter}

Using the highly stable output from the FTS, the digitally controlled
oscillator (DCO), through digitally controlled frequency multipliers,
generates the Track Synthesizer Frequency (TSF) of $\sim$~22~MHz. This
is then sent to the Exciter Assembly. The Exciter Assembly multiplies
the TSF by 96 to produce the S-band carrier signal at
$\sim$~2.2~GHz. The signal power is amplified by Traveling Wave Tubes
(TWT) for transmission. If ranging data are required, the Exciter
Assembly adds the ranging modulation to the carrier. The DSN tracking
system has undergone many upgrades during the 34 years of tracking
Pioneer~10. During this period internal frequencies have changed (see
Section~\ref{sec:tracking-stations}).

This S-band frequency is sent to the antenna where it is amplified and
transmitted to the spacecraft. The onboard receiver tracks the up-link
carrier using a phase lock loop. To ensure that the reception signal
does not interfere with the transmission, the spacecraft (e.g.,
Pioneer) has a turnaround transponder with a ratio of 240/221 in the
S-band. The spacecraft transmitter's local oscillator is phase locked
to the up-link carrier. It multiplies the received frequency by the
above ratio and then re-transmits the signal to Earth.

\subsubsection{Receiver and Doppler extractor}

When the signal reaches the ground, the receiver locks on to the
signal and tunes the Voltage Control Oscillator (VCO) to null out the
phase error. The signal is sent to the Doppler Extractor. At the
Doppler Extractor the current transmitter signal from the Exciter is
multiplied by 240/221 (or 880/241 in the X-band) and a bias of 1~MHz
for S-band or 5~MHz for X-band is added to the Doppler. The Doppler
data is no longer modulated at S-band but has been reduced as a
consequence of the bias to an intermediate frequency of 1 or 5~MHz.

The transmitter frequency of the DSN is a function of time, due to
ramping and other scheduled frequency changes. When a two-way or
three-way (see Section~\ref{sec:pio-doppler}) Doppler measurement is
performed, it is necessary to know the precise frequency at which the
uplink signal was transmitted. This, in turn, requires knowledge of
the exact light travel time to and from the spacecraft, which is
available only when the position of the spacecraft is determined with
precision. For this reason, DSN transmitter frequencies are recorded
separately so that they can be accounted for in the orbit
determination programs that we discuss in Section~\ref{sec:anomaly}.

\subsubsection{Metric data assembly}

The intermediate frequency (IF) of 1 or 5~MHz with a Doppler
modulation is sent to the Metric Data Assembly (MDA). The MDA consists
of computers and Doppler counters where continuous count Doppler data
are generated. From the FTS a 10 pulse per second signal is also sent
to the MDA for timing. At the MDA, the IF and the resulting Doppler
pulses are counted at a rate of 10 pulses per second. At each tenth of
a second, the number of Doppler pulses is recorded. A second counter
begins at the instant the first counter stops. The result is
continuously-counted Doppler data. (The Doppler data is a biased
Doppler of 1~MHz, the bias later being removed by the analyst to
obtain the true Doppler counts.) The range data (if present) together
with the Doppler data are sent separately to the Ranging Demodulation
Assembly. The accompanying Doppler data is used to ``rate-aid'' (i.e.,
to ``freeze'' the range signal) for demodulation and cross
correlation.

\subsubsection{Radiometric Doppler data}
\label{sec:pio-doppler}

Doppler data is the measure of the cumulative number of cycles of a
spacecraft's carrier frequency received during a user-specified count
interval. The exact precision to which these measurements can be
carried out is a function of the received signal strength and station
electronics, but it is a small fraction of a cycle. Raw Doppler data
is generated at the tracking station and delivered via a DSN interface
to customers.

When the measured signal originates on the spacecraft, the resulting
Doppler data is called one-way or F1 data. In order for such data to
be useful for precision navigation, the spacecraft must be equipped
with a precision oscillator on board. The Pioneer~10 and 11 spacecraft
had no such oscillator. Therefore, even though a notable amount of F1
Doppler data was collected from these spacecraft, these data are not
usable for precision orbit determination.

An alternative to one-way Doppler involves a signal generated by a
transmitter on the Earth, which is received and then returned by the
spacecraft. The Pioneer~10 and 11 spacecraft were equipped with a
radio communication subsystem that had the capability to operate in
``coherent'' mode, a mode of operation in which the return signal from
the spacecraft is phase-locked to a signal received by the spacecraft
from the Earth. In this mode of operation, the precision of the
frequency measurement is not limited by the stability of equipment on
board the spacecraft, only by the frequency stability of ground-based
DSN stations.

When the signal is transmitted from, and received by, the same
station, the measurement is referred to as a two-way or F2 Doppler
measurement; if the transmitting and receiving stations differ, the
measurement is a three-way (F3) Doppler measurement.

Knowing the frequency of a signal received at a precise time at a
precisely known location is only half the story in the case of two-way
or three-way Doppler data: information must also be known about the
time and location of transmission and the frequency of the transmitted
signal.

The frequency of the transmitter, or the frequency of the receiver's
reference oscillator may not be fixed. In order to achieve better
quality communication with spacecraft the velocity of which varies
with respect to ground stations, a technique called ramping has been
implemented at the DSN. When a frequency is ramped, it is varied
linearly starting with a known initial frequency, at a known rate of
frequency change over unit time.

Thus, a Doppler data point is completely characterized by the
following:

\begin{itemize}
\item The location of the receiving station,
\item \vskip -6pt The time of reception,
\item \vskip -6pt The reference frequency of the receiver,
\item \vskip -6pt Receiving station ramp information,
\item \vskip -6pt The length of the Doppler count interval,
\item \vskip -6pt The beat frequency (Doppler) count,
\item \vskip -6pt Transmission frequency,
\end{itemize}

and, for two- and three-way signals only,

\begin{itemize}
\item The location of the transmitting station,
\item \vskip -6pt Transmitting station amp information.
\end{itemize}


\nocite{JPL32-1526-II-C,
JPL32-1526-III-C, JPL32-1526-IV-D, JPL32-1526-V-B, JPL32-1526-V-W, JPL32-1526-VI-D, JPL32-1526-VII-B,
JPL32-1526-VIII-C, JPL32-1526-IX-E, JPL32-1526-IX-AA,
JPL32-1526-IX-FF, JPL32-1526-X-C, JPL32-1526-X-G,
JPL32-1526-X-KK, JPL32-1526-XI-C, JPL32-1526-XI-D,
JPL32-1526-XV-D, JPL32-1526-XV-E, JPL32-1526-XVI-E,
JPL32-1526-XVI-BB, JPL32-1526-XVII-D, JPL32-1526-XVIII-C,
JPL32-1526-XIX-D, JPL32-1526-XIX-I, JPL42-20-CC, JPL42-21-D,
JPL42-21-G, JPL42-22-C, JPL42-22-H, JPL42-24-D, JPL42-26-D,
JPL42-27-Y, JPL42-30-E, JPL42-31-J, JPL42-33-D, JPL42-35-D, JPL42-35-P,
JPL42-36-R, JPL42-37-E, JPL42-39-C, JPL42-41-F, JPL42-44-E,
JPL42-45-G, JPL42-46-E, JPL42-47-D, JPL42-52-B, JPL42-53-C,
JPL42-54-D, JPL42-55-C, JPL42-58-I, JPL42-61-C, JPL42-63-D,
JPL42-65-F, JPL42-91-O, JPL42-93-F, JPL42-106-W, JPL42-111-O,
JPL42-61-B}

Below we discuss the radiometric Doppler data formats that were used
to support navigation of the Pioneer~10 and 11 spacecraft.

\subsubsection{Pioneer Doppler data formats}

The Pioneer radiometric data was received by the DSN in
``closed-loop'' mode, i.e., it was tracked with phase lock loop
hardware. (``Open loop'' data is recorded to tape but not tracked by
phase lock loop hardware.) There are basically two types of data:
Doppler (frequency) and range (time of flight), recorded at the
tracking sites of the DSN as a function of UT Ground Received
Time~\cite{DSN810-5}. During their missions, the raw radiometric
tracking data from Pioneer~10 and 11 were received originally in the
form of Intermediate Data Record (IDR) tapes, which were then
processed into special binary files called Archival Tracking Data
Files (ATDF, format TRK-2-25~\cite{TRK-2-25}), containing Doppler data
from the standard DSN tracking receivers
(Figure~\ref{fig:pio-data-format-flow})\epubtkFootnote{After the
  installation of Block~V receivers, there is high time resolution
  Doppler data available, called ``open loop'' data. This data is
  stored in files called ODRs (Original Data Records), which became
  available only at the very end of the Pioneer missions and were not
  used as a standard format for navigating these missions in deep
  space~\cite{JPL42-93-F}.}. Note that the ``closed-loop'' data
constitutes the ATDFs that were used
in~\cite{pioprd}. Figure~\ref{fig:pio-data-format-flow} shows a typical
tracking configuration for a Pioneer-class mission and corresponding
data format flow.

\epubtkImage{}{%
  \begin{figure}[t!]
    \centerline{\includegraphics[width=\linewidth]{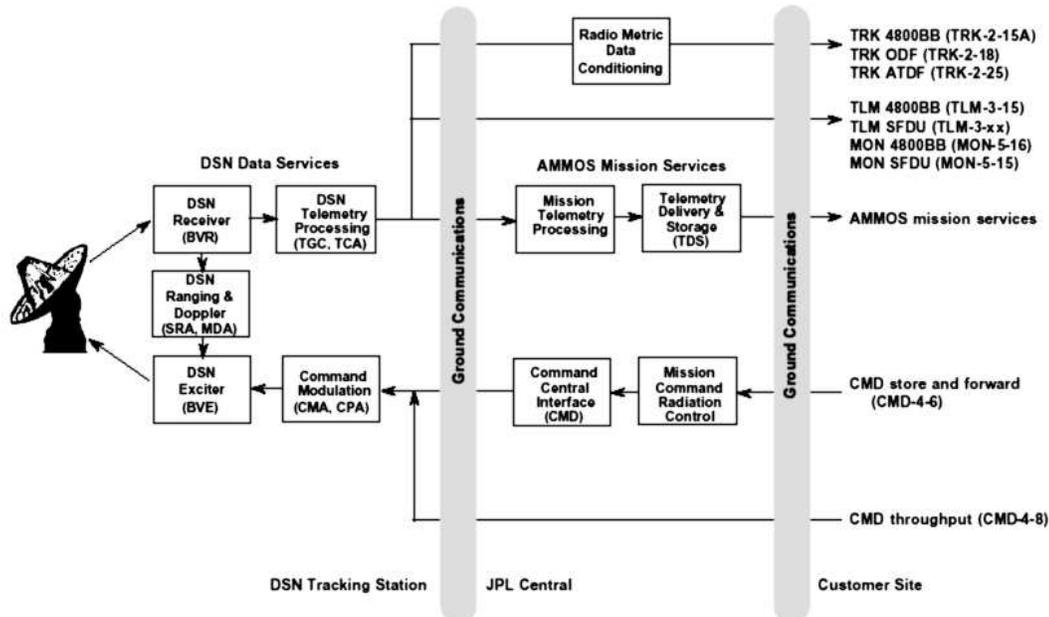}}
    \caption{Typical tracking configuration for a Pioneer-class
    mission and corresponding data format flow~\cite{MDR2005}.}
    \label{fig:pio-data-format-flow}
\end{figure}}

ATDFs are files of radiometric data produced by the Network Operations
Control Center (NOCC) Navigation Subsystem (NAV) (see
Figure~\ref{fig:dsn_block}). They are derived from Intermediate Data
Records by NAV and contain all radiometric measurements received from
the DSN station including signal levels (AGC = automatic gain control
in dB), antenna pointing angles, frequency (often referred to simply
as ``Doppler''), range, and residuals. Each ATDF consists of all
tracking data types used to navigate a particular spacecraft (Pioneers
had only Doppler data type) and typically include Doppler, range and
angular types (in S-, X-, and L-frequency bands), differenced range
versus integrated Doppler, programmed frequency data,
pseudo-residuals, and validation data. (Unfortunately there was no
range capability implemented on Pioneer~10 and
11~\cite{pioprl,pioprd}. Early in the mission, JPL successfully
simulated range data using the ``ramped-range''
technique~\cite{JPL32-1526-XIX-I}. In this method, the transmitter
frequency is ramped up and down with a known pattern. One round trip
light time later, this same pattern appears at the DSN receiver, and
can be used to solve for the round trip light travel time. Later
in the mission, this technique became unusable because the carrier
tracking loop bandwidth required to detect the carrier became too
narrow to track the ramped frequency changes.) Also, ATDFs contain
data for a single spacecraft, for one or more ground receiving
stations, and for one or more tracking passes or days.

After a standard processing at the Radio Metric Data Conditioning
group (RMDC) of JPL's Navigation and Mission Design Section, ATDFs are
transformed into Orbit Data Files (ODF, format
TRK-2-18~\cite{TRK-2-18}). A program called {\tt STRIPPER} is used to
produce the ODFs that are, at this point, the main product that is
distributed to the end users for their orbit determination needs (see
more discussion on the conversion process in~\cite{pioprd}). At JPL,
after yet additional processing, these ODFs are used to produce
sequentially formatted input/output files in {\tt NAVIO} format that
is used by navigators while working with the JPL Orbit Determination
Program. (Note that the {\tt NAVIO} input/output file format is used
only at JPL; other orbit determination programs convert ODFs to their
particular formats.)

Each ODF physical record is 2016 32-bit words in length and consists
of 224 9-word logical records per data block. The ODF records are
arranged in a sequence that consists of one file label record, one
file identifier logical record, orbit data logical records in time
order, ramp data logical records in time order, clock offset data
logical records in time order, data summary logical records in time
order and software/hardware end-of-file markers. Bit lengths of data
fields are variable and cross word boundaries. An ODF usually contains
most types of records, but may not have them all. The first record in
each of the 7 primary groups is a header record; depending on the
group, there may be from zero to many data records following each
header. For further details, see Appendix~\ref{app:ODF}.

\subsection{Available Doppler data}
\label{sec:new-data}

The Pioneer~10 and 11 spacecraft were in space for more than three
decades. The original 2002 study~\cite{pioprd} of the Pioneer anomaly
(see Section~\ref{sec:anomaly}) used approximately 11.5~years of
Pioneer~10 and 3.5 years of Pioneer~11 Doppler data. These data points
were obtained from the late phases of the respective missions, when
the spacecraft were far from the Earth and the Sun.

An effort was launched at JPL in June 2005 to recover more Pioneer
Doppler data, preferably the entire Doppler mission records, if
possible: almost 30 years of Pioneer~10 and 20 years of Pioneer~11
Doppler data, most of which was never used previously in the
investigation of the anomalous acceleration. This task proved much
harder than anticipated, due to antiquated file formats, missing data,
and corrupted files (see discussion below and~\cite{MDR2005}).

Below we discuss these two sets of Pioneer~10 and 11 Doppler data.

\subsubsection{Pioneer Doppler data available in 2002}

The anomalous acceleration of the Pioneer spacecraft was first
reported in 1998~\cite{pioprl}. This effort utilized Pioneer~10 data
from 1987 to 1995, and a shorter span of Pioneer~11 data to obtain
accelerations of $a_{P10}=(8.09\pm 0.20)\times 10^{-10}\mathrm{\ m/s}^2$, and
$a_{P11}=(8.56\pm 0.11)\times 10^{-10}\mathrm{\ m/s}^2$, respectively, using
JPL's Orbit Determination Program (ODP). The Pioneer~10 result was
also verified by the Aerospace Corporation's Compact High Accuracy
Satellite Motion Program (CHASMP), which yielded $a_{P10}=(8.65\pm
0.03)\times 10^{-10}\mathrm{\ m/s}^2$. The errors quoted are the statistical
formal errors produced by the fitting procedure.

In 2002, JPL published the results of a study that, to this date,
remains the definitive result on the Pioneer anomaly~\cite{pioprd}
(see Section~\ref{sec:anomaly}). In this study, a significantly
expanded set of Doppler data was utilized. For Pioneer~10, the data
set covered the period between January~3, 1987 and July~22, 1998. This
corresponds to heliocentric distances between 40 and 70.5~AU. The data
set contained 19,403 Doppler data points. For Pioneer~11, the data set
was smaller: the trajectory of the spacecraft between January~5, 1987
and October~1, 1990 was covered, corresponding to heliocentric
distances between 22.42 and 31.7~AU, with 10,252 Doppler data points.

The 2002 JPL study was also the first to combine Pioneer~10 and
Pioneer~11 results. The study attempted to estimate realistic errors,
e.g., errors due to physical or computational systematic effects. This
approach resulted in the value of $a_P=(8.74\pm 1.33)\times
10^{-10}\mathrm{\ m/s}^2$, which is now the widely quoted ``canonical'' value
of the anomalous acceleration of Pioneer~10 and 11.

\subsubsection{The extended Pioneer Doppler data set}
\label{sec:extend-Doppler-data}

The recovery of radiometric Doppler data for a mission with such a
long duration was never attempted before. It presents unique
challenges, as a result of changing data formats, changes in
navigational software and supporting hardware, changes in the
configuration of the DSN (new stations built, old ones demolished,
relocated, or upgraded), and the loss of people~\cite{MDR2005}. Even
physically locating the data proved to be a difficult task, as
incomplete holdings were scattered among various
archives. Nonetheless, as of November 2009, the transfer of the
available Pioneer Doppler data to modern media formats has been
completed.

Initially, the following sources for Doppler data were considered:

\begin{itemize}
\item Archived tracking files at JPL and the Deep Space Network;
\item \vskip -6pt Files archived by JPL researchers at the NSSDC
  (i.e., National Space Science Data Center);
\item \vskip -6pt Data segments available from individual researchers
  at JPL.
\end{itemize}

During these data collection efforts, multiple serious problems were
encountered (see details in~\cite{MDR2005}), including

\begin{itemize}
\item The files were in several different formats, including old
  (e.g., ``Type 66''~\cite{TYPE66}) formats that are no longer used,
  nor very well documented, and for which no data conversion tools
  exist;
\item \vskip -6pt The files were often missing critical information;
  for instance, tracking data may have been recorded for the
  spacecraft, but critically important ramp (see
  Section~\ref{sec:dsn-receivers} and also
 ~\cite{JPL32-1526-XIX-I}) data for the corresponding
  transmitting stations may have been lost;
\item \vskip -6pt The files were not processed with a consistent
  strategy; for instance, some files contained Doppler frequency data
  that was corrected for the spacecraft's spin, whereas other files
  did not include such corrections;
\item \vskip -6pt Some files were corrupted but recoverable; for
  instance, writing fixed size records using inappropriate software
  tools may have introduced spurious bytes into the data in a manner
  that can be corrected;
\item \vskip -6pt Some files were corrupted in an unrecognizable
  manner and had to be discarded.
\end{itemize}

Despite the unanticipated complexities, as of late 2009 the transfer
of the available Pioneer Doppler data to modern media formats has been
completed. Initial analysis of these data is under way, and it appears
that while Pioneer~10 data prior to February 1980 is not usable,
coverage is nearly continuous from that data until the end of
mission. Similarly, for Pioneer~11, good data is available from
mid-1978 until the loss of coherent mode in late 1990. Further details
will be reported as appropriate upon the conclusion of this initial
data analysis.

To summarize, there exists more than 20 years of Pioneer~10 and more
than 10~years of Pioneer~11 data, a significant fraction of which had
never been well studied for the purposes of anomaly
investigation. This new, expanded data set may make it possible to
answer questions concerning the constancy and direction of the
anomalous acceleration.

\subsection{Doppler observables and data preparation}
\label{sec:doppler_tech}

The Doppler observable can be predicted if the spacecraft's orbit is
known. Given known initial conditions and the ephemerides of
gravitating sources, a dynamic model can be constructed that yields
predictions of the spacecraft's position as a function of time. An
observational model can account for the propagation of the signal,
allowing a computation of the received signal frequency at a ground
station of known terrestrial location. The difference between the
calculated and observed values of the received frequency is known as
the Doppler residual. If this residual exceeds acceptable limits, the
dynamic model or observational model must be adjusted to account for
the discrepancy. Once the model is found to be sufficiently accurate,
it can also be used to plan the spacecraft's future trajectory (see
Figure~\ref{fig:radio-nav} and relevant discussion in
Section~\ref{sec:navigation}).

Various radio tracking strategies are available for determining the
trajectory parameters of interplanetary spacecraft. However, radio
tracking Doppler and range techniques are the most commonly used
methods for navigational purposes. (Note that Pioneers did not have a
range observable; all the navigational data is in the form of
Doppler observations.) The position and velocities of the DSN tracking
stations must be known to high accuracy. The transformation from a
Earth fixed coordinate system to the International Earth Rotation
Service (IERS) Celestial System is a complex series of rotations that
includes precession, nutation, variations in the Earth's rotation
({\tt UT1-UTC}) and polar motion.

Calculations of the motion of a spacecraft are made on the basis of
the range time-delay and/or the Doppler shift in the signals. This
type of data was used to determine the positions, the velocities, and
the magnitudes of the orientation maneuvers for the Pioneer
spacecraft.

Theoretical modeling of the group delays and phase delay rates are
done with the orbit determination software we describe in
Section~\ref{sec:navigation}.

\subsubsection{Doppler experimental techniques and strategy}
\label{sec:doppler}

In Doppler experiments a radio signal transmitted from the Earth to
the spacecraft is coherently transponded and sent back to the
Earth. Its frequency change is measured with great precision, using
the hydrogen masers at the DSN stations. The observable is the DSN
frequency shift\epubtkFootnote{The JPL and DSN convention for Doppler
  frequency shift is $(\Delta \nu)_{\tt DSN} = \nu_0 - \nu$, where
  $\nu$ is the measured frequency and $\nu_0$ is the reference
  frequency. It is positive for a spacecraft receding from the
  tracking station (red shift), and negative for a spacecraft
  approaching the station (blue shift), just the opposite of the usual
  convention, $(\Delta \nu)_{\tt usual} = \nu - \nu_0$. In
  consequence, the velocity shift, $\Delta v = v - v_0$, has the same
  sign as $(\Delta \nu)_{\tt DSN}$ but the opposite sign to $(\Delta
  \nu)_{\tt usual}$. Unless otherwise stated, we use the DSN frequency
  shift convention in this document.}
\begin{equation}
 \Delta \nu(t)={\nu_0}\,\frac{1}{c}\frac{d \ell}{dt},
\label{eq:doppler}
\end{equation}
where $\ell$ is the overall optical distance (including diffraction
effects) traversed by a photon in both directions. (In the Pioneer
Doppler experiments, the stability of the fractional drift at the
S-band is on the order of $\Delta \nu/\nu_0\simeq10^{-12}$, for
integration times on the order of 10\super{3}~s.) Doppler measurements
provide the ``range rate'' of the spacecraft and therefore are
affected by all the dynamical phenomena in the volume between the
Earth and the spacecraft.

Expanding upon what was discussed in
Section~\ref{sec:data-acquisition}, the received signal and the
transmitter frequency (both are at S-band) as well as a 10 pulse per
second timing reference from the FTS are fed to the Metric Data
Assembly (MDA). There the Doppler phase (difference between
transmitted and received phases plus an added bias) is counted. That
is, digital counters at the MDA record the zero crossings of the
difference (i.e., Doppler, or alternatively the beat frequency of the
received frequency and the exciter frequency). After counting, the
bias is removed so that the true phase is produced.

The system produces ``continuous count Doppler'' and it uses two
counters. Every tenth of a second, a Doppler phase count is recorded
from one of the counters. The other counter continues the counts. The
recording alternates between the two counters to maintain a continuous
unbroken count. The Doppler counts are at 1~MHz for S-band or 5~MHz
for X-band. The wavelength of each S-band cycle is about
13~cm. Dividers or ``time resolvers'' further subdivide the cycle into
256 parts, so that fractional cycles are measured with a resolution of
0.5~mm. This accuracy can only be maintained if the Doppler is
continuously counted (no breaks in the count) and coherent frequency
standards are kept throughout the pass. It should be noted that no
error is accumulated in the phase count as long as lock is not
lost. The only errors are the stability of the hydrogen maser and the
resolution of the ``resolver.''

Consequently, the JPL Doppler records are not frequency
measurements. Rather, they are digitally counted measurements of the
Doppler phase difference between the transmitted and received S-band
frequencies, divided by the count time.

Therefore, the Doppler observables to which we will refer have units
of cycles per second or Hz. Since total count phase observables are
Doppler observables multiplied by the count interval $T_c$,
they have units of cycles. The Doppler integration time refers to the
total counting of the elapsed periods of the wave with the reference
frequency of the hydrogen maser. The usual Doppler integrating times
for the Pioneer Doppler signals refers to the data sampled over
intervals of 10~s, 60~s, 600~s, 660~s, or 1980~s.

In order to acquire Doppler data, the user must provide a reference
trajectory and information concerning the spacecraft's RF system to
JPL's Deep Space Mission System (DSMS), to allow for the generation of
pointing and frequency predictions. The user specified count interval
can vary from 0.1~s to tens of minutes.  Absent any systematic
errors, the precision improves as the square root of the count
interval. Count times of 10 to 60 seconds are typical~\cite{dsms-cat},
as well as intervals of $\sim$~2000~s, which is an averaging interval
located at the minimum of the Allan variance curve for hydrogen
masers. The average rate of change of the cycle count over the count
interval expresses a measurement of the average velocity of the
spacecraft in the line between the antenna and the spacecraft. The
accuracy of Doppler data is quoted in terms of how accurate this
velocity measurement is over a 60 second count.

It is also possible to infer the position in the sky of a spacecraft
from the Doppler data. This is accomplished by examining the diurnal
variation imparted to the Doppler shift by the Earth's rotation. As
the ground station rotates underneath a spacecraft, the Doppler shift
is modulated by a sinusoid. The sinusoid's amplitude depends on the
declination angle of the spacecraft and its phase depends upon the
right ascension. These angles can therefore be estimated from a record
of the Doppler shift that is (at least) of several days duration. This
allows for a determination of the distance to the spacecraft through
the dynamics of spacecraft motion using standard orbit theory
contained in the orbit determination programs.

\subsubsection{Data preparation}
\label{sec:data_edit}

In an ideal system, all scheduled observations would be used in
determining parameters of physical interest. However, there are
inevitable problems that occur in data collection and processing that
corrupt the data. So, at various stages of the signal processing one
must remove or ``edit'' corrupted data. Thus, the need arises for
objective editing criteria. Procedures have been developed, which
attempt to excise corrupted data on the basis of objective
criteria. There is always a temptation to eliminate data that is not
well explained by existing models, to thereby ``improve'' the
agreement between theory and experiment. Such an approach may, of
course, eliminate the very data that would indicate deficiencies in
the \textit{a~priori} model. This would preclude the discovery of
improved models.

In the processing stage that fits the Doppler samples, checks are made
to ensure that there are no integer cycle slips in the data stream
that would corrupt the phase. This is done by considering the
difference of the phase observations taken at a high rate (10 times a
second) to produce Doppler. Cycle slips often are dependent on
tracking loop bandwidths, the signal-to-noise ratios, and predictions
of frequencies. Blunders due to out-of-lock can be determined by
looking at the original tracking data. In particular, cycle slips due
to loss-of-lock stand out as a 1~Hz blunder point for each cycle
slipped.

If a blunder point is observed, the count is stopped and a Doppler
point is generated by summing the preceding points. Otherwise the
count is continued until a specified maximum duration is
reached. Cases where this procedure detected the need for cycle
corrections were flagged in the database and often individually
examined by an analyst. Sometimes the data was corrected, but
nominally the blunder point was just eliminated. This ensures that the
data is consistent over a pass. However, it does not guarantee that
the pass is good, because other errors can affect the whole pass and
remain undetected until the orbit determination is done.

To produce an input data file for an orbit determination program, JPL
has a software package known as the Radio Metric Data Selection,
Translation, Revision, Intercalation, Processing and Performance
Evaluation Reporting (RMD-STRIPPER) program. As we discussed in
Section~\ref{sec:doppler}, this input file has data that can be
integrated over intervals with different durations: 10~s, 60~s, 600~s,
660~s, and 1980~s. This input orbit determination file obtained from
the RMDC group is the data set that can be used for
analysis. Therefore, the initial data file already contained some
common data editing that the RMDC group had implemented through
program flags, etc. The data set we started with had already been
compressed to 60~s. So, perhaps there were some blunders that had
already been removed using the initial {\tt STRIPPER} program.

The orbit analyst manually edits the remaining corrupted data
points. Editing is done either by plotting the data residuals and
deleting them from the fit or plotting weighted data residuals. That
is, the residuals are divided by the standard deviation assigned to
each data point and plotted. This gives the analyst a realistic view
of the data noise during those times when the data was obtained while
looking through the solar plasma. Applying an ``$N$-$\sigma$''
($\sigma$ is the standard deviation) test, where $N$ is the choice of
the analyst (usually 4\,--\,10) the analyst can delete those points that
lie outside the $N$-$\sigma$ rejection criterion without being biased
in his selection.

A careful analysis edits only very corrupted data; e.g., a blunder due
to a phase lock loss, data with bad spin calibration, etc. If needed
or desired, the orbit analyst can choose to perform an additional data
compression of the original navigation data.

\subsubsection{Data weighting}
\label{sec:data_weight}

The Pioneers used S-band ($\sim$~2.2~GHz) radio signals to communicate
with the DSN. The S-band data is available from 26~m, 70~m, and some
34~m antennas of the DSN complex (see baseline DSN configuration in
the Figure~\ref{fig:dsn_block}). The dominant systematic error that
can affect S-band tracking data is ionospheric transmission
delays. When the spacecraft is located angularly close to the Sun,
with Sun-Earth-spacecraft angles of less than 10~degrees, degradation
of the data accuracy will occur. S-band data is generally unusable for
Sun-Earth-spacecraft angles of less than 5~degrees.

Therefore, considerable effort has gone into accurately estimating
measurement errors in the observations. These errors provide the data
weights necessary to accurately estimate the parameter adjustments and
their associated uncertainties. To the extent that measurement errors
are accurately modeled, the parameters extracted from the data will be
unbiased and will have accurate sigmas assigned to them. Typically,
for S-band Doppler data one assigns a standard 1-$\sigma$ uncertainty
of 1~mm/s over a 60~s count time after calibration for transmission
media effects.

A change in the DSN antenna elevation angle also directly affects the
Doppler observables due to tropospheric refraction. Therefore, to
correct for the influence of the Earth's troposphere the data can also
be deweighted for low elevation angles. The phenomenological range
correction used in JPL's analysis technique is given as
\begin{equation}
\sigma=\sigma_\mathrm{nominal} \bigg(1+\frac{18}{(1+\theta_E)^2}\bigg),
\label{eq:sig_aer0}
\end{equation}
where $\sigma_\mathrm{nominal}$ is the basic standard deviation (in
Hz) and $\theta_E$ is the elevation angle in degrees. Each leg is
computed separately and summed. For Doppler the same procedure is
used. First, Equation~(\ref{eq:sig_aer0}) is multiplied by
$\sqrt{60\mathrm{\ s}/T_c}$, where $T_c$ is the count time. Then a
numerical time differentiation of Equation~(\ref{eq:sig_aer0}) is
performed. That is, Equation~(\ref{eq:sig_aer0}) is differenced and
divided by the count time, $T_c$. (For more details on this standard
technique see~\cite{Moyer:1971, Moyer-2003}.)

There is also the problem of data weighting for data influenced by the
solar corona. This is discussed in Section~\ref{sec:effect_solplasm}.

\subsubsection{Spin calibration of the data}
\label{sec:spin-calibrate}

The radio signals used by DSN to communicate with spacecraft are
circularly polarized. When these signals are reflected from spinning
spacecraft antennas a Doppler bias is introduced that is a function of
the spacecraft spin rate. Each revolution of the spacecraft adds one
cycle of phase to the up-link and the down-link. The up-link cycle is
multiplied by the turn around ratio 240/221 so that the bias equals
(1+240/221) cycles per revolution of the spacecraft.

For the Pioneer~10 and 11 spacecraft, high-accuracy spin data is
available from the spacecraft telemetry. Due to the star sensor
failure on board Pioneer~10 (see Section~\ref{sec:pio-project}), once
the spacecraft was more than $\sim$~30~AU from the Sun, no on-board
roll reference was available. Until mid-1993, a science instrument
(the Infrared Photo-Polarimeter) was used as a surrogate star sensor,
which allowed the accurate determination of the spacecraft spin rate;
however, due to the lack of available electrical power on board, this
instrument could not be used after 1993. However, analysts still could
get a rough spin determination approximately every six months using
information obtained from the conscan maneuvers. No spin
determinations were made after 1995. However, the archived conscan
data could still yield spin data at every maneuver time if such work
was approved. Further, as the phase center of the main antenna is
slightly offset from the spin axis, a very small (but detectable)
sine-wave signal appears in the high-rate Doppler data. In principle,
this could be used to determine the spin rate for passes taken after
1993, but it has not been attempted.

The changing spin rates of Pioneer~10 and 11 can be an indication of
gas leaks, which can also be a source of unmodeled accelerations. We
discussed this topic in more detail in Section~\ref{sec:spin}.

\subsection{Pioneer telemetry data}
\label{sec:telemetry}

Telemetry received from the Pioneer~10 and 11 spacecraft was supplied
to the Pioneer project by the DSN in the form of Master Data Records
(MDRs). MDRs contained all information sent by the spacecraft to the
ground, and some information about the DSN station that received the
data. The information in the MDRs that was sent by the spacecraft
included engineering telemetry as well as scientific observations.

The Pioneer project used engineering data extracted from the MDRs to
monitor and control the spacecraft, while scientific data, also
extracted from the MDRs, was converted into formats specific to each
experiment and supplied to the experimenter groups.

Far beyond the original expectations\epubtkFootnote{Normally, MDRs are
  seen to be of little use once the relevant information is extracted
  from them. Scientific measurements are extracted, packaged in the
  appropriate form, and sent to the corresponding experimenters for
  further processing and evaluation. Engineering telemetry is used by
  the spacecraft operations team to control and guide the
  spacecraft. MDRs are not necessarily considered worth preserving
  once the scientific data has been extracted, and the engineering
  telemetry is no longer needed for spacecraft operations. Indeed, the
  MDR retention schedule prescribed that the tapes be destroyed after
  7~years. Fortunately, most of the MDRs for the Pioneer~10 and 11
  projects have been preserved nevertheless (see discussion
  in~\cite{MDR2005}). }, telemetry is now seen to be of value for the
investigation of the Pioneer anomaly, as the MDRs, specifically the
telemetry data contained therein, are helpful in the construction of
an accurate model of the spacecraft during their decades long journey,
including a precise thermal profile, the time history of propulsion
system activation and usage, and many other potential sources of
on-board disturbances. After recent recovery efforts~\cite{MDR2005},
this data is available for investigation.

\subsubsection{MDR data integrity and completeness}

The total amount of data stored in these MDR files is approximately
40~GB~\cite{MDR2005}. According to the original log sheets that record
the transcription from tape to magneto-optical media, only a few days
worth of data is missing, some due to magnetic tape damage. One
notable exception is the Jupiter encounter period of
Pioneer~10. According to the transcription log sheets, DOY 332--341
from 1973 were not available at the time when the magnetic tapes on
which the MDRs were originally stored were transcribed to more durable
magneto-optical media.

Other significant periods of missing data are listed in
Table~\ref{tb:missing_mdr}. It is not known why these records are not
present, except that we know that very few days are missing due to
unreadable media (i.e., the cause is missing, not damaged, tapes.)

\begin{table}[h!]
  \caption{Pioneer~10 and 11 missing MDRs (periods of missing data
  shorter than 1\,--\,2 days not shown.)}
  \label{tb:missing_mdr}
  \centering
  \begin{tabular}{lll}
    \toprule
    Spacecraft & Year & Days of the Year (DOY)\\
    \midrule
    Pioneer-10 & 1972 & 133--149\\
             ~ & 1973 & 004--008, 060--067, 332--341\\
             ~ & 1974 & 034--054\\
             ~ & 1979 & 025--032, 125--128, 137--157, 171--200\\
             ~ & 1980 & 173--182, 187--199, 248--257\\
             ~ & 1983 & 329--348\\
             ~ & 1984 & 346--359\\
    \midrule
    Pioneer-11 & 1973 & 056--064, 067--080, 082--086, 088--094\\
             ~ & 1980 & 309--330, 337--365\\
             ~ & 1982 & 318--365\\
             ~ & 1983 & 001--050\\
             ~ & 1984 & 343--357\\
             ~ & 1990 & 081--096\\
    \bottomrule
  \end{tabular}
\end{table}

So, the record is fairly complete. But how good is the data? Over
forty billion bytes were received by the DSN, processed, copied to
tape, copied from tape to magneto-optical disks, then again copied
over a network connection to a personal computer. It is not
inconceivable that the occasional byte was corrupted by a transmission
or storage error. There are records that contain what is apparently
bogus data, especially from the later years of operation. This, plus
the fact that the record structure (e.g., headers, synchronization
sequences) is intact suggests reception errors as the spacecraft's
signal got weaker due to increasing distance, and not copying and/or
storage errors.

The MDRs contain no error detection or error correction code, so it is
not possible to estimate the error rate. However, it is likely to be
reasonably low, since the equipment used for storage and copying is
generally considered very reliable. Furthermore, any errors would
likely show up as random noise, and not as a systemic bias. In this
regard, the data should generally be viewed to be of good quality
insofar as the goal of constructing an engineering profile of the
spacecraft is considered.

\subsubsection{Interpreting the data}

MDRs are a useless collection of bits unless information is available
about their structure and content. Fortunately, this is the case in
the case of the Pioneer~10 and Pioneer~11 MDRs.

The structure of an MDR is shown in Appendix~\ref{app:mdr} (see
also~\cite{ARC221}). The frame at the beginning and at the end of each
1344-bit record contained information about the DSN station that
received the data, and included a timestamp, data quality and error
indicators, and the strength of the received signal. The middle of the
record was occupied by as many as four consecutive data frames
received by the spacecraft.

The MDR header is followed by four data frames (not all four frames
may be used, but they are all present) of 192~bits each. Lastly, an
additional 8 words of DSN information completes the record. The total
length of an MDR is thus 42 words of 32~bits each.

The 192-bit data frames are usually interpreted as 64 3-bit words or,
alternatively, as 32 6-bit words. The Pioneer project used many
different data frame formats during the course of the mission. Some
formats were dedicated to engineering telemetry (accelerated
formats). Other formats are science data formats, but still contain
engineering telemetry in the form of a subcommutator: a different
engineering telemetry value is transmitted in each frame, and
eventually, all telemetry values are cycled through.

The Pioneer spacecraft had a total of 128 6-bit words reserved for
engineering telemetry. Almost all these values are, in fact, used. A
complete specification of the engineering telemetry values can be
found in Section~3.5 (``Data Handling Subsystem'')
of~\cite{PC202}. When engineering telemetry was accelerated to the
main frame rate, four different record formats (C-1 through C-4) were
used to transmit telemetry information. When the science data formats
were in use, an area of the record was reserved for a subcommutator
identifier and value.

The formats are further complicated by the fact that some engineering
telemetry values appear only in subcommutators, whereas others only
appear at the accelerated (main frame) rate.

In the various documentation packages, engineering data words are
identified either by mnemonic, by the letter `C' followed by a
three-digit number that runs from 1 through 128, or most commonly, by
the letter `C' followed by a digit indicating which `C' record (C-1
through C-4) the value appears in, and a two-digit number between 1
and 32: for instance, C-201 means the first engineering word in the
C-2 record.

\subsubsection{Available telemetry information}
\label{sec:mdr-summary}

\begin{table}
  \caption{Available parameter set that may be useful for the Pioneer
  anomaly investigation.}
  \label{tab:telemetry}
  \centering
{\linespread{0.95}\small
\begin{tabular}{|l|l|l|}\hline
Parameters & Subsystem & Telemetry words\\\hline\hline
\multicolumn{3}{|l|}{\tt TEMPERATURES}\\\hline
RTG fin root temps & Thermal & C\sub{201}, C\sub{202}, C\sub{203}, C\sub{204}\\
RTG hot junction temps & Thermal & C\sub{220}, C\sub{219}, C\sub{218}, C\sub{217}\\
TWT temperatures & Communications & C\sub{205}, C\sub{206}, C\sub{207}, C\sub{228}, C\sub{223}, C\sub{221}\\
Receiver temperatures & Communications & C\sub{222}, C\sub{227}\\
Platform temperatures & Thermal & C\sub{301}, C\sub{302}, C\sub{304}, C\sub{318}, C\sub{319}, C\sub{320}\\
PSA temperatures & Thermal & C\sub{225}, C\sub{226}\\
Thruster cluster temps & Propulsion & C\sub{309}, C\sub{326}, C\sub{310}, C\sub{311}, C\sub{312}, C\sub{328}, C\sub{325}\\
SRA/SSA temperatures & ACS & C\sub{303}, C\sub{317}\\
Battery temperature & Power & C\sub{115}\\
Propellant temperature & Propulsion & C\sub{327}\\
N2 tank temperature & Propulsion & C\sub{130}\\
Science instr temps & Science & E\sub{101}, E\sub{102}, E\sub{109}, E\sub{110}, E\sub{117}, E\sub{118}, E\sub{125}, E\sub{128},\\
~ & ~ &  E\sub{201}, E\sub{209}, E\sub{213}, E\sub{221}\\\hline
\multicolumn{3}{|l|}{\tt VOLTAGES}\\\hline
Calibration voltages & Data handling & C\sub{101}, C\sub{102}, C\sub{103}\\
RTG voltages & Power & C\sub{110}, C\sub{125}, C\sub{131}, C\sub{113}\\
Battery/Bus voltages & Power & C\sub{106}, C\sub{107}, C\sub{117}, C\sub{118}, C\sub{119}\\
TWT voltages & Communications & C\sub{224}, C\sub{230}\\
Science instr voltages & Science & E\sub{119}, E\sub{129}, E\sub{210}, E\sub{211}, E\sub{217}, E\sub{220}\\\hline
\multicolumn{3}{|l|}{\tt CURRENTS}\\\hline
RTG currents & Power & C\sub{127}, C\sub{105}, C\sub{114}, C\sub{123}\\
Battery/Bus currents & Power & C\sub{109}, C\sub{126}, C\sub{129}\\
Shunt current & Power & C\sub{122}, C\sub{209}\\
TWT currents & Communications & C\sub{208}, C\sub{211}, C\sub{215}, C\sub{216}\\
Science instr currents & Science & E\sub{111}, E\sub{112}, E\sub{113}\\\hline
\multicolumn{3}{|l|}{\tt PRESSURE}\\\hline
Propellant pressure & Propulsion & C\sub{210}\\\hline
\multicolumn{3}{|l|}{\tt OTHER ANALOG}\\\hline
TWT power readings & Communications & C\sub{231}, C\sub{214}\\
Receiver readings & Communications & C\sub{111}, C\sub{212}, C\sub{232}, C\sub{121}, C\sub{229}, C\sub{213}\\\hline
\multicolumn{3}{|l|}{\tt BINARY/BIT FIELDS}\\\hline
Conscan & Communications & C\sub{313}, C\sub{314}, C\sub{315}, C\sub{316}\\
Stored commands & Electrical & C\sub{305}, C\sub{306}, C\sub{307}\\
Thruster pulse counts & Propulsion & C\sub{329}, C\sub{321}, C\sub{322}, C\sub{330}\\
Status bits & Data handling & C\sub{104}\\
~ & Power & C\sub{128}\\
~ & Electrical & C\sub{120}, C\sub{132}, C\sub{324}, C\sub{332}\\
~ & Communications & C\sub{308}\\
Power switches & Electrical & C\sub{108}, C\sub{124}\\
Roll attitude & Data handling & C\sub{112}, C\sub{116}\\
Precession & ACS & C\sub{403}, C\sub{411}, C\sub{412}, C\sub{415}, C\sub{416}, C\sub{422}, C\sub{423}, C\sub{424}, \\
~ & ~ &  C\sub{425}, C\sub{428}, C\sub{429}, C\sub{430}\\
Spin/roll & Data handling & C\sub{405}, C\sub{406}, C\sub{407}, C\sub{408}, C\sub{417}\\
Delta v & ACS & C\sub{413}, C\sub{414}, C\sub{426}\\
ACS status & Propulsion & C\sub{409}\\
~ & ACS & C\sub{410}, C\sub{427}, C\sub{431}, C\sub{432}\\
Star sensor & ACS & C\sub{404}, C\sub{419}, C\sub{420}, C\sub{421}\\\hline
\multicolumn{3}{|l|}{\tt SCIENCE INSTRUMENTS}\\\hline
Status/housekeeping & Science & E\sub{108}, E\sub{124}, E\sub{130}, E\sub{202}, E\sub{224}, E\sub{131}, E\sub{132}, E\sub{208}\\
JPL/HVM readings & Science & E\sub{103}, E\sub{104}, E\sub{105}, E\sub{106}, E\sub{107}, E\sub{203}, E\sub{204}, E\sub{205}\\
UC/CPI readings & Science & E\sub{114}, E\sub{115}, E\sub{116}, E\sub{206}, E\sub{212}, E\sub{214}, E\sub{215}, E\sub{216}\\
GE/AMD readings & Science & E\sub{122}, E\sub{123}, E\sub{222}, E\sub{223}\\
GSFC/CRT readings & Science & E\sub{126}, E\sub{127}\\
LaRC/MD readings & Science & E\sub{207}\\\hline
\end{tabular}}
\end{table}

Pioneer~10 and 11 telemetry data is very relevant for a study of
on-board systematics. The initial studies of the Pioneer
anomaly~\cite{pioprl, pioprd, 2002gr.qc.....8046M, moriond} and
several subsequent papers~\cite{2001gr.qc.....7022A, Turyshev:2005zm,
  Turyshev:2005zk, Turyshev:2005vj} had emphasized the need for a very
detailed investigation of the on-board systematics. Other researchers
also focused their work on the study of several on-board generated
mechanisms that could contribute to an anomalous acceleration of the
spacecraft~\cite{1999PhRvL..83.1892K, 1999PhRvL..83.1890M,
  2003PhRvD..67h4021S}. Most of these investigations of on-board
systematics were not very precise. This was due to a set of several
reasons, one of them is insufficient amount of actual telemetry data
from the vehicles. In 2005, this picture changed dramatically when
this critical information became available.

Table~\ref{tab:telemetry} summarizes all available telemetry values in
the C (engineering) and E (science) telemetry formats. The MDRs also
contain a complete set of science readings that were telemetered in
the A, B, and D formats.

As this table demonstrates, telemetry readings can be broadly
categorized as temperature, voltage, current readings; other analog
readings; various binary counters, values, and bit fields; and
readings from science instruments. Temperature and electrical readings
are of the greatest use, as they help to establish a detailed thermal
profile of the spacecrafts' major components. Some binary readings are
useful; for instance, thruster pulse count readings help to understand
maneuvers and their impact on the spacecrafts' trajectories. It is
important to note, however, that some readings may not be available
and others may not be trusted. For example, thruster pulse count
readings are only telemetered when the spacecraft is commanded to send
readings in accelerated engineering formats; since these formats were
rarely used late in the mission, we may not have pulse count readings
for many maneuvers. Regarding reliability, we know from mission status
reports about the failure of the sun and star sensors; these failures
invalidate many readings from that point onward. Thus it is important
to view telemetry readings in context before utilizing them as source
data for our investigation.

On-board telemetry not only gives a detailed picture of the spacecraft
and its subsystems, but this picture is redundant: electrical,
thermal, logic state and other readings provide means to examine the
same event from a multitude of perspectives.

In addition to telemetry, there exists an entire archive of the Pioneer
Project documents for the period from 1966 to 2003. This archive
contains all Pioneer~10 and 11 project documents discussing the
spacecraft and mission design, fabrication of various components,
results of various tests performed during fabrication, assembly,
pre-launch, as well as calibrations performed on the vehicles; and
also administrative documents including quarterly reports, memoranda,
etc. Information on most of the maneuver records, spin rate data,
significant events of the craft, etc.\ is also available.

\newpage
\section{Navigation of the Pioneer Spacecraft}
\label{sec:navigation}

Modern radio tracking techniques made it possible to explore the
gravitational environment in the solar system up to a level of
accuracy never before possible. The two principal forms of celestial
mechanics experiments that were used involve planets (e.g., passive
radar ranging) and Doppler and range measurements with interplanetary
spacecraft~\cite{anderson74, dsn86}. This work was motivated by the
desire to improve the ephemerides of solar system bodies and knowledge
of solar system dynamics.

The main objective of spacecraft navigation is to determine the
present position and velocity of a spacecraft and to predict its
future trajectory. This is usually done by measuring changes in the
spacecraft's radio signal and then, using those measurements,
correcting (fitting and adjusting) the predicted spacecraft
trajectory.

In this section we discuss the theoretical foundations that are used
for the analysis of tracking data from interplanetary spacecraft. We
describe the basic physical models used to determine a trajectory.

\subsection{Models for gravitational forces acting on a spacecraft}
\label{sec:gravity}

The primary force acting on a spacecraft in deep space is the force of
gravity, specifically, the gravitational attraction of the Sun and, to
a lesser extent, planetary bodies. When the spacecraft is in deep
space, far from the Sun and not in the vicinity of any planet, these
sources of gravity can be treated as Newtonian point sources. However,
when the spacecraft is in the vicinity of a massive body, corrections
due to general relativity as well as the finite extent and mass
distribution of the body in question must be considered.

Of particular interest is the possibility that a celestial mechanics
experiment might help distinguish between different theories of
gravity. There exists a method, the parametrized post-Newtonian (PPN)
formalism, that allows one to describe various metric theories of
gravity up to ${\cal O}(c^{-5})$ using a common parameterized
metric. As this formalism forms the basis of modern spacecraft
navigational codes, in the following, we provide a brief summary.

\subsubsection{The parametrized post-Newtonian formalism}
\label{sec:ppn}

In 1922, Eddington~\cite{Eddington} developed the first parameterized
generalization of Einstein's theory, expressing components of the
metric tensor in the form of a power series of the Newtonian
potential. This phenomenological parameterization has since been
developed into what is known as the parametrized post-Newtonian (PPN)
formalism~\cite{MTW, Nordtvedt_1968a, Nordtvedt_1968b, Nordtvedt_1969,
  Will-Nordtvedt-72-II, Thorne-Will-1971-I, Will-1971-I, Will-1971-II,
  Thorne-Will-1971-II, Thorne-Will-1971-III, Will-1973, Will93,
  WillNordtvedt72}. This formalism represents the metric tensor for
slowly moving bodies and weak gravitational fields. The formalism is
valid for a broad class of metric theories, including general
relativity as a unique case. The parameters that appear in the PPN
formalism are individually associated with various symmetries and
invariance properties of the underlying theory (see~\cite{Will93} for
details).

The full PPN formalism has 10
parameters~\cite{Will-lrr-2006-3}. However, if one assumes that
Lorentz invariance, local position invariance and total momentum
conservation hold, the metric tensor for a system of $N$ point-like
gravitational sources in four dimensions at the spacetime position of
body $i$ may be written as (see~\cite{Turyshev-UFN-2008})
\begin{eqnarray}
g_{00}&=&1-\frac{2}{c^2}\sum_{j\ne i}\frac{\mu_j}{r_{ij}}+\frac{2\beta}{c^4}\bigg[\sum_{j\ne i}\frac{\mu_j}{r_{ij}}\bigg]^2-\frac{1+2\gamma}{c^4}\sum_{j\ne i}\frac{\mu_j{\dot r}^2_j}{r_{ij}}\nonumber\\
&+&\frac{2(2\beta-1)}{c^4}\sum_{j\ne i}\frac{\mu_j}{r_{ij}}\sum_{k\ne j}\frac{\mu_k}{r_{jk}}-\frac{1}{c^4}\sum_{j\ne i}\mu_j\frac{\partial^2 r_{ij}}{\partial t^2}+{\cal O}(c^{-5}),\nonumber
\end{eqnarray}
\begin{eqnarray}
g_{0\alpha}&=&\frac{2(\gamma+1)}{c^3}\sum_{j\ne i}\frac{\mu_j\dot{\vec{r}}^\alpha_j}{r_{ij}}+{\cal O}(c^{-5}),~~~(\alpha=1...3)\nonumber\\
g_{\alpha\beta}&=&-\delta_{\alpha\beta}\bigg(1+\frac{2\gamma}{c^2}\sum_{j\ne i}\frac{\mu_j}{r_{ij}}\bigg)+{\cal O}(c^{-5}),~~~(\alpha,\beta=1...3)
\label{eq:metric}
\end{eqnarray}
where the indices $1\le i,j\le N$ refer to the $N$ bodies, $r_{ij}$ is
the distance between bodies $i$ and $j$ (calculated as
$|\vec{r}_j-\vec{r}_i|$ where $\vec{r}_i$ is the spatial position
vector of body $i$), $\mu_i$ is the gravitational constant for body $i$
given as $\mu_i=Gm_i$, where $G$ is the Newtonian gravitational
constant and $m_i$ is the body's rest mass. The Eddington-parameters
$\beta$ and $\gamma$ have, in this special case, clear physical
meaning: $\beta$ represents a measure of the nonlinearity of the law
of superposition of the gravitational fields, while $\gamma$
represents the measure of the curvature of the spacetime created by a
unit rest mass.

The Newtonian scalar gravitational potential in Equation~(\ref{eq:metric})
is given by the $1/c^2$ term in $g_{00}$. Corrections of order
$1/c^4$, parameterized by $\beta$ and $\gamma$, are post-Newtonian
terms. In the case of general relativity, $\beta=\gamma=1$. One of the
simplest generalizations of general relativity is the theory of Brans
and Dicke~\cite{Brans-Dicke-1961} that contains, in addition to the
metric tensor, a scalar field and an undetermined dimensionless
coupling constant $\omega$. Brans--Dicke theory yields the values
$\beta=1$, $\gamma=(1+\omega)/(2+\omega)$ for the Eddington
parameters. The value of $\beta$ may be different for other
scalar-tensor theories~\cite{Damour_Nordtvedt_93b,
  Damour_Nordtvedt_93a, Turyshev-UFN-2008}.

The PPN formalism is widely used in studies of relativistic
gravitation~\cite{Brumberg-book-1972, Brumberg-book-1991,
  Moyer-1981-1, Moyer-1981-2, Standish_etal_92, Turyshev:1996,
  Will93}. The relativistic equations of motion for an $N$-body system
are derived from the PPN metric using a Lagrangian
formalism~\cite{Turyshev:1996, Will93}, discussed below.

\subsubsection{Relativistic equations of motion}
\label{sec:eqs-motion}

Navigation of spacecraft in deep space requires computing a
spacecraft's trajectory and the compilation of spacecraft ephemeris: a
file containing the position and velocity of the spacecraft as
functions of time~\cite{Turyshev-UFN-2008}. Spacecraft ephemerides are
computed by orbit determination codes that utilize a numerical
integrator in conjunction with various input parameters. These include
an estimate of the spacecraft's initial state vector (comprising its
position and velocity vector), adopted constants ($c$, $G$, planetary
mass ratios, etc.) and parameters that are estimated from fits to
observational data (e.g., corrections to the ephemerides of solar
system bodies).

The principal equations of motion used by orbit determination codes
describe the relativistic gravitational acceleration in the presence
of the gravitational field of multiple point sources that include the
Sun, planets, major moons and larger asteroids
~\cite{Standish_Williams_2008}. These equations are derived from the
metric Equation~(\ref{eq:metric}) using a Lagrangian formalism. The point
source model is adequate in deep space when a spacecraft is traveling
far from those sources. When the spacecraft is in the vicinity of a
planet, ephemeris programs also compute corrections due to deviations
from spherical symmetry in the planetary body, as well as the
gravitational influences from the planet's moons, if any.

The acceleration of body $i$ due to the gravitational field of point
sources, including Newtonian and relativistic perturbative
accelerations~\cite{Einstein-Infeld-Hoffmann-1938, Turyshev:1996,
  Will93}, can be derived in the solar system barycentric frame in the
form~\cite{pioprd,estabrook69, Moyer:1971, Moyer-1981-1, Moyer-2003,
  Newhall83}:
\begin{align}
\ddot{\vec{r}}_i&=\sum_{j\not=i}\frac{\mu_j(\vec{r}_j-\vec{r}_i)}{r_{ij}^3}\bigg\{1-\frac{2(\beta+\gamma)}{c^2}\sum_{l\ne i}\frac{\mu_l}{r_{il}}-\frac{2\beta-1}{c^2}\sum_{k\ne j}\frac{\mu_k}{r_{jk}}+\gamma\left(\frac{{\dot r}_i}{c}\right)^2+(1+\gamma)\left(\frac{{\dot r}_j}{c}\right)^2\nonumber\\
&~-\frac{2(1+\gamma)}{c^2}\dot{\vec{r}}_i \dot{\vec{r}}_j-\frac{3}{2c^2}\left[\frac{(\vec{r}_i-\vec{r}_j)\dot{\vec{r}}_j}{r_{ij}}\right]^2+\frac{1}{2c^2}(\vec{r}_j-\vec{r}_i)\ddot{\vec{r}}_j\bigg\}+\frac{3+4\gamma}{2c^2}\sum_{j\ne i}\frac{\mu_j\ddot{\vec{r}}_j}{r_{ij}}\nonumber\\
&~+\frac{1}{c^2}\sum_{j\not=i}\frac{\mu_j}{r_{ij}^3}\Big\{\left[\vec{r}_i-\vec{r}_j\right]\cdot\left[(2+2\gamma)\dot{\vec{r}}_i-(1+2\gamma)\dot{\vec{r}}_j\right]\Big\}(\dot{\vec{r}}_i-\dot{\vec{r}}_j)+{\cal O}(c^{-4}),
\label{eq:rdotdot}
\end{align}
where the indices $1\le j,k,l\le N$ refer to the $N$ bodies and where
$k$ includes body $i$, whose motion is being investigated.

These equations can be integrated numerically to very high precision
using standard techniques in numerical codes that are used to
construct solar system ephemerides, for spacecraft orbit
determination~\cite{Moyer-2003, Standish_etal_92, Turyshev:1996}, and
for the analysis of gravitational experiments in the solar
system~\cite{Turyshev:2008dr, Turyshev_etal_acfc_2003, Will93,
  Will-lrr-2006-3, Williams:2004qba}.

In the vicinity of a celestial body, one must also take into account
that a celestial body is not spherically symmetric. Its gravitational
potential can be modeled in terms of spherical harmonics. As
Pioneers~10 and 11 both flew by Jupiter and Pioneer~11 visited Saturn,
of specific interest to the navigation of theses spacecraft is the
gravitational potential due to the oblateness of a planet, notably a
gas giant. The gravitational potential due to the oblateness of
planetary body $i$ can be expressed using zonal harmonics in the
form~\cite{Moyer-2003}:
\begin{equation}
U_{i\mathrm{obl}}=-\frac{\mu_i}{|\vec{r}_i-\vec{r}|}\sum_{k=1}^\infty\frac{J_k^{(i)}a_i^kP_k(\sin\theta)}{|\vec{r}_i-\vec{r}|^k},
\label{eq:Ui}
\end{equation}
where $P_k(x)$ is the $k$-th Legendre polynomial in $x$, $a_i$ is the
equatorial radius of planet $i$, $\theta$ is the latitude of the
spacecraft relative to the planet's equator, and $J_k^{(i)}$ is the
$k$-th spherical harmonic coefficient of planet $i$.

In order to put Equation~(\ref{eq:Ui}) to use, first it must be translated
into an expression for force by calculating its gradient. Second, it
is also necessary to express the position of the spacecraft in a
coordinate system that is fixed to the planet's center and equator.

\subsection{Light times and time scales}
\label{sec:time_scales}

The complex gravitational environment of the solar system manifests
itself not just through its effects on the trajectory of a spacecraft
or celestial body: the propagation of electromagnetic signals to or
from an observing station on the Earth must also be
considered. Additionally, proper timekeeping becomes an issue of
significance: clocks that are in relative motion do not tick at the
same rate, and changing gravitational potentials may also affect them.

\subsubsection{Light time solution}

The time it takes for a signal to travel between two locations in space
in the gravitational environment of a massive point source with
gravitational constant $\mu$ can be derived from the PPN metric
Equation~(\ref{eq:metric}) in the form~\cite{Moyer-2003}:
\begin{equation}
t_2-t_1=\frac{r_{12}}{c}+(1+\gamma)\frac{\mu}{c^3}\ln\frac{r_1+r_2+r_{12}+(1+\gamma)\mu/c^2}{r_1+r_2-r_{12}+(1+\gamma)\mu/c^2}+{\cal O}(c^{-5}),
\label{eq:light-time}
\end{equation}
where $t_1$ refers to the signal transmission time, and $t_2$ refers
to the reception time, $r_{1,2}$ represent the distance of the point
of transmission and point of reception, respectively, from the massive
body, and $r_{12}$ is the spatial separation of the points of
transmission and reception. The terms proportional to $\mu/c^2$ are
important only for the Sun and are negligible for all other bodies in
the solar system.

\subsubsection{Standard time scales}

The equations of motion Equation~(\ref{eq:rdotdot}) and the light time
solution Equation~(\ref{eq:light-time}) are both written in terms of an
independent time variable, which is called the ephemeris time, or
ET. Ephemeris time is simply coordinate time in the chosen coordinate
frame, such as a solar system barycentric frame. As such, the
ephemeris time differs from the standard International Atomic Time
(TAI, Temps Atomique International), measured in SI (Syst\`eme
International) seconds relative to a given epoch, namely the beginning
of the year 1958.

When a solar system barycentric frame of reference is used to
integrate the equations of motion, the relationship between ET and TAI
can be expressed, to an accuracy that is sufficient for the purposes
of the Pioneer project\footnote{JPL's Orbit Determination Program, ODP,
uses a higher precision conversion algorithm, not the simplified
formula presented here.}, as
\begin{equation}
\mathrm{ET}-\mathrm{TAI}=(32.184+1.657\times 10^{-3}\sin{E})~\mathrm{seconds},
\end{equation}
where
\begin{eqnarray}
E&=&M+0.01671\sin{M},\\
M&=&6.239996 + 1.99096871\times 10^{-7}\,t,
\end{eqnarray}
and $t$ is ET in seconds since the J2000 epoch (noon, January 1,
2000). For further details, including higher accuracy time conversion
formulae, see the relevant literature~\cite{DSX-I, DSX-II, DSX-III,
  DSX-IV, MG2005, Moyer-2003} (in particular, see Eqs.~(2--26) through
(2--28) in~\cite{Moyer-2003}.)

There exist alternate expressions with up to several hundred
additional periodic terms, which provide greater accuracies. The use of
these extended expressions provide transformations of
$\mathrm{ET}-\mathrm{TAI}$ to accuracies of 1~ns~\cite{Moyer-2003}.

For the purposes of the investigation of the Pioneer anomaly, the
Station Time (ST) is especially significant. The station time is the
time kept by the ultrastable oscillators of DSN
stations, and it is measured in Universal Coordinated Time (UTC). All
data records generated by DSN stations are timestamped using ST, that
is, UTC as measured by the station's clock.

UTC is a discontinuous time scale; it is similar to TAI, except for
the regular insertion of leap seconds, which are used to account for
minute variations in the Earth's rate of rotation. Converting from UTC
to international atomic time (TAI) requires the addition or
subtraction of the appropriate number of leap seconds (ranging between
10 and 32 during the lifetime of the Pioneer missions.) For more
details see~\cite{Moyer-2003,exp_cat}.

\subsection{Nongravitational forces external to the spacecraft}

Even in the vacuum of interplanetary space, the motion of a spacecraft
is governed by more than just gravity. There are several
nongravitational forces acting on a spacecraft, many of which must be
taken into account in order to achieve an orbit determination accuracy
at the level of the Pioneer anomaly. (For a general introduction to
nongravitational forces acting on spacecraft, consult~\cite{longuski} and p.~125 in~\cite{milani}.) To determine the
Pioneer orbits to sufficient precision, orbit determination programs
must take into account these nongravitational accelerations unless a
particular force can be demonstrated to be too small in magnitude to
have a detectable effect on the spacecraft's orbit.

In the presentation below of the standard modeling of small
nongravitational forces, we generally follow the discussion in
~\cite{pioprd}, starting with nongravitational forces that
originate from sources external to the spacecraft, and followed by a
review of forces of on-board origin. We also discuss effects acting on
the radio signal sent to, or received from, the spacecraft.

\subsubsection{Solar radiation pressure}
\label{sec:radpress}

Most notable among the sources for the forces external to the Pioneer
spacecraft is the solar pressure. This force is a result of the
exchange of momentum between solar photons and the spacecraft, as
solar photons are absorbed or reflected by the spacecraft. This force
can be significant in magnitude in the vicinity of the Earth, at
$\sim$~1~AU from the Sun, especially when considering spacecraft with a
large surface area, such as those with large solar panels or
antennas. For this reason, solar pressure models are usually developed
before a spacecraft is launched. These models take into account the
effective surface areas of the portions of the spacecraft exposed to
sunlight, and their thermal and optical properties. These models offer
a computation of the acceleration of the spacecraft due to solar
pressure as a function of solar distance and spacecraft orientation.

The simplest way of modeling solar pressure is by using a ``flat
plate'' model. In this case, the spacecraft is treated as a flat
surface, oriented at same angle with respect to incoming solar
rays. The surface absorbs some solar heat, while it reflects the rest;
reflection can be specular or diffuse. A flat plate model is fully
characterized by three numbers: the area of the plate, its specular
and its diffuse reflectivities. This model is particularly applicable
to Pioneer~10 and 11 throughout most of their mission, as the
spacecraft were oriented such that their large parabolic dish antennas
were aimed only a few degrees away from the Sun, and most of the
spacecraft body was behind the antenna, not exposed to sunlight.

In the case of a flat plate model, the force produced by the solar
pressure can be described using a combination of several force
vectors. One vector, the direction of which coincides with the
direction of incoming solar radiation, represents the force due to
photons from solar radiation intercepted by the spacecraft. The
magnitude of this vector $\vec{F}_\mathrm{intcpt}$ is proportional to
the solar constant at the spacecraft's distance from the Sun,
multiplied by the projected area of the flat plate surface:
\begin{equation}
\vec{F}_\mathrm{intcpt}=\frac{f_\odot A}{cr^2}(\vec{n}\cdot\vec{k})\vec{n},
\label{eq:srp}
\end{equation}
where $\vec{r}$ is the Sun-spacecraft vector, $\vec{n}=\vec{r}/r$ is
the unit vector in this direction with $r=|\vec{r}|$, $A$ is the
effective area of the spacecraft (i.e., flat plate), $\vec{k}$ is a
unit normal vector to the flat plate, $f_\odot$ is the solar radiation
constant at 1~AU from the Sun and $c$ is the speed of light. The
standard value of the solar radiation constant is $f_\odot\simeq
1367~\mathrm{AU}^2\mathrm{\ Wm}^{-2}$ when $A$ is measured in
units of m$^2$ and $\vec{r}$, in units of AU. According
to Equation~(\ref{eq:srp}), approximately 65~W of intercepted sunlight can
produce a force comparable in magnitude to that of the Pioneer
anomaly; in contrast, in the vicinity of the Earth, the Pioneer~10 and
11 spacecraft intercepted $\sim$~7~kW of sunlight, indicating that
solar pressure is truly significant, even as far away from the Sun as
Saturn, for precision orbit determination.

Equation~(\ref{eq:srp}) reflects the amount of momentum carried by
solar photons that are intercepted by the spacecraft body. However,
one must also account for the amount of momentum carried away by
photons that are reflected or re-emitted by the spacecraft body. These
momenta depend on the material properties of the spacecraft exterior
surfaces. The absorptance coefficient $\alpha$ determines the amount
of sunlight absorbed (i.e., not reflected) by spacecraft
materials. The emittance coefficient $\epsilon$ determines the
efficience with which the spacecraft radiates (absorbed or internally
generated) heat relative to an idealized black body. Finally, the
specularity coefficient $\sigma$ determines the direction in which
sunlight is reflected: a fully specular surface reflects sunlight like
a mirror, whereas a diffuse (Lambertian) surface reflects light in the
direction of its normal. Together, these coefficients can be used in
conjunction with basic vector algebra to calculate the force acting on
the spacecraft due to specular reflection:

\begin{equation}
\vec{F}_\mathrm{spec}=(1-\alpha)\sigma[\vec{F}_\mathrm{intcpt}-2(\vec{F}_\mathrm{intcpt}\cdot\vec{k})\vec{k}],
\end{equation}
and the force due to diffuse reflection:
\begin{equation}
\vec{F}_\mathrm{diffuse}=(1-\alpha)(1-\sigma)|\vec{F}_\mathrm{intcpt}|\vec{k}.
\end{equation}

Lastly, the force due to solar heating (i.e., re-emission of absorbed
solar heat) can be computed in conjunction with the recoil force due
to internally generated heat, which is discussed later in this
section.

\subsubsection{Solar wind}
\label{sec:force_solwind}

The solar wind is a stream of charged particles, primarily protons and
electrons with energies of $\sim$~1~keV, ejected from the upper
atmosphere of the Sun. Solar wind particles intercepted by a
spacecraft transfer their momentum to the spacecraft. The acceleration
caused by the solar wind has the same form as Equation~(\ref{eq:srp}), with
$f_\odot$ replaced by $m_pv^3n$, where $n\approx5\mathrm{\ cm}^{-3}$ is the
proton density at 1~AU and $v\approx400\mathrm{\ km/s}$ is the speed of the wind
(electrons in the solar wind travel faster, but due to their smaller
mass, their momenta are much smaller than the momenta of the
protons). Thus,
\begin{equation}
\vec{F}_\mathrm{solar~wind}=\frac{m_pv^3nA}{cr^2}(\vec{n}\cdot\vec{k})\vec{n}\simeq 7\times 10^{-4}\frac{A}{r^2}(\vec{n}\cdot\vec{k})\vec{n}.
\label{eq:sw}
\end{equation}
Because the density can change by as much as 100\%, the exact
acceleration is unpredictable. Nonetheless, as confirmed by actual
measurement\epubtkFootnote{See
  \url{http://www.ngdc.noaa.gov/stp/SOLAR/IRRADIANCE/irrad.html}.},
the magnitude of Equation~(\ref{eq:sw}) is at least 10\super{5} times smaller
than the direct solar radiation pressure. This contribution is
completely negligible~\cite{pioprd}, and therefore, it can be safely
ignored.

\subsubsection{Interaction with planetary environments}
\label{sec:force_planrad}

When a spacecraft is in the vicinity of a planetary body, it interacts
with that body in a variety of ways. In addition to the planet's
gravity, the spacecraft may be subjected to radiation pressure from
the planet, be slowed by drag in the planet's extended atmosphere, and
it may interact with the planet's magnetosphere.

For instance, for Earth orbiting satellites, the Earth's optical
albedo of~\cite{MG2005}:
\begin{equation}
\alpha_\mathrm{Earth}\simeq 0.34
\end{equation}
yields typical albedo accelerations of 10\,--\,35\% of the
acceleration due to solar radiation pressure. On the other hand, when
the spacecraft is in the planetary shadow, it does not receive direct
sunlight.

Atmospheric drag can be modeled as follows~\cite{MG2005}:
\begin{equation}
\ddot{\vec{r}}=-\frac{1}{2}C_D\frac{A}{m}\rho|\dot{\vec{r}}|\dot{\vec{r}},
\end{equation}
where $\vec{r}$ is the spacecraft's position, $\dot{\vec{r}}$ its
velocity, $A$ is its cross-sectional area, $m$ its mass, $\rho$ is the
atmospheric density, and the coefficient $C_D$ has typical values
between 1.5 and 3.

The Lorentz force acting on a charged object with charge $q$ traveling
through a magnetic field with field strength $\vec{B}$ at a velocity
$\vec{v}$ is given by
\begin{equation}
\vec{F}=q(\vec{v}\times\vec{B}).
\label{eq:Lorentz-force}
\end{equation}
Considering the velocity of the Pioneer spacecraft relative to a
planetary magnetic field during a planetary encounter (up to 60~km/s
during Pioneer~11's encounter with Jupiter) and a strong planetary
magnetic field (up to 1~mT for Jupiter near the poles), if the
spacecraft carries a net electric charge, the resulting force can be
significant: up to 60~N per Coulomb of charge. In actuality, the
maximum measured magnetic field by the two Pioneers at Jupiter was
113.5~$\mu$T~\cite{1976AJ.....81.1153N}. An upper bound of 0.1~$\mu$C
exists for any positive charge carried by the
spacecraft~\cite{1976AJ.....81.1153N}, but a possible negative charge
cannot be excluded~\cite{1976AJ.....81.1153N} and a negative charge as
high as 10\super{-4}~C cannot be ruled out~\cite{1976AJ.....81.1153N}.

The long-term accelerations of Pioneer~10 and 11, however, remain
unaffected by planetary effects, due to the fact that except for brief
encounters with Jupiter and Saturn, the two spacecraft traveled in
deep space, far from any planetary bodies.

\subsubsection{Interplanetary magnetic fields}

The interplanetary magnetic field strength is less than
1~nT~\cite{pioprd}. Considering a spacecraft velocity of 10\super{4}~m/s
and a charge of 10\super{-4}~C, Equation~(\ref{eq:Lorentz-force}) gives a force of
10\super{-9}~N or less, with a corresponding acceleration (assuming a
spacecraft mass of $\sim$~250~kg) of $4\times 10^{-12}\mathrm{\ m/s}^2$ or
less. This value is two orders of magnitude smaller than the anomalous
Pioneer acceleration of $a_P=(8.74\pm1.33)\times10^{-10}\mathrm{\ m/s}^2$ (see
Section~\ref{sec:anomaly_sum}).

\subsubsection{Drag forces}
\label{sec:force_drag}

While there have been attempts to explain the anomalous acceleration
as a result of a drag force induced by exotic forms of matter (see
Section~\ref{sec:new-physics}), no known form of matter (e.g., gas,
dust particles) in interplanetary space produces a drag force of
significance.

The drag force on a sail was estimated as~\cite{Nieto:2003qy}:
\begin{equation}
\vec{F}_\mathrm{sail}=-{\cal K}_d\rho Av^2,
\end{equation}
where $v$ is the spacecraft's velocity relative to the interplanetary
medium, $A$ its cross sectional area, $\rho$ is the density of the
interplanetary medium, and ${\cal K}_d$ is a dimensionless coefficient
that characterizes the absorptance, emittance, and transmittance of
the spacecraft with respect to the interplanetary medium.

Using the Pioneer spacecraft's 2.74~m high-gain antenna as a sail and
an approximate velocity of 10\super{-4}~m/s relative to the interplanetary
medium, and assuming ${\cal K}_d$ to be of order unity, we can
estimate a drag force of
\begin{equation}
\vec{F}_\mathrm{sail}\simeq -5.9\times 10^{8}\rho,
\end{equation}
with $\vec{F}$ and $\rho$ measured in SI units. According to this
result, a density of $\rho\sim 3\times 10^{-16}\mathrm{\ kg/m}^3$ or higher
can produce accelerations that are comparable in magnitude to the
Pioneer anomaly~\cite{Marmet2003}.

The density $\rho_\mathrm{ISD}$ of dust of interstellar origin has
been measured by the Ulysses
probe~\cite{2000JGR...10510317M,2005PhLB..613...11N} at
$\rho_\mathrm{ISD}\lesssim 3\times 10^{-23}\mathrm{\ kg/m}^3$. The average
interplanetary dust density, which also contains orbiting dust, is
believed to be almost two orders of magnitude higher according to the
consensus view~\cite{2005PhLB..613...11N}. However, higher dust
densities are conceivable.

On the other hand, if one {\em presumes} a model density, the
constancy of the observed anomalous acceleration of the Pioneer
spacecraft puts upper limits on the dust density. For instance, an
isothermal density model $\rho_\mathrm{isoth}\propto r^{-2}$ yields
the limit $\rho_\mathrm{isoth}\lesssim 5\times 10^{-17}
\,(20~\mathrm{AU}/r)^2\mathrm{\ kg/m}^3$.

\subsection{Nongravitational forces of on-board origin}
\label{sec:nongrav}

Perhaps the most fundamental question concerning the anomalous
acceleration of Pioneer~10 and 11 is whether or not the acceleration
is due to an on-board effect: i.e., is the ``anomalous'' acceleration
simply a result of our incomplete understanding of the engineering
details of the two spacecraft? Therefore, it is essential to analyze
systematically any possible on-board source of acceleration that may
be present.

In the broadest terms, momentum conservation dictates that in order
for an on-board effect to accelerate the spacecraft, the spacecraft
must eject mass or emit radiation. As no significant anomaly occurred
in the Pioneer~10 and 11 missions, it is unlikely that either
spacecraft lost a major component during their cruise. In any case,
such an occurrence would have resulted in a one time change in the
spacecraft's velocity, not any long-term acceleration. Therefore, it
is safe to consider only the emission of volatiles as a means of mass
ejection. Such emissions can be intentional (as during maneuvers) or
due to unintended leaks of propellant or other volatiles on
board. Radiation emitted as radiative energy is produced by
on-board processes.

The spin of the Pioneer spacecraft makes it possible to apply a
simplified treatment of forces of on-board origin that change slowly
with time. Let us denote the unit vector normal to the spacecraft's
plane of rotation (i.e., the spin axis, which we assume to remain
constant in time) by $\vec{s}$. Then, considering a force $\vec{F}$
that is a linear function of time in a co-rotating reference frame
that is attached to the spacecraft, it can be described in a
co-moving (nonrotating) inertial frame as
\begin{equation}
\vec{F}(t)=\vec{F}_\parallel(t)+\vec{F}_\perp(t)=\vec{F}_\parallel(t)+\mathbb{R}(\omega t)\cdot[\vec{F}_\perp(t_0)+\dot{\vec{F}}_\perp \cdot (t-t_0)],
\end{equation}
where $\vec{F}_\parallel(t)=[\vec{F}(t)\cdot\vec{s}]\vec{s}$ is the
component of $\vec{F}(t)$ parallel with the spin axis and
$\vec{F}_\perp(t)=\vec{F}(t)-\vec{F}_\parallel(t)$ is the
perpendicular component of $\vec{F}(t)$. The component
$\vec{F}_\parallel$ accelerates the spacecraft in the spin axis
direction. The displacement due to the perpendicular component can be
obtained by double integration with integration limits of $t=t_0$ and
$t=t_0+2\pi n/\omega$:
\begin{equation}
\Delta\vec{x}_\perp=\frac{1}{\omega^2m}\mathbb{R}(\omega t_0-\pi)\cdot\dot{\vec{F}}_\perp\Delta t,
\end{equation}
describing an arithmetic spiral around the spin axis. This spiral
vanishes (i.e., the displacement of the spacecraft remains confined
along the spin axis) if $\dot{\vec{F}}=0$, and even for nonzero
$\dot{\vec{F}}$ its radius increases only linearly with time, and thus
in most cases, it can be ignored safely.

\subsubsection{Modeling of maneuvers}
\label{sec:force_maneuvers}

There were several hundred\epubtkFootnote{The exact number of maneuvers is
  unknown due to incomplete records, although most maneuvers can be
  reconstructed from the preserved telemetry.} Pioneer~10 and
Pioneer~11 maneuvers during their entire missions. The modeling of
maneuvers entails significant uncertainty due to several
reasons. First, the duration of a thruster firing is known only
approximately and may vary between maneuvers due to thermal and
mechanical conditions, aging, manufacturing deficiencies in the
thruster assembly, and other factors. Second, the thrust can vary as a
result of changing fuel temperature and pressure. Third, imperfections
in the mechanical mounting of a thruster introduce uncertainties in
the thrust direction. Lastly, after a thruster has fired, leakage may
occur, producing an additional, small amount of slowly decaying
thrust. When combined, these effects result in a velocity change of
several mm/s.

By the time Pioneer~11 reached Saturn, the behavior of its thrusters
was believed to be well understood~\cite{pioprd}. The effectively
instantaneous velocity change caused by the firing of a thruster was
followed by several days of decaying acceleration due to gas
leakage. This acceleration was large enough to be observable in the
Doppler data~\cite{null81}.

The Jet Propulsion Laboratory's analysis of Pioneer orbits included
either an instantaneous velocity increment at the beginning of each
maneuver (instantaneous burn model) or a constant acceleration over
the duration of the maneuver (finite burn model)~\cite{pioprd}. In
both cases, the burn is characterized by a single unknown
parameter. The gas leak following the burn was modeled by fitting to
the post-maneuver residuals a two-parameter exponential model in the
form of
\begin{equation}
\Delta v(t)=v_0\exp (-t/\tau),
\end{equation}
with $v_0$ and $\tau$ being the unknown parameters. The typical
magnitude of $v_0$ is several mm/s, while the time constant $\tau$ is
of order $\sim$~10~days. Due to the spin of the spacecraft, only
acceleration in the direction of the spin axis needs to be accounted
for, as accelerations perpendicular to the spin axis are averaged out
over several resolutions~\cite{pioprd, Toth2009}.

\subsubsection{Other sources of outgassing}
\label{sec:gas_leaks}

Regardless of the source of a leak, the effects of outgassing on the
spacecraft are governed by the rocket equation~\cite{pioprd}:
\begin{equation}
a=-v_e\frac{\dot m}{m},\label{eq:rocket}
\end{equation}
where the dot denotes differentiation with respect to time, and $v_e$
is the exhaust velocity. The Pioneer spacecraft mass is approximately
$m\simeq 250\mathrm{\ kg}$. For comparison, the anomalous acceleration,
$a_P=8.74\times 10^{-10}\mathrm{\ m/s}^2$, requires an outgassing rate of
$\sim$~6.89~g/yr at an exhaust velocity of 1~km/s.

The exhaust velocity $v_e$ of a hot gas, according to the rocket
engine nozzle equation, can be calculated
as~\cite{1992wi...book.....S}\epubtkFootnote{See also
  \url{http://en.wikipedia.org/wiki/Rocket_engine_nozzles}}:
\begin{equation}
v_e^2=\frac{2kRT}{(k-1)M_\mathrm{mol}}\bigg[1-\bigg(\frac{P_e}{P_i}\bigg)^{(k-1)/k}\bigg],\label{eq:nozzle}
\end{equation}
where $k$ is the isentropic expansion factor (or heat capacity ratio,
$k=C_p/C_v$ where $C_p$ and $C_v$ are the heat capacities at constant
pressure and constant volume, respectively) of the exhaust gas, $T$ is
its temperature, $R=8314\mathrm{\ JK}^{-1}\mathrm{\ kmol}^{-1}$ is the
gas constant, $m_\mathrm{mol}$ is the molecular weight of the exhaust gas in
kg/kmol, $P_i$ is its pressure at the nozzle intake, and $P_e$ is the
exhaust pressure. At room temperature, $k=1.41$ for H\sub{2}, $k=1.66$
for He, and $k=1.40$ for O\sub{2}. Typical values for liquid
monopropellants are $1.7~\mathrm{\ km/s}~<v_e<2.9\mathrm{\ km/s}$.

A review of the Pioneer~10 and 11 spacecraft design reveals only three
possible sources of outgassing: the propulsion system (fuel leaks),
the radioisotope thermoelectric generators, and the battery.

The propulsion system carried $\sim$~30~kg of hydrazine propellant and
N$_2$ pressurant. Loss of either due to a leak could produce a
constant or slowly changing acceleration term. Propellant and
pressurant can be lost due to a malfunction in the propulsion system,
and also due to the regular operation of thruster valves, which are
known to have small, persistent leaks lasting days or even weeks after
each thruster firing event, as described above in
Section~\ref{sec:force_maneuvers}. While the possibility of additional
propellant leaks cannot be ruled out, in order for such leaks to be
responsible for a constant acceleration like the anomalous
acceleration of Pioneer~10 and 11, they would have had to be
\begin{inparaenum}[i)]
\item constant in time;
\item the same on both spacecraft;
\item not inducing any detectable changes in the spin rate or precession.
\end{inparaenum}
Given these considerations, Anderson et al. \textit{conservatively}
estimate that undetected gas leaks introduce an uncertainty not
greater than
\begin{equation}
\sigma_\mathrm{gl}=\pm 0.56\times 10^{-10}~\mathrm{m}/\mathrm{s}^2.
\end{equation}

Outgassing can also occur in the radioisotope thermoelectric
generators as a result of alpha decay. Each kg of \super{238}Pu produces
$\sim$~0.132~g of helium annually; the total amount of helium produced
by the approx. 4.6~kg of radioisotope fuel on board\epubtkFootnote{The fuel
  inventory quoted in~\cite{pioprd} is 5.8~kg. This much \super{238}Pu
  would produce significantly more heat at the rate of 0.57~W/g than
  the known thermal power of the Pioneer RTGs. Likely, 5.8~kg was the
  total mass of the plutonium-molybdenum cermet pucks on board, which
  contained approximately 4.6~kg plutonium metal.} is, therefore,
0.6~g/year. Exterior temperatures of the RTGs at no point exceeded
320~\textdegree~F=433~K. According to Equation~(\ref{eq:nozzle}), the
corresponding exhaust velocity is 2.13~km/s, resulting in an
acceleration of $1.62\times 10^{-10}\mathrm{\ m/s}^2$. (This is slightly
larger than the corresponding estimate in~\cite{pioprd}, where the
authors adopted the figures of $\dot{m}=0.77\mathrm{\ g/year}$ and
$v_e=1.16\mathrm{\ km/s}$.) However, the circumstances required to achieve this
acceleration are highly unrealistic, requiring all the helium to be
expelled at maximum efficiency and in the spin axis direction. Using a
more realistic (but still conservative) scenario, Anderson et
al. estimate the bias and error in acceleration due to He-outgassing
as
\begin{equation}
a_\mathrm{He}=(0.15\pm 0.16)\times 10^{-10}~\mathrm{m}/\mathrm{s}^2.
\end{equation}

Another source of possible outgassing not previously considered may be
the spacecraft's battery. According to Equation~(\ref{eq:nozzle}), H\sub{2} gas
leaving the battery system at a temperature of 300~K can acquire an
exhaust velocity of 92.6~m/s. For O\sub{2} at 300~K, the exhaust velocity
is 23.2~m/s. At these velocities, an outgassing of $\sim$~74~g/year of
H\sub{2} or 298~g/year of O\sub{2} can produce an acceleration equal to
$a_P$, so the battery cannot be ruled out in principle as a source of
a near constant acceleration term. However, no realistic
construction~\cite{BATTERY} for a 5~A, 11.3~V AgCd battery would
provide near enough volatile electrolites for such outgassing to
occur, and in any case, the nominal performance of the battery system
for a far longer time period than designed indicates that no
significant loss of volatiles from the battery has taken place. A
conservative (but still generous) estimate using a battery of maximum
weight, 2.35~kg, assuming a loss of 10\% of its mass over 30 years,
and a thrust efficiency of 50\% yields
\begin{equation}
\sigma_\mathrm{bat}=\pm 0.14\times 10^{-10}~\mathrm{m}/\mathrm{s}^2.
\end{equation}

\subsubsection{Thermal recoil forces}
\label{sec:force_recoil}

The spacecraft carried several on-board energy sources that produced
waste heat (see Section~\ref{sec:heat}). Most notably among these are
the RTGs; additional heat was produced by electrical
instrumentation. Further heat sources include Radioisotope Heater
Units and the propulsion system.

As the spacecraft is in an approximate thermal steady state, heat
generated on board must be removed from the
spacecraft~\cite{Toth2009}. In deep space, the only mechanism of heat
removal is thermal radiation: the spacecraft can be said to be
radiatively coupled to the cosmic background, which can be modeled by
surrounding the spacecraft with a large, hollow spherical black body
at the temperature of $\sim$~2.7~K.

As the spacecraft emits heat in the form of thermal photons, these
also carry momentum $p_\gamma$, in accordance with the well known law
of $p_\gamma=h\nu/c$, where $\nu$ is the photon's frequency, $h$ is
Planck's constant, and $c$ is the velocity of light. This results in a
recoil force in the direction opposite to that of the path of the
photon. For a spherically symmetric body, the net recoil force is
zero. However, if the pattern of radiation is not symmetrical, the
resulting anisotropy in the radiation pattern yields a net recoil
force.

The magnitude of this recoil force is a subject of many factors,
including the location and thermal power of heat sources, the
geometry, physical configuration, and thermal properties of the
spacecraft's materials, and the radiometric properties of its external
(radiating) surfaces.

Key questions concerning the thermal recoil force that have been
raised during the study of the Pioneer anomaly
include~\cite{1999PhRvL..83.1892K, 1999PhRvL..83.1890M,
  2003PhRvD..67h4021S}:

\begin{itemize}
\item How much heat from the RTGs is reflected by the spacecraft,
  notably the rear of its high-gain antenna, and in what direction?
\item \vskip -6pt Was there a fore-aft asymmetry in the radiation
  pattern of the RTGs due to differential aging?
\item \vskip -6pt How much electrical heat generated on-board was
  radiated through the spacecraft's louver system?
\end{itemize}

The recoil force due to on-board generated heat that was emitted
anisotropically was recognized early as a possible origin of the
Pioneer anomaly. The total thermal inventory on board the Pioneer
spacecraft exceeded 2~kW throughout most of their mission
durations. The spacecraft were in an approximate steady state: the
amount of heat generated on-board was equal to the amount of heat
radiated by the spacecraft.

The mass of the Pioneer spacecraft was $\sim$~250~kg. An acceleration
of $8.74\times 10^{-10}\mathrm{\ m/s}^2$ is equivalent to a force of $\sim
0.22~\mu$N acting on a $\sim$~250~kg object. This is the amount of
recoil force produced by a 65~W collimated beam of photons. In
comparison with the available thermal inventory of 2500~W, a fore-aft
anisotropy of less than 3\% can account for the anomalous acceleration
in its entirety. Given the complex shape of the Pioneer spacecraft, it
is certainly conceivable that an anisotropy of this magnitude is
present in the spacecrafts' thermal radiation pattern.

The issue of the thermal recoil force remains a subject of on-going
study, as estimates of the actual magnitude of this force may require
significant revision in the light of new data and new
investigations~\cite{2006CaJPh..84.1063T,2007arXiv0710.2656T,
  Toth2009, MDR2005}.

\subsubsection{The radio beam recoil force}
\label{sec:force_radio}

Throughout most of their missions, the Pioneer~10 and 11 spacecraft
were transmitting continuously in the direction of the Earth using a
highly focused microwave radio beam that was emitted by the high gain
antenna (HGA; see Section~\ref{sec:radiobeam}). The recoil force due
to the radio beam can be readily calculated.

A naive calculation uses the nominal value of the radio beam's power
(8~W), multiplied by the reciprocal of the velocity of light,
$c^{-1}$, to obtain the radio beam recoil force. This is a useful way
to estimate the recoil force, but it may need refinement.

The spacecraft's radio transmission is concentrated into a very narrow
beam: signal attenuation exceeds 20~dB at only 3.75\textdegree\ deviation
from the antenna centerline (see Figure~3.6-13 in~\cite{PC202}). The
projected transmitter power $P_0$ in the beam direction can be
computed using the integral $P_0=\int d\theta\sin{\theta}{\cal
  P}(\theta)$ where $\cal{P}(\theta)$ is the angular power
distribution of the antenna. Given the antenna power distribution, we
find that $P_0=P$, where $P$ is the total power radiated by the
antenna, to an accuracy much better than 1\%. For this reason, the
shape of the transmission beam needs not be taken into account when
computing the recoil force.

However, as discussed in Section~\ref{sec:radiobeam}, the power of the
spacecraft's transmitter was not constant in time: if the telemetry
readings are accepted as reliable, transmitter power may have
decreased by as much as 3~W or more near the end of Pioneer~10's
mission. Furthermore, some (estimated $\sim$~10\%) of the radio beam
may have missed the antenna dish altogether, resulting in a reduced
efficiency with which the energy of the spacecraft's transmitter is
converted into momentum.

Note that the navigational model that was used to navigate the
Pioneers did not include this effect. It became clear only recently
leading to the need to include this model as a part of the on-going
efforts to re-analyze the Pioneer data.

\subsection{Effects on the radio signal}

The radio signal to or from the Pioneer spacecraft travels several
billion kilometers in interplanetary space. Unsurprisingly, the
interplanetary environment, notably charged particles emitted by the
Sun and the gravitational fields in the solar system all affect the
length of the path that the radio signal travels and its frequency.

The communication antennas of the DSN complex that are used to
exchange data with the Pioneer spacecraft are located on the Earth's
surface. This introduces many corrections to the modeling of the
uplinked or downlinked radio signal due to the orbital motion,
rotation, internal dynamics and atmosphere of our home planet.

\subsubsection{Plasma in the solar corona and weighting}
\label{sec:effect_solplasm}

The interplanetary medium in the solar system is dominated by the
solar wind, i.e., charged particles originating from the Sun. Although
their density is low, the presence of these particles has a noticeable
effect on a radio frequency signal, especially when the signal passes
relatively close to the Sun.

Delay due to solar plasma is a function of the electron density in the
plasma. Although this can vary significantly as a result of solar
activity, the propagation delay $\Delta t$ (in microseconds) can be
approximated using the formula~\cite{DSMS106}:
\begin{equation}
\Delta t=-\frac{1}{2cn_\mathrm{crit}(f)}\int_\oplus^{SC}d\ell n_e(t,\vec{r}),
\label{eq:sol_plasma}
\end{equation}
where $f$ is the signal frequency, $n_{\mathrm{crit}}(f)$ is is the
critical plasma density for frequency $f$ that is given by
\begin{equation}
n_\mathrm{crit}(f)=1.240\times 10^4\left(\frac{f}{1~\mathrm{MHz}}\right)^2~\mathrm{cm}^{-3},
\end{equation}
and $n_e$ is the electron density as a function of time $t$ and
position $\vec{r}$, which is integrated along the propagation path
$\ell$ between the spacecraft and the Earth.

We write the electron density as a sum of a static, steady-state
part, $n_e(\vec{r})$ and fluctuation $\delta
n_e(t,\vec{r})$~\cite{Turyshev:2005zk}:
\begin{equation}
n_e(t,\vec{r})=n_e(\vec{r})+\delta n_e(t,\vec{r}).
\end{equation}
The second term, which is difficult to quantify, has only a small
effect on the Doppler observable~\cite{Turyshev:2005zk}, except at
conjunction, when noise due to the solar corona dominates the Doppler
observable. In contrast, the steady-state behavior of the solar corona
is well known and can be approximated using the
formula~\cite{anderson74, DSMS106, MuhlemanAnderson81,
  MuhlemanEspositoAnderson77}:
\begin{equation}
n_e(t,\vec{r})=A\Big(\frac{R_\odot}{r}\Big)^2+B\Big(\frac{R_\odot}{r}\Big)^{2.7}e^{-\left[\frac{\phi}{\phi_0}\right]^2}+C\Big(\frac{R_\odot}{r}\Big)^6.
\label{eq:corona_model_content}
\end{equation}
where $R_0=6.96\times10^8\mathrm{\ m}$ is the solar radius, and $r$ is the
distance from the Sun along the propagation path.

Using Equation~(\ref{eq:corona_model_content}) in Equation~(\ref{eq:sol_plasma}), we
obtain the range model~\cite{pioprd}:
\begin{equation}
\Delta\mathrm{range}=\pm\Big(\frac{f_0}{f}\Big)^2\Big[A\Big(\frac{R_\odot}{\rho}\Big)F+B\Big(\frac{R_\odot}{\rho}\Big)^{1.7}e^{-\left[\frac{\phi}{\phi_0}\right]^2}+C\Big(\frac{R_\odot}{\rho}\Big)^{5}\Big],
\end{equation}
where $f_0=2295\mathrm{\ MHz}$ is the reference frequency used for the analysis
of Pioneer~10, $\rho$ is the impact parameter with respect to the Sun,
and $F$ is a light-time correction factor, which is given for distant
spacecraft as
\begin{equation}
F=F(\rho,r_T,r_E)=\frac{1}{\pi}\Big[\arctan\Big(\frac{\sqrt{r_T^2-\rho^2}}{\rho}\Big)+\arctan\Big(\frac{\sqrt{r_E^2-\rho^2}}{\rho}\Big)\Big],
\label{eq:weight_doppler*}
\end{equation}
where $r_T$ and $r_E$ are the heliocentric radial distances to the
target and to the Earth, respectively. The sign of the solar corona
range correction is negative for Doppler measurements (positive for
range).

The values of the parameters $A$, $B$, and $C$ are: $A=6.0\times 10^3,
B= 2.0\times 10^4, C= 0.6\times 10^6$, all in meters~\cite{pioprd}.

\subsubsection{Effects of the ionosphere}
\label{sec:effect_iono}

As the radio signal to or from the spacecraft travels through the
Earth's ionosphere, it suffers an additional propagation delay due to
the presence of charged particles. This delay $\Delta t$ can be
modeled as~\cite{MEYER2005, Schaer1999}
\begin{equation}
\Delta t_{\tt iono}=-\frac{1}{cN\sin\theta}\int_0^{h_{\tt max}} N_{\mathrm{iono}}~dh,
\end{equation}
where $\theta$ is the antenna elevation, $N$ is the atmospheric
refractivity index, $N_{\mathrm{iono}}$ is the ionospheric refractivity
index, and $h_{\tt max}$ is the height of the ionosphere. $N$ can be
approximated at 10\super{6}, while $N_{\mathrm{iono}}$ is well approximated
by the formula
\begin{equation}
N_\mathrm{iono}=-40.28\times 10^6\frac{n_e}{f^2},
\end{equation}
where $n_e$ is the electron density and $f$ is the signal frequency in
Hz. We introduce the total electron content,
\begin{equation}
N_e=\int_0^{h_{\tt max}} n_e~dh,
\end{equation},
which allows us to express the propagation delay in the form,
\begin{equation}
\Delta t_{\tt iono}=\frac{40.28}{cf^2}N_e,
\end{equation}
with $c=3\times 10^8\mathrm{\ m/s}$. For European latitudes, the total
electron content may vary from very few electrons at night to
(20\,--,100)~\texttimes~10\super{16} electrons during the day at
various stages during the solar cycle.

\subsubsection{Effects of the troposphere}
\label{sec:effect_tropo}

Chao (\cite{SOVERS1996}; see also~\cite{JPL9424, IERS32, TDA42122,
  KIS94, Schuler2001}) estimates the delay due to signal propagation
through the troposphere using the following formula:
\begin{equation}
\Delta l_{\tt tropo}=\frac{1}{\sin\theta+A/(\tan\theta+B)},
\end{equation}
where $\Delta l_{\tt tropo}$ is the additional propagation path,
$\theta$ is the elevation angle, and $A=A_\mathrm{dry}+A_\mathrm{wet}$
and $B=B_\mathrm{dry}+B_\mathrm{wet}$ are coefficients defined as
\begin{eqnarray}
A_\mathrm{dry}&=&0.00143,\\
A_\mathrm{wet}&=&0.00035,\\
B_\mathrm{dry}&=&0.0445,\\
B_\mathrm{wet}&=&0.017.
\end{eqnarray}

Unfortunately, historical weather data going back over 30 years may
not be available for most DSN stations. In the absence of such data,
C.B.~Markwardt suggests that seasonal weather data or historical
weather data from nearby weather stations can be used to achieve good
modeling accuracy.\epubtkFootnote{C.B.~Markwardt, private communication.}

\subsubsection{The effect of spin}
\label{sec:effect_spin}

The radio signal emited by the DSN and the radio signal returned by
the Pioneer~10 and 11 spacecraft are circularly polarized. The
spacecraft themselves are spinning, and the spin axis coincides with
the axis of the HGA. Therefore, every revolution of the spacecraft
adds a cycle to both the radio signal received by, and that
transmitted by the spacecraft.

At a nominal rate of 4.8 revolutions per minute, the spacecraft spin
adds 0.08~Hz to the radio signal frequency in each direction.

The sign of the spin contribution to the spacecraft frequency depends
on whether or not the radio signal is left or right circularly
polarized, and the direction of the spacecraft's rotation.

The rotation of the spacecraft is clockwise~\cite{PC202} as viewed
from a direction behind the spacecraft, facing towards the Earth. This
implies that the spacecraft spin would contribute to the frequency of
a right circularly polarized (as seen from the transmitter) signal's
frequency with a positive sign. The assumption that the DSN signal is
right circularly polarized is consistent with the explanation provided
in~\cite{Moyer-2003}. This interpretation of the spacecraft's spin in
relation to the radio signal agrees with what one finds when
comparing orbit data files with or without previously applied spin
correction.

The total amount of spin correction, therefore, must be written as
\begin{equation}
\Delta_\mathrm{spin}f=\left(1+\frac{240}{221}\right)\frac{\omega}{2\pi},
\end{equation}
where $\omega$ is the angular velocity of the spacecraft, and we
accounted for the Pioneer communication system turnaround ratio of
240/221.

\subsubsection{Station locations}
\label{sec:force_stations}

Accurate estimation of the amount of time it takes for a signal to
travel between a DSN station and a distant spacecraft, and the
frequency shift due to the relative motion of these, requires precise
knowledge of the position and velocity of not just the spacecraft
itself, but also of any ground stations participating in the
communication.

DSN transmitting and receiving stations are located on the surface of
the Earth. Therefore, their coordinates in a solar system barycentric
frame of reference are determined primarily by the orbital motion,
rotation, precession and nutation of the Earth.

In addition to these motions of the Earth, station locations also
change relative to a geocentric frame of reference due to tidal
effects and continental drift.

Information about station locations is readily available for stations
presently in operation; however, for stations that are no longer
operating, or for stations that have been relocated, it is somewhat
more difficult to obtain (see Section~\ref{sec:ant}).

The transformation of station coordinates from a terrestrial reference
frame, such as ITRF93, to a celestial (solar system barycentric) reference
frame can be readily accomplished using publicly available algorithms
or software libraries, such as NASA's SPICE library\epubtkFootnote{See also
\url{http://naif.jpl.nasa.gov/}.} \cite{1996PSS...44...65A}.

\subsection{Modeling the radiometric Doppler observable}

The Pioneer spacecraft were navigated using radiometric Doppler
data\epubtkFootnote{Some ranging observations were also performed for the
  Pioneer spacecraft, by varying the frequency of the transmitted
  signal and observing the corresponding changes in the signal
  received from the spacecraft.}. The Doppler observable is defined as
the difference between the number of cycles received by a receiving
station and the number of cycles produced by a (fixed or ramped) known
reference frequency, during a specific count interval.

The expected value of the Doppler observable can be calculated
accurately if the trajectory of the transmitting station (e.g., a
spacecraft) and receiving station (e.g., a ground-based tracking
station) are known accurately, along with information about the
transmission medium along the route of the received signal.

The trajectory of the spacecraft is determined using the spacecraft's
initial position and velocity according to Section~\ref{sec:gravity},
in conjunction with a model of nongravitational forces, as detailed
in Section~\ref{sec:nongrav}.

The signal propagation delay due to the gravitational field of solar
system can be calculated using Equation~(\ref{eq:light-time}). Afterwards,
from the known arrival times of the first and last cycle during a
Doppler count interval, the times of their transmission can be
obtained. Given the known frequency of the transmitter, one can then
calculate the actual number of cycles that were transmitted during
this interval. Comparing the two figures gives the difference known as
the Doppler residual.

\epubtkImage{}{%
  \begin{figure}[h!]
    \centerline{\includegraphics[width=\linewidth]{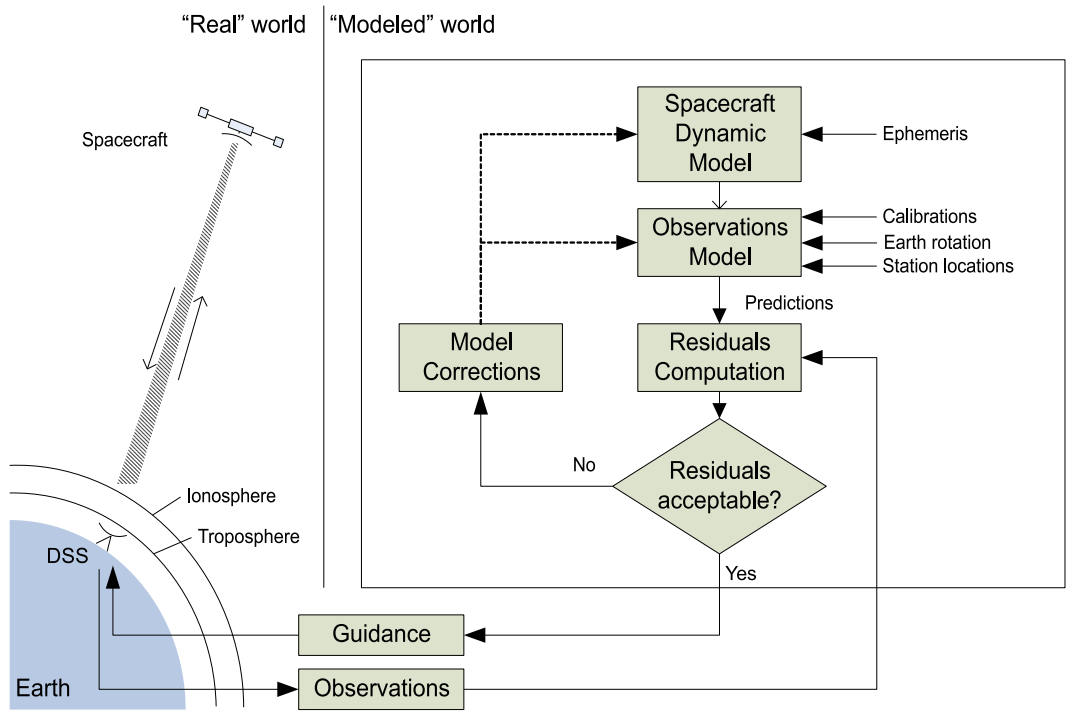}}
    \caption{Schematic overview of the radio navigation
    process. Adapted from~\cite{THORNTON2000}.}
    \label{fig:radio-nav}
\end{figure}}

The model parameters, which include an estimate of the spacecraft's
initial state vector and other factors, can be adjusted to achieve a
``best fit'' between model and observation. A commonly used method to
achieve a best fit is the use of a least squares estimator. The
``solve-for'' parameters can include orbital parameters of solar
system bodies; the visible light and infrared radiometric properties
of the spacecraft; or properties of the Earth's atmosphere or the
interplanetary medium.

While a spacecraft is in flight, the revised model can be used to make
navigational predictions and provide guidance, as depicted in
Figure~\ref{fig:radio-nav}. In the case of the historical Doppler data
of Pioneer~10 and 11, the purpose is not to navigate a live
spacecraft, but to provide a model of as large a segment of the
spacecraft's trajectory as possible, using a consistent set of
parameters and minimizing the model residual. Several, independent
efforts to analyze the trajectories of the two spacecraft have
demonstrated that this can be accomplished using multi-year spans of
data with a root-mean-square model residual of no more than a few mHz.

\subsection{Orbit determination and parameter estimation}
\label{sec:odp}

The Pioneer anomaly has been verified using a variety of independent
orbit determination codes. The code that was used for the initial
discovery of the anomaly, JPL's Orbit Determination Program (ODP), is
probably also the most comprehensive and best tested among these, as
it is the primary code that is being used to navigate US and many
international spacecraft anywhere in the solar system, especially in
very deep space. ODP is a complex engineering achievement that
includes many thousands of lines of code that were built during the last
50 years of space exploration.  The physical models in ODP draw on
fundamental principles and practices developed during decades of deep
space exploration (see~\cite{Gelb-1974, Melbourne-2004, MG2005,
  Moyer-2003, Simon-2001, THORNTON2000}).  In its core, ODP relies on
a program called ``{\tt Regress}'' that calculates the computed values
of Doppler (and other) observables obtained at the tracking stations
of the DSN.  {\tt Regress} also calculates media corrections and
partial derivatives of the computed values of the observables with
respect to the solve-for-parameter vector-state.

An orbit determination procedure first determines the spacecraft's
initial position and velocity in a data interval. For each data
interval, we then estimate the magnitudes of the orientation
maneuvers, if any. The analysis uses models that include the effects
of planetary perturbations, radiation pressure, the interplanetary
media, general relativity, and bias and drift in the Doppler and range
(if available). Planetary coordinates and solar system masses are
obtained using JPL's Export Planetary Ephemeris DE{\tt nnn}, where DE
stands for the Development Ephemeris and {\tt nnn} is the current
number. (Earlier in the study, DE200 and DE405 were used, presently
  DE412 is available.)

Current versions of ODP implement computations in the J2000.0 epoch.
Past versions used B1950.0. (See~\cite{1982AA...115...20S} for
details on the conversion of positions and proper motions between
these two epochs.)

Standard ODP modeling includes a number of solid-Earth effects, namely
precession, nutation, sidereal rotation, polar motion, tidal effects,
and tectonic plates drift (see discussion in~\cite{pioprd}).  Model
values of the tidal deceleration, nonuniformity of rotation, polar
motion, Love numbers, and Chandler wobble are obtained
observationally, by means of lunar and satellite laser ranging (LLR,
SLR) techniques and very long baseline interferometry
(VLBI). Currently this information is provided by way of the International
Celestial Reference Frame (ICRF). JPL's Earth Orientation Parameters
(EOP) is a major source contributor to the ICRF.

Since the previous analysis~\cite{pioprl, pioprd}, physical models for
the Earth's interior and the planetary ephemeris have greatly
improved. This is due to progress in GPS, SLR, LLR and VLBI
techniques, Doppler spacecraft tracking, and new radio science data
processing algorithms. ODP models have been updated using these latest
Earth models (adopted by the IERS) and also are using the latest
planetary ephemeris. This allows for a better characterization of not
only the constant part of any anomalous acceleration, but also of the
annual and diurnal terms detected in the Pioneer~10 and 11 Doppler
residuals~\cite{pioprd, 2007AA...463..393O, moriond}.

During the last few decades, the algorithms of orbital analysis have
been extended to incorporate a Kalman-filter estimation procedure that
is based on the concept of ``process noise'' (i.e., random,
nonsystematic forces, or random-walk effects).  This was motivated by
the need to respond to the significant improvement in observational
accuracy and, therefore, to the increasing sensitivity to numerous
small perturbing factors of a stochastic nature that are responsible
for observational noise. This approach is well justified when one
needs to make accurate predictions of the spacecraft's future behavior
using only the spacecraft's past hardware and electronics state
history as well as the dynamic environmental conditions in the distant
craft's vicinity. Modern navigational software often uses
Kalman-filter estimation since it more easily allows determination of
the temporal noise history than does the weighted least-squares
estimation.

ODP also enables the use of batch-sequential filtering and a smoothing
algorithm with process noise~\cite{pioprd}. Though the name may imply
otherwise, batch-sequential processing does not involve processing the
data in batches. Instead, in this approach any small anomalous forces
may be treated as stochastic parameters affecting the spacecraft
trajectory. As such, these parameters are also responsible for the
stochastic noise in the observational data. To characterize these
noise sources, we split the data interval into a number of constant or
variable size batches with respect to the stochastic parameters, and
make assumptions on the possible statistical properties of the noise
factors. We then estimate the mean values of the unknown parameters
within the batch and their second statistical moments. (More details
on this ``batch-sequential algorithm with smoothing filter''
in~\cite{Gelb-1974, Moyer-2003}). ODP is permanently being updated to
suit the needs of precision navigation; the progress in the estimation
algorithms, programming languages, models of small forces and new
navigation methods have strongly supported its recent upgrades.

There have been a number of new models developed that are needed for
the analysis of tracking data from interplanetary spacecraft that are
now an integral part of the latest generation of the JPL's ODP. These
include an update to the relativistic formulation of the planetary and
spacecraft motion, relativistic light propagation and relevant
radiometric observables (i.e., Doppler, range, VLBI, and Delta
Differential One-way Ranging or $\Delta$DOR), coordinate
transformation between relativistic reference frames, and several
models for nongravitational forces. Details on other models and their
application for the analysis of the Pioneer anomaly are
in~\cite{pioprd, Turyshev:2005zk}.

\newpage
\section{The Original 1995\,--\,2002 Study of the Pioneer Anomaly}
\label{sec:anomaly}

The Pioneer~10 and 11 spacecraft have been described informally as the
most precisely navigated deep space vehicles to date. Such precise
navigation~\cite{pioprl, pioprd, 2004CQGra..21.4005N,
  2005PhLB..613...11N,moriond, Turyshev:2005zm, Turyshev:2005zk,
  Turyshev:2005vj} was made possible by many factors, including a
conservative design (see Figure~\ref{fig:pio-craft} for a design drawing
of the spacecraft) that placed the spacecraft's RTGs at the end of
extended booms, providing added stability and reducing thermal
effects. For attitude control, the spacecraft were spin-stabilized,
requiring a minimum number of attitude correction maneuvers, further
reducing navigation noise. As a result, precision navigation of the
Pioneer spacecraft was possible across multi-year stretches spanning a
decade or more~\cite{1976AJ.....81.1153N}.

Due in part to these excellent navigational capabilities, NASA
supported a proposal to extend the Pioneer~10 and 11 missions beyond
the originally planned mission durations, and use the spacecraft in an
attempt to perform deep space celestial mechanics experiments, as
proposed by J.D.~Anderson from the Jet Propulsion Laboratory
(JPL). Starting in 1979, the team led by Anderson began a systematic
search for unmodeled accelerations in the trajectories of the two
spacecraft. The principal aim of this investigation was the search for
a hypothetical tenth planet, Planet~X. Later, Pioneer~10 and 11 were
used to search for trans-Neptunian objects; the superior quality of
their Doppler tracking results also yielded the first ever limits on
low frequency gravitational radiation~\cite{pioprd}.

The acceleration sensitivity of the Pioneer~10 and 11 spacecraft was
at the level of $\sim 10^{-10}\mathrm{\ m/s}^2$. At this level of sensitivity,
however, a small, anomalous, apparently constant Doppler frequency
drift was detected~\cite{pioprl,pioprd,moriond}.

\subsection{The early evidence for the anomaly and the original study}

By 1980, when Pioneer~10 had already passed a distance of  $\sim$~20
AU from the Sun and the acceleration contribution from solar radiation
pressure on the spacecraft had decreased to less than $4\times
10^{-10}$ m/s$^2$, the radiometric data started to show the presence of
the anomalous sunward acceleration. Figure~\ref{fig:early} shows these
early unmodeled accelerations of Pioneer~10 (from about 1981 to 1989)
and Pioneer~11 (from 1977 to 1989).

The JPL team continued to monitor the unexpected anomalous
accelerations of Pioneer~10 and 11. Eventually, a proposal was made to
NASA to initiate a formal study. The proposal argued that the anomaly
is evident in the data of both spacecraft; that no physical model
available can explain the puzzling behavior; and that, perhaps, an
investigation with two independent software codes is needed to exclude
the possibility of a systematic error in the navigational
software. NASA supported the proposed investigation and, in 1995, the
formal study was initiated at JPL and, independently, at the Aerospace
Corporation, focusing solely on the acceleration anomaly detected in
the radiometric Doppler data of both spacecraft Pioneer~10 and 11.

\epubtkImage{}{%
  \begin{figure}[t]
    \centerline{\includegraphics[width=0.65\linewidth]{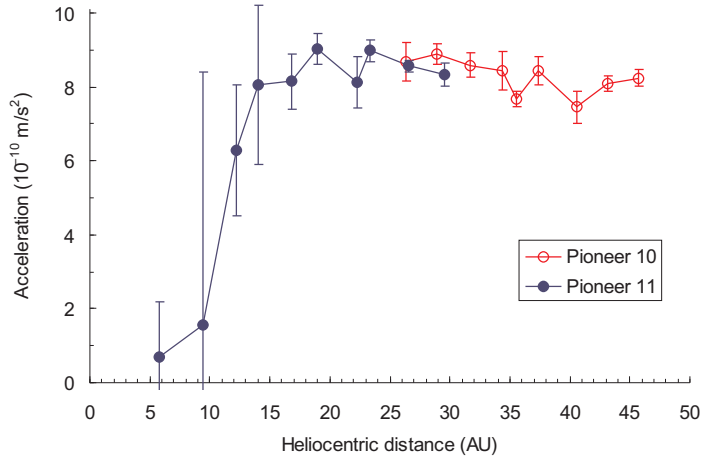}}
    \caption{Early unmodeled sunward accelerations of Pioneer~10 (from
    about 1981 to 1989) and Pioneer~11 (from 1977 to 1989). Adapted
    from~\cite{pioprd}, which contained this important footnote:
    ``Since both the gravitational and radiation pressure forces
    become so large close to the Sun, the anomalous contribution close
    to the Sun in [this figure] is meant to represent only what
    anomaly can be gleaned from the data, not a measurement.''}
    \label{fig:early}
\end{figure}}

Acceleration estimates for Pioneer~10 and 11 were developed at the JPL
using the Orbit Determination Program (ODP). The independent analysis
performed by The Aerospace Corporation utilized a different software
package called Compact High Accuracy Satellite Motion Program
(CHASMP). This effort confirmed the presence of the anomalous
acceleration, excluded computational systematics as a likely cause,
and also indicated a possible detection of a sunward acceleration
anomaly in the Galileo and Ulysses spacecrafts' signals.

Standard navigational models account for a number of post-Newtonian
perturbations in the dynamics of the planets, the Moon, and
spacecraft. Models for light propagation are correct to order
$(v/c)^2$. The equations of motion of extended celestial bodies are
valid to order  $(v/c)^4$. Nongravitational effects, such as solar
radiation pressure and precessional attitude-control maneuvers, make
small contributions to the apparent acceleration we have observed. The
solar radiation pressure decreases as $r^{-2}$; at distances
$>$~10\,--\,15~AU it produces an  acceleration in the case of the
Pioneer~10 and 11 spacecraft that is much less than $8\times
10^{-10}\mathrm{\ m/s}^2$, directed \textit{away} from the Sun. (The
acceleration due to the solar wind is roughly a hundred times smaller
than this.)

The initial results of both teams (JPL and The Aerospace Corporation)
were published in 1998~\cite{pioprl}. The JPL group concluded that
there is indeed an unmodeled acceleration, $a_P$, towards the Sun, the
magnitude of which is $(8.09\pm0.20)\times 10^{-10}\mathrm{\ m/s}^2$ for
Pioneer~10 and $(8.56\pm 0.15) \times 10^{-10}\mathrm{\ m/s}^2$ for
Pioneer~11. The formal error is determined by the use of a five-day batch
sequential filter with radial acceleration as a stochastic parameter,
subject to white Gaussian noise ($\sim$~500 independent five-day
samples of radial acceleration). No magnitude variation of $a_P$ with
distance was found, within a sensitivity of $2\times 10^{-10}\mathrm{\ m/s}^2$
over a range of 40 to 60~AU.

To determine whether or not the anomalous acceleration is specific to
the spacecraft, an attempt was made to detect any anomalous
acceleration signal in the tracking data of the Galileo and Ulysses
spacecraft. It soon became clear that in the case of Galileo, the
effects of solar radiation and an anomalous acceleration component
cannot be separated. For Ulysses, however, a possible sunward
anomalous acceleration was seen in the data, at $(12\pm 3)\times
10^{-10}\mathrm{\ m/s}^2$. Thus, the data from the Galileo and Ulysses
spacecraft yielded ambiguous results for the anomalous
acceleration. Nevertheless, the analysis of data from these two
additional spacecraft was useful in that it ruled out the possibility
of a systematic error in the DSN Doppler system that could easily have
been mistaken as a spacecraft acceleration.

The systematic error found in the Pioneer~10/11 post-fit residuals
could not be eliminated by taking into account all known gravitational
and nongravitational forces, both internal and external to the
spacecraft. A number of potential causes have been ruled
out. Continuing the search for an explanation, the authors considered the
following forces and effects:
\begin{itemize}
\item gravity from the Kuiper belt (see Section~\ref{sec:kuiper});
\item \vskip -6pt gravity from the galaxy (see Section~\ref{sec:misconceptions});
\item \vskip -6pt spacecraft gas leaks (see Section~\ref{sec:gas_leaks});
\item \vskip -6pt errors in planetary ephemerides (see Section~\ref{sec:solar-system-data});
\item \vskip -6pt errors in accepted values of the Earth's orientation, precession and nutation (see Section~\ref{sec:solar-system-data});
\item \vskip -6pt solar radiation pressure (see Section~\ref{sec:radpress});
\item \vskip -6pt precession attitude control maneuvers (see Section~\ref{sec:gasmaneuvers});
\item \vskip -6pt radio beam recoil force (see Section~\ref{sec:force_radio});
\item \vskip -6pt anisotropic thermal radiation (see Section~\ref{sec:force_recoil}).
\end{itemize}
Other possible sources of error were considered but none found to be
able to explain the puzzling behavior of the two Pioneer spacecraft.

The availability of further data (the data spanned January 1987 to
July 1998) from the then-still-active Pioneer~10 spacecraft allowed
the collaboration to publish a revised solution for $a_P$. In 1999, based
partially on this extended data set, they published a new estimate of
the average Pioneer~10 acceleration directed towards the Sun, which was
found to be $\sim 7.5 \times 10^{-10}\mathrm{\ m/s}^{2}$~\cite{moriond}. The
analyses used JPL's Export Planetary Ephemeris DE200, and modeled
planetary perturbations, general relativistic corrections, the Earth's
nonuniform rotation and polar rotation, and effects of radiation
pressure and the interplanetary medium.

A possible systematic explanation of the anomalous residuals is
nonisotropic thermal radiation. The thermal power of the spacecrafts'
radioisotope thermoelectric generators was in excess of 2500~W at
launch with a half-life for the \super{238}Pu fuel of 87.74~years, and
most of this power was thermally radiated into space. The power needed
to explain the anomalous acceleration is $\sim$~65~W. Nonetheless,
anisotropically emitted thermal radiation was not seen as a likely
explanation, for two reasons: first, it was assumed, after an initial
analysis of the spacecraft's geometry, that the thermal radiation
would be largely isotropic, and further, the observed acceleration did
not appear to be consistent with the decay rate of the radioactive
fuel.

As a result of this work, it became clear that a detailed
investigation of the Pioneer anomaly was needed.

\subsection{The 2002 formal solution for the anomalous acceleration}
\label{sec:formal_2002}

The most definitive study to date of the Pioneer anomaly~\cite{pioprd}
used Pioneer~10 data from January~3, 1987 to July~22, 1998, and
Pioneer~11 data from January~5, 1987 to October~1, 1990 (at this time,
Pioneer~11 lost coherent mode capability, as described in
Section~\ref{sec:comm}). The data were again analyzed with two
independently developed software packages, JPL's ODP and The Aerospace
Corporation's CHASMP.

Following an analysis of the anomalous spin behavior of Pioneer~10
(see Section~\ref{sec:spin}, and also Figure~\ref{fig:spin}), the
Pioneer~10 data set was further divided into three
intervals. Interval~I contained data January~3, 1987 to July~17, 1990;
Interval~II, from July~17, 1990 to July~12, 1992; and Interval~III,
from July~12, 1992 to July~22, 1998 (Table~\ref{tb:2002results}).

Analysis of results shown in Table~\ref{tb:2002results} let the
collaboration develop their estimate for the baseline ``experimental''
values for Pioneer~10 and 11~\cite{pioprd}. They found the
optimally weighted least-squares solution ``experimental'' number for
Pioneer~10:

\begin{equation}
a^{\mathrm{Pio10}}_{\mathrm{exp}}=(7.84\pm
0.01)\times~10^{-10}\mathrm{\ m/s}^2.
\label{pio10lastresult}
\end{equation}
Similarly, the  experimental value for Pioneer~11 was found to be:
\begin{equation}
a^{\mathrm{Pio11}}_{\mathrm{exp}}=(8.55\pm 0.02)\times
10^{-10}\mathrm{\ m/s}^2.
\label{pio11lastresult}
\end{equation}

\clearpage

\epubtkImage{}{%
  \begin{figure}[htbp]
    \centerline{
      \includegraphics[width=0.52\textwidth]{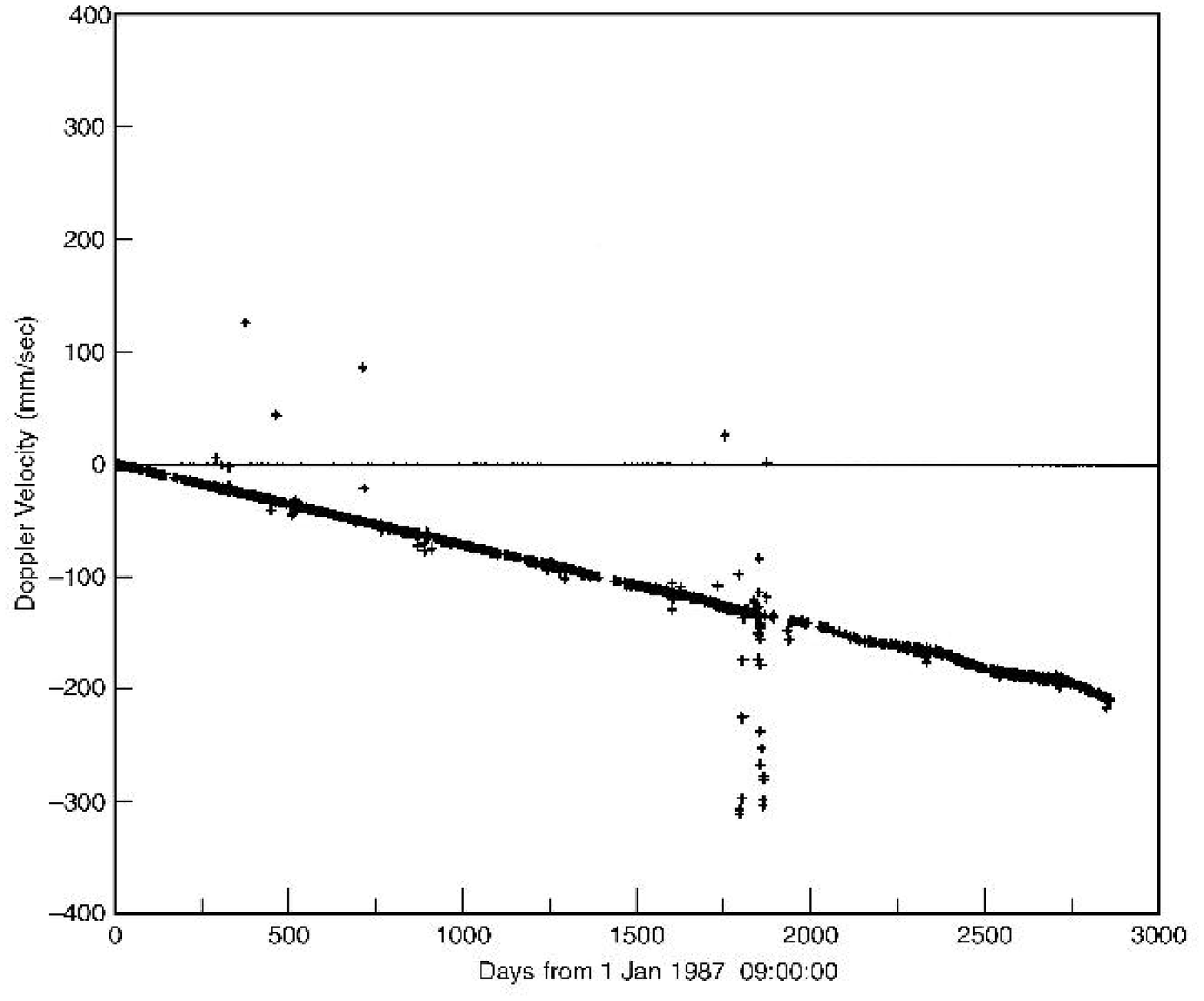}
      \includegraphics[width=0.51\textwidth]{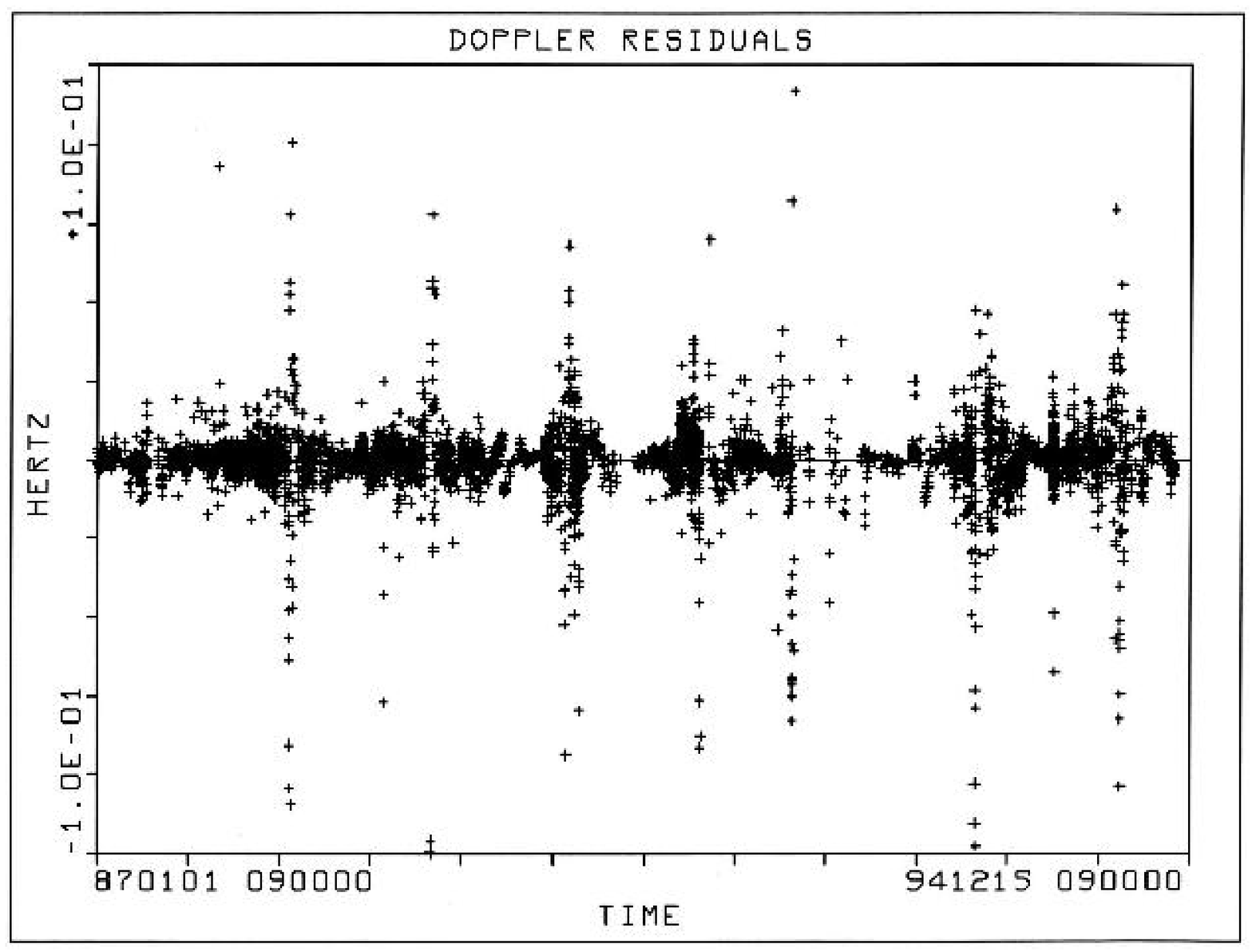}
    }
    \caption{\emph{Left:} Two-way Doppler residuals (observed Doppler
    velocity minus model Doppler velocity) for Pioneer~10. On the
    vertical axis, 1~Hz is equal to 65~mm/s range change per
    second. \emph{Right:} The best fit for the Pioneer~10 Doppler residuals
    with the anomalous acceleration taken out. After adding one more
    parameter to the model (a constant radial acceleration of
    $a_P=(8.74\pm 1.33)\times 10^{-10}\mathrm{\ m/s}^2$) the residuals are
    distributed about zero Doppler velocity with a systematic
    variation $\sim$~3.0~mm/s on a time scale of
    $\sim$~3~months~\cite{pioprd}.}
    \label{fig:residuals}
\end{figure}}

\begin{table}[htbp]
  \caption[Acceleration estimates (in units of
  10\super{-10}~m/s\super{2}). Two programs (JPL's ODP and The Aerospace
  Corporation's CHASMP) were used to obtain weighted least squares
  (WLS) and batch-sequential filtering (BSF, 1-day batch)
  estimates. CHASMP could also incorporate corrections based on
  10.7~cm solar flux observations, called F10.7
  corrections.]{Acceleration estimates (in units of
  10\super{-10}~m/s\super{2}) published in~\cite{pioprd}. Two programs (JPL's
  ODP and The Aerospace Corporation's CHASMP) were used to obtain
  weighted least squares (WLS) and batch-sequential filtering (BSF,
  1-day batch) estimates. CHASMP could also incorporate corrections
  based on 10.7~cm solar flux observations, called F10.7 corrections.}
  \label{tb:2002results}
  \centering
  \begin{tabular}{lllcccc}
    \toprule
    Program & Method & Corona & P10 (I) & P10 (II) & P10 (III) & P11\\
    \midrule
    ODP    & WLS       & no  & 8.02~\textpm~0.01 & 8.65~\textpm~0.01 & 7.83~\textpm~0.01 & 8.64~\textpm~0.04\\
    ODP    & WLS       & yes & 8.00~\textpm~0.01 & 8.66~\textpm~0.01 & 7.84~\textpm~0.01 & 8.44~\textpm~0.04\\
    ODP    & BSF       & yes & 7.82~\textpm~0.29 & 8.16~\textpm~0.40 & 7.59~\textpm~0.22 & 8.49~\textpm~0.33\\
    CHASMP & WLS       & no  & 8.25~\textpm~0.02 & 8.86~\textpm~0.02 & 7.85~\textpm~0.01 & 8.71~\textpm~0.03\\
    CHASMP & WLS       & yes & 8.22~\textpm~0.02 & 8.89~\textpm~0.02 & 7.92~\textpm~0.01 & 8.69~\textpm~0.03\\
    CHASMP & WLS+F10.7 & yes & 8.25~\textpm~0.03 & 8.90~\textpm~0.03 & 7.91~\textpm~0.01 & 8.91~\textpm~0.04\\
    \bottomrule
  \end{tabular}
\end{table}

The main conclusions of the 2002 study~\cite{pioprd} can be summarized
as follows:

\begin{itemize}
\item The tracking data of Pioneer~10 and 11 are consistent with
  constant sunward acceleration;
\item \vskip -6pt The magnitude of acceleration is several times noise
  level;
\item \vskip -6pt No known source of error or systematic bias can
  account for the anomalous acceleration.
\end{itemize}

Initial announcement of the anomalous acceleration
(e.g., \cite{pioprl,moriond}) triggered many proposals that invoked
various conventional physics mechanisms, all aimed at explaining the
origin of the anomaly. Finding a systematic origin of the proper
magnitude and behavior was the main focus of these proposals. Although
the most obvious explanation would be that there is a systematic
origin to the effect, perhaps generated by the spacecraft themselves
from anisotropic heat rejection or propulsive gas leaks, the analysis
did not find evidence for either mechanism: That is, no unambiguous,
on-board systematic has been discovered.

This initial search was summarized
in~\cite{pioprd,2001gr.qc.....7022A}, where possible contributions of
various mechanisms to the final solution for $a_P$ were given. The
entire error budget was subdivided in three main types of effects,
namely
\begin{inparaenum}[i)]
\item effects due to sources external to the spacecraft;
\item the contribution of on-board systematics; and
\item computational systematic errors
\end{inparaenum}
(see Table~\ref{tab:error_budget}.) These three categories are
detailed in the following sections.

\subsection{Sources of systematic error external to the spacecraft}
\label{sec:ext-systema}

External forces can contribute to all three vector components of
spacecraft acceleration (in contrast, as detailed in Section~\ref{sec:theory-explain},
forces generated on board contribute primarily along the axis of
rotation). However, nonradial spacecraft accelerations are difficult
to observe by the Doppler technique, which measures the velocity along
the Earth-spacecraft line of sight, which approximately coincides with
the spacecraft spin axis.

Following \cite{pioprl,pioprd,moriond}, we first consider forces
that affect the spacecraft motion, such as those due to
\begin{inparaenum}[i)]
\item solar-radiation pressure, and
\item solar wind pressure. We then discuss the effects on the propagation of the radio signal that are from
\item the solar corona and its mismodeling,
\item electro-magnetic Lorentz forces,
\item the influence of the Kuiper belt,
\item the phase stability of the reference atomic clocks, and
\item the mechanical and phase stability of the DSN antennae, together with influence of the station locations and troposphere and ionosphere contributions.
\end{inparaenum}
Although some of the mechanisms detailed below are near the limit for
contributing to the final error budget, it was found that none of them
could explain the behavior of the detected signal.  Moreover, some were
three orders of magnitude or more too small.

\subsubsection{Direct solar radiation pressure and mass}
\label{sec:solarP}

\cite{pioprd} estimated the systematic error from solar radiation
pressure on the Pioneer~10 spacecraft over the interval from 40 to
70.5~AU, and for Pioneer~11 from 22.4 to 31.7~AU. Using
Equation~(\ref{eq:srp}) they  estimated that when the spacecraft reached
10~AU, the solar radiation acceleration was $18.9\times
10^{-10}\mathrm{\ m/s}^2$ decreasing to $0.39\times 10^{-10}\mathrm{\ m/s}^2$ at
70~AU. Because this contribution falls off with the inverse square of
the spacecraft's heliocentric distance, it can bias the Doppler
determination of a constant acceleration. By taking the average of the
inverse square acceleration curve over the Pioneer distance,
\cite{pioprd} estimated the error in the acceleration of the
spacecraft due to solar radiation pressure. This error, in units of
$10^{-10}\mathrm{\ m/s}^2$, is $\sigma_{\mathrm{sp}}=0.001$ for Pioneer~10 over
the interval from 40 to 70.5~AU, and six times this amount for
Pioneer~11 over the interval from 22.4 to 31.7~AU. In addition the
uncertainty in the spacecraft's mass for the studied data interval
also introduced a bias of $b_{\mathrm{sp}}=0.03\times 10^{-10}\mathrm{\ m/s}^2$
in the acceleration value. These estimates resulted in the error
estimates
\begin{eqnarray}
\delta a^\mathrm{Pio10}_{\mathrm{sp}} &=&
b_{\mathrm{sp}}\pm\sigma^\mathrm{Pio10}_{\mathrm{sp}} = (0.03\pm0.001)\times 10^{-10}\mathrm{\ m/s}^2,\\
\delta a^\mathrm{Pio11}_{\mathrm{sp}} &=&
b_{\mathrm{sp}}\pm\sigma^\mathrm{Pio11}_{\mathrm{sp}} = (0.03\pm0.006)\times 10^{-10}\mathrm{\ m/s}^2.
\end{eqnarray}

\subsubsection{The solar wind}
\label{solarwind}

The acceleration caused by solar wind particles intercepted by the
spacecraft can be estimated, as discussed in
Section~\ref{sec:force_solwind}. Due to variations with a magnitude of
up to 100\%, the exact acceleration is unpredictable, but its
magnitude is small, therefore its contribution to the Pioneer
acceleration is completely negligible. Based on these arguments, the
authors of \cite{pioprd} concluded that the total uncertainty in
$a_P$ due to solar wind can be limited as

\begin{equation}
\sigma_{\mathrm{sw}}\leq 10^{-15}\mathrm{\ m/s}^2.
\end{equation}

\subsubsection{The effects of the solar corona}
\label{sec:corona}

Given Equation~(\ref{eq:weight_doppler*}) derived in
Section~\ref{sec:effect_solplasm} and the values of the parameters $(A,
B, C)=(6.0\times 10^3,~2.0\times 10^4,~0.6\times 10^6)$, all in
meters, \cite{pioprd} estimated the acceleration error due to the
effect of the solar corona on the propagation of radio waves between
the Earth and the spacecraft.

The correction to the Doppler frequency shift is obtained from
Equation~(\ref{eq:weight_doppler*}) by simple time differentiation. (The
impact parameter depends on time as $\rho=\rho(t)$ and may be
expressed in terms of the relative velocity of the spacecraft with
respect to the Earth, $v\approx 30\mathrm{\ km/s}$).

The effect of the solar corona is expected to be small on the Doppler
frequency shift, which is our main observable. This is due to the fact
that most of the data used for the Pioneer analysis were taken with
large Sun-Earth-spacecraft angles. Further, the solar corona effect on
the Doppler observable has a periodic signature, corresponding to the
Earth's orbital motion, resulting in variations in the
Sun-Earth-spacecraft angle. The time-averaged effect of the corona on
the propagation of the Pioneers' radio-signals is of order

\begin{equation}
\sigma_{\mathrm{corona}}=\pm 0.02\times 10^{-10}\mathrm{\ m/s}^2.
\end{equation}

\subsubsection{Electro-magnetic Lorentz forces}

The authors of~\cite{Turyshev:2005zk} considered the possibility that
the Pioneer spacecraft can hold a charge and be deflected in its
trajectory by Lorentz forces. They noted that this was a concern
during planetary flybys due to the strength of Jupiter's and Saturn's
magnetic fields (see Figure~\ref{fig:pioneer_inner_path}). The
magnetic field strength in the outer solar system, $\le10^{-5}$~Gauss,
is five orders of magnitude smaller than the magnetic field strengths
measured by the spacecraft at their nearest approaches to Jupiter:
0.185~Gauss for Pioneer~10 and 1.135~Gauss for Pioneer~11. Data from
the Pioneer~10 plasma analyzer can be interpreted as placing an upper
bound of $0.1 \mu\mathrm{C}$ on the positive charge during its Jupiter
encounter~\cite{1976AJ.....81.1153N}.

These bounds allow us to estimate the upper limit of the contribution
of the electromotive force on the motion of the Pioneer spacecraft in
the outer solar system. This was accomplished in \cite{pioprd}
using the standard formula for the Lorentz-force,
$\vec{F}=q\vec{v}\times\vec{B}$, and found that the greatest force
would be on Pioneer~11 during its closest approach to Jupiter,
$<\,20\times 10^{-10}\mathrm{\ m/s}^2$. However, once the spacecraft reached
the interplanetary medium, this force would decrease to
\begin{equation}
\sigma_{\mathrm{Lorentz}}\lesssim 2\times 10^{-14}\mathrm{\ m/s}^2,
\end{equation},
which is negligible.

\subsubsection{The Kuiper belt's gravity}
\label{sec:kuiper}

\cite{pioprd} specifically studied three distributions of matter
in the Kupier belt, including a uniform distribution and resonance
distributions that were hypothesized in \cite{malhotra}. The
authors assumed a total mass of one Earth mass, which is significantly
larger than standard estimates. Even so, the resulting accelerations
are only on the order of 10\super{-11}~m/s\super{2}, two orders of magnitude
less than the observed effect. The calculated accelerations vary with
time, increasing as Pioneer~10 approaches the Kuiper belt, even with a
uniform density model. For these reasons, \cite{pioprd} excluded
the dust belt as a source for the Pioneer effect.

More recent infrared observations established an upper limit of 0.3
Earth masses of Kuiper Belt dust in the trans-Neptunian
region~\cite{backman,stern1996,teplitzinfra}. Therefore, for the
contribution of Kuiper belt gravity, the authors of \cite{pioprd}
placed a limit of

\begin{equation}
\sigma_\mathrm{KB} = 0.03 \times 10^{-10}~\mathrm{m/s}^{2}.
\end{equation}

\subsubsection{Stability of the frequency references}
\label{sec:clocks}

Reliable detection of a precision Doppler observable requires a very
stable frequency reference at the observing stations. High precision
Pioneer~10 and 11 Doppler measurements were made using 2-way and 3-way
Doppler observations. In this mode, there was no on-board frequency
reference at the spacecraft; the received frequency was converted to a
downlink frequency using a fixed frequency ratio (240/221), and this
signal was returned to the Earth. As the round-trip light time was
many hours, the stability of the frequency reference over such
timescales is essential. Further, in the case of 3-way Doppler
measurements when the transmitting and receiving stations were not the
same, it was essential to have stable frequency references that were
synchronized between ground stations of the DSN.

The stability of a clock or frequency reference is usually measured by
its Allan deviation. The Allan deviation $\sigma_y(\tau)$, or its
square, the Allan-variance, are defined as the variance of the
frequency departure $y_n=\left<\delta\nu/\nu\right>_n$ (where $\nu$ is
the frequency and $\delta\nu$ is its variance during the measurement
period) over a measurement period $\tau$:
\begin{equation}
\sigma_y^2(\tau)=\frac{1}{2}\left<(y_{n+1}-y_n)^2\right>,
\end{equation}
where angle brackets indicate averaging.

The S-band communication systems of the DSN that were used for
communicating with the Pioneer spacecraft had Allan deviations that
are of order $\sigma_y\sim 1.3\times 10^{-12}$ or less for
$\sim\,10^3\mathrm{\ s}$ integration times~\cite{Turyshev:2005zk}.
Using the Pioneer S-band transmission frequency as $\nu\simeq
2.295\mathrm{\ GHz}$, we obtain
\begin{equation}
\delta\nu=\sigma_y\nu\simeq 2.98~\mathrm{mHz}
\end{equation}
over a Doppler integration time of $\sim 10^3$~s. Applying this figure
to the case of a steady frequency drift, the corresponding
acceleration error over the course of a year was
estimated~\cite{pioprd} as
\begin{equation}
\sigma_\mathrm{freq}=0.0003\times 10^{-10}\mathrm{\ m/s}^2.
\end{equation}

\subsubsection{Stability of DSN antenna complexes}
\label{sec:dsn_complex}

The measurement of the frequency of a radio signal is affected by the
stability of physical antenna structures. The large antennas of the
DSN complexes are not perfectly stable. Short term effects include
thermal expansion, wind loading, tides and ocean loading. Long term
effects are introduced by continental drift, gravity loads and the
aging of structures.

All these effects are well understood and routinely accounted for as
part of DSN operations. DSN personnel regularly assess the performance
of the DSN complex to ensure that operational limits are
maintained~\cite{SOVERS1998, SOVERS1996}.

The authors of~\cite{pioprd} found that none of these effects can
produce a constant drift comparable to the observed Pioneer Doppler
acceleration. Their analysis, which included errors due to imperfect
knowledge of DSN station locations, to troposphere and ionosphere
models at different stations, and to Faraday rotation effects of the
atmosphere, shows a negligible contribution to the observed
acceleration:

\begin{equation}
\sigma_{\tt DSN} \leq 10^{-14}\mathrm{\ m/s}^2.
\end{equation}

\subsection{Sources of systematic error internal to the spacecraft}
\label{sec:on-board-systematics}

There exist several on-board mechanisms that can contribute to the
acceleration of the Pioneer spacecraft. For spinning spacecraft, like
Pioneer~10 and 11, the contribution of these forces to the
spacecraft's acceleration will be primarily in the direction of the
spin axis. The reason for this is that for any force that is constant
in a co-rotating coordinate system, the force component that is
perpendicular to the spin axis will average to zero over the course of
a full revolution. Consequently, for an arbitrary force, its
contribution to lateral accelerations will be limited to its
time-varying component (see Section~\ref{sec:nongrav}).

There are several known forces of on-board origin that can result in
unmodeled accelerations. These forces, in fact, represent the most
likely sources of the anomaly, in particular because previously
published magnitudes of several of the considered effects are subject
to revision, in view of the recently recovered telemetry data and
newly developed thermal models.

On-board mechanisms that we consider in this section include:
\begin{inparaenum}[i)]
\item thruster gas leaks,
\item nonisotropic radiative cooling of the spacecraft body,
\item heat from the RTGs,
\item the radio beam reaction force, and
\item the expelled helium produced within the RTG and other gas emissions.
\end{inparaenum}

We also review the differences in experimental results between the two spacecraft.

\subsubsection{Propulsive mass expulsion}
\label{sec:propulsive_mass_expulsion}

The attitude control subsystems on board Pioneer~10 and 11 were used
frequently to ensure that the spacecrafts' antennas remained oriented
in the direction of the Earth. This raises the possibility that the
observed anomalous acceleration is due to mismodeling of these
attitude control maneuvers, or inadequate modeling of the inevitable
gas leaks that occur after thruster firings.

The characteristics of propulsive gas leaks are well understood and
routinely modeled by trajectory estimation software. Typical gas leaks
vary in magnitude after each thruster firing, and usually decrease in
time, until they become negligible.

The placement of thrusters (see Section~\ref{sec:propsubsys}) makes it
highly likely that any leak would also induce unaccounted-for changes
in the spacecraft's spin and attitude.

In contrast, to produce the observed acceleration, any propulsion
system leaks would have had to be
\begin{inparaenum}[i)]
\item constant in time;
\item the same on both spacecraft;
\item not inducing any detectable changes in the spin rate or precession.
\end{inparaenum}
Given these considerations, \cite{pioprd} \textit{conservatively}
estimates that undetected gas leaks introduce an uncertainty no
greater than
\begin{equation}
\sigma_\mathrm{gl}=\pm 0.56\times 10^{-10}\mathrm{\ m/s}^2.
\end{equation}

\subsubsection{Heat from the RTGs}

The radioisotope thermoelectric generators of the Pioneer~10 and 11
spacecraft emitted up to $\sim$~2500~W of heat at the beginning of the
mission, slowly decreasing to $\sim$~2000~W near the end. Even a small
anisotropy ($<$~2\%) in the thermal radiation pattern of the RTGs can
account, in principle, for the observed anomalous
acceleration. Therefore, the possibility that the observed
acceleration is due to anisotropically emitted RTG heat has been
considered~\cite{pioprd, Turyshev:2005zk}.

The cylindrical RTG packages (see Section~\ref{sec:RTG}) have
geometries that are fore-aft symmetrical. Two mechanisms were
considered that would nonetheless lead to a pattern of thermal
radiation with a fore-aft asymmetry.

According to one argument, heat emitted by the RTGs would be reflected
anisotropically by the spacecraft itself, notably by the rear of the
HGA.

\cite{pioprd} used the spacecraft geometry and the resultant RTG
radiation pattern to estimate the contribution of the RTG heat
reflecting off the spacecraft to the Pioneer anomaly. The solid angle
covered by the antenna as seen from the RTG packages was estimated at
$\sim$~2\% of $4\pi$ steradians. The equivalent fraction of RTG heat
is $\sim$~40~W. This estimate was further reduced after the shape of
the RTGs (cylindrical with large radiating fins) and the resulting
anisotropic radiation pattern of the RTGs was considered. Thus,
\cite{pioprd} estimated that this mechanism could produce only
4~W of directed power.

The force from 4\, W of directed power suggests a systematic bias of
$\approx 0.55 \times 10^{-10}\mathrm{\ m/s}^2$. The authors also add an
uncertainty of the same size, to obtain a contribution from heat
reflection of
\begin{equation}
a_\mathrm{hr}=(-0.55\pm 0.55)\times 10^{-10}\mathrm{\ m/s}^2.
\end{equation}

Another mechanism may also have contributed to a fore-aft asymmetry in
the thermal radiation pattern of the RTGs. Especially during the early
part of the missions, one side of the RTGs was exposed to continuous
intense solar radiation, while the other side was in permanent
darkness. Furthermore this side, facing deep space, was sweeping
through the dust contained within the solar system. These two
processes may have led to different modes of surface degradation,
resulting in changing emissivities~\cite{A634203}.

To obtain an
estimate of the uncertainty, \cite{pioprd} considered the
case when one side (fore or aft) of the RTGs has its emissivity
changed by only 1\% with respect to the other side.\epubtkFootnote{A
  fore-aft difference of 1\% is possible, but not supported by any
  available data.} In a simple cylindrical model of the RTGs, with
2000~W power (only radial emission is assumed with no loss out of the
sides), the ratio of the power emitted by the two sides would be
$995/1005=0.99$, or a differential emission between the half cylinders
of 10~W. Therefore, the fore/aft asymmetry toward the normal would be
$10~\mathrm{\ W}\times\frac{1}{\pi}\int_0^\pi d\phi\sin\phi\approx
6.37\mathrm{\ W}$. A more sophisticated model of the fin structure
resulted in the slightly smaller estimate of 6.12~W, which the authors
of \cite{pioprd} took as the uncertainty from the differential
emissivity of the RTGs, to obtain an acceleration uncertainty of
\begin{equation}
\sigma_\mathrm{de}=0.85\times 10^{-10}\mathrm{\ m/s}^2.
\end{equation}

\subsubsection{Nonisotropic radiative cooling of the spacecraft}
\label{subsec:mainbus}

It has also been suggested that the anomalous acceleration seen in the
Pioneer~10/11 spacecraft can be, ``explained, at least in part, by
nonisotropic radiative cooling of the
spacecraft~\cite{1999PhRvL..83.1890M}.'' Later this idea was modified,
suggesting that ``most, if not all, of the unmodeled acceleration'' of
Pioneer~10 and 11 is due to an essentially constant supply of heat
coming from the central compartment, directed out the front of the
craft through the closed louvers~\cite{Scheffer:2001we}.

To address the original proposal~\cite{1999PhRvL..83.1890M} and
several later
modifications~\cite{Scheffer:2001we,2003PhRvD..67h4021S},
\cite{1999PhRvL..83.1891A,2001gr.qc.....7022A,2001gr.qc.....8054S}
developed a bound on the constancy of $a_P$. This bound came from
first noting the 11.5 year 1-day batch-sequential result, sensitive to
time variation: $a_P=(7.77\pm 0.16)\times 10^{-10}\mathrm{\ m/s}^2$. It is
conservative to take three times this error to be our systematic
uncertainty for radiative cooling of the craft,
\begin{equation}
\sigma_\mathrm{rc}=\pm 0.48 \times 10^{-10}\mathrm{\ m/s}^2.
\end{equation}

\subsubsection{Radio beam reaction force}
\label{sec:radio-beam}

The emitted radio-power from the spacecraft's HGA produces a recoil
force, which is responsible for an acceleration bias, $b_\mathrm{rp}$,
on the spacecraft away from the Earth. If the spacecraft were equipped
with ideal antennas, the total emitted power of the spacecrafts' radio
transmitters would be in the form of a collimated beam aimed in the
direction of the Earth. In reality, the antenna is less than 100\%
efficient: some of the radio frequency energy from the transmitter may
miss the antenna altogether, the radio beam may not be perfectly
collimated, and it may not be aimed precisely in the direction of the
Earth.

Therefore, using $\beta$ to denote the efficiency of the antenna, we
can compute an acceleration bias as
\begin{equation}
b_\mathrm{rp}=\frac{1}{mc}\beta P_\mathrm{rp},
\end{equation}
where $P_\mathrm{rp}$ is the transmitter's power. The nominal
transmitted power of the spacecraft is 8~W. Given the
$m=241\mathrm{\ kg}$ as the mass of a spacecraft with half its fuel
gone, and using the 0.4~dB antenna error as a means to estimate the
uncertainty, we obtain the acceleration figure of
\begin{equation}
a_\mathrm{rp}=b_\mathrm{rp}\pm\sigma_\mathrm{rp}=-(1.10\pm 0.10)\times 10^{-10}\mathrm{\ m/s}^2,
\end{equation}
where the negative sign indicates that this acceleration is in the
direction \emph{away} from the Earth (and thus from the Sun), i.e.,
this correction actually \emph{increases} the amount of anomalous
acceleration required to account for the Pioneer Doppler
observations~\cite{pioprd}.

\subsubsection{Expelled helium produced within the RTGs}

Another possible on-board systematic error is from the expulsion of
the He being created in the RTGs from the $\alpha$-decay of
\super{238}Pu. According to the discussion presented in
Section~\ref{sec:gas_leaks}, Anderson et al. estimate the bias and
error in acceleration due to He-outgassing as
\begin{equation}
a_\mathrm{He}=(0.15\pm 0.16)\times 10^{-10}\mathrm{\ m/s}^2.
\end{equation}

\subsubsection{Variation between determinations from the two spacecraft}
\label{sec:two-craft}

Section~\ref{sec:formal_2002} presented two experimental results for
the Pioneer anomaly from the two spacecraft: $7.84\times
10^{-10}\mathrm{\ m/s}^2$ (Pioneer~10) and $8.55 \times 10^{-10}\mathrm{\ m/s}^2$
(Pioneer~11). The first result represents the entire 11.5~year data
period for Pioneer~10; Pioneer~11's result represents a 3.75~year data
period.

The difference between the two craft could be due to differences in
gas leakage. It also could be due to heat emitted from the RTGs. In
particular, the two sets of RTGs have had different histories and so
might have different emissivities. Pioneer~11 spent more time in the
inner solar system (absorbing radiation). Pioneer~10 has swept out
more dust in deep space. Further, Pioneer~11 experienced about twice
as much Jupiter/Saturn radiation as Pioneer~10.

\cite{pioprd} estimated the value for the Pioneer anomaly based
on the two independent determinations derived from the two spacecraft,
Pioneer~10 and 11. They calculated the time-weighted average of the
experimental results from the two craft:
$[(11.5)(7.84)+(3.75)(8.55)]/(15.25)=8.01$ in units of
$10^{-10}\mathrm{\ m/s}^2$. This result implies a bias of
$b_\mathrm{2~craft}=0.17\times 10^{-10}\mathrm{\ m/s}^2$ with respect to the
Pioneer~10 experimental result $a_{P(\mathrm{exp})}$ (see
Equation~(\ref{pio10lastresult})). We can take this number to be a measure
of the uncertainty from the separate spacecraft measurements, so the
overall quantitative measure is
\begin{equation}
a_\mathrm{2~craft}=b_\mathrm{2~craft}\pm\sigma_\mathrm{2~craft}=(0.17\pm 0.17)\times 10^{-10}\mathrm{\ m/s}^2.
\end{equation}

\subsection{Computational systematics}

The third group of effects was composed of contributions from
computational errors (see Table~\ref{tab:error_budget}). The effects
in this group dealt with
\begin{inparaenum}[i)]
\item the numerical stability of least-squares estimations,
\item accuracy of consistency/model tests,
\item mismodeling of maneuvers, and that of
\item the solar corona model used to describe the propagation of radio
  waves. It has also been demonstrated that the influence of
\item annual/diurnal terms seen in the data on the accuracy of the
  estimates was small.
\end{inparaenum}

\subsubsection{Numerical stability of least-squares estimation}

The authors of \cite{pioprd} looked at the numerical stability of
the least squares estimation algorithms and the derived solutions.

Common precision orbit determination algorithms use double precision
arithmetic. The representation uses a 53-bit mantissa, equivalent to
more than 15 decimal digits of precision~\cite{IEEE754}. Is this
accuracy sufficient for precision orbit determination within the solar
system? At solar system barycentric distances between 1 and 10 billion
kilometers (10\super{12}\,--\,10\super{13}~m), 15 decimal digits of accuracy
translates into a positional error of 1~cm or less. Therefore, we can
conclude that double precision arithmetic is adequate in principle for
modeling the orbits of Pioneer~10 and 11 in the outer solar system in
a solar system barycentric reference frame. However, one must still be
concerned about cumulative errors and the stability of the employed
numerical algorithms.

The leading source for computational errors in finite precision
arithmetic is the addition of quantities of different magnitudes,
causing a loss of least significant digits in the smaller quantity. In
extreme cases, this can lead to serious instabilities in numerical
algorithms. Software codes that perform matrix operations are
especially vulnerable to such stability issues, as are algorithms that
use finite differences for solving systems of differential equations
numerically.

While it is difficult to prove that a particular solution is not a
result of a numerical instability, it is extremely unlikely that two
independently-developed programs could produce compatible results that are
nevertheless incorrect, as a result of computational error. Therefore,
verifying a result using independently-developed software codes is a
reliable way to exclude numerical instabilities as a possible error
source, and also to put a limit on any numerical errors.

In view of the above, given the excellent agreement in various
implementations of the modeling software, the authors of
\cite{pioprd} concluded that differences in analyst choices
(parameterization of clocks, data editing, modeling options, etc.)
give rise to coordinate discrepancies only at the level of
0.3~cm. This number corresponds to an uncertainty in estimating the
anomalous acceleration on the order of
8~\texttimes~10\super{-14}~m/s\super{2}, which was found to be
negligible for the investigation.

Analysis identified, however, a slightly larger error to contend
with. After processing, Doppler residuals at JPL were rounded to 15
and later to 14 significant figures. When the Block~5 receivers came
online in 1995, Doppler output was further rounded to 13 significant
digits. According to~\cite{pioprd}, this roundoff results in the
estimate for the numerical uncertainty of
\begin{equation}
\sigma_\mathrm{num}=\pm 0.02\times 10^{-10}\mathrm{\ m/s}^2.
\end{equation}

\subsubsection{Model consistency}

The accuracy of navigational codes that are used to model the motion
of spacecraft is limited by the accuracy of the mathematical models
employed by the programs to model the solar system. The two programs
used in the investigation -- JPL's ODP/{\em Sigma} modeling software
and The Aerospace Corporation's POEAS/CHASMP software package -- used
different parameter estimation procedures, employed different
realizations of the Earth's orientation parameters, used different
planetary ephemerides, and different data editing strategies. While it
is possible that some of the differences were partially masked by
maneuver estimations, internal consistency checks indicated that the
two solutions were consistent at the level of one part in
$10^{15}$, implying an acceleration error $\leq
10^{-4}a_P$~\cite{pioprd}.

The consistency of the models was verified by comparing separately the
Pioneer~11 results and the Pioneer~10 results for the three intervals
studied in~\cite{pioprd}. The models differed, respectively, by
(0.25, 0.21, 0.23, 0.02)~m/s\super{2}. Assuming that these errors are
uncorrelated, \cite{pioprd} computed the combined effect on
anomalous acceleration $a_P$ as
\begin{equation}
\sigma_\mathrm{consist/model}=\pm 0.13\times 10^{-10}\mathrm{\ m/s}^2.
\end{equation}

\subsubsection{Error due to mismodeling of maneuvers}

The velocity change that results from a propulsion maneuver cannot be
modeled exactly. Mechanical uncertainties, fuel properties and
impurities, valve performance, and other factors all contribute
uncertainties. The authors of~\cite{pioprd} found that for a typical
maneuver, the standard error in the residuals is $\sigma_0\sim
0.095\mathrm{\ mm/s}$. Given 28 maneuvers during the Pioneer~10 study
period of 11.5~years, a mismodeling of this magnitude would contribute
an error to the acceleration solution with a magnitude of $\delta
a_{\mathrm{man}}=\sigma_0/\tau=0.07\times 10^{-10}\mathrm{\ m/s}^2$. Assuming a
normal distribution around zero with a standard deviation of $\delta
a_{\mathrm{man}}$ for each single maneuver, a total of $N=28$ maneuvers
yields a total error of
\begin{equation}
\sigma_{\mathrm{man}}=\frac{\delta a_{\mathrm{man}}}{\sqrt{N}}=0.01\times 10^{-10}\mathrm{\ m/s}^2,
\end{equation}
due to maneuver mismodeling.

\subsubsection{Annual/diurnal mismodeling uncertainty}
\label{sec:annualdiurnal}

In addition to the constant anomalous acceleration term, an annual
sinusoid has been reported~\cite{pioprd,moriond}. The peaks of the
sinusoid occur when the spacecraft is nearest to the Sun in the
celestial sphere, as seen from the Earth, at times when the Doppler
noise due to the solar plasma is at a maximum. A parametric fit to
this oscillatory term~\cite{pioprd, Turyshev:2005zk} modeled this
sinusoid with amplitude $v_{\mathrm{at}}=(0.1053\pm 0.0107)\mathrm{\ mm/s}$,
angular velocity $\omega_{\mathrm{at}}=(0.0177\pm 0.0001)\mathrm{\ rad/day}$, and
bias $b_{\mathrm{at}}=(0.0720\pm 0.0082)\mathrm{\ mm/s}$, resulting in post-fit
residuals of $\sigma_T=0.1\mathrm{\ mm/s}$, averaged over the data interval $T$.

The obtained amplitude and angular velocity can be combined to form an
acceleration amplitude:
$a_{\mathrm{at}}=v_\mathrm{at}\omega_{\mathrm{at}}=(0.215\pm 0.022)\times
10^{-10}\mathrm{\ m/s}^2$. The likely cause of this apparent acceleration is a
mismodeling of the orbital inclination of the spacecraft to the
ecliptic plane~\cite{pioprd, Turyshev:2005zk}.

\cite{pioprd} estimated the annual contribution to the error
budget for $a_P$. Combining $\sigma_T$ and the magnitude of the annual
sinusoidal term for the entire Pioneer~10 data span, they calculated
\begin{equation}
\sigma_\mathrm{at}=0.32\times 10^{-10}\mathrm{\ m/s}^2.
\end{equation}
This number is assumed to be the systematic error from the annual
term.

\cite{pioprd} also indicated the presence of a significant
diurnal term, with a period that is approximately equal to the
sidereal rotation period of the Earth,
23\super{h}56\super{m}04\super{s}.0989. The magnitude of
the diurnal term is comparable to that of the annual term, but the
corresponding angular velocity is much larger, resulting in large
apparent accelerations relative to $a_P$. These large accelerations,
however, average out over long observational intervals, to less than
$0.03\times 10^{-10}\mathrm{\ m/s}^2$ over a year. The origin of the annual
and diurnal terms is likely the same modeling problem~\cite{pioprd}.

These small periodic modeling errors are effectively masked by
maneuvers and plasma noise. However, as they are uncorrelated with the
observed anomalous acceleration (characterized by an essentially
linear drift, not annual/diurnal sinusoidal signatures), they do not
represent a source of systematic error.

\subsection{Error budget and the final 2002 result}
\label{sec:anomaly_sum}

\begin{table}[h!]
  \caption[Error budget: a summary of biases and uncertainties as
  known in 2002.]{Error budget: a summary of biases and uncertainties as
  known in 2002~\cite{pioprd}. Values that are the subject of on-going
  study are marked by an asterisk.}
  \label{tab:error_budget}
  \centering
      {\small
    \begin{tabular}{rlll}\\
      \toprule
Item & Description of error budget constituents & Bias & Uncertainty\\
 ~ & ~ & 10\super{-10}~m/s\super{2} & 10\super{-10}~m/s\super{2}\\
\midrule
1 & {\sf Systematics generated external to the spacecraft:} & ~ & \\
 & a) Solar radiation pressure and mass & +0.03 & \textpm~0.01\\
 & b) Solar wind & ~ & \textpm~\textless~10\super{-5}\\
 & c) Solar corona & ~ & \textpm~0.02\\
 & d) Electro-magnetic Lorentz forces & ~ & \textpm~\textless~10\super{-4}\\
 & e) Influence of the Kuiper belt's gravity & ~ & \textpm~0.03\\
 & f) Influence of the Earth's orientation &  & \textpm~0.001\\
 & g) Mechanical and phase stability of DSN antennae & ~ & \textpm~\textless~0.001\\
 & h) Phase stability and clocks & ~ & \textpm~\textless~0.001\\
 & i) DSN station location & ~ & \textpm~\textless~10\super{-5}\\
 & j) Troposphere and ionosphere & ~ & \textpm~\textless~0.001\\
2 &  {\sf On-board generated systematics:} & ~ & \\
 & a) Radio beam reaction force & +1.10\super{*} & \textpm~0.11\\
 & b) RTG heat reflected off the craft & --0.55\super{*} & \textpm~0.55\\
 & c) Differential emissivity of the RTGs & ~ &\textpm~0.85\\
 & d) Nonisotropic radiative cooling of the spacecraft  & ~~0.00\super{*} & \textpm~0.48\\
 & e) Expelled Helium  produced within the RTGs & +0.15  & \textpm~0.16\\
 & f) Gas leakage & ~ & \textpm~0.56\\
 & g) Variation between spacecraft determinations & +0.17 & \textpm~0.17\\
3 & {\sf Computational systematics:} & ~ & \\
 & a) Numerical stability of least-squares estimation &  & \textpm~0.02\\
 & b) Accuracy of consistency/model tests & ~ & \textpm~0.13\\
 & c) Mismodeling of maneuvers & ~ & \textpm~0.01\\
 & d) Mismodeling of the solar corona & ~ & \textpm~0.02\\
 & e) Annual/diurnal terms & ~ & \textpm~0.32\\
\midrule
 ~ & Estimate of total bias/error & +0.90 & \textpm~1.33\\
\bottomrule
\end{tabular}
}
\end{table}

The results of the 2002 study~\cite{pioprd} are summarized in
Table~\ref{tab:error_budget}. Sources that contribute to the overall
bias and error budget are grouped depending on their origin: external
to the spacecraft, generated on-board, or computational in
nature. Sources of error are treated as uncorrelated; the combined
error is the root sum square of the individual error values.

The contribution of effects in the first group in
Table~\ref{tab:error_budget}, that is, effects external to the
spacecraft to the overall error budget is negligible:
$\sigma_{\mathrm{external}}\sim 0.04\times 10^{-10}\mathrm{\ m/s}^2$. The second
group (on-board effects) yields the largest error contribution:
$\sigma_{\mathrm{on\text{-}board}}\sim 1.29\times
10^{-10}\mathrm{\ m/s}^2$. Lastly, computational systematics amount to
$\sigma_{\mathrm{comp}}\sim 0.35\times 10^{-10}\mathrm{\ m/s}^2$.

Similarly, the largest contribution to bias comes from on-board
effects: $b_{\mathrm{on\text{-}board}}\sim 0.87\mathrm{\ m/s}^2$, a value that
is dominated by the radio beam reaction force. External effects
contribute a bias of $b_{\mathrm{external}}\sim 0.03\mathrm{\ m/s}^2$, while
computational systematics contribute no bias.

Note that several items in Table~\ref{tab:error_budget} are marked
with an asterisk, indicating that these items are the subject of an
on-going new investigation of the Pioneer anomaly (discussed in
Section~\ref{sec:current-status}).

The bias (third column) and error (fourth column) in
Table~\ref{tab:error_budget} give the final acceleration result in the
form
\begin{equation}
a_P = a_{P(\mathrm{exp)}}+b_P\pm \sigma_P,
\label{eq:tot}
\end{equation}
where
\begin{equation}
a_{P(\mathrm{exp)}}=(7.84\pm 0.01)\times 10^{-10}\mathrm{\ m/s}^2
\label{eq:apexp}
\end{equation}
is the reported formal solution for the Pioneer anomaly that was
obtained with the data set available prior to
2002~\cite{pioprd}. Specifically, after accounting for the systematics
listed in Table~\ref{tab:error_budget} and using Equations~(\ref{eq:tot})
and~(\ref{eq:apexp}), the authors of~\cite{pioprd} presented the final
result of their study as
\begin{equation}
a_P=(8.74\pm1.33)\times 10^{-10}\mathrm{\ m/s}^2.
\end{equation}
This 6-$\sigma$ effect is clearly significant and, as of 2009, still
remains unexplained.

The 2002 analysis demonstrated that after accounting for the
gravitational and other large forces included in standard orbit
determination programs~\cite{pioprl,pioprd, Turyshev:2005zk}, the
anomaly in the Doppler frequency blue shift drift is uniformly
changing with a rate of $\dot{f}_P = (5.99\pm0.01)\times
10^{-9}\mathrm{\ Hz/s}$~\cite{Turyshev:2005zm} (see
Figure~\ref{fig:residuals}). Let us denote the frequency of the signal
observed by a DSN antenna as $f_{\mathrm{obs}}$, and the predicted
frequency of that signal after modeling conventional forces and other
signal propagation effects as $f_{\mathrm{model}}$. Then, for a one-way
signal, the observed anomalous effect to first order in $v/c$ is given
by $f_{\mathrm{obs}}-f_{\mathrm{model}}=-\dot f_Pt$. This translates to
\begin{equation}
\left[f_{\mathrm{obs}}(t)-f_{\mathrm{model}}(t)\right]_{\mathrm{DSN}}=-f_{0}\frac{a_Pt}{c},
\end{equation}
where $f_{0}$ is the DSN reference frequency~\cite{pioprd,
  Turyshev:2005zm, Turyshev:2005vj} (for a discussion of the DSN sign
conventions, see Endnote~38 of~\cite{pioprd}).

Since the publication of the 2002 study~\cite{pioprd}, many proposals
have been put forth offering theoretical explanations of the
anomaly. These are reviewed in the next section
(Section~\ref{sec:theory-explain}). On the other hand, our knowledge
of the anomaly also improved. The existence and magnitude of the
anomalous acceleration has been confirmed by several independent
researchers. Others have attempted to model the thermal behavior of
the spacecraft, arguing that the magnitude of thermal recoil forces
might have been underestimated by the 2002 study. The recovery of
essentially all Pioneer~10 and 11 telemetry, as well as large
quantities of archived project documentation, raised hope that it
might be possible to construct a sufficiently accurate thermal model
of the spacecraft using modeling software, and properly estimate the
magnitude of the thermal recoil force. This remains one of several
open, unresolved questions that, hopefully, will be answered in the
near future as a result of on-going study, as detailed in
Section~\ref{sec:current-status}.

\newpage
\section{Efforts to Explain and Study the Anomaly}
\label{sec:theory-explain}

Since the initial announcement of the anomalous acceleration of the
Pioneer~10 and 11 spacecraft, a significant number of proposals have
been made in an attempt to explain the nature of the discovered
effect. The explanations targeted the effect with the properties
presented in~\cite{pioprl, pioprd, moriond} and summarized in
Section~\ref{sec:anomaly_sum}. These key properties include
\begin{inparaenum}[i)]
\item the magnitude and the apparent constancy of the anomalous acceleration,
\item its nearly Sun-pointing direction, and
\item the apparent ``onset'' of the anomaly.
\end{inparaenum}
This set of ``known'' properties was used to analyze the mechanisms
that were put forward in numerous attempts to identify the origin of
the effect. Although the proposals are all very different and include
conventional and new physics ideas, it is possible to place them into
several broad categories. In this section we review some of these
proposed mechanisms.

First, there are attempts to explain the anomaly using unmodeled
conventional forces with an origin external to the spacecraft
(Section~\ref{sec:external-forces}), which may be both gravitational
or nongravitational in nature. Some authors considered the
possibility that the anomalous effect may be due to a new physics
mechanism indicating, for instance, modification of gravity
(Section~\ref{sec:new-physics}), or may have a cosmological origin
(Section~\ref{sec:cosmol-mech}). On the other hand, the fact that the
Pioneer anomaly was observed in the radiometric Doppler signal opens
up the possibility that the anomaly is not a dynamical effect on the
trajectories of the probes but instead is due to an unmodeled effect
on their radio signal (Section~\ref{sec:forces-radio-signal}). We also
consider proposals that attempt to explain the anomaly using
unmodeled forces of on-board origin
(Section~\ref{sec:on-board-forces}). Lastly, we review miscellaneous
mechanisms (Section~\ref{sec:othermechanisms}) and some common
misconceptions before moving on to a discussion of independent
observational confirmations (Section~\ref{sec:experiments}) and
proposals for dedicated space experiments (Section~\ref{sec:mission}).

\subsection{Unmodeled forces external to the spacecraft}
\label{sec:external-forces}

The trajectory of the Pioneer spacecraft, while governed primarily by the
gravity of the solar system, is nevertheless a result of a complex
combination of gravitational and nongravitational forces, all of
which must be taken into account for a precision orbit
determination. What if some of those forces were not properly
accounted for in the model, resulting in an unmodeled acceleration of
the observed magnitude? Several authors considered this possibility.

\subsubsection{Gravitational forces due to unknown mass distributions
  and the Kuiper belt}

Of course, one of the most natural mechanisms to generate a putative
physical force is the gravitational attraction due to a known mass
distribution in the outer solar system; for instance, due to Kuiper
belt objects or interplanetary dust. Anderson et al.~\cite{pioprd}
have considered such a possibility by studying various known density
distributions for the Kuiper belt and concluded these density
distributions are incompatible with the discovered properties of the
anomaly. Even worse, these distributions cannot circumvent the
constraint from the undisturbed orbits of Mars and Jupiter.

The possibility of a gravitational perturbation on the Pioneer paths
has also been considered by the authors
of~\cite{2006CQGra..23.4625B, 2006IJMPD..15..533D,
  2007MNRAS.375.1311I, 2005PhRvD..72h3004N},
who studied the possible effects produced by different Kuiper Belt
mass distributions, and concluded that the Kuiper Belt cannot produce
the observed acceleration.

Nieto~\cite{2005PhRvD..72h3004N} studied several models for
3-dimensional rings and wedges whose densities are either constant or
vary as the inverse of the distance, as the inverse-squared distance,
or according to the Boss--Peale model. It was demonstrated that
physically viable models of this type can produce neither the
magnitude nor the constancy of the Pioneer anomaly. In fact, the
results emphasized the difficulty in achieving a constant acceleration
within a finite cylindrically-symmetric distribution of matter. The
difficulties are even stronger if one considers the amount of mass
that would be needed to mimic the Pioneer anomaly.

The density of dust is not large enough to produce a gravitational
acceleration on the order of
$a_P$~\cite{pioprd,2006CQGra..23.4625B,2005PhRvD..72h3004N} and also
it varies greatly within the Kuiper belt, precluding any constant
acceleration. In particular, Bertolami and
Vieira~\cite{2006CQGra..23.4625B} obtained the largest acceleration
when the Kuiper belt was represented by a two-ring model. In
this case, the following magnitude of a radial acceleration
$a_\mathrm{rad}$ could be obtained (using spherical coordinates
$(r,\theta,\phi)$):
\begin{equation}
a_\mathrm{rad}(r,\theta)=-\frac{GM}{2\pi(R_1+R_2)}\int_0^{2\pi}\sum\limits_{i=1}^2R_i\frac{r-R_i\cos\theta\cos\phi}{(r^2+R_i^2-2rR_i\cos\theta\cos\phi)^{3/2}}d\phi,
\label{eq:KB-bv}
\end{equation}
where $R_1$~=39.4~AU (3:2 resonance) and $R_2$~=47.8~AU (2:1 resonance)
are the radii of the two rings, $M$ is the total mass of the Kuiper
belt, and $G$ is the gravitational constant. Equation~(\ref{eq:KB-bv}) can
be evaluated numerically, yielding a nonuniform acceleration that is
at least an order of magnitude smaller than the Pioneer anomaly. Other
dust distributions, such as those represented by a uniform disk model,
a nonuniform disk model, or a toroidal model yield even smaller
values. Hence, a gravitational attraction by the Kuiper belt can, to a
large extent, be ruled out.

\subsubsection{Drag forces due to interplanetary dust}

Several nongravitational, conventional forces have been proposed by
different authors to explain the anomaly. In particular, the drag
force due to interplanetary dust has been investigated by the authors
of~\cite{2006CQGra..23.4625B,2005PhRvD..72h3004N}. The acceleration
$a_\mathrm{drag}$ due to drag can be modeled as
\begin{equation}
a_\mathrm{drag}=-\frac{\kappa\rho v^2A}{m},\label{eq:adrag}
\end{equation}
where $\rho$ is the density of the interplanetary medium, $v$ is the
velocity of the spacecraft, $A$ its effective cross section, $m$ its
mass, while $\kappa$ is a dimensionless coefficient the value of which
is 2 for reflection, 1 for absorption, and 0 for transmission of the
dust particles through the spacecraft.

Using Equation~(\ref{eq:adrag}) as an \textit{in situ} measurement of the
``apparent'' density of the interplanetary medium, one obtains
$\rho\simeq 2.5\times 10^{-16}\mathrm{\ kg/m}^3$. This is several orders of
magnitude larger than the interplanetary dust density ($\sim\,
10^{-21}\mathrm{\ kg/m}^3$) reported by other spacecraft (see discussion
in~\cite{2005PhLB..613...11N}).

The analysis of data from the inner parts of the solar system taken by
the Pioneer~10 and 11 dust detectors strongly favors a spherical
distribution of dust over a disk. Ulysses and Galileo measurements in
the inner solar system find very few dust grains in the
$10^{-18}\mbox{\,--\,}10^{-12}\mathrm{\ kg}$ range~\cite{2005PhLB..613...11N}. The density
of dust is not large enough to produce a gravitational acceleration on
the order of $a_P$~\cite{pioprd}. The resistance caused by the
interplanetary dust is too small to provide support for the
anomaly~\cite{2005PhLB..613...11N}, so is the dust-induced frequency
shift of the carrier signal.

The mechanism of drag forces due to interplanetary dust as the origin
of the anomaly was discussed in detail in
\cite{2005PhLB..613...11N}. In particular, the authors considered
this idea by taking into account the known properties of dust in the
solar system, which is composed of  thinly scattered matter with two
main contributions:
\begin{itemize}[i)]
\item \textit{Interplanetary Dust (IPD)}: a hot-wind plasma (mainly
  $p$ and $e^-$) within the Kuiper Belt, from 30 to 100~AU with a
  modeled density of
  $\rho_{\mathrm{IPD}}\lesssim10^{-21}\mathrm{\ kg/m}^3$; and
\item \textit{Interstellar Dust (ISD)}: composed of fractions of
  interstellar dust (characterized by greater impact velocity). The
  density of ISD was directly measured by the Ulysses spacecraft,
  yielding
  $\rho_{\mathrm{ISD}}\lesssim3\times10^{-23}\mathrm{\ kg/m}^3$.
\end{itemize}
In~\cite{2005PhLB..613...11N} these properties were used to estimate
the effect of dust on Pioneer~10 and 11 and it was found that one
needs an axially-symmetric dust distribution within 20\,--\,70~AU with a
constant, uniform, and unreasonably high density of $\sim3\times
10^{-16}\mathrm{\ kg/m}^3\simeq3\cdot 10^5\,
(\rho_{\mathrm{IPD}}+\rho_{\mathrm{ISD}})$. Therefore, interplanetary
dust cannot explain the Pioneer anomaly.

One may argue that higher densities are present within the Kuiper
belt. IR observations rule out more than 0.3 Earth mass from Kuiper
Belt dust in the trans-Neptunian region. Using this figure, the
authors of~\cite{2006CQGra..23.4625B} have noted that the Pioneer
measurement of the interplanetary dust density is comparable to the
density of various Kuiper belt models. Nonetheless, the density varies
greatly within the Kuiper belt, precluding any constant acceleration.

\subsection{Possibility for new physics? Modified gravity theories}
\label{sec:new-physics}

Many authors investigated the possibility that the origin of the
anomalous signal is ``new physics''~\cite{pioprl,pioprd}. This is an
interesting conjecture, even though the probability is that some
standard physics or some as-yet-unknown systematic will be found to
explain this acceleration. Being more specific, one may ask the
question, ``Is it dark matter or a modification of gravity?''
Unfortunately, as we discuss below, it is not easy for either of these
solutions to provide a satisfactory answer.

\subsubsection{Dark matter}

Various distributions of dark matter in the solar system have been
proposed to explain the anomaly, e.g., dark matter distributed in the
form of a disk in the outer solar system with a density of
$\sim\,4\times10^{-16}\mathrm{\ kg/m}^3$, yielding the wanted effect. However,
it would have to be a special kind of dark matter that was not seen in
other nongravitational processes. Dark matter in the form of a
spherical halo of a degenerate gas of heavy neutrinos around the
Sun~\cite{1999astro.ph.10566M} and a hypothetical class of dark matter
that would restore the parity symmetry, called the \emph{mirror
  matter}~\cite{FootVolkas01}, have also been discussed. However, it
would have to be a special smooth distribution of dark matter that is
not gravitationally modulated as normal matter so obviously is.

It was suggested that the observed deceleration in the Pioneer probes
can be explained by the gravitational pull of a distribution of
undetected dark matter in the solar
system~\cite{2006IJMPD..15..533D}. Explanations of the Pioneer anomaly
involving dark matter depend on the small scale structure of
Navarro--Frenk--White (NFW) haloes, which are not known. {\it N}-body
simulations to investigate solar system size subhalos would require on
the order of 10\super{12} particles~\cite{2005PhRvD..72h3513N}, while the
largest current simulations involve around 10\super{8}
particles~\cite{2007ApJ...667..859D}. As a consequence of this lack of
knowledge about the small scale structure of dark matter, the existence of
a dark matter halo around the Sun is still an open question.

It has been proposed that dark matter could become trapped in the
Sun's gravitational potential after experiencing multiple
scatterings~\cite{1985ApJ...296..679P}, perhaps combined with
perturbations due to planets~\cite{1998PhRvL..81.5726D}. Moreover, the
birth of the solar system itself may be a consequence of the existence
of a local halo. The existence of dark matter streams crossing the
solar system, perhaps forming ring-shaped caustics analogous to the
dark matter ring postulated in \cite{2006IJMPD..15..533D}, has also
been considered by
Sikivie~\cite{1998PhLB..432..139S}. Considering an NFW dark matter
distribution~\cite{1997ApJ...490..493N}, de Diego et
al.~\cite{2006IJMPD..15..533D} show that there should be several
hundreds of earth masses of dark matter available in the solar system.

Gor'kavyi et al.~\cite{1998astro.ph.12480G} have shown that the solar
system dust distributes in two dust systems and four resonant belts
associated with the orbits of the giant planets. The density profile
of these belts approximately follows an inverse heliocentric distance
dependence law [$\rho \propto (R-k)^{-1}$, where $k$ is a
  constant]. As in the case of dark matter, dust is usually modeled as
a collisionless fluid as pressure, stresses, and internal friction are
considered negligible. Although dust is subjected to radiation
pressure, this effect is very small in the outer solar
system. Gravitational pull by dark matter has also recently been considered
also by Nieto~\cite{2008PhLB..659..483N}, who also mentioned the
possibility of searching for the Pioneer anomaly using the  New Horizons
spacecraft when the probe crosses the orbit of Saturn.

\subsubsection{Modified Newtonian Dynamics (MOND)}

The anomalous behavior of galaxy rotational curves led to an extensive
search for dark matter particles. Some authors considered the
possibility that a modification of gravity is needed to address this
challenge. Consequently, there were many attempts made at constructing
a theory that modifies Newton's laws of gravity in the regime of weak
gravitational fields. Presently, these efforts aim at constructing a
consistent and stable theory that would also be able to account for a
range of puzzling phenomena -- such as flat galaxy rotational curves,
gravitational lensing observations, and recent cosmological
data -- without postulating the existence of nonbaryonic dark matter
or dark energy of yet unknown origin. Some of these novel theories
were used to provide a cause of the Pioneer anomaly.

One approach to modify gravity, called Modified Newtonian Dynamics
(MOND), is particularly well studied in the literature. MOND is a
phenomenological modification that was proposed by
Milgrom~\cite{1984ApJ...286....7B, 1983ApJ...270..371M, 1983ApJ...270..365M,
  Milgrom01, 2007LNP...720..375S} to explain the ``flat'' rotation
curves of galaxies  by inducing a long-range modification of
gravity. In this approach, the Newtonian force law for a test particle
with mass $m$ and acceleration $\vec{a}$ is modified as follows:
\begin{equation}
m\vec{a}=\vec{F}~\rightarrow~\mu(|\vec{a}|/a_0)m\vec{a}=\vec{F}, \qquad {\rm with} \qquad \mu(x)\simeq\Bigg\{
\begin{matrix}
1~\mathrm{if}~|x|\gg 1,\\
x~\mathrm{if}~|x|\ll 1,\\
\end{matrix}
\label{eq:mond-F}
\end{equation}
where $\mu(x)$ is an unspecified function (a frequent, particularly
simple choice is $\mu(x)=x/(x+1)$; other forms of $\mu$ are also used)
and $a_0$ is some constant acceleration.

It follows from Equation~(\ref{eq:mond-F}) that a test particle separated
by $\vec{r}={\vec n}r$ from a large mass $M$, instead of the standard
Newtonian expression $\vec{a}=-GM\vec{n}/r^2$ (which still holds when
$|\vec{a}|\gg {a}_0$), is subject to an acceleration that is given
phenomenologically by the rule
\begin{equation}
|\vec{a}|~\rightarrow~\mu(|\vec{a}|/a_0)\vec{a}\simeq\Bigg\{
\begin{matrix}
\,~~~GM/r^{2}\propto {1}/{r^2}~~~\mathrm{if}~a\gg a_0~~\mathrm{(or\ large\ forces)},\\
\sqrt{a_0GM}/r\propto~{1}/{r}~~~\mathrm{if}~a\ll a_0~~\mathrm{(or\ small\ forces)}.\\
\end{matrix}
\label{eq:mond}
\end{equation}
Such a modification of Newtonian law produces a very distinct
modification of galactic rotational curves.  The velocities of
circular orbits are modified by Equation~(\ref{eq:mond}) as
\begin{equation}
a_\mathrm{centrifugal}=\frac{v^2}{r}\simeq\Bigg\{
\begin{matrix}
~~~~GM/r^{2}\\
\sqrt{a_0GM}/r\\
\end{matrix}
\quad\Rightarrow\quad
v^2\simeq\Bigg\{
\begin{matrix}
\,~~~GM/r~~~\mathrm{if}~a\gg a_0~~\mathrm{(or\ small\ distances)},\\
\sqrt{a_0GM}~~~\mathrm{if}~a\ll a_0~~\mathrm{(or\ large\ distances)}. \\
\end{matrix}
\label{eq:mond-v}
\end{equation}
With a value of $a_0\simeq 1.2\times 10^{-10}\mathrm{\ m/s}^2$. MOND
reproduces many galactic rotation curves.

Clearly, the original MOND formulation is purely phenomenological,
which drew some criticism toward the approach. However, recently a
relativistic theory of gravitation that reduces to MOND in the
weak-field approximation was proposed by Bekenstein in the form of the
tensor-vector-scalar (TeVeS) gravity
theory~\cite{Bekenstein:2004ne}. As the exact form of $\mu(x)$ remains
unspecified in both MOND and TeVeS, it is conceivable that an
appropriately chosen $\mu(x)$ might reproduce the Pioneer anomaly even
as the theory's main result, its ability to account for galaxy
rotation curves, is not affected.

As far as the Pioneer anomaly is concerned, considering the strong
Newtonian regime (i.e., $a_0\ll GM/r^2$) and choosing $\mu(x)=1+\xi
x^{-1}$, one obtains a modification of Newtonian acceleration in the
form $a=-GM/r^2-\xi a_0$, which reproduces the qualitative behavior
implied by the observed anomalous acceleration of the
Pioneers. However, Sanders~\cite{2006MNRAS.370.1519S} concludes that
if the effects of a MONDian modification of gravity are not observed
in the motion of the outer planets in the solar system (see
Section~\ref{sec:solar-system-data} for discussion), the acceleration
cannot be due to MOND. On the other hand, Bruneton and
Esposito-Far\`ese~\cite{Bruneton:2007si} demonstrate that while it may
require model choices that are not justified by underlying symmetry
principles, it is possible to simultaneously account for the Pioneer
anomalous acceleration and for the tests of general relativity in the
solar system within a consistent field theory.

Laboratory experiments have recently reached new levels of precision in
testing the proportionality of force and acceleration in Newton's
second law, $\vec{F}=m\vec{a}$, in the limit of small forces and
accelerations~\cite{Abramovichi-Vager-PRD-1986,
  Gundlach-etal-2007}. The tests were motivated to explore the
acceleration scales implied by several astrophysical puzzles, such as
the observed flatness of galactic rotation curves (with MOND-implied
acceleration of $a_0=1.2\times 10^{-10}\mathrm{\ m/s}^2$), the Pioneer anomaly
(with $a_P\sim 9\times 10^{-10}\mathrm{\ m/s}^2$) and the natural scale set by
the Hubble acceleration ($a_H=cH\simeq
7\times10^{-10}\mathrm{\ m/s}^2$). Gundlach et
al.~\cite{Gundlach-etal-2007} reported no violation of Newton's
second law at accelerations as small as $5\times
10^{-14}\mathrm{\ m/s}^2$. The obtained result does not invalidate
MOND directly as the formalism requires that the measurement must be
carried out in the absence of any other large accelerations (i.e.,
those due to the Earth and our solar system). However, the test
constrains theoretical formalisms that seek to derive MOND from
fundamental principles by requiring that formalism to reproduce
$\vec{F}=m\vec{a}$ under laboratory conditions similar to those used
in the experiment.

Finally, there were suggestions that rather then modifying laws of
gravity in order to explain the Pioneer effect, perhaps we needs to
modify laws of inertia instead~\cite{Milgrom:2005mc}. To that extent,
modified-inertia as a reaction to Unruh radiation has been considered
in \cite{2007MNRAS.376..338M}.

\subsubsection{Large distance modifications of Newton's potential}

Motivated by the puzzle of the anomalous galactic rotation curves,
many phenomenological models of modified Newtonian potential (leading
to the changes in the gravitational inverse-square law) were
considered, a Yukawa-like modification being one of the most popular
scenario. Following Sanders~\cite{Sanders-1984AA}, consider the
ansatz:
\begin{equation}
U(r)=U_\mathrm{Newton}(r)\left(1+\alpha e^{-r/\lambda}\right),
\label{eq:yukawa}
\end{equation}
which, as was shown in~\cite{Sanders-1984AA}, is able to successfully
explain many galactic rotation curves. The same expression may also be
used to study the physics of the Pioneer
anomaly~\cite{2006NewA...12..142M}. Indeed, Equation~(\ref{eq:yukawa}) implies that a
body moving in the gravitational field of the Sun is subject to the
acceleration law:
\begin{equation}
a(r)=-G_0\frac{M_\odot}{r^2}+\frac{\alpha}{1+\alpha}G_0\frac{M_\odot}{2\lambda^2}- \frac{\alpha}{1+\alpha}G_0\frac{M_\odot}{3\lambda^2}\frac{r}{\lambda}+...,
\label{eq:yukawa-accel}
\end{equation}
where $G_0=(1+\alpha)G$ is the observed gravitational constant in the
limit $r\rightarrow 0$. Identifying the second term in
Equation~(\ref{eq:yukawa-accel}) with the Pioneer acceleration one can
solve, for instance, for the parameter $\alpha=\alpha(a_P,\lambda)$,
and obtain:
\begin{equation}
a_P=\frac{\alpha}{1+\alpha}G_0\frac{M_\odot}{2\lambda^2} \qquad \Rightarrow\qquad
\alpha=\frac{2\lambda^2 a_P}{G_0M_\odot-2\lambda^2 a_P}.
\label{eq:yukawa-aP}
\end{equation}
The denominator in Equation~(\ref{eq:yukawa-aP}) implies that
$\lambda>\sqrt{GM_\odot/(2a_P)}$, or $\lambda\ge 2.8 \times
10^{14}\mathrm{\ m}$.

A combination of $\log \lambda >16$ with $\log |\alpha|\approx 0$ is
compatible with the existing solar system data and the Yukawa
modification in the form of Equation~(\ref{eq:yukawa}) may provide a viable
model for the Pioneer anomaly~\cite{Laemmerzahl-2009}. Furthermore,
after rearranging the terms in Equation~(\ref{eq:yukawa-accel}) as
\begin{equation}
a(r)=-G_0\frac{M_\odot}{r^2}+a_P- \frac{2}{3}a_P\frac{r}{\lambda}+...,
\label{eq:yukawa-accel-rear}
\end{equation}
one can note that the third term in Equation~(\ref{eq:yukawa-accel-rear})
is smaller than $a_P$ by a factor of
$\frac{2}{3}\frac{r}{\lambda}\le0.06$ and may account for the small
decrease in the observed acceleration. The values for the parameters
$\alpha$ and $\lambda$ obtained for the Pioneer anomaly are also
compatible with the analysis of the galactic rotation curves. Indeed,
for the case of $\log \lambda >16$ the Pioneer anomalous acceleration
implies $\alpha+1\le 10^{-5}$ while the galactic curves data yields a
weaker limit of  $\alpha+1\le 10^{-1}$.

A modification of the gravitational field equations for a metric
theory of gravity, by introducing a momentum-dependent linear relation
between the Einstein tensor and the energy-momentum tensor, has been
developed by Jaekel and Reynaud~\cite{2005MPLA...20.1047J,
Jaekel-Reynaud:2005,2006CQGra..23..777J,2006CQGra..23.7561J,mark_serge_05}
and was shown to be able to account for $a_P$. The authors identify two
sectors, characterized by the two potentials
\begin{equation}
g_{00}\simeq 1+2\Phi_N,~~~~~g_{00}g_{rr}\simeq 1+2\Phi_P,
\end{equation}
where $g_{00}$ and $g_{rr}$ are components of the metric written in
Eddington isotropic coordinates. The Pioneer anomaly could be
accounted for by an anomaly in the Newtonian potential,
$\delta\Phi_N\simeq (r-r_1)/\ell_P$ with a characteristic length scale
given by $\ell_P^{-1}\equiv a_P/c^2$. However, this model is likely
excluded by measurements such as Viking Mars radio ranging. On the
other hand, an anomaly due to the potential in the second sector in
the form
\begin{equation}
\delta\Phi_P\simeq -\frac{(r-r_1)^2+\mu_P(r-r_1)}{3\kappa\ell_P},
\end{equation}
with $\kappa$ given by $1/3\kappa\ell_P\simeq 4\times 10^8\mathrm{\ AU}^{-2}$
and $\mu_P$ being a further characteristic length representing the
radial derivative of the metric anomaly at the Earth's orbit, could
account for the Pioneer anomaly, and the conflict with Viking ranging
data can be resolved~\cite{2006CQGra..23.7561J}.

Other related proposals include Yukawa-like or higher-order
corrections to the Newtonian potential~\cite{pioprd} and Newtonian
gravity as a long wavelength excitation of a scalar condensate
inducing electroweak symmetry breaking~\cite{1999hep.ph...10372C}.

\subsubsection{Scalar-tensor extensions of general relativity}
\label{sec:scalar-tensor}

There are many proposals that attempt to explain the Pioneer anomaly
by invoking scalar fields. In scalar-tensor theories of gravity, the
gravitational coupling strength exhibits a dependence on a scalar
field $\varphi$. A general action for these theories can be written as
\begin{equation}
S={c^3\over 4\pi G}\int d^4x \sqrt{-g}\left[\frac{1}{4}f(\varphi)R-\frac{1}{2}g(\varphi)\partial_{\mu}\varphi\partial^{\mu}\varphi+V(\varphi)\right]+\sum_{i}q_{i}(\varphi)\mathcal{L}_{i}, \label{eq:sc-tensor}
\end{equation}
where $f(\varphi)$, $g(\varphi)$, and $V(\varphi)$ are generic
functions, $q_i(\varphi)$ are coupling functions, and
$\mathcal{L}_{i}$ is the Lagrangian density of matter fields, as
prescribed by the Standard Model of particles and fields.

Effective scalar fields are prevalent in supersymmetric field theories
and string/M-theory. For example, string theory predicts that the
vacuum expectation value of a scalar field, the dilaton, determines
the relationship between the gauge and gravitational couplings. A
general, low energy effective action for the massless modes of the
dilaton can be cast as a scalar-tensor theory (as in
Equation~(\ref{eq:sc-tensor})) with a vanishing potential, where
$f(\varphi)$, $g(\varphi)$, and $q_{i}(\varphi)$ are the dilatonic
couplings to gravity, the scalar kinetic term, and the gauge and
matter fields, respectively, which encode the effects of loop effects
and potentially nonperturbative corrections.

Brans--Dicke theory~\cite{Brans-Dicke-1961} is the best known
alternative scalar theory of gravity. It corresponds to the choice
\begin{equation}
\label{eq:BD}
f(\varphi)=\varphi,\qquad g(\varphi)=\frac{\omega}{\varphi},\qquad V(\varphi)=0.
\end{equation}
In Brans--Dicke theory, the kinetic energy term of the field $\varphi$
is noncanonical and the latter has a dimension of energy squared. In
this theory, the constant $\omega$ marks observational deviations from
general relativity, which is recovered in the limit
$\omega\to\infty$. In the context of Brans--Dicke theory, one can
operationally introduce Mach's principle, which states that the
inertia of bodies is due to their interaction with the matter
distribution in the Universe. Indeed, in this theory the gravitational
coupling is proportional to $\varphi^{-1}$, which depends on the
energy-momentum tensor of matter through the field equation
$\Box^2\varphi=8\pi/(3+2\omega)T$ where $T$ is the trace of the matter
stress-energy tensor defined as the variation of $\mathcal{L}_i$ with
respect to the metric tensor.

The $\omega$ parameter can be directly related to the
Eddington--Robertson (PPN) parameter $\gamma$ by the
relation~\cite{Will-lrr-2006-3}: $\gamma=(\omega+1)/(\omega+2)$. The
stringent observational bound resulting from the 2003 experiment with
the Cassini spacecraft require that $|\omega| \gtrsim
40\,000$~\cite{2003Natur.425..374B, Will-lrr-2006-3}. On the other
hand, $\omega=-3/2$ may be favored by cosmological observations and
also offer a resolution of the Pioneer
anomaly~\cite{2007AnP....16..237D}. A possible resolution can be
obtained by incorporating a Gauss--Bonnet term in the form of ${\cal
  L}_\mathrm{GB}=R_{\mu\nu\rho\sigma}R^{\mu\nu\rho\sigma}-4R_{\mu\nu}R^{\mu\nu}+R^2$
into the Brans--Dicke version of the Lagrangian
Equation~(\ref{eq:sc-tensor}) with the choice of Equation~(\ref{eq:BD}), which may
allow the Eddington parameter $\gamma$ to be arbitrarily close to 1,
while choosing an arbitrary value for
$\omega$~\cite{2008PhRvD..78h4009A}. Another scalar-tensor model,
proposed by Novati et al.~\cite{2000GrCo....6..173C}, was also
motivated in part by the observed anomalous acceleration of the two
Pioneer spacecraft.

Other scalar-tensor approaches using different forms of the Lagrangian
Equation~(\ref{eq:sc-tensor}) were used to investigate the
anomaly. Capozziello et al. ~\cite{2001MPLA...16..693C} developed a
proposal based on flavor oscillations of neutrinos in Brans--Dicke
theory; Wood~\cite{Wood:2001ve} proposed a theory of conformal gravity
with dynamical mass generation, including the Higgs
scalar. Cadoni~\cite{2004GReGr..36.2681C} studied the coupling of
gravity with a scalar field with an exponential potential, while
Bertolami and P\'aramos~\cite{2004CQGra..21.3309B} also applied a
scalar field in the context of the braneworld scenarios. In
particular, Bertolami and Par\'amos~\cite{2004CQGra..21.3309B} have
shown that a generic scalar field cannot explain $a_P$; on the other
hand, a non-uniformly-coupled scalar could produce the wanted
effect. In addition, although braneworld models with large extra
dimensions may offer a richer phenomenology than standard
scalar-tensor theories, it seems difficult to find a convincing
explanation for the Pioneer anomaly~\cite{2005PhRvD..71b3521B}.

\subsubsection{Scalar-tensor-vector modified gravity theory (MOG)}

Moffat~\cite{2006JCAP...03..004M} attempted to explain the anomaly in
the framework of Scalar-Tensor-Vector Gravity (STVG) theory. The
theory originates from investigations of a nonsymmetric
generalization of the metric tensor, which gives rise to a
skew-symmetric field. Endowing this field with a mass led to the
Metric-Skew-Tensor Gravity (MSTG) theory, while the further step of
replacing the skew-symmetric field with the curl of a vector field
yields STVG. The theory successfully accounts for observed galactic
rotation curves, galaxy cluster mass profiles, gravitational lensing
in the Bullet Cluster (1E0657-558), and cosmological observations.

The STVG Lagrangian takes the form,
\begin{align}
{\cal L}&=\sqrt{-g}\bigg\{-\frac{1}{16\pi G}\left(R+2\Lambda\right)
-\frac{1}{4\pi}\omega\bigg[\frac{1}{4}B^{\mu\nu}B_{\mu\nu}-\frac{1}{2}\mu^2\phi_\mu\phi^\mu+V_\phi(\phi)\bigg]\nonumber\\
&-\frac{1}{G}\bigg[\frac{1}{2}g^{\mu\nu}\bigg(\frac{\nabla_\mu G\nabla_\nu G}{G^2}+\frac{\nabla_\mu\mu\nabla_\nu\mu}{\mu^2}-\nabla_\mu\omega\nabla_\nu\omega\bigg)+\frac{V_G(G)}{G^2}+\frac{V_\mu(\mu)}{\mu^2}+V_\omega(\omega)\bigg]\bigg\},\label{eq:MOG}
\end{align}
where $g$ is the determinant of the metric tensor, $R$ is the
Ricci-scalar, $\Lambda$ is the cosmological constant, $\phi_\nu$ is a
massive vector field with (running) mass $\mu$,
$B_{\mu\nu}=\partial_\mu\phi_\nu-\partial_\nu\phi_\mu$, $G$ is the
(running) gravitational constant, $\omega$ is the (running) vector
field coupling constant, and $V_\phi$, $V_G$, $V_\mu$ and $V_\omega$
are the potentials associated with the vector field and the three
running scalar fields.

The spherically symmetric, static vacuum solution of
Equation~(\ref{eq:MOG}) yields, in the weak field limit, an effective
gravitational potential that is a combination of a Newtonian and a
Yukawa-like term, and can be written as
\begin{equation}
G_\mathrm{eff}=G_N\Big(1+\alpha-\alpha(1+\mu r)e^{-\mu r}\Big).
\end{equation}
In earlier papers, the values of $\alpha$ and $\mu$ were treated as
fitted parameters. This allowed Brownstein and
Moffat~\cite{2006CQGra..23.3427B, Moffat:2004ud} to reproduce an
anomalous acceleration of the correct magnitude and also account for
the anomaly's apparent ``onset'' at a distance of $\sim$~10~AU from the
Sun. More recently, the values of $\alpha$ and $\mu$ were derived
successfully as functions of the gravitational source
mass~\cite{Moffat:2007nj}. This later approach results in the
prediction of Newtonian behavior within the solar system, indeed
within all self-gravitating systems with a mass below several times
$10^6 M_{\odot}$.

\subsection{Cosmologically-motivated mechanisms}
\label{sec:cosmol-mech}

There have been many attempts to explain the anomaly in terms of the
expansion of the Universe, motivated by the numerical coincidence
$a_P\simeq cH_0$, where $c$ the speed of light and $H_0$ is the Hubble
constant at the present time (see Section~\ref{sec:misconceptions} for
details). These attempts were also stimulated by the fact that the
initial announcement of the anomaly~\cite{pioprl} came almost
immediately after reports on the luminosity distances of type Ia
supernovae~\cite{Perlmutter-1998Natur.391...51P, Riess_supernovae98}
that were followed by measurements of the angular structure of the
cosmic microwave background (CMB)~\cite{deBernardis_CMB2000},
measurements of the cosmological mass densities of large-scale
structures~\cite{Peacock_LargeScale01} that have placed stringent
constraints on the cosmological constant $\Lambda$ and led to a
revolutionary conclusion: The expansion of the universe is
accelerating. These intriguing numerical and temporal coincidences led
to heated discussion (see, e.g., \cite{2007CQGra..24.2735L,
2007IJTP...46.3193O} for contrarian views) of the
possible cosmological origin of the Pioneer anomaly.

Below we discuss cosmologically-motivated mechanisms used to explain
the Pioneer anomaly.

\subsubsection{Cosmological constant as the origin of the Pioneer anomaly}

An inverse time dependence for the gravitational constant $G$ produces
effects similar to that of an expanding universe. So does a length or
momentum scale-dependent cosmological term in the gravitational action
functional~\cite{1999NuPhB.556..397M, Rosales:1998mj}. It was claimed
that the anomalous acceleration could be explained in the frame of a
quasi-metric theory of relativity~\cite{Ostvang:1999dm}. The possible
influence of the cosmological constant on the motion of inertial
systems leading to an additional acceleration has been
discussed~\cite{Rosales:1998mj}.  Gravitational coupling resulting in
an increase of the constant $G$ with scale is analyzed by Bertolami
and Garc\'ia-Bellido~\cite{1996IJMPD...5..363B}. A 5-dimensional
cosmological model with a variable extra dimensional scale factor in a
static external space~\cite{1999gr.qc.....3016B, Belayev:2004vy} was
also proposed. It was suggested that the coupling of a cosmological
``constant'' to matter ~\cite{Massa:2008zz} may provide a connection
with the Pioneer anomaly.

Kagramanova et al.~\cite{Kagramanova:2006ax} (see
also~\cite{Kerr-etal-2003,2006PhRvD..73f3004S}) have studied the
effect of the cosmological constant on the outcome of the various
gravitational experiments in the solar system by taking the metric of
the Schwarzschild--de~Sitter spacetime:
\begin{equation}
ds^2=\alpha(r) dt^2-\alpha(r)^{-1} dr^2-r^2(d \theta^2+\sin^2\theta d\phi^2),
\end{equation}
where
\begin{equation}
 \alpha(r) = 1-\frac{2M}{r}-\frac{1}{3}\Lambda r^2
\end{equation}
with $\Lambda$ being the cosmological constant and $M$ the mass of the
source. (Note that $\Lambda<0$ would result in attraction, while
$\Lambda>0$ will lead to repulsion.)

The authors of~\cite{pioprd} have shown that if the cosmological
expansion would be at the origin of the Pioneer anomaly, such a
mechanism would produce an opposite sign for the effect. Taken at face
value, the anomaly would imply a negative cosmological constant of
$\Lambda\sim -3\times 10^{-37}\mathrm{\ m}^{-2}$, which contradicts both the
solar system data and the data on cosmological expansion. Indeed, the
highest limit on $\Lambda$ allowable by the solar system tests is set
by the data on the perihelion advance, which limit the value of the
cosmological constant to
$\Lambda\leq3\times10^{-42}\mathrm{\ m}^{-2}$~\cite{2006PhRvD..73f3004S}.
However, the data on the cosmological accelerated expansion yields the value of
$\Lambda\approx10^{-52}\mathrm{\ m}^{-2}$, leading Kerr et
al.~\cite{Kerr-etal-2003} to conclude that the cosmological effects
are  too small to be measured in the solar system dynamical experiments.

Hackmann and L\"ammerzahl~\cite{Hackmann:2008zza, Hackmann:2008zz}
developed the analytic solution of the geodesic equation in
Schwarzschild--(anti-)de~Sitter spacetimes and show that the influence
of the cosmological constant on the orbits of test masses is
negligible. They concluded that the cosmological constant cannot be
held responsible for the Pioneer anomaly.

Thus, there is now a consensus that the Pioneer anomaly cannot be of a
cosmological origin and, specifically, $\Lambda$ cannot be responsible
for the observed anomalous acceleration of the Pioneer~10 and 11
spacecraft. However, the discussion is still ongoing.

\subsubsection{The effect of cosmological expansion on local systems}

The effect of cosmological expansion on local systems had been studied
by a number of authors~\cite{Anderson-1995PhysRevLett.75.3602,
  Cooperstock:1998ny, Einstein-Strauss-1945RevModPhys.17.120,
  Gautreu-1984PhysRevD.29.198, McVittie-1931MNRAS}, (for reviews,
see~\cite{Carrera-Giulini-2008, Ferreira:2009cx, LaemmerzahlDittus05,
  Wilson:2009rg}). To study the behavior of small isolated mass in
expanding universe, one starts with the weak field
ansatz~\cite{Carrera-Giulini-2008, LaemmerzahlDittus05}:
\begin{equation}
g_{\mu\nu}=b_{\mu\nu}+h_{\mu\nu}, \qquad h_{\mu\nu}\ll b_{\mu\nu},
\end{equation}
and derives the linearized Einstein equations for $h_{\mu\nu}$:
\begin{equation}
b^{\rho\sigma}D_\rho D_\sigma{\bar h}_{\mu\nu}+b^{\rho\sigma}R^\kappa_{\mu\rho\nu}{\bar h}_{\kappa\sigma}=16G\pi T_{\mu\nu}.
\end{equation}
The relevant solution with modified Newtonian potential is given below
\begin{equation}
h_{00}=\frac{2GM}{R}\frac{\cos(\sqrt{6}|{\dot R}|r)}{r}=\frac{2GM}{Rr}\Big(1-3H^2(Rr)^2+...\Big).
\label{eq:h00}
\end{equation}
The first part in Equation~(\ref{eq:h00}) is the standard Newtonian
potential with the measured distance $R(t)r$ in the
denominator. L\"emmerzahl et al.~\cite{LaemmerzahlDittus05} observed
that the additional acceleration towards the gravitating body is of
the second order in the Hubble constant $H$.  As such, this potential
practically does not participate in the cosmic expansion; thus, there
is no support for the cosmological origin of $a_P$.

Oliveira~\cite{2007IJTP...46.3193O} conjectured that the solar system
has escaped the gravity of the Galaxy, as evidenced by its orbital
speed and radial distance and by the visible mass within the solar
system radius. Spacecraft unbound to the solar system would also be
unbound to the galaxy and subject to the Hubble law. However, this
hypothesis produces practically unnoticeable effects.

\subsubsection{The cosmological effects on planetary orbits}

The cosmological effects on the planetary orbits has been addressed in
many papers recently (for instance, \cite{Ferreira:2009cx}). To study
the effect of cosmological modification of planetary orbit, one
considers the action
\begin{equation}
S=-mc^2\int\Big(1-2\frac{U(x)}{c^2}-R^2(t)\frac{{\dot x}^2}{c^2}\Big)^\frac{1}{2}dt \approx \int\Big(-mc^2+mU(r)+\frac{m}{2}R^2(t)({\dot r}^2+r^2{\dot\varphi}^2)\Big)dt.
\label{eq:S}
\end{equation}
In the case of weak time dependence of $R(t)$ the action above has two
adiabatic invariants:
\begin{equation}
I_\varphi =L, \qquad I_r=-L+\frac{GmM}{R}\sqrt{\frac{m}{2|E|}},
\label{eq:ai}
\end{equation}
which determine the energy, $E$, momentum, $p$ and the eccentricity,
$e$, of an orbit:
\begin{equation}
E=-\frac{m^3G^2M^2}{2(I_r+I_\varphi)}, \qquad p=\frac{I_\varphi^2}{m^2MR(t)}, \qquad e^2=1-\Big[\frac{I_\varphi}{I_r+I_\varphi}\Big]^2.
\label{eq:po}
\end{equation}
L\"ammerzahl~\cite{Laemmerzahl-2009} noted that in this scenario $E,
e$ and $R(t)p$ stay nearly constant. In addition, $R(t)p$ and $e$ of
the planetary orbits practically do not participate in cosmic
expansion. Remember that the stability of adiabatic invariants is
governed by the factor of $e^{-\tau/T}$, where $\tau$ is the
characteristics time of change of external parameter (here:
$\tau=1/H$) and $T$ is the characteristic time of the periodic motion
(here: $T$ = periods of planets). In the the case of periodic bound
orbits $\tau/T\sim10^{10}$, thus, any change of the adiabatic
invariants is truly negligible.

\subsubsection{Gravitationally bound systems in an expanding universe}

The question of whether or not the cosmic expansion has an influence
on the size of the Solar system was addressed in conjunction to
the study of the Pioneer anomaly. In particular, is there a difference
between locally bound and escape orbits? If the former are proven to
be practically immune to cosmic expansion, what about the latter? In
fact, the properties of bound (either electrically or gravitationally)
systems in an expanding universe have been discussed controversially
in many papers, notably
by~\cite{Anderson-1995PhysRevLett.75.3602,1933MNRAS..93..325M}.

Effects of cosmological expansion on local systems were addressed by a
number of authors (for reviews see~\cite{Carrera-Giulini-2008,
  Laemmerzahl-2009}). Gautreu~\cite{Gautreu-1984PhysRevD.29.198}
studied the behavior of a spherical mass with the energy-stress tensor
taken in the form of an ideal fluid. The obtained results show that
the outer planets would tend to out-spiral away from the solar
system. Anderson~\cite{Anderson-1995PhysRevLett.75.3602} has studied
the behavior of local systems in the cosmologically-curved
background. He obtained cosmological modifications of local
gravitational fields with an additional drag term for escape orbits and
demonstrated that $Rr=const$ for bound orbits. Cooperstock et
al.~\cite{Cooperstock:1998ny} used the Einstein--de Sitter universe
$ds^2=c^2dt^2-R^2(t)(dr^2+d\Omega^2)$ and derived a geodesic deviation
equation in Fermi normal coordinates $\ddot x-({\ddot
  R}/R)x=0$. Clearly, the additional terms are too small be observed
in the solar system.

As far as the Pioneer anomaly is concerned, the papers above are
consistent in saying that planetary orbits in the solar system should see
the effects of the  anomaly and $a_P$ may not be of gravitational
origin. However, Anderson~\cite{Anderson-1995PhysRevLett.75.3602}
found an interesting result that suggests that the expansion couples
to escape orbits, while it does not couple to bound orbits.

To study this possibility L\"ammerzahl~\cite{Laemmerzahl-2009} used a
PPN-inspired spacetime metric:
\begin{equation}
g_{00}=1-2\alpha\frac{U}{c^2}+2\beta\frac{U^2}{c^4}, \qquad
g_{ij}=-\Big(1+2\gamma \frac{U}{c^2}\Big)R^2(t)\delta_{ij},
\end{equation}
that led to the following equation of motion (written in terms of
measured distances and times):
\begin{equation}
\frac{d^2X^i}{dT^2}=\frac{\partial U}{\partial X^i}\bigg[\alpha+\gamma\frac{1}{c^2}\left(\frac{dX}{dT}\right)^2+\big(2\alpha^2-\gamma\alpha-2\beta\big)\frac{U}{c^2}\bigg]-\bigg[\big(\alpha+\gamma\big)\frac{1}{c^2}\frac{\partial U}{\partial X^j}\frac{dX^j}{dT}+\frac{\dot R}{R}\bigg]\frac{dX^i}{dT}.
\label{eq:PPN-cosm}
\end{equation}
Using Equation~(\ref{eq:PPN-cosm}) L\"ammerzahl~\cite{Laemmerzahl-2009}
concludes that in cosmological context the behavior of bound orbits is
different from that of unbound ones. However, cosmological expansion
results in a decelerating drag term, which is a factor $v/c$ too small
to account for the Pioneer effect.

\subsubsection[Dark-energy-inspired $f(R)$ gravity models]{Dark-energy-inspired \boldmath$f(R)$ gravity models}

The idea that the cosmic acceleration of the Universe may be caused by a
modification of gravity at very large distances, and not by a dark
energy source, has recently received a great deal of attention
(see~\cite{Rubakov:2008nh, Sotiriou:2008rp}). Such a modification
could be triggered by extra space dimensions, to which gravity spreads
over cosmic distances. Recently, models involving inverse powers of
the curvature or other invariants have been proposed as an alternative
to dark energy. Although such theories can lead to late-time
acceleration, they typically result in one of two problems: either
they are in conflict with tests of general relativity in the solar
system, due to the existence of additional dynamical degrees of
freedom~\cite{Chiba:2003ir}, or they contain ghost-like degrees of
freedom that seem difficult to reconcile with fundamental theories.

Gravity theories that supplement the Einstein--Hilbert Lagrangian with
a nonlinear function $f(R)$ of the curvature scalar have attracted
interest in the context of inflationary cosmological scenarios, and
because these theories may explain the accelerated expansion of the
universe without the need to postulate scalar fields or a cosmological
constant (for a review, see~\cite{Sotiriou:2008rp}).

As an upshot of these efforts, Bertolami et
al.~\cite{2007PhRvD..75j4016B} explored a particular family of $f(R)$
Lagrangians that gives rise to an extra acceleration in the form
\begin{equation}
a_E=\frac{f^2r^2}{GM}+2f.
\end{equation}
Assuming that $f\sim\alpha GM/r$ where $\alpha$ is a constant leads to
$a_E\propto \alpha^2=const$ in the limit of large $r$. For a galaxy,
$a_E$ is naturally associated with the MOND acceleration $a_0$, while
in the solar system, it may acquire the Pioneer acceleration value,
$a_P$.

A similar result was obtained by Saffari and
Rahvar~\cite{2008PhRvD..77j4028S} who used a metric formalism in
$f(R)$ modified gravity to study the dynamics of various systems from
the solar system to the cosmological scales. Replacing $f(R)$ with
$F(R)=df/dR$, the authors postulated the ansatz
$F(r)=(1+r/d)^{-\alpha}$ where $\alpha\ll 1$ is a dimensionless
constant and $\alpha r/d\ll 1$. They obtained an anomalous
acceleration term of $-\alpha/2d$, which agrees with the Pioneer
anomaly for $\alpha/d\simeq 10^{-26}\mathrm{\ m}^{-1}$.

On the other hand, Exirifard \cite{2007arXiv0708.0662E} demonstrated that
a correction to the Einstein-Hilbert Lagrangian in the form of
$\Delta{\cal L}=R_{\kappa\lambda\mu\nu}R^{\kappa\lambda\mu\nu}$
cannot offer a covariant resolution to the Pioneer anomaly.

Capozziello et al.,~\cite{Capozziello:2007ms} developed a general
analytic procedure to investigate the Newtonian limit of $f(R)$
gravity. The authors discussed the Newtonian and the post-Newtonian
limits of these gravity models, including the investigation of their
possible relevance to the Pioneer anomaly. Capozziello et
al.~\cite{Capozziello:2009nq} mention that, by treating the Pioneer
anomaly as a correction to the Newtonian potential, it could be
studied in the general theoretical scheme of $f(R)$ gravity theories.

\subsection{Effects on the radio signal}
\label{sec:forces-radio-signal}

It has been argued~\cite{2003EL.....63..653R} that the cosmic
expansion influences the measurement process via a change in the
frequency of the traveling electromagnetic signals. However, one
expects that taking all effects of the cosmic expansion on the
frequency as well as on the Pioneer motion into account, the resulting
acceleration is $-vH_0$ and, thus, has the correct sign but is too
small by a factor $v/c$~\cite{LaemmerzahlDittus05}. The ways in which
the cosmic expansion might be responsible for $a_P$ vary considerably
between the approaches. It is known~\cite{pioprl,pioprd} that the very
presence of the Pioneer anomalous acceleration contradicts the
accurately known motion of the inner planets of our solar system. This
motivated focusing on the effect of cosmic acceleration on the radio
communication signal rather than on the spacecraft themselves. This
mechanism might be able to overcome the apparent conflict that $a_P$
presents to modern solar system planetary
ephemerides~\cite{pioprl,pioprd}.

\subsubsection{Helicity-rotation coupling}

The radio signal used for communicating with Pioneer~10 and 11 (and
indeed, used routinely with other spacecraft) is circularly
polarized. Therefore, the question naturally arises as to whether the
coupling between the helicity of the radio signal and the rotation of
either the transmitter or the receiver could contribute to the
observed Doppler anomaly in the Pioneer radio
signal~\cite{2002PhLA..306...66M}. To first order, this coupling can
increase or decrease the frequency of a radio signal by the rotational
frequency of the transmitter (or receiver):
\begin{equation}
\omega'=\omega\mp\Omega,
\end{equation}
where $\omega$ is the signal frequency that would be observed in the
absence of rotation, $\Omega$ is the rotational frequency, while
$\omega'$ is the observed frequency. The sign is positive for a
negative helicity wave, and vice versa, and the formula must be
applied to each leg of the communication separately (e.g., to the
uplink and downlink leg in the case of two-way or three-way Doppler
data.)

For the Pioneer spacecraft, $\Omega\simeq 0.076$~Hz (Pioneer~10) and
$\Omega\simeq 0.13$~Hz (Pioneer~11). However, Anderson and
Mashhoon~\cite{2003PhLA..315..199A} note that this effect cannot
account for the Pioneer anomaly. The effect of rotation on the radio
signal is already phenomenologically incorporated into the Doppler
data analysis (see Section~\ref{sec:effect_spin}).

\subsubsection{Clock acceleration}

The fact that the anomaly was discovered using Doppler techniques
leaves duality in the nature of the detected signal -- it is either
true physical acceleration $a_P$ or a time acceleration $a_t$ that is
connected with the former by the relationship  $a_P = c~ a_t$. This
fact motivated Anderson et al.~\cite{pioprd} to try to look for
purely phenomenological ``time'' distortions  that might fit the
Pioneer data.  The question was ``is there any evidence that some kind
of `time acceleration' is being seen?''

A number of models were investigated and discarded for various reasons
(see~\cite{pioprd} for discussion), but there was one model that was
especially interesting.  This model adds a term that is quadratic in
time to the light time, as seen by the DSN station as $t\rightarrow t
+\frac{1}{2}a_tt^2$. In particular, let any  labeled time $t_a$ be
given as
\begin{equation}
t_a-t_0\rightarrow t_a-t_0+\frac{1}{2}a_t\left(t_a^2-t_0^2\right),
\end{equation}
then the light time is
\begin{equation}
\Delta t=t_\mathrm{received}-t_\mathrm{sent}\rightarrow \Delta t+\frac{1}{2}a_t\left(t_\mathrm{received}^2-t_\mathrm{sent}^2\right).
\label{eq:a-quadratic}
\end{equation}

Expression~(\ref{eq:a-quadratic}) mimics a line of sight acceleration
of the spacecraft, and could be thought of as an \textit{expanding
  space}  model.  Note that $a_t$ affects only the data, not the
trajectory. It was pointed out by Anderson et al. that this model fits
both Doppler and range very well for several spacecraft used in their
study~\cite{pioprd}. This fact motivated the discussion on the nature
of the implied numerical relationship between the Hubble constant and
$a_P$ (Section~\ref{sec:misconceptions}). To investigate further the
nature of this relation one would need to check the data of other
spacecraft, compare modern clocks with accuracy much higher then that
used in the navigation of the Pioneers, as well as the data on
millisecond binary pulsars. Presently, not all of these venues are yet
properly explored.

Ra{\~n}ada~\cite{Ranada:2004mf} investigated the effect of a background
gravitational potential that pervades the universe and is increasing
because of the expansion, provoking a drift of clocks~\cite{pioprd};
however, such an effect should also be observed in the radio signals
from pulsars~\cite{1997AA...326..924M, wex01}, which is not the
case. Further refining their argument, Ra{\~n}ada and and
Tiemblo~\cite{2008FoPh...38..458R} investigated the nonequivalence
of atomic and astronomical times and concluded that these times
could be accelerating adiabatically with respect to one another.

Ostvang~\cite{Ostvang:1999dm} proposes that cosmic expansion
applies directly to gravitationally bound systems according to the
{\it quasi-metric} framework. According
to~\cite{Rosales:2004kb,2005ESASP.588..263R}, the scale factor of the
spacetime background would cause an anomaly in the frequency. The
cosmological constant has also been invoked to produce
acceleration~\cite{Nottale:2003zj} or a gravitational frequency
shift~\cite{2004gr.qc.....7023M,2007EL.....7719001M}.

\subsubsection{Cosmological effects on radio signals}

L\"ammerzahl~\cite{Laemmerzahl-2009} considered the possibility that an
expanding universe may have an effect on the Doppler microwave signals
traveling in the solar system. The basic question is whether or not,
if it exists, the coupling of the expansion of the universe to light has
an observable effect. It was shown that for a spacecraft moving with
velocity $v$, the cosmologically-induced acceleration $a_H$ would have
the following form:
\begin{equation}
a_H=vH=\frac{v}{c} cH.
\end{equation}
Clearly, this acceleration is a factor of $v/c$ smaller that the
observed Pioneer acceleration, for which $a_P\simeq cH$. Therefore,
Doppler tracking in an expanding universe cannot account for the
observed Pioneer anomaly~\cite{Carrera-Giulini-2008,
  Laemmerzahl-2009}.

\subsection{Unmodeled forces of on-board origin}
\label{sec:on-board-forces}

The trajectory of the Pioneer~10 and 11 spacecraft may be affected by
an unmodeled force of on-board origin. The Pioneer spacecraft can
accelerate as a result of ejecting mass or radiation. Here we consider
these mechanisms as the possible sources for the anomaly.

\subsubsection{Thermal recoil forces}

Immediately after publication of the initial report on the Pioneer
anomaly~\cite{pioprl}, independent researchers raised questions
concerning the estimated magnitude of the thermal recoil force due to
internally generated heat. These ideas are summarized below.

Katz~\cite{1999PhRvL..83.1893A,1999PhRvL..83.1892K} pointed out that
although the electrical power generated on board Pioneer~10 and 11 was
less than 100~W during the period that was investigated, the waste
heat produced by each spacecraft's four RTGs amounted to $\sim
2.5$~kW. Therefore, even a small anisotropy in the spacecraft's
thermal radiation pattern may be sufficient to produce the observed
acceleration. Further, the amount of heat generated on-board decreased
slowly, consistent with the quoted error bounds of the
acceleration. In their reply, Anderson et
al.~\cite{1999PhRvL..83.1893A} reasoned that due to the geometrical
features of the Pioneer spacecraft such a mechanism would produce an
acceleration much smaller than that observed in the Pioneer Doppler
data.

Murphy~\cite{1999PhRvL..83.1891A,1999PhRvL..83.1890M} brought
attention to the fact that the louver system on the back of the
Pioneer spacecraft is not a Lambertian emitter, and the electrical
heat produced on-board may not have been declining steadily with time,
as the spacecraft body contained mostly components that had to be
powered at all times. In their reply, Anderson et
al.~\cite{1999PhRvL..83.1891A} argued that by the time Pioneer~10
reached the distances of $\sim$40~AU (i.e., the beginning of the data
interval used in~\cite{pioprd}), all louvers on the spacecraft
were closed, thus forcing the remaining and decreasing heat to escape
nonpreferentially.

Scheffer~\cite{2001gr.qc.....8054S,2003PhRvD..67h4021S} asserted that
the 2002 study may have seriously underestimated the role of the
thermal recoil force, and that up to 70~W of collimated thermal
radiation, corresponding to an acceleration of $9.3\times
10^{-10}$~m/s$^2$, may have been present on board the two Pioneer
spacecraft. Using the data on the electrical status of the Pioneer~10
spacecraft available prior to 2002, Anderson et
al.~\cite{1999PhRvL..83.1891A} argued that such a mechanism would
produce a significant decrease in the magnitude of the anomalous
acceleration as a response to the decreasing supply of the
radio-isotopic fuel. In fact, for the Pioneer~10 data interval used in
the 2002 study, the amount of electrical energy decreased by nearly
30\%, implying a similar decrease in the magnitude of the anomalous
acceleration. Such a decrease was not observed in the data.

Given the later recovery of
\begin{inparaenum}[i)]
\item the telemetry records of Pioneers 10 and 11 for the entire duration of their respective missions and
\item documentation detailing the design of the Pioneer F/G spacecraft (both completed in 2005~\cite{MDR2005}),
\end{inparaenum}
further investigation into the contribution of the thermal recoil
force as the source of the Pioneer anomaly is well founded. Indeed,
the recoil force due to on-board generated thermal radiation is the
subject of on-going study, discussed in
Section~\ref{sec:therml-recoil-force}.

\subsubsection{Propulsive gas maneuvers}
\label{sec:gasmaneuvers}

Among the possible mechanisms for the anomaly Anderson et
al.~\cite{pioprd} emphasized that propulsive gas maneuvers could
contribute to the anomalous acceleration (see discussion in
Sections~\ref{sec:force_maneuvers} and
\ref{sec:propulsive_mass_expulsion}). (The topic of fuel leaks was
addressed in Section~\ref{sec:gas_leaks}.) The errors associated with
mismodeling of the maneuvers could be large enough to seriously affect
the acceleration estimates. The number of maneuvers, their magnitude
and the details of the trajectories of both Pioneers make it unlikely
that similar acceleration estimates for both spacecraft would result, thus
undermining this idea. Recently this mechanism was again put forward
by Tortora et al.~\cite{Tortora-ISSI-2006-talk}. The recovered
telemetry record, however, shows no indication of a loss of fuel or
propulsion system activity that would support this
notion~\cite{MDR2005}.

\subsection{Miscellaneous mechanisms}
\label{sec:othermechanisms}

Some proposed mechanisms do not easily fit into the preceding
categories, yet represent potentially promising avenues of
research. In contrast, several proposed solutions to the Pioneer
anomaly are based on simple misconceptions and numerical coincidences.

\subsubsection{Other proposed mechanisms}

Cherubini and Mashhoon~\cite{2004PhRvD..70d4020B} discuss the
possibility of  nongravitational acceleration of the Sun, orthogonal
to the ecliptic plane, but they found that it is necessary for the Sun
to emit all electromagnetic radiation in the opposite direction.

The possibility of an unknown interaction of the Pioneer radio signals
with the solar wind was considered in~\cite{pioprl}. In addition,
there were ideas to invoke a model for superstrong interaction of
photons or massive bodies with the graviton
background~\cite{2001GReGr..33..479I}.

Wilson and Blome~\cite{Wilson:2009rg} derived the equations of motion
for an accelerated, rotating observer in a G\"odel universe, and
calculated the contribution of the universal cosmic rotation or
vorticity. However, they found that this term cannot account for the
observed Pioneer acceleration.

Trencevski~\cite{Trencevski2004} considered a time-dependent
gravitational potential that could explain the Pioneer anomaly without
causing planetary perihelion precessions that are in conflict with
observation. Mansouri, Nasseri, and
Khorrami~\cite{1999PhLA..259..194M} argued in favor of an effective
time variation of the gravitational constant; a similar analysis was
performed by Sidharth~\cite{2000NCimB.115..151S}.

Kowalski-Glikman and Smolin~\cite{KowalskiGlikman:2004kp} proposed
triply special relativity: a generalization of Einstein's theory of
special relativity with three invariant scales, the speed of light
$c$, a mass $\kappa$, and a length $R$, as a means to address several
observational puzzles, including the Pioneer anomaly.

Wiltshire~\cite{2007NJPh....9..377W} considered the effects of
decoupling bound systems from the global expansion of the
universe. While he found that this cannot be responsible for the
Pioneer effect, he also proposed an intriguing test case: a cosmic
microwave background imager flown on a trajectory similar to that of
Pioneer~10 and 11, with sufficient sensitivity to determine if an
anomaly, if observed, is due to a clock effect related to
gravitational energy.

Capozziello and Lambiase~\cite{Capozziello:1999qm} argued that
neutrino oscillations in Brans--Dicke theory could produce a phase
shift with observable effects on astronomical or cosmological time
scales.

\subsubsection{Common misconceptions and numerical coincidences}
\label{sec:misconceptions}

Attempting to make sense of the Pioneer anomaly, one can easily
discover several numerical coincidences that could mislead the
discussion. We review some common misconceptions in the hope that this
helps researchers navigate more easily through a formidable volume of
literature.

The numerical value of the anomalous acceleration gave rise to much
speculation. First, it was noticed that $a_P\simeq cH_0$ where $c$ is
the speed of light and $H_0$ is Hubble's constant at the present
epoch. However, an anomalous acceleration that is somehow connected to
the cosmic \textit{expansion} rate would necessarily point away from,
not toward, the Sun. Furthermore, an acceleration with the magnitude
$cH_0$ would only be observed for distant objects at a cosmic redshift
of $z=1$. Additionally, the numerical coincidence is only approximate,
not exact; good agreement with the observed value of $a_P$ requires
$H_0\simeq 95\pm 14\mathrm{\ km/s/Mpc}$, well above the presently
accepted value of $H_0\simeq
73.2\mathrm{\ km/s/Mpc}$~\cite{Spergel:2006hy}.

Then there were suggestions that the Pioneer anomaly may be related to
galactic gravity. This claim is not true. The Milky Way, as most
spiral galaxies, if approximated as a point mass using a nominal value
of $r_{\mathrm{gal}}\simeq8,000\mathrm{\ pc}$ as the distance from the Earth to the
galactic bulge, and a nominal mass of $M_{\mathrm{gal}}\simeq 4\times
10^{11} M_{\odot}$, yields an acceleration of $8.75\times
10^{-10}\mathrm{\ m/s}^2$, a number that is almost identical to $a_P$. Yet
this coincidence is misleading. Galactic gravity acts on all bodies in
the solar system, including the Sun, the Earth, and flying
spacecraft. The observed value of $a_P$ is an acceleration relative to
the Sun, so if it is caused by galactic gravity, it would have to be a
result of a difference between the galactic gravitational force on the
Sun vs. the spacecraft, i.e., a tidal force. Even at {\it d}~=~100~AU, this
tidal acceleration amounts to only
$2GM_{\mathrm{gal}}d/r^3_{\mathrm{gal}}\simeq10^{-16}\mathrm{\ m/s}^2$,
which is about seven orders of magnitude less than $a_P$.

Another numerical coincidence exploited in some proposals that have
been communicated to the authors concerns the value $v^4/c^3$ or some
variation thereof, where $v$ is the heliocentric velocity of the
Pioneer spacecraft. Evaluated numerically using the heliocentric
velocity $v\simeq 12\mathrm{\ km/s}$ of Pioneer~10, we obtain $v^4/c^3\simeq
7.68\times 10^{-10}\mathrm{\ m/s}$. While this value is very close numerically
to $a_P$, it has the dimensions of a velocity, not acceleration, and
the numerical coincidence is valid only so long as seconds are used as
the unit of time.

M\"akel\"a~\cite{2007arXiv0710.5460M} notes that $(\ln
2/16\pi)a_P\simeq Gm_P/l_P^2\simeq 8.83\times 10^{-10}\mathrm{\ m/s}^2$, where
$m_P\simeq 1.67\times 10^{-27}\mathrm{\ kg}$ is the proton mass and
$l_P=2\pi\bar h/cm_P$ is the proton's Compton wavelength, is very
close in value to the Pioneer acceleration. The factor of $\ln
2/16\pi$ may arise as the result of an unspecified quantum field
theoretical calculation.

As described in Section~\ref{sec:anomaly}, early data indicates that
there was an ``onset'' of the anomalous acceleration of Pioneer~11 at
around the time of its encounter with Saturn, which also marks
this spacecraft's transition from an elliptic orbit to an hyperbolic
escape trajectory. Some authors~\cite{2008PhLB..659..483N,
  Nieto:2009ve} attempted to draw far reaching conclusions based
solely on this fact. However, this observation cannot be used to draw
any firm conclusions, as analysis of the early data is highly suspect:
it was not developed using consistent methods and instead is an
amalgamation of results derived by different analysts who used to work
on Pioneer navigation at a particular time. So at best, it is only an
indication of the temporal dependence of the anomaly, and quite
likely, the ``onset'' is simply an artifact that will vanish when data
is re-analyzed using a consistent strategy.

Lastly, the {\it flyby anomaly}~\cite{anderson:091102, AG1998}
must be mentioned, as it has been related to the Pioneer anomaly by
some authors~\cite{anderson:091102,2007NewA...12..383A}. Several
spacecraft, including Galileo, NEAR and Rosetta, demonstrated a small,
unexplained change in kinetic energy during a gravity-assist Earth
flyby. The origin of this flyby anomaly remains unknown, although the
possibility of a systematic origin cannot be
excluded~\cite{Turyshev:2009mt}.

\subsection{Search for independent confirmation}
\label{sec:experiments}

So far, the Pioneer anomaly has been observed unambiguously using two
spacecraft of nearly identical design. The question naturally arises
whether the anomalous acceleration is observed in the motion of solar
system bodies or in the motion of other spacecraft.

\subsubsection{Solar-system planetary data analysis}
\label{sec:solar-system-data}

The biggest challenge to the mechanisms that use to explain the
anomaly either the unseen matter distribution in the outer solar
system or modifications of gravitational theory (see in
Section~\ref{sec:new-physics}) is the existence of a large volume of
high-precision solar system data. Any naive modification of
gravitational theory, for instance, that is designed to account for
the anomalous acceleration of the Pioneer spacecraft would also likely
induce changes in the predicted orbits of the outer planets.

A gravitational source in the solar system as a possible origin for
the anomaly has been considered by~\cite{pioprd}. According to the
equivalence principle, such a gravitational source would also affect
the orbits of the planets. In the case of the inner planets, which
have orbits determined with great accuracy, they show no evidence of
the expected anomalous motion if the source of the anomaly were
located in the inner solar system. In fact, the anomalous acceleration
is too large to have gone undetected in planetary orbits, particularly
for Earth and Mars.  Indeed, the authors of~\cite{pioprd} observed
that if a planet experiences a small, anomalous, radial acceleration,
$a_A$,  its orbital radius $r$ is perturbed  by
\begin{equation}
\Delta r =-\frac{{\it l}^6 a_A}{(GM_\odot)^4}
         \rightarrow  - \frac{r~ a_A}{a_N} ,
  \label{deltar}
\end{equation}
where \textit{l} is the orbital angular momentum per unit mass and
$a_N$ is the Newtonian acceleration at distance $r$.

For Earth and Mars, $\Delta r$ is about --21~km and --76~km. Take the
orbit of Mars, for example, for which range data provided by the Mars
Global Surveyor and Mars Odyssey missions have yielded measurements of
the Mars system center-of-mass relative to the Earth to an accuracy of
one meter~\cite{2006Icar..182...23K}. However, the anomaly has been
detected beyond 20~AU (i.e., beyond Uranus, 19~AU), and the orbits of
the outer planets have been determined only by optical methods,
resulting in much less accurate planet ephemerides.

The idea of looking at the impact of a Pioneer-like acceleration on
the orbital dynamics of the solar system bodies was originally put
forth in~\cite{pioprl,pioprd}. These papers, however, considered the
motion of the Earth and Mars finding no evidence of any effect induced
by an extra acceleration like the Pioneer anomaly. In particular, a
perturbation in $r$ produces a perturbation to the orbital angular
velocity of
\begin{equation}
\Delta \omega = \frac{2la_A}{GM_\odot}
   \rightarrow \frac{2 \dot{\theta}~ a_A}{a_N}.
\end{equation}
The determination of the synodic angular velocity $(\omega_E -
\omega_M)$ is accurate to 7 parts in $10^{11}$, or to about 5~ms
accuracy in synodic period. The only parameter that could possibly
mask the spacecraft-determined $a_R$ is  $(GM_\odot)$. But a large
error here would cause inconsistencies with the overall planetary
ephemeris~\cite{pioprd}.

More interestingly, the authors of~\cite{2002AAS...201.4509A} have
investigated the effect of an ever-present, uniform Pioneer-like force
on the long-period comets. The authors of~\cite{2005AAS...20715402P}
proposed to use comets and asteroids to assess the gravitational field
in the outer regions of the solar system and thereby investigate the
Pioneer anomaly.

Furthermore, \cite{pioprd} concluded that available planetary
ranging data limit any unmodeled radial acceleration acting on Earth
and Mars to less than $0.1 \times 10^{-10}\mathrm{\ m/s}^2$. Consequently, if
the anomalous radial acceleration acting on spinning spacecraft is
gravitational in origin, it is \textit{not} universal.  That is, it
must affect bodies in the 1000~kg range more than bodies of planetary
size by a factor of 100 or more. As the authors of \cite{pioprd}
said: ``This would be a strange violation of the Principle of
Equivalence''.


Attempts to detect observable evidence of unexpected gravitational
effects acting on the orbits of the outer planets have not yielded any
positive results yet. Hence, the authors of~\cite{2005astro.ph..4634I}
used parametric constraints to the orbits of Uranus and Neptune and
found that the reduced solar mass to account for the Pioneer anomaly
would not be compatible with the measurements. A similar result was
obtained in~\cite{2006NewA...11..600I} based on the Gauss equations to
estimate the effect of a gravitational perturbation in terms of the
time rate of change on the osculating orbital elements. These authors
argue that the perturbation would produce long-period, secular rates
on the perihelion and the mean anomaly, and short-period effects on
the semimajor axis, the eccentricity, the perihelion and the mean
anomaly large enough to be detected. \cite{2007PhRvD..76d2005T}
also considers the effect on the path of the outer planets of a
disturbance on a spherically-symmetric spacetime metric, and rules
out any model of the anomaly that implies that the Pioneer spacecraft
move geodesically in a perturbed spacetime metric. A recent test of
the orbits of 24 Trans-Neptunian Objects using bootstrap analysis also
failed to find evidence of the anomaly in these
objects~\cite{2007ApJ...666.1296W}.

Nevertheless, the absence of support for a perturbation of the
planetary orbits in the outer solar system is weak and inconclusive
(primarily due to the lack of precision range measurements). For
example, the authors of~\cite{2006ApJ...642..606P} conclude that such
anomalous gravitational disturbances would not be detected in the
orbits of the outer planets.
Therefore, efforts to find a gravitational explanation
continue, as in the case of a recent paper~\cite{nyambuya-2008} that
proposes an azimuthally symmetric solution to Poisson's equation for
empty space to explain qualitatively the Pioneer anomaly. This
solution results in a gravitational potential dependent on the
distance and the polar angle, and it also has implications for the
planetary orbits, albeit not yet tested with ephemeris data yet.

Many of these considerations are based on simplified models. As such,
they must be contrasted with efforts that incorporate a radial
acceleration into computations of planetary ephemerides.

Using the planetary ephemerides, Standish~\cite{2010IAUS..261..179S,2008AIPC..977..254S}
investigated modifications to the laws of gravitation that can explain the
anomaly and still be compatible with the known motion of the planets
from Saturn and outward. Out of five different acceleration models,
four were shown to be not viable.

Using radiometric tracking data from the Cassini spacecraft,
Folkner~\cite{2010IAUS..261..155F} established an upper bound on Saturn's
radial acceleration as less than $10^{-14}\mathrm{\ m/s}^2$. This is around just
one tenth the anomalous acceleration of Pioneer 10 and 11.

Using the latest
INPOP08 planetary ephemerides, Fienga et~al~\cite{2010IAUS..261..159F}
determined that a radial acceleration greater than 1/4 times the observed
Pioneer anomaly at distances beyond 20~AU is not consistent with planetary
orbits.

These investigations represent serious challenges to any attempt to
explain the Pioneer anomaly using unseen mass distributions in the solar
system or modifications of the theory of gravity. Such attempts, in
addition to accounting for the anomalous acceleration of the two spacecraft,
must also be able to explain why an acceleration of the same magnitude is
\emph{not} readily seen in the orbit of Saturn and in the known ephemerides
of the outer planets.

\subsubsection{Attempts to confirm the anomaly using other spacecraft}

One reason that the Pioneer anomaly generated so much interest is that
two spacecraft on very different trajectories yielded similar
anomalous acceleration values. However, the two spacecraft are nearly
identical in design, and although their trajectories are different,
they both follow hyperbolic escape trajectories. This raises the
obvious question: is it possible to confirm the presence of an
anomalous acceleration using
\begin{inparaenum}[i)]
\item different spacecraft, or
\item spacecraft on different trajectories?
\end{inparaenum}

The Pioneer~10 and 11 spacecraft are sensitive to such a small,
anomalous acceleration term because they are very far from the Sun,
and therefore, the effects of solar radiation pressure are much
smaller than the anomalous acceleration. For spacecraft located in the
inner parts of the solar system (e.g., within the orbit of Jupiter),
solar radiation pressure is significant; for Pioneer~10 and 11 at
5~AU, solar radiation results in an acceleration that exceeds the
anomalous acceleration by nearly a full order of magnitude. This is a
clear indication that the only candidates for investigating
accelerations as small as the Pioneer anomaly are spacecraft in deep
space, preferably at or beyond the orbit of Jupiter.

The most obvious choice that comes to mind are the twin
Voyagers. Voyager~1 and 2, like Pioneer~10 and 11 before them, are
traveling on hyperbolic escape trajectories after encounters with the
gas giants in the outer solar system. Unfortunately, unlike Pioneer~10
and 11, Voyager 1~and 2 are not spin stabilized. The three-axis
stabilization employed by these spacecraft requires the use of small
attitude stabilization thrusters several times a day. The noise
produced by unmodeled small accelerations that arise as a result of
uncertainties in the magnitude and exact duration of thruster firings
and possible leakage afterwards reduce the sensitivity of these
spacecraft to small accelerations such that an anomalous acceleration
of order 10\super{-9}~m/s\super{2} is completely undetectable.

Next on the list was Galileo. This spacecraft, designed to orbit
Jupiter, featured an innovative design comprising a spinning and a
nonspinning section. However, due to mechanical problems between the
two sections, the spacecraft was often configured to use the
all-spinning mode. The malfunction of Galileo's high gain antenna
opening mechanism put this antenna out of commission, and therefore,
Galileo communicated with the Earth using an omnidirectional low-gain
antenna at much lower than planned data rates, at a very low
signal-to-noise ratio. Despite these difficulties, navigational data
was obtained from this spacecraft with sufficient accuracy for
precision orbit estimation. A combined analysis of Doppler and ranging
data by The Aerospace Corporation from a 113-day period beginning
shortly before Galileo's second Earth encounter (Galileo flew a
complicated series of flyby orbits before it achieved a transfer orbit
that took it to Jupiter) does show the presence of a possible
acceleration term of the correct magnitude~\cite{pioprd}.

Ulysses, a joint project between NASA and the European Space Agency
(ESA), flew on a highly elliptical heliocentric orbit tilted about
80\textdegree\ from the plane of the elliptic. Although Ulysses flew
relatively close to the Sun and therefore, solar radiation had a
significant effect on its trajectory, its varying distance from the
Sun, combined with variations in the Earth-spacecraft-Sun angle make
it possible in principle to separate the effects of solar radiation
from any anomalous acceleration. This is how an estimate of
(12~\textpm~3)~\texttimes~10\super{-10}~m/s\super{2} was obtained
using JPL's ODP software. Although this value remains highly
correlated with solar radiation pressure ~\cite{pioprd}, it is
compatible with the value of $a_P$ measured for Pioneer~10 and 11.

Yet another deep-space craft, Cassini, is three-axis stabilized like
Voyager~1 and 2. Nevertheless, the sophisticated tracking capabilities
of this spacecraft could offer a potential contribution to the
research of the anomalous acceleration~\cite{pioprd}.

The recently (2006) launched New Horizons mission to Pluto is
potentially a better candidate for researching the anomalous
acceleration. Like Pioneer~10 and 11, New Horizons is spin stabilized,
using only infrequent thruster maneuvers for attitude
stabilization. Unfortunately, current mission plans call for New
Horizons to spend much of the time between Jupiter and Pluto in
``hibernation'' mode, with only infrequent communications with the
Earth. Therefore, during the flight to Pluto, little or no precision
Doppler data may be collected~\cite{2008PhLB..659..483N, Toth2009,
  Turyshev:2005zm}.

As a result, attempts to verify the anomaly using other spacecraft
have proven disappointing~\cite{pioprd, 2001gr.qc.....7022A,
  2002IJMPD..11.1545A, 2004CQGra..21.4005N}, because the Voyager,
Galileo, Ulysses, and Cassini spacecraft navigation data all have
their own individual difficulties for use as an independent test of
the anomaly. In addition, many of the deep space missions that are
currently being planned will not provide the needed
navigational accuracy and trajectory stability of under
10\super{-10}~m/s\super{2}.

\subsubsection{Possible future spacecraft searches}

The acceleration regime in which the anomaly was observed diminishes
the value of using modern disturbance compensation systems for a
test. For example, the systems that are currently being developed for
the LISA (Laser Interferometric Space Antenna) and LISA Pathfinder
missions are designed to operate in the presence of a very low
frequency acceleration noise (at the mHz level), while the Pioneer
anomalous acceleration is a strong constant bias in the Doppler
frequency data. In addition, currently available accelerometers are a
few orders of magnitude less sensitive than is needed for a
test~\cite{2002IJMPD..11.1545A, 2004CQGra..21.4005N,
  2005AIPC..758..113N, Turyshev:2005zm, 2007AdSpR..39..291T}. To
enable a clean test of the anomaly there is also a requirement to have
an escape hyperbolic trajectory. This makes a number of other
currently proposed missions less able to directly test the anomalous
acceleration.

A number of alternative ground-based verifications of the anomaly have
also been considered; for example, using VLBI astrometric
observations. However, the trajectories of Pioneers, with small proper
motions in the sky, make it presently impossible to use VLBI in
accurately isolating an anomalous sunward acceleration of the size of
$a_P$.

\subsection{A mission to explore the Pioneer anomaly}
\label{sec:mission}

The apparent inability to explain the anomalous behavior of the
Pioneers with conventional physics has contributed to growing
discussion about its origin. A number of researchers emphasized the
need for a new experiment to explore the detected signal. As  a
result, \cite{Turyshev:2005zm} advocated for a program to study the
Pioneer anomaly. This program effectively includes three phases:
\begin{enumerate}[i)]
\item Analysis of the entire set of existing Pioneer data, obtained
  from launch to the last useful data received from Pioneer~10 in
  April 2002.  This data provides critical new information about the
  anomaly~\cite{2007arXiv0710.2656T, Turyshev:2005zm}. If the anomaly
  is confirmed,
\item Development of an instrument, to be carried on another deep
  space mission, to provide an independent confirmation for the
  anomaly.  If further confirmed,
\item Development of a dedicated deep-space experiment to explore the
  Pioneer anomaly with an accuracy for acceleration resolution at the
  level of 10\super{-12}~m/s\super{2} in the extremely low frequency (or
  nearly DC) range.
\end{enumerate}

Significant progress was accomplished concerning the first and second
phases~\cite{2004CQGra..21.4005N, Rathke:2004gv, 2007arXiv0710.2656T,
  2007AdSpR..39..291T}. Work on the third phase was also
initiated~\cite{2007IJMPD..16.1611B,ESLAB2005_Pioneer,2006gr.qc....10160R,
2007arXiv0710.2656T,Turyshev:2005zm}. Mission studies conducted in
2004\,--\,2005~\cite{ESLAB2005_Pioneer} have identified two options:
\begin{inparaenum}[i)]
\item an experiment on a major mission to deep space capable of
  reaching acceleration sensitivity similar to that demonstrated by
  the Pioneers and
\item a dedicated mission to explore the Pioneer anomaly that offers
  full characterization of the anomaly.
\end{inparaenum}
Below we discuss both of these proposals.

\subsubsection{A Pioneer instrument package}

A way to confirm independently the anomaly is to fly an instrumental
package on a mission heading to the outer regions of the solar
system. The primary goal is to provide an independent experimental
confirmation of the anomaly. One can conceive of an instrument placed
on a major mission to deep space. The instrument must be able to
compensate for systematic effects to an accuracy below the level of
10\super{-10}~m/s\super{2}.  Another concept is a simple autonomous probe that
could be jettisoned from the main vehicle, such as the proposed
Interstellar Probe\epubtkFootnote{see \url{http://interstellar.jpl.nasa.gov/interstellar/probe/}
  for details.}, presumably further out than at least the orbit of
Jupiter or Saturn. The probe would then be navigated from the ground
yielding a navigational accuracy below the level of
10\super{-10}~m/s\super{2}.  The data collected could provide an independent
experimental verification of the anomaly's existence.

Dittus et al.~\cite{ESLAB2005_Pioneer} emphasized that the option of
an instrument on a major mission to deep space would have a major
impact on spacecraft and mission designs with limited improvement in
measuring $a_P$. Nevertheless, a highly-accurate accelerometer has
been proposed as part of the Gravity Advanced Package, which is a fundamental physics
experiment that is being considered by the ESA for the future Jupiter
Ganymede Orbiter Mission\epubtkFootnote{See presentation by
  B. Christophe, ``Gravity Advanced Package, a fundamental physics
  experiment for Jupiter Ganymede Orbiter Mission'', at the recent
  \textit{Gravitation and Fundamental Physics in Space}, the GPhyS
  ``Kick-Off'' Colloquium, Les Houches, 20\,--\,22 October, 2009, at
    \url{http://gphys.obspm.fr/LesHouches2009/Program.html}.}. At the same
time, it is clear that to explore the anomaly one needs to travel
beyond the orbit of Jupiter. Furthermore, an acceleration sensitivity
at the level of $\sim$~10\super{-12}~m/s\super{2} would be preferable,
which can be done only with a dedicated mission, as discussed in
Section~\ref{sec:dsgp} below.

\subsubsection{A dedicated mission concept}
\label{sec:dsgp}

\epubtkImage{}{%
  \begin{figure}[t!]
    \centerline{\includegraphics[width=0.72\textwidth]{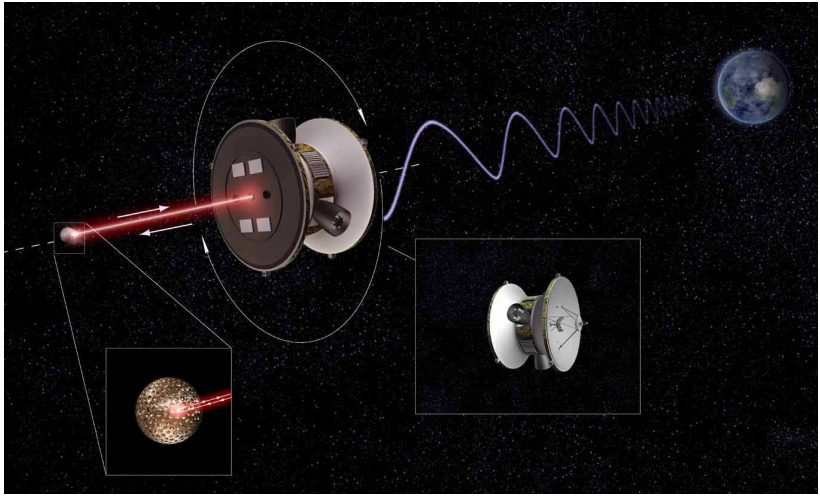}}
    \caption{A drawing for the measurement concept chosen of the Deep
    Space Gravity Probe (from~\cite{ESLAB2005_Pioneer}, drawing
    courtesy of Alexandre D.\ Szames). The formation-flying approach
    relies on actively controlled spacecraft and a set of passive
    test-masses. The main objective is to accurately determine the
    heliocentric motion of the test-mass by utilizing the 2-step
    tracking needed for common-mode noise rejection purposes. The
    trajectory of the spacecraft will be determined using standard
    methods of radiometric tracking, while the motion of the test
    mass relative to the spacecraft will be established by laser
    ranging technology. The test mass is at an environmentally quiet
    distance from the craft, $\geq$~250~m. With occasional maneuvers to
    maintain formation, the concept establishes a flexible craft to
    test mass formation.}
    \label{fig:pae-concept}
    \label{fig:dsgp-concept}
\end{figure}}

The available knowledge of the Pioneer anomaly lead to the following
science objectives for a dedicated mission to explore the Pioneer
anomaly:
\begin{inparaenum}[i)]
\item investigate the origin of the anomaly with an improvement by a factor of 1,000;
\item improve spatial, temporal, and directional resolution;
\item identify and measure all possible disturbing and competing effects;
\item test Newtonian gravity potential at large distances;
\item discriminate amongst candidate theories explaining $a_P$, and
\item study the deep-space environment in the outer solar system.
\end{inparaenum}

A viable concept would utilize a spacecraft pair capable of flying in
a flexible formation (see Figure~\ref{fig:dsgp-concept}).  The main
craft would have a precision star-tracker and an accelerometer and
would be capable of precise navigation, with disturbances, to a level
less than $\sim$~10\super{-10}~m/s\super{2} in the low-frequency acceleration
regime.  Mounted on the front would be a container holding a probe -- a
spherical test mass covered with corner cubes.  Once the configuration
is on its solar system escape trajectory and will undergo no further
navigation maneuvers, and is at a heliocentric distance of
$\sim$~5\,--\,20~AU, the test mass would be released from the primary
craft. The probe will be passively laser-ranged from the primary craft
with the latter having enough capabilities to maneuver with respect to
the probe, if needed.  The distance from the Earth to the primary
would be determined with either standard radiometric methods
operating at Ka-band or with optical communication.  Note that any
dynamical noise at the primary would be a common mode contribution to
the Earth-primary and primary-probe distances. This design satisfies
the primary objective, which would be accomplished by the two-staged
accurate navigation of the probe with sensitivity down to the
10\super{-12}~m/s\super{2} level in the DC of extremely low frequency
bandwidth.

Since the small forces affecting the motion of a craft in four
possible directions all have entirely different characteristics
(i.e., sunward, earthward, along the velocity vector or along the
spin-axis~\cite{2004CQGra..21.4005N, Turyshev:2005zm}), it is clear
that an antenna with a highly directional radiation pattern along with
star sensors will create even better conditions for resolving the true
direction of the anomaly when compared to standard navigation
techniques. On a craft with these additional capabilities, all
on-board systematics will become a common mode factor contributing to
all the attitude sensors and antennas. The combination of all the
attitude measurements will enable one to clearly separate the effects
of the on-board systematics referenced to the direction towards the
Sun.

To enable fast orbital transfer to distances greater than 20~AU,
hyperbolic escape trajectories enabled by solar sail propulsion
technology were considered as an attractive candidate.  Among other
options is a standard chemical rocket and nuclear electric propulsion,
as  was successfully demonstrated recently~\cite{2004CQGra..21.4005N}.
The proposed combination of a formation-flying system aided by solar
sail propulsion for fast trajectory transfer, leads to a technology
combination that will benefit many missions in the
future~\cite{Turyshev-UFN-2008, Turyshev-etal-2007}.

Two missions were recently proposed to explore the Pioneer anomaly in
a dedicated space experiment. The Solar System Odyssey mission will
use modern-day high-precision experimental techniques to test the laws
of fundamental physics, which determine dynamics in the solar
system~\cite{Christophe:2007uj}. The mission design is similar to the
one proposed for the Deep Space Gravity Probe
(DSGP)~\cite{ESLAB2005_Pioneer}. Also, the proposed SAGAS (Search for
Anomalous Gravitation using Atomic Sensors) mission~\cite{Wolf:2007sh}
aims at flying highly sensitive atomic sensors (optical clock, cold
atom accelerometer, optical link) on a solar system escape
trajectory. SAGAS has numerous science objectives in solar system
exploration and fundamental physics, including an accurate test of the
Pioneer anomaly.

The extraordinary nature of the Pioneer anomalous acceleration led to
serious questions concerning the possible origin of the
effect. Answering these questions requires further in-depth
analysis. This is especially true before any serious discussion of a
dedicated experiment can take place. In fact, prior to the development
of any dedicated mission to investigate the Pioneer anomaly, it is
absolutely essential to analyze the complete Pioneer Doppler data in
order to rule out, as much as possible, any engineering cause. This
study is on-going and will be reviewed in
Section~\ref{sec:current-status}.

\newpage
\section{Current Status of the Study of the Pioneer Anomaly}
\label{sec:current-status}

With the publication of the JPL's definitive 2002 study of the Pioneer
anomaly~\cite{pioprd} the efforts to study the discovered effect have
intensified. After some hesitation there was an agreement among the
researchers that a thorough investigation of the anomaly is
needed. The main question remained: What is the nature of the Pioneer
anomaly? Is this a manifestation of new physics? Or a
representation of some unmodeled  conventional physics mechanism?
Furthermore, it became clear that before the Pioneer anomaly can be
fully accepted as a novel physical effect, the issue of the on-board
heat redistribution must be fully addressed.

Ultimately, the search for the origin of the Pioneer anomaly came down
to one principal question: ``Is it the heat or not the heat?''
Multiple community efforts are now guided by this question. In
addition to the teams from JPL and The Aerospace Corporation, several
independent teams and researchers have confirmed the existence of the
anomalous acceleration in the Pioneer Doppler data; the new analyses
also studied temporal behavior of the anomaly, establishing limits on
the temporal change of anomalous acceleration. As a result of these
extensive multi-year efforts, our understanding of the Pioneer anomaly
improved significantly (see Section~\ref{sec:indep-verify}).

Meanwhile, a team at JPL along with C.B.~Markwardt at the NASA Goddard
Space Flight Center have been working on the recovery of a much
extended radiometric Doppler data set, covering a significantly
longer time period for both spacecraft than previously available data
(Section~\ref{sec:new-data}). At the same time, thanks mainly to the
efforts of L.R.~Kellogg at the NASA Ames Research Center, a near
complete record of telemetry from Pioneer~10 and 11
(Section~\ref{sec:telemetry}) has been recovered, as well as
significant quantities of original project documentation
(Section~\ref{sec:pio-docs}). This made it possible to construct a
detailed thermal model of the spacecraft, investigating the extent to
which heat generated on-board and radiated away anisotropically may
contribute to the anomalous acceleration
(Section~\ref{sec:therml-recoil-force}).

As of early 2010, some of these efforts are finished, such as the
recovery of the Doppler data, while others, notably, the construction
of a comprehensive thermal model, are still under way. This new effort
to investigate the Pioneer anomalous acceleration, when completed,
will for the first time use all available Doppler data to establish
the acceleration profile of the two spacecraft, and it will also
accurately account for any acceleration of on-board thermal origin.

In this section, we review these recent and on-going efforts and
summarize our current knowledge of the Pioneer anomaly.

\subsection{Independent verifications}
\label{sec:indep-verify}

By now several studies of the Pioneer~10 and 11 radiometric Doppler
data have demonstrated that the anomaly is unambiguously present in
the trajectory solutions for both spacecraft. These studies were
performed with six independent (and different!) navigational computer
programs (see~\cite{pioprl,pioprd,levy-2008,
  2002gr.qc.....8046M, 2007AA...463..393O, Toth2008}), namely:

\begin{itemize}
\item The most detailed analysis of the Pioneer anomaly to date, the
  2002 study by JPL~\cite{pioprd} (which is discussed in depth in
  Section~\ref{sec:anomaly}), used various versions of the JPL's Orbit
  Determination Program (ODP), developed between
  1980\,--\,2005~\cite{pioprl,pioprd},
\item The Aerospace Corporation's Compact High Accuracy Satellite
  Motion Program (CHASMP) code, extended for deep space
  navigation~\cite{pioprl,pioprd},
\item Code written at the Goddard Space Flight Center (GSFC) by
  C.B.~Markwardt~\cite{2002gr.qc.....8046M} that was used to analyze
  Pioneer~10 data for the period 1987\,--\,1994 obtained from the
  National Space Science Data Center (NSSDC)\epubtkFootnote{The
    National Space Science Data Center (NSSDC), see details at
    \url{http://nssdc.gsfc.nasa.gov/}},
\item The HELIOSAT orbit determination program that was co-developed
  at the Institute for Theoretical Astrophysics, University of Oslo,
  Norway, was recently used by one of the code's authors,
  \O.~Olsen~\cite{2007AA...463..393O}, to analyze the set of the
  Pioneer~10 data set identical to the one used in the JPL's 2002
  study,
\item An orbit determination code that was independently developed by
  V.T.~Toth~\cite{Toth2008} for the purposes of studying the Pioneer
  anomaly, and finally
\item A dedicated software package called ODYSSEY that has been
  developed by Groupe Anomalie Pioneer (GAP)\epubtkFootnote{Groupe
    Anomalie Pioneer (GAP), a French collaboration on Pioneer Anomaly
    supported by The Centre National d'Etudes Spatiales (CNES),
    France, which includes researchers from LKB, ONERA, OCA, IOTA and
    SYRTE laboratories, see details at
  \url{http://www.lkb.ens.fr/-GAP-?lang=fr}} for the purposes of
  investigating the Pioneer anomaly~\cite{levy-2008}.
\end{itemize}

These recent independent analyses of the Pioneer~10 and 11
radiometric Doppler data confirmed the existence of the Pioneer
anomaly at the level reported by the JPL's 2002 study and they also
provided new knowledge of the effect. Below we review these analyses
in some detail.

\subsubsection{Independent verification by Markwardt}
\label{sec:markwardt}

Shortly after publication of the 2002 JPL result,
Markwardt~\cite{2002gr.qc.....8046M} published an independent analysis
that was unique in the sense that it utilized a separately obtained
data set. Rather than using data in the form of JPL-supplied Orbit
Determination Files, Markwardt obtained Pioneer~10 tracking data from
the National Space Science Data Center (NSSDC) archive. This data was
in the Archival Tracking Data File (ATDF) format, which Markwardt
processed using tools developed for the purposes of this specific
study~\cite{MARKWARDTIDL, MARKWARDTATDF}.

The Pioneer~10 data used in Markwardt's investigation spanned the
years 1987 through 1994, and his result, $a_{P10}=(7.70\pm 0.02)\times
10^{-10}\mathrm{\ m/s}^2$ (see Figure~\ref{fig:markwardt}), is consistent with
the JPL result. Markwardt was also the first to investigate explicitly
the possible presence of a jerk (i.e., the rate of change of
acceleration\epubtkFootnote{See definition of the jerk term at
    \url{http://en.wikipedia.org/wiki/Jerk_(physics)}.}, defined as
$j\equiv\dot a=da/dt$) term, and found that a term
$|j_{P10}|<0.18\times 10^{-10}\mathrm{\ m/s}^2/\mathrm{year}$ is
consistent with the data. Based on the studied Pioneer~10 data set,
Markwardt found that the anomaly is nearly constant with time, with a
characteristic variation time scale of over 70~yr, which is still too
short to rule out on-board thermal radiation effects.

\epubtkImage{}{%
  \begin{figure}[t!]
    \centerline{\includegraphics[width=0.70\textwidth]{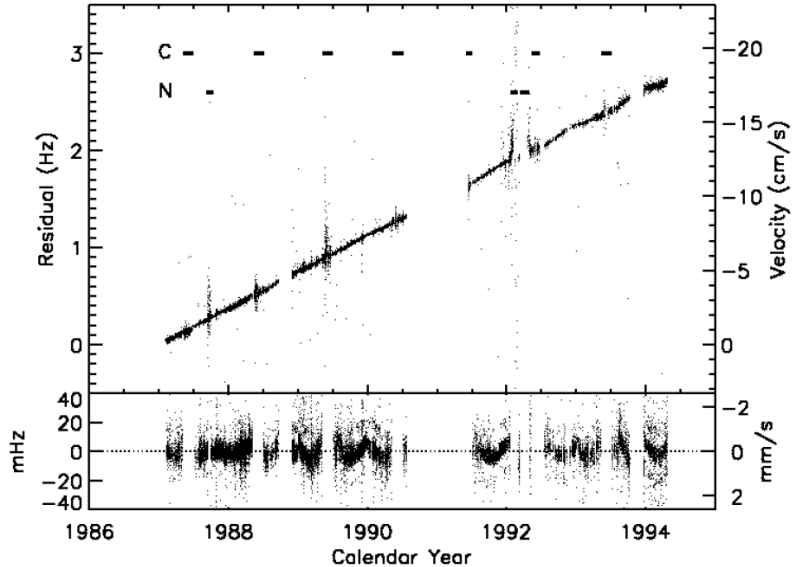}}
    \caption{Results of Markwardt's
    analysis~\cite{2002gr.qc.....8046M} show Doppler residuals as a
    function of time of the best fit model. The top panel shows the
    residuals after setting $a_P = 0$, and demonstrates the linear
    increase with time. The top panel shows all of the data, including
    segments that were filtered out because of interference due to the
    solar corona (designated by a horizontal bar with ``C'') or due to
    general noise (designated ``N''). The bottom panel shows the
    filtered residuals, including the best fit value of the anomalous
    acceleration. The equivalent spacecraft velocity is also shown.}
    \label{fig:markwardt}
\end{figure}}

\subsubsection{Analysis by Olsen using HELIOSAT}

Olsen~\cite{2007AA...463..393O} focused on the constancy of the
anomalous acceleration using the HELIOSAT orbit determination program
that was independently  developed by him at the University of Oslo,
Norway. Analysis confirmed the acceleration at the levels reported
by~\cite{pioprd} for the same segments of Pioneer~10 and 11 data that
were used by JPL (see Table~\ref{tb:olsen}). The study found that
systematic variations in the anomalous acceleration are consistent
with solar coronal mass ejections and that the Doppler data alone
cannot distinguish between constant acceleration and slowly decreasing
acceleration. Specifically, the study concluded that heat dissipation
cannot be excluded as a source of the anomaly.

\begin{table}[h!]
  \caption[The Pioneer anomalous acceleration in units of
  10\super{-10}~m/s\super{2}. This table compares the results from JPL's ODP
  and the Aerospace Corporation's CHASMP codes to results obtained
  using the HELIOSAT program developed by Olsen.]{The Pioneer
  anomalous acceleration in units of 10\super{-10}~m/s\super{2}. This table
  compares the results from JPL's ODP and the Aerospace Corporation's
  CHASMP codes from~\cite{pioprd} (see Table~\ref{tb:2002results}) to
  results obtained using the HELIOSAT program developed by
  Olsen~\cite{2007AA...463..393O}.}
  \label{tb:olsen}
  \centering
  \begin{tabular}{ccccc}
    \toprule
    Software & Pioneer~10 (I) &  Pioneer~10 (II) & Pioneer~10 (III) & Pioneer~11\\
    \midrule
    ODP/Sigma  & 8.00~\textpm~0.01  & 8.66~\textpm~0.01  & 7.84~\textpm~0.01  & 8.44~\textpm~0.04\\
    CHASMP  & 8.22~\textpm~0.02  & 8.89~\textpm~0.01  & 7.92~\textpm~0.01  & 8.69~\textpm~0.03\\
    HELIOSAT  & 7.85~\textpm~0.02  & 8.78~\textpm~0.01  & 7.75~\textpm~0.01 & 8.10~\textpm~0.01\\
    \bottomrule
  \end{tabular}
\end{table}

\subsubsection{Independent analysis by Toth}
\label{sec:toth}

\epubtkImage{}{%
\begin{figure}[t!]
\begin{raggedright}
\vskip -6pt
\begin{minipage}{.45\linewidth}a)\\\vskip -12pt\includegraphics[width=\linewidth]{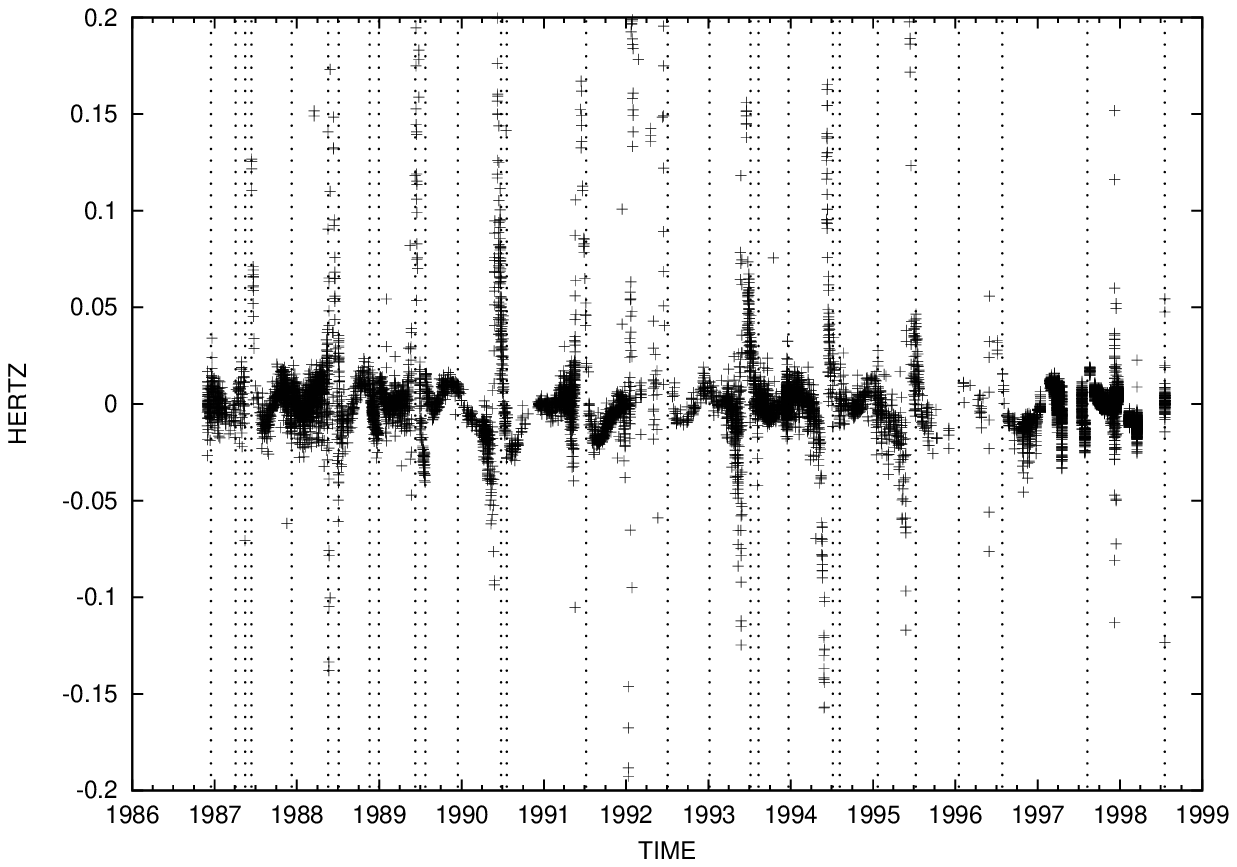}\end{minipage}
\hskip 0.05\linewidth
\begin{minipage}{.45\linewidth}c)\\\vskip -12pt\includegraphics[width=\linewidth]{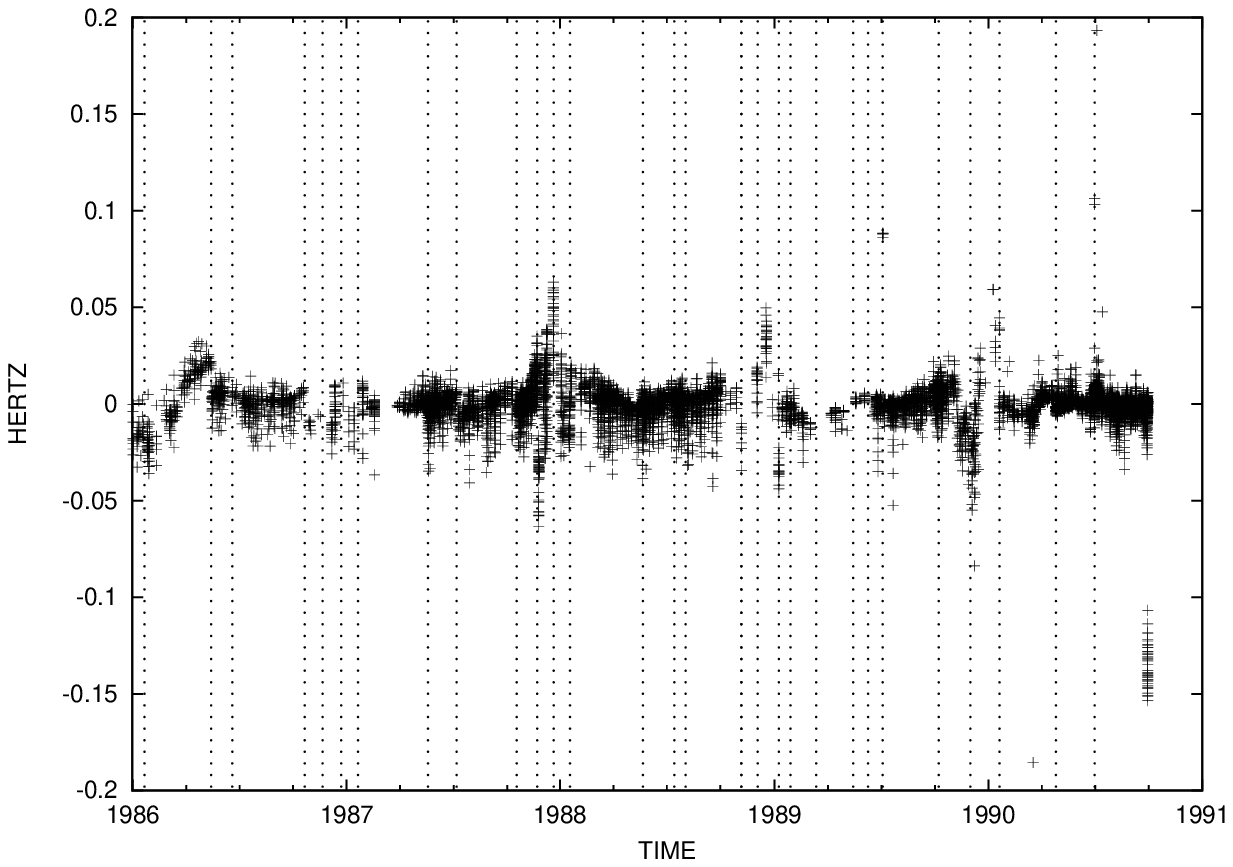}\end{minipage}
\vskip -6pt
\begin{minipage}{.45\linewidth}b)\\\vskip -12pt\includegraphics[width=\linewidth]{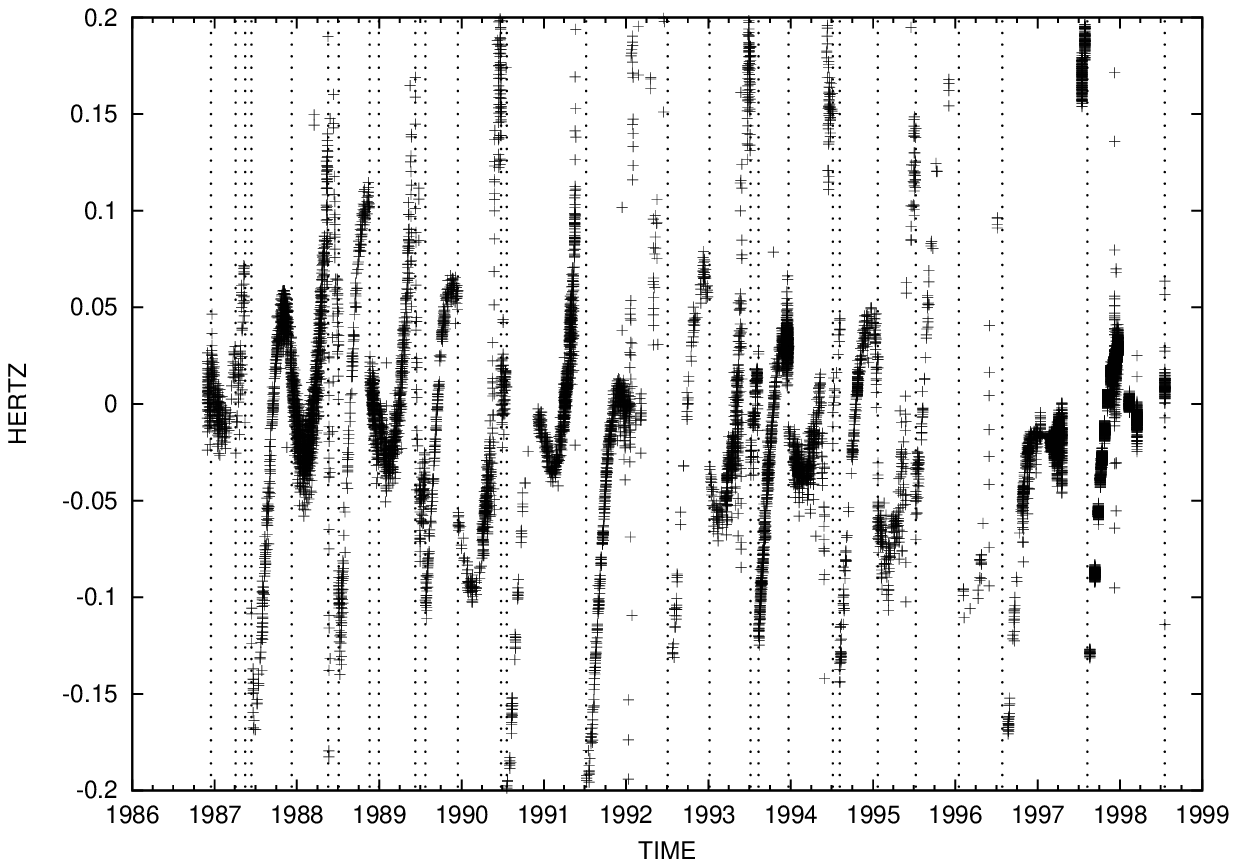}\end{minipage}
\hskip 0.05\linewidth
\begin{minipage}{.45\linewidth}d)\\\vskip -12pt\includegraphics[width=\linewidth]{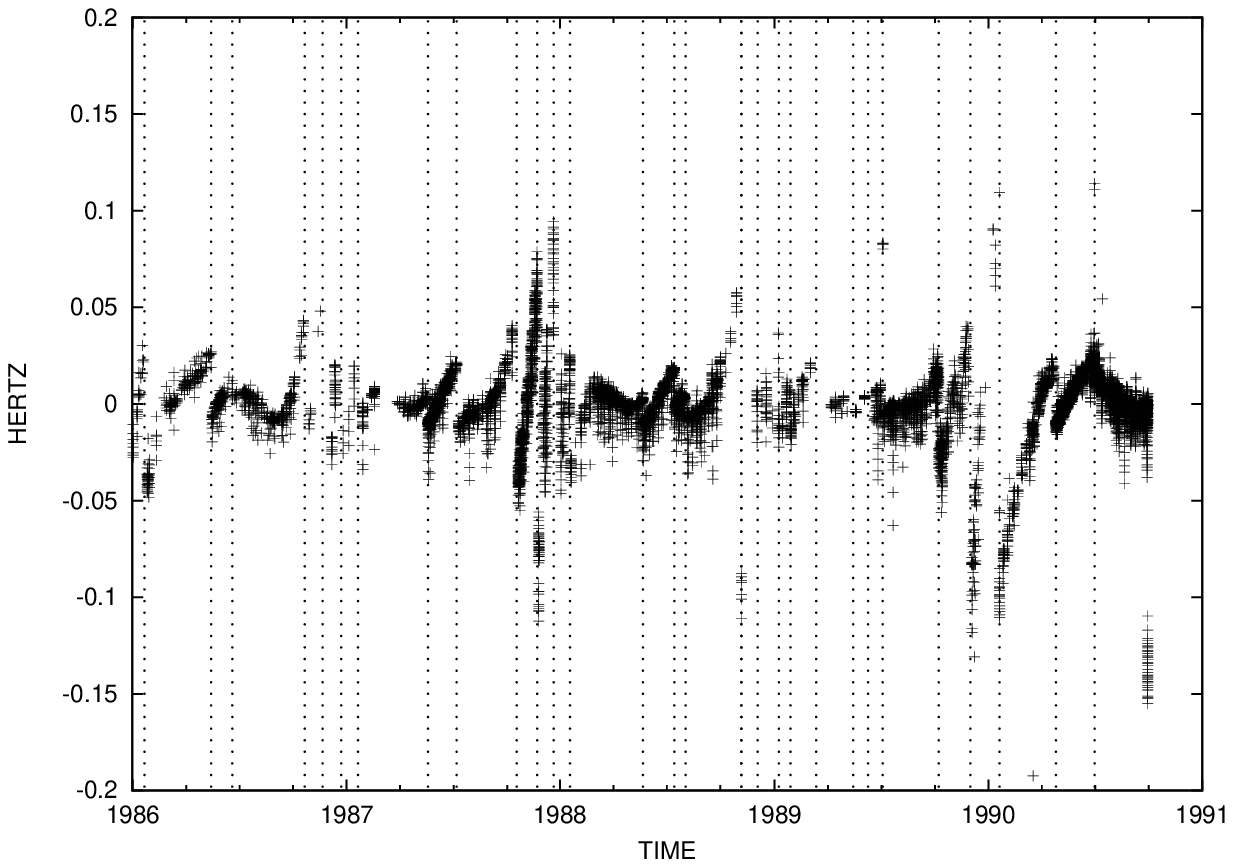}\end{minipage}
\caption{Results of Toth's analysis~\cite{Toth2008}: a) best-fit
  residuals for Pioneer~10; b) best-fit residuals for Pioneer~10 with
  no anomalous acceleration term; c--d) same as a--b, for Pioneer~11.}
\label{fig:Toth}
\end{raggedright}
\end{figure}}

Toth~\cite{Toth2008} also studied the anomalous acceleration using
independently developed orbit determination software, and confirmed
that the introduction of a constant acceleration term significantly
improves the post-fit residuals (Figure~\ref{fig:Toth}). Toth
determined the anomalous accelerations of Pioneers 10 and 11 as
$a_{P10}=(9.03\pm 0.86)\times 10^{-10}\mathrm{\ m/s}^2$ and $a_{P11}=(8.21\pm
1.07) \times 10^{-10}\mathrm{\ m/s}^2$ correspondingly, where the error terms
were taken from~\cite{pioprd} (excluding terms related to thermal
modeling, which is the subject of on-going effort). Studying the
temporal behavior of the anomalous acceleration, he was able to find a
best fit for the acceleration and jerk terms of both spacecraft:
$a_\mathrm{P10}=(10.96\pm 0.89)\times
10^{-10}\mathrm{\ m/s}^2$ and $j_\mathrm{P10}=(-0.21\pm
0.04)\times 10^{-10}\mathrm{\ m/s}^2/\mathrm{year}$
(Pioneer~10) and $a_\mathrm{P11}=(9.40\pm 1.12)\times
10^{-10}\mathrm{\ m/s}^2$ and $j_\mathrm{P11}=(-0.34\pm
0.12)\times 10^{-10}\mathrm{\ m/s}^2/\mathrm{year}$
(Pioneer~11). Toth's study demonstrated that a moderate jerk term is
consistent with the Doppler data and, therefore, an anomalous
acceleration that is a slowly changing function of time cannot be
excluded at present.

Toth's orbit determination software also has the capability to utilize
telemetry data. In particular, the code can be used to estimate the
thermal recoil force as a function of the heat generated on-board, or
conversely, to fit thermal recoil force coefficients to radiometric
Doppler measurements, as discussed in
Section~\ref{sec:thermal_force}.

\subsubsection{Analysis by Levy et al. using ODYSSEY}

\epubtkImage{}{%
\begin{figure}[t!]
\hskip -6pt
\begin{minipage}[b]{.5\linewidth}
\centering \includegraphics[width=\linewidth]{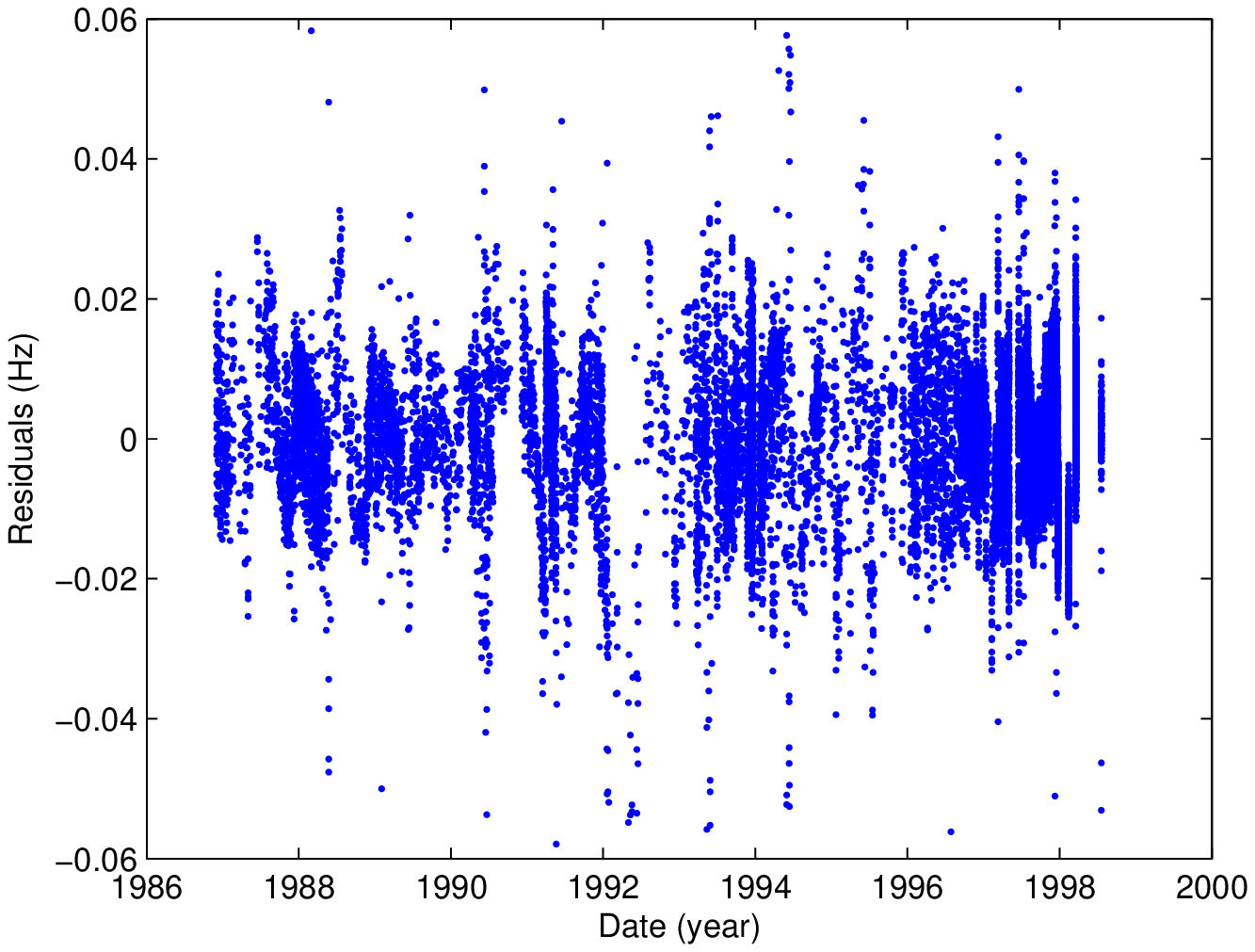}
\end{minipage}
\hskip 0.001\linewidth
\begin{minipage}[b]{.5\linewidth}
\centering \includegraphics[width=\linewidth]{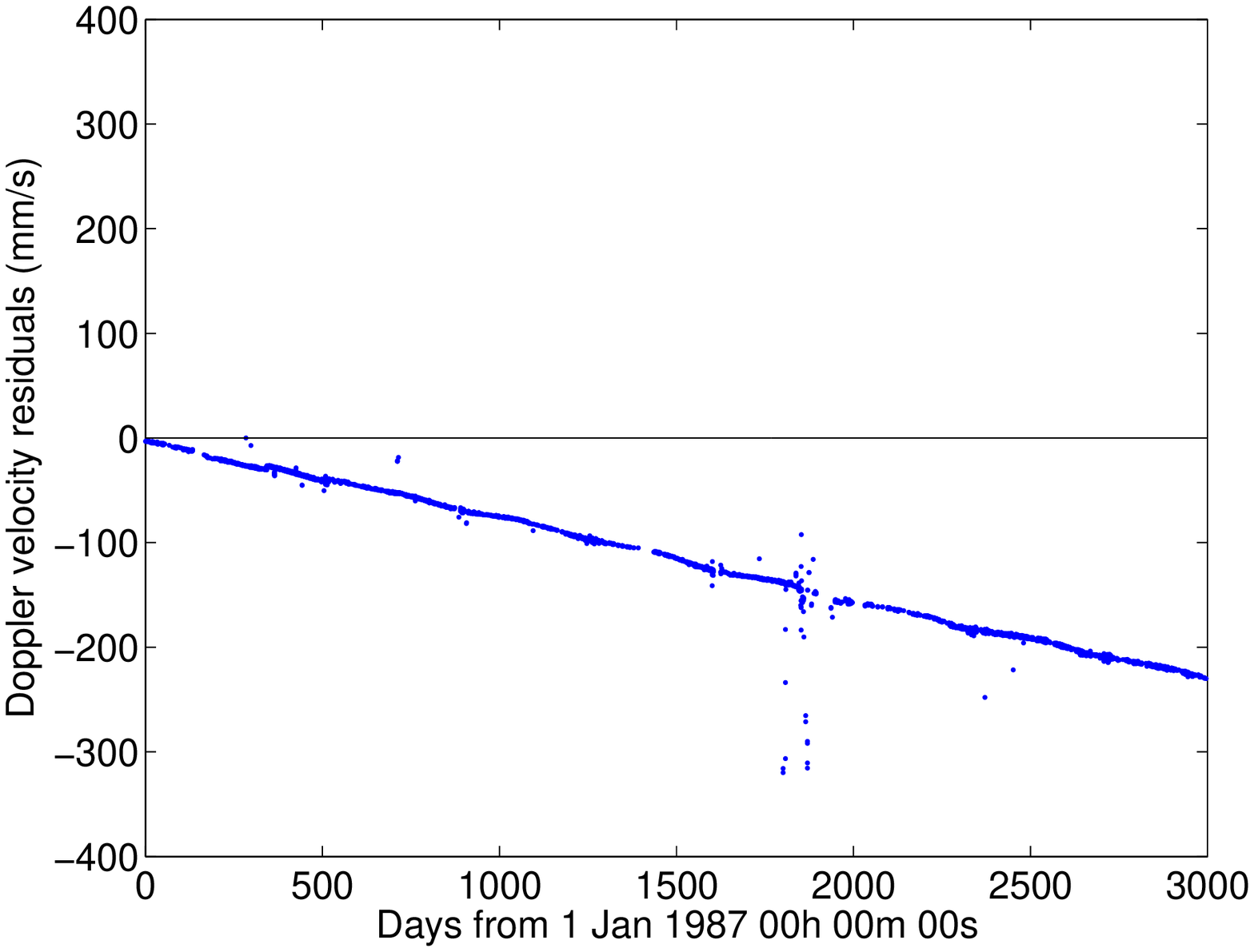}
\end{minipage}
\caption{Best-fit Pioneer~10 residuals using the ODYSSEY orbit
  determination program~\cite{levy-2008}. Left: residuals after a
  best-fit constant acceleration of $a_P=(8.40\pm 0.01)\times
  10^{-10}\mathrm{\ m/s}^2$. Right: reconstruction of the anomalous
  acceleration contribution.}
\label{fig:levy}
\vskip -5pt
\end{figure}}

Levy et al.~\cite{levy-2008} also performed an analysis of the Pioneer
data using the independently developed orbit determination program
ODYSSEY. The team confirmed the presence of an acceleration signal
consistent with that found in other studies: for Pioneer~10, they
obtained an anomalous acceleration of $a_P=(8.40\pm 0.01)\times
10^{-10}\mathrm{\ m/s}^2$ (see Figure~\ref{fig:levy}). Their study shows the
presence in the residual of periodic terms with periods consistent
with half a sidereal day, one sidereal day, and half a year, and they
investigate the possibility that these variations may be due to
perturbations of unknown origin that modify the propagation of the
signal.

In view of all these studies, the existence of the Pioneer anomaly in
the Pioneer~10 and 11 radiometric Doppler data is established beyond
doubt. Furthermore, the
analyses~\cite{levy-2008,2002gr.qc.....8046M,2007AA...463..393O,
  Toth2008} brought new knowledge about the effect, especially insofar
as the temporal behavior of the anomaly is concerned. As a result, the
anomalous acceleration can no longer be characterized as having a
constant magnitude. Instead, the effect clearly shows temporal
decrease -- perhaps consistent with the decay of the radioactive fuel
on board -- the conjecture that needs further investigation. This
recently-gained knowledge serves as a guide for new study of the
effect (discussed in Section~\ref{sec:strategy}); it also points out the
unresolved questions that we summarize below.

\subsection{Unresolved questions}
\label{sec:unresolved-questions}

Although JPL's 2002 study~\cite{pioprd} offered a very comprehensive
review of the Pioneer anomaly, in some cases it left some questions
unanswered. In other cases, the report's conclusions were put into
question, either by independent verifications or by the analysis of
subsequently recovered data and spacecraft documentation.

Specifically, the following open questions are important for
understanding the physical nature of the Pioneer anomaly and are a
subject of on-going investigation:

\begin{itemize}
\item \textit{Direction}: The true direction of the anomalous
  acceleration is yet unclear. Within the typical angular uncertainty
  of Doppler navigation in the S-band and a spacecraft HGA pointing
  accuracy of 3\textdegree\ (set within 10~dB antenna gain bandwidth),
  $a_P$ behaves as a constant acceleration of the craft  pointing in
  the innermost region of the solar system. Is it possible to
  determine, on the basis of available data, if the anomalous
  acceleration points towards the Sun, the Earth, the spin-axis
  direction, or the direction of the spacecraft's motion?
\item \textit{Constancy}: The earlier analysis concluded that both
  temporal and spatial variations of the anomaly's magnitude are of
  order 10\% for each craft, while formal errors are significantly
  smaller. But what is the true temporal behavior of the anomaly? Is
  it really possible to exclude on-board heat (which decays with time)
  as its source?
\item \textit{Distance}: It is unclear how far out the anomaly
  extends, but the Pioneer~10 data supports its presence at distances
  up to $\sim$~70~AU from the Sun. The Pioneer~11 data shows the
  presence of the anomaly as close in as $\sim$~20 AU. What is the
  true range of the anomaly, does it extend beyond these limits?
\item \textit{Onset}: Early data suggests that the anomalous
  acceleration may not have been present when the spacecraft were much
  nearer to the Sun. Can this be confirmed using the extended Doppler
  data set?
\item \textit{Annual/diurnal terms}: Even after a best fit analysis is
  completed, the resulting residual is not completely random: both
  annual and diurnal variations are clearly
  visible~\cite{pioprd,levy-2008,moriond}. Is it possible to pinpoint
  the source of these variations?
\item \textit{Pioneer spin anomaly}: The Pioneer~10 spacecraft
  exhibited anomalous spin behavior, and corresponding anomalous
  behavior in its fuel system. Similar issues exist with the spin of
  Pioneer~11. Does this spin anomaly have any bearing on the anomalous
  acceleration?
\item \textit{The thermal recoil force}: Since the publication of the
  2002 study~\cite{pioprd}, it has become clear that the magnitude of
  the recoil force due to anisotropically emitted heat has been
  underestimated. To what extent is this heat responsible for the
  anomalous acceleration?
\end{itemize}

In the following subsections, we review these unresolved questions in
detail, including information that has become available since 2002 as
a result of the on-going efforts of several teams.

\subsubsection{Direction}
\label{sec:direction}

The direction of the anomalous acceleration vector has been the
subject of many discussions. If this direction was precisely known, it
would allow one to establish the possible cause(s) of the anomaly; at
present, all causes must be considered.

The canonical value of $a_P=(8.74\pm 1.33) \times 10^{-10}\mathrm{\ m/s}^2$
was developed using the \textit{hypothesis} that the anomalous
acceleration of the two spacecraft is Sun-pointing, and it finds that
the hypothesis is consistent with the data. However, this does not
exclude the possibility that equally good solutions can be obtained by
postulating an hypothetical acceleration vector in some other
direction.

Thus, we must consider at least four possible directions for the
anomaly (see Figure~\ref{fig:directions}), all indicating a different
physical mechanism.  Because these directions differ by at most a few
degrees during the period of time from which Doppler data was studied,
they cannot be distinguished easily. However, each of these four
directions implies very different physics.

Specifically, if the acceleration was:
\begin{inparaenum}[i)]
\item in the direction \textit{towards the Sun}, this would indicate a
  force, likely gravitational, originating from the Sun, likely
  signifying a need for gravity modification;
\item in the direction \textit{towards the Earth}, this would indicate
  a time signal anomaly originating in the DSN hardware or
  introduced by the space flight-control methods;
\item in the direction \textit{of the velocity vector}, this would
  indicate an inertial force or a drag force providing support for a
  media-dependent origin; or, finally,
\item in \textit{the spin-axis direction}, this would indicate an
  on-board systematic, which is the most plausible explanation for the
  effect.
\end{inparaenum}
As seen on the left of Figure~\ref{fig:directions}, the corresponding
directional signatures of these four directions are distinct and could
be easily extracted from the data~\cite{2004CQGra..21.4005N,
  Turyshev:2005zm, MDR2005}.

\epubtkImage{}{%
  \begin{figure}[t!]
    \centerline{\includegraphics[width=0.70\textwidth]{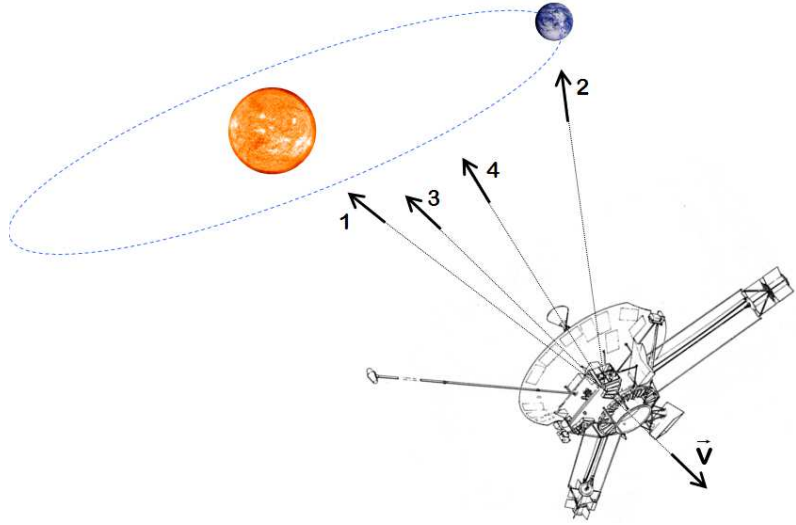}}
    \caption{The four possible different directions for the Pioneer
    anomaly: (1)~toward the Sun, (2)~toward the Earth, (3)~along the
    velocity vector, and (4)~along the spin axis. These directions
    would offer different signal
    modulations~\cite{2004CQGra..21.4005N, Turyshev:2005zm} that could
    be detected in the new study.}
    \label{fig:directions}
\end{figure}}

The navigation of the Pioneer spacecraft relied on an S-band
radio-Doppler observable (no ranging capabilities!), which is not very
accurate for the purposes of a 3-dimensional orbit reconstruction. To
maintain communications with the Earth using the narrow beam of their
HGA, an Earth-pointing attitude was necessary. During the flight
through the inner solar system, this required frequent attitude
correction maneuvers to re-orient the spin axis. At this time, the
Sun-craft-Earth angle was relatively large, as was the angle between
these and the direction of motion (Figure~\ref{fig:directions}). Even
so, the Pioneer data from distances up to 27~AU indicate an
approximately sunward directional anomaly for both craft. However, at
distances farther than 40~AU, both the Sun and the Earth were within
the 3\textdegree\ of the antenna radiation pattern (set by 10~dbm range in
the antenna gain), thus limiting accuracy in directional
reconstruction.

Given the importance of the direction of the anomalous acceleration,
it is perhaps surprising that this direction has not yet been
established. This is due to the fact that the data that has been
investigated to date is from the late cruise phase of the Pioneer~10
and 11 missions, when Pioneer~10 was over 40, and Pioneer~11 was over
22~AU from the Sun. From this distance, the Earth-spacecraft-Sun angle
is less than 3\textdegree. The angle between the spacecraft spin axis and
the Earth-spacecraft line is also small, as the spacecraft was always
oriented towards the Earth in order to maintain continuous radio
communication. Lastly, at these distances from the Sun the hyperbolic
orbits of the spacecraft are nearly asymptotic, and the direction of
motion corresponds closely with the Sun-spacecraft direction.

The discussion above indicates that the true direction of the Pioneer
anomalous acceleration still remains poorly determined. Based on the
Pioneer data analyzed to date, a more correct statement about the
direction would be that the anomaly is directed towards the
\textit{inner} part of the solar system -- the region that includes
both the Sun and the Earth. This question will be re-examined with the
extended data set that is now available for new investigation (see
Section~\ref{sec:new-data}).

\subsubsection{The magnitude and temporal behavior of the effect}

The magnitude of the anomalous acceleration has been confirmed by
several
studies~\cite{levy-2008,2002gr.qc.....8046M, 2007AA...463..393O,
  Toth2008}, all of which show results that are consistent with JPL's
``canonical'' value of $a_P=(8.74\pm 1.33)\times
10^{-10}\mathrm{\ m/s}^2$. However, the constancy of the Pioneer anomalous
acceleration remains a subject of serious debate.

For example, if the anomaly was of thermal origin, its magnitude would
decay with time, consistent with the decreasing amount of heat
generated on-board. Initially~\cite{pioprd} it was assumed that the
decay of a thermal recoil force would be consistent with the half-life
of the \super{238}Pu fuel (87.74~years), but later it was
realized~\cite{1999PhRvL..83.1892K, 1999PhRvL..83.1890M, 2003PhRvD..67h4021S}
that other effects, such as regulation of the electric power on board,
can mask at least some of this decay.

Meanwhile, Markwardt (\cite{2002gr.qc.....8046M}; see also
Section~\ref{sec:markwardt}) and later, Toth (\cite{Toth2008},
see Section~\ref{sec:toth}) demonstrated that the Doppler data
are, in fact, consistent with a change-of-acceleration (jerk) term for
both spacecraft, and that the magnitude of this jerk term is
consistent with the decay of the on-board thermal inventory.

\subsubsection{Onset of the anomaly}
\label{sec:onset}

Early Pioneer~10 and 11 data (before 1987) were never analyzed in
detail, especially with regard to systematics. However, by about
1980 the Doppler navigational data had began to indicate the presence
of an anomaly. At first this was considered to be only an interesting
navigational curiosity. But even so, samples of data a few months long
in duration were periodically examined by different analysts. By 1992
an interesting string of data-points had been obtained; they were
gathered in a JPL memorandum~\cite{jda-memo}, and are shown in
Figure~\ref{fig:early}. (More details on this issue are
in~\cite{pioprl, pioprd, Turyshev:2005zk}.)

For Pioneer~10, an approximately constant anomalous acceleration seems
to exist in the data as close in as 27~AU from the Sun. For
Pioneer~11, beginning just after Jupiter flyby, the early navigational
data show a small value for the anomaly during the Jupiter-Saturn
cruise phase in the interior of the solar system. But right at Saturn
encounter, when the craft passed into an hyperbolic escape orbit,
there was an apparent fast increase in the anomaly, whereafter it
settled into the canonical value.

The data, therefore, indicate the possibility that there was an
``onset'' of the anomaly at around this time, as the anomalous
acceleration component was much smaller prior to Saturn
encounter. However, this is likely a premature conclusion. The
observations shown in Figure~\ref{fig:early} do not represent a
systematic set of measurements obtained using a consistent, common
editing strategy. The apparent onset of the anomaly that is seen in
Figure~\ref{fig:early} has been used to justify theoretical work that
predicted a modification of gravity or other nongravitational forces
that affect only objects at sufficient distance from the Sun, or
objects in hyperbolic orbits. Yet it must be emphasized (as indeed,
the authors of the 2002 study~\cite{pioprd} emphasized in a footnote)
that this early detection data cannot be viewed as a measurement;
confirmation of the onset of the anomaly requires re-analysis of early
Doppler data using a consistent editing strategy.

Other factors, such as the greater frequency of maneuvers, also put
into question the extent to which one can rely on the apparent pattern
represented by the first few data points in
Figure~\ref{fig:early}. Also, during this time period, solar radiation
pressure produced an acceleration that was several times larger than
the anomaly. Scheffer~\cite{2001gr.qc.....8054S,2003PhRvD..67h4021S}
pointed out that the onset that is seen in the data may, in fact, be
an artifact of solar model calibration. If an anomalous sunward
acceleration was present but not accounted for at the time the solar
model coefficients were measured when the spacecraft were still near
the Sun (i.e., if the anomalous acceleration was absorbed into
estimates of solar model coefficients), later, as solar radiation
decreased, one would observe an apparent onset as a result of this
miscalibration.

\subsubsection{Annual/diurnal terms}

As indicated in Section~\ref{sec:annualdiurnal}, even after a best fit
solution is obtained, the resulting residuals contain clearly
discernible annual and diurnal signatures. These small, approximately
sinusoidal contributions do not affect the determination of the
(approximately constant) anomalous acceleration, as they are
uncorrelated with it. However, the origin of these sinusoidal terms
remains unknown.

The authors of~\cite{pioprd,moriond} expressed the belief that these terms
are due to orbital mismodeling, notably mismodeling of the orbital
inclination of the spacecraft to the ecliptic plane. Other
possibilities also exist: for instance, the annual term may be related
to mismodeling of the effects of solar plasma on the radio signal,
whereas the diurnal term may be related to atmospheric effects on the
signal.

\subsubsection{Radio beam reaction force}

Among the effects considered by the authors of~\cite{pioprd}, the
radio beam reaction force produced the largest bias to the result (see
Section~\ref{sec:radio-beam}). This bias is due to the fact that the
spacecraft are continuously transmitting a $P_{\mathrm{radio}}=\, \sim
8\mathrm{\ W}$ highly collimated radio beam in the direction of the Earth, and
as a result, experience a proportional recoil force, resulting in an
acceleration of $a_{\mathrm{radio}}=P_\mathrm{radio}/mc\simeq
-1.10\times 10^{-10}\mathrm{\ m/s}^2$. As the force exerted by the
radio beam necessarily points away from the Earth and the Sun (as
indicated by the negative sign in the preceding equation), the
correction of the measured data increases the amount of the observed
anomalous attractive force and makes the Pioneer effect larger.

Two open questions remain concerning the actual magnitude of the radio
beam recoil force. First, according to the recovered flight telemetry,
the radio beam reaction force might not have been constant.  Flight
telemetry indicates that during much of the mission, the radio beam
was more powerful than the nominal 8~W, exceeding the nominal value by
1~W and more. Near the end of Pioneer~10's mission, however, the
transmitter power may have decreased by as much as 3~W (see
Figure~\ref{fig:TWT}). As this apparent decrease coincides with a drop
in the main bus voltage on board (due to the depletion of the
spacecraft's \super{238}Pu power supply), the decrease may be an artifact
of a failing telemetry system (see also Section~\ref{sec:radiobeam}).

Second, Scheffer~\cite{2003PhRvD..67h4021S} argues that assuming a
typical antenna design, as much as 10\% of the power emitted by the
high gain antenna (HGA) feed would have missed the parabolic dish
altogether, and would have produced a reaction force in the opposite
direction, at an approximately 45\textdegree\ angle. Although the HGA is
discussed in detail in the recovered project documentation, no attempt
has yet been made to establish more precise estimates on the
efficiency with which the transmitter's power is converted into a
recoil force.

\subsubsection{The Pioneer spin anomaly}

The spin rates of the Pioneer~10 and 11 spacecraft, nominally 4.8
revolutions per minute (rpm), were in fact changing with time. As we
showed in Section~\ref{sec:spin}, the two spacecraft exhibited
markedly different behavior, with unique features.

Pioneer~10 was slowly spinning down, with three discernible phases in
its spin history, described in detail in Section~\ref{sec:spin}. The
changes in Pioneer~10's spin rate approximately coincide with
unexplained readings from its propulsion tank (see
Section~\ref{sec:propulsion_syst}).

Meanwhile, Pioneer~11 was spinning up, although a detailed examination
of its spin history reveals that spin-up events coincided with
attitude correction maneuvers, and between maneuvers the spacecraft
was spinning down, albeit at varying rates (see
Section~\ref{sec:spin}).

It is possible that a constant or near constant rate of spin change is
of thermal origin: if thermal radiation is emitted by the spacecraft
not just anisotropically but also asymmetrically with respect to the
spacecraft's center-of-gravity, a torque acts on the spacecraft.

No convincing explanation has yet been offered for the anomalous
change in the spin change rate during the history of Pioneer~10. The
approximate coincidence of this change with the anomalous change in
fuel tank pressure readings may be significant; on the other hand, it
must be emphasized that this coincidence is only approximate, and may
very well be accidental.

In contrast, the spin rate change of Pioneer~11 exhibits no dramatic
changes, and appears to be a combination of three effects: A spindown
that may be similar to that of Pioneer~10 and may be of thermal
origin, instantaneous spinups that occur at each maneuver and may be a
result of a small thruster misalignment, and a third effect that
changes the spindown rate after each maneuver, and may be related to
thruster leaks and outgassing.

It should be noted that these considerations are qualitative in
nature, and no detailed study of the Pioneer~10 and 11 spin history
has taken place to date. In particular, we do not know if the spin
rate change is related to the anomalous acceleration of these
spacecraft.

\subsubsection{Thermal recoil forces}

The effect of rejected thermal radiation was the second largest
bias/uncertainty that has been the most critical systematic bias to
quantify (see Section~\ref{sec:force_recoil}). If heat generated by
the on-board power sources was asymmetrically reflected by the body of
the craft, an acceleration along the spin axis could be produced
causing the measured anomaly.

The Pioneer spacecraft were powered by SNAP-19 (Space Nuclear
Ancillary Power) RTGs mounted on long extended booms (designed to
protect the on-board electronics from heat and radiation
impact)~\cite{pioprl,pioprd,2001gr.qc.....7022A}. It was recognized
early that, in principle, there was more than enough heat available on
the craft to cause the anomaly (see Section~\ref{sec:heat}). In
addition to heat from the RTGs, additional waste heat was produced by
electrical instrumentation, Radioisotope Heater Units (RHUs), and the
propulsion system.

However, it was assumed that the spacecraft's spin-stabilized attitude
control, special design of the RTGs and the length of the RTG booms
that resulted in a relatively small spacecraft surface available for
the preferential heat rejection significantly minimized the amount of
heat for the mechanism to work. These considerations led to an
estimated acceleration not exceeding $a_{\mathrm{hr}}=(-0.55\pm
0.55)\times 10^{-10}\mathrm{\ m/s}^2$. This result is now being reconsidered,
in view of on-going work on Pioneer thermal modeling, which suggests
that the acceleration due to asymmetrically reflected heat from the
RTGs may have been several times this value.

As the spacecraft is in an approximate thermal steady state, heat
generated on board must be removed from the
spacecraft~\cite{Toth2009}. In deep space, the only mechanism of heat
removal is thermal radiation: the spacecraft can be said to be
radiatively coupled to the cosmic background, which can be modeled by
surrounding the spacecraft with a large, hollow spherical black body
at the temperature of $\sim$~2.7~K.

The spacecraft emits heat in the form of thermal photons, which also
carry momentum $p_\gamma$, in accordance with the well known law of
$p_\gamma=h\nu/c$, where $\nu$ is the photon's frequency, $h$ is
Planck's constant, and $c$ is the velocity of light. This results in a
recoil force in the direction opposite to that of the path of the
photon. For a body that emits radiation in a spherically symmetric
pattern, the net recoil force is zero. However, if the pattern of
radiation is not symmetrical, the resulting anisotropy in the
radiation pattern yields a net recoil force.

The magnitude of this recoil force is a subject of many factors,
including the location and thermal power of heat sources, the
geometry, physical configuration, and thermal properties of the
spacecraft's materials, and the radiometric properties of its external
(radiating) surfaces.

The total thermal inventory on board the Pioneer spacecraft exceeded
2~kW throughout most of their mission durations. The spacecraft were
in an approximate steady state: the amount of heat generated on-board
was equal to the amount of heat radiated by the spacecraft.

The mass of the Pioneer spacecraft was approximately 250~kg. An
acceleration of $8.74\times 10^{-10}\mathrm{\ m/s}^2$ is equivalent to a force
of $\sim 0.22~\mu\mathrm{N}$ acting on a $\sim$~250~kg object. This is the
amount of recoil force produced by a 65~W collimated beam of
photons. In comparison with the available thermal inventory of 2500~W,
a fore-aft anisotropy of less than 3\% can account for the anomalous
acceleration in its entirety. Given the complex shape of the Pioneer
spacecraft, it is certainly conceivable that an anisotropy of this
magnitude is present in the spacecrafts' thermal radiation pattern.

The possibility that heat from the RTGs can be responsible for the
Pioneer anomaly was first proposed by
Katz~\cite{1999PhRvL..83.1893A,1999PhRvL..83.1892K}.
Murphy~\cite{1999PhRvL..83.1891A,1999PhRvL..83.1890M} pointed out the
potential significance of anisotropic rejection of electrically
generated
heat. Scheffer~\cite{2001gr.qc.....8054S,2003PhRvD..67h4021S}
attempted to account for all possible sources of the thermal recoil
force and estimated that the total recoil force is more than
sufficient to produce an anomalous acceleration of the observed
magnitude. Unfortunately, none of these studies benefited from
detailed information about the spacecrafts' design or from thermal and
electrical telemetry. These data became available for study in 2005.

Key questions concerning the thermal recoil force that have been
raised during the study of the Pioneer anomaly
include~\cite{1999PhRvL..83.1892K, 1999PhRvL..83.1890M,
  2003PhRvD..67h4021S}:

\begin{itemize}
\item How much heat from the RTGs is reflected by the spacecraft,
  notably the rear of its high-gain antenna, and in what direction?
\item \vskip -6pt Was there a fore-aft asymmetry in the radiation
  pattern of the RTGs due to differential aging?
\item \vskip -6pt How much electrical heat generated on-board was
  radiated through the spacecraft's louver system?
\end{itemize}

A presently (2009) on-going effort is to build a comprehensive thermal
model of the Pioneer spacecraft from design documentation. The model
is to be validated by telemetry, and evaluated at different
heliocentric distances at different times during the lifetime of the
spacecraft. If successful, this effort will yield a high-accuracy
estimate of the thermal recoil force, which can then be incorporated
into future trajectory models.

\subsection{An approach to finding the origin of the Pioneer anomaly}
\label{sec:strategy}

With the availability of the new Pioneer~10 and 11 radiometric
Doppler data (see Section~\ref{sec:new-data}), a new study of the
Pioneer anomaly became possible. This much extended set of Pioneer
Doppler data is the primary source for the new ongoing investigation of
the anomaly. In addition, the entire record of flight telemetry files
received from Pioneer~10 and 11 is also available (see
Section~\ref{sec:telemetry}). Together with original project
documentation (see Section~\ref{sec:pio-docs}) and newly developed
software codes and trajectory analysis tools, this additional
information is now used to reconstruct the engineering history of both
spacecraft with the aim to establish the nature of the Pioneer
anomaly.  Below we review the current status of these efforts.

The primary objective of the new investigation is to determine the
physical origin of the Pioneer anomaly and identify its properties.
To achieve this goal, a study of the recently recovered radiometric
Doppler and telemetry data has begun, focusing in particular on
improving our understanding of the thermal behavior of the spacecraft
and the extent to which radiated heat can be responsible for the
acceleration anomaly.

The objectives of this new investigation of the Pioneer anomaly are
sixfold:
\begin{enumerate}[i)]
\item To analyze the early mission data; the goal would be  to
  determine the true direction of the anomaly and thus, its origin;
\item To study the physics of the planetary encounters; the goal would
  be to learn more about the onset of the anomaly (e.g., Pioneer~11's Saturn flyby),
\item To study the temporal evolution of $a_P$ with the entire data
  set; the goal would be a better determination of the temporal
  behavior of the anomaly,
\item To perform a comparative analysis of individual anomalous
  accelerations for the two Pioneers with the data taken from similar
  heliocentric distances, which could highlight properties of $a_P$,
  and
\item To investigate the on-board systematics with recently recovered
  MDRs; the goal here would be to investigate the effect of on-board
  systematics on the solution for the Pioneer anomaly obtained with
  the Doppler data, and, finally
\item To build a model of the thermal, electrical and dynamical
  behavior of the Pioneer vehicles and verify it with the actual data
  from the MDRs; the goal here would be to develop a model to be used
  to calibrate the Doppler anomaly with respect to the on-board
  sources of dynamical noise.
\end{enumerate}
These objectives are not entirely independent of each other; by
putting them on this list, we are identifying the main areas that are
the focus of the on-going new investigation of the anomaly. Below we
will discuss these objectives in more detail.

\subsubsection{Analysis of the earlier trajectory phases}

One objective of the new investigation is the study of the early parts
of the trajectories of the Pioneers with the goal of determining the
true direction of the Pioneer anomaly and possibly its
origin~\cite{2004CQGra..21.4005N, Turyshev:2005zm,
  Turyshev:2005vj}. The much longer data span is expected to improve
the ability to determine the source of the acceleration. In
particular, with data from much closer to the Earth and Sun, one
should be able to better determine whether the acceleration is
\begin{inparaenum}[i)]
\item in the sunward direction,
\item in the Earth-pointing direction,
\item in the direction along the velocity vector, or
\item along the spin axis direction (see Section~\ref{sec:direction}).
\end{inparaenum}
Analysis of the earlier data is critical in helping to establish a
precise 3-dimensional time history of the effect, and therefore to
find out whether it is due to a systematic or new physics.

\epubtkImage{}{%
  \begin{figure}[!ht]
    \centerline{\includegraphics[width=0.6\linewidth]{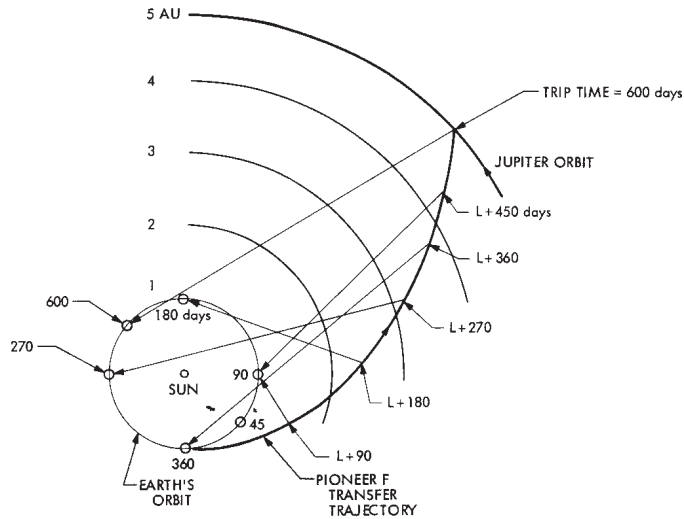}}
    \caption{Proposed directions (along the spin and antenna axes)
    from the Pioneer~F spacecraft (to become Pioneer~10) toward the
    Earth~\cite{JPL32-1526-II-C}.}
    \label{fig:antenna-point}
\end{figure}}

\epubtkImage{}{%
  \begin{figure}[!ht]
    \centerline{\includegraphics[width=0.8\linewidth]{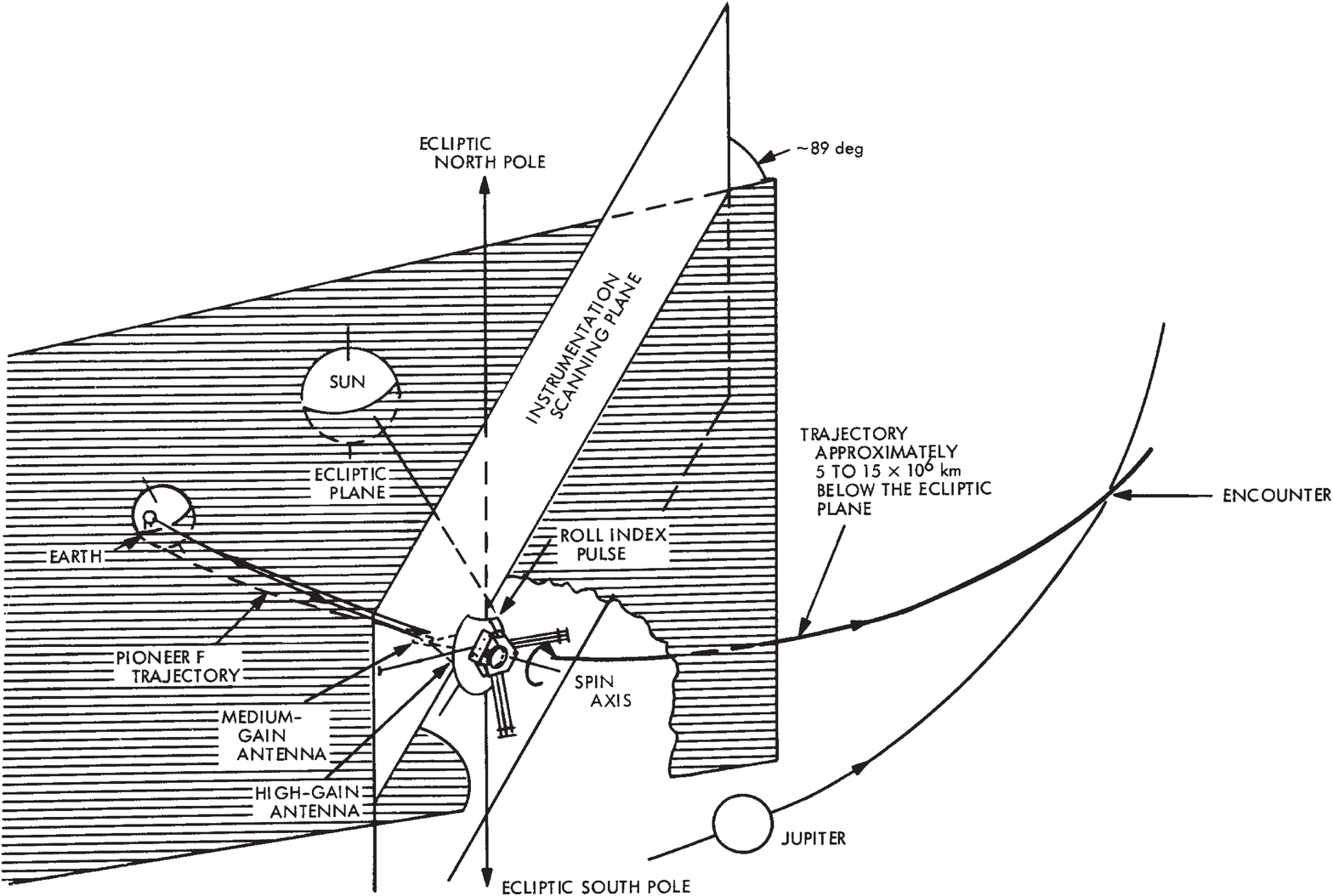}}
    \caption{The earlier part of the Pioneer~10 trajectory before Jupiter
    encounter, the part of the trajectory when antenna articulation
    was largest~\cite{JPL32-1526-II-C}.}
    \label{fig:pio-inner-orient}
\end{figure}}

\subsubsection{Study of the planetary encounters}

An approximately constant anomalous acceleration seems to exist in the
data of Pioneer~10 as close in as 27~AU from the
Sun~\cite{pioprl,pioprd,1976AJ.....81.1153N}. Navigational data
collected for Pioneer~11, beginning just after Jupiter flyby, show a
small value for the anomaly during the Jupiter-Saturn cruise phase in
the interior of the solar system. However, right at Saturn encounter,
when Pioneer~11 passed into a hyperbolic escape orbit, there was an
apparent fast increase in the anomalous
acceleration~\cite{pioprd,2005CQGra..22.5343N, Turyshev:2005vj}, but
this has not yet been confirmed by rigorous analysis (see
Section~\ref{sec:onset}).

Doppler data covering Pioneer~11's encounter with Saturn are
available. A successful study of the data surrounding the
encounter~\cite{Turyshev:2005zm} would lead to an improved
understanding of the apparent onset of the anomalous acceleration. The
encounters of both spacecraft with Jupiter may also be of interest
(see~\cite{MDR2005}), although that close to the sun, much larger
contributions to the acceleration noise are present.

While early data may improve our understanding of the direction of the
anomaly, a difficult obstacle exists along the way towards this
goal~\cite{pioprd,2004CQGra..21.4005N}. Radiometric observables,
notably Doppler, are sensitive in the line-of-sight direction, but are
insensitive to small changes in the spacecraft's orbit in a direction
that is perpendicular to the line-of-sight. The lack of a range
observable on Pioneer~10 and 11 also reduces the accuracy with which
the orbit can be determined in three dimensions. Nevertheless, these
problems can be addressed and the on-going analysis should be able to
yield the true direction of the anomaly and its
origin~\cite{2005CQGra..22.5343N, Turyshev:2005zm, Turyshev:2005vj}.

\subsubsection{Study of the temporal evolution of the anomaly}

JPL's 2002 analysis~\cite{pioprd} found that the anomalous
acceleration is approximately constant. On the other hand, any
explanation involving the on-board thermal inventory of the spacecraft
must necessarily take into account this inventory's decay with
time. It was on this basis that the authors of~\cite{pioprd} rejected
the hypothesis that the acceleration is due to collimated thermal
emission.

While JPL's study of 11.5~years of Pioneer~10
data~\cite{pioprd, 2001gr.qc.....7022A} found no change in the
anomalous acceleration, Markwardt~\cite{2002gr.qc.....8046M},
Olsen~\cite{2007AA...463..393O} and Toth~\cite{Toth2008} were not
able to rule out this possibility. The now available extended data
set, which includes over 20~years of usable Pioneer~10 data, may be
sufficient to demonstrate unambiguously whether or not a jerk term is
present in the signal, and if it is compatible with the temporal
behavior of the on-board thermal
inventory~\cite{2006CaJPh..84.1063T}. We note, however, the difficulty
of the task of disentangling such a jerk term from the effects of
solar radiation pressure.

\subsubsection{Analysis of the individual trajectories for both Pioneers}

The trajectories of Pioneer~10 and 11 were profoundly different. After
its encounter with Jupiter, Pioneer~10 continued on a hyperbolic
escape trajectory, leaving the solar system while remaining close to
the plane of the ecliptic. Pioneer~11, in contrast, proceeded from
Jupiter to Saturn along a trajectory that took it closer to the Sun,
while outside the ecliptic plane. After its encounter with Saturn,
Pioneer~11 also proceeded along a hyperbolic escape trajectory, but
once again it was flying outside the plane of the ecliptic. In the
end, the two spacecraft were flying out of the solar system in
approximately opposite directions.

Nonetheless, the limited data set that was available previously
precluded a meaningful comparison. The individual solutions for the
two spacecraft were obtained from data segments that not only differed in
length (11.5 and 3.75~years), but were also taken from different
heliocentric distances (see Section~\ref{sec:anomaly_sum}).

From the recovered telemetry~\cite{MDR2005} we now also know that the
actual thermal and electrical behavior of the two spacecraft was
different~\cite{2006CaJPh..84.1063T, 2007arXiv0710.2656T,
  MDR2005}. These facts underline the importance of studying and
comparing the behavior of both spacecraft, as this may help determine
if the anomaly is of on-board origin or extravehicular in nature.

\subsubsection{Investigation of on-board systematics}
\label{sec:system-mdr}

The availability of telemetry information makes it possible to conduct
a detailed investigation of the on-board systematic forces as a source
of the anomalous acceleration.

Previously, all known mechanisms of on-board systematic forces were
examined~\cite{pioprl,1999PhRvL..83.1893A, 1999PhRvL..83.1891A,
  pioprd, 1999PhRvL..83.1892K, 1999PhRvL..83.1890M,
  2003PhRvD..67h4021S, MDR2005} (see
Table~\ref{tab:error_budget}). Current efforts are designed to improve
our understanding of the contribution of on-board heat -- notably, heat
from the RTGs reflecting off the spacecraft, and electrical heat
generated within the spacecraft -- to the anomalous acceleration. The
available telemetry also helps refine estimates of the radio beam
reaction force. (Other effects, such as the differential emissivity of
the RTGs, helium expulsion from the RTGs, propulsive gas leaks, were
also analyzed~\cite{2007arXiv0710.2656T, MDR2005} but were found to be
insignificant.)

As pointed out in~\cite{2001gr.qc.....7022A}, any thermal explanation
should clarify why either the radioactive decay (if the heat is
directly from the RTGs/RHUs) or electrical power decay (if the heat is
from the instrument compartment) is not seen. One reason could be that
previous analyses used only a limited data set of only 11.5~years when
the thermal signature was hard to disentangle from the Doppler
residuals or the fact that the actual data on the performance of the
thermal and electrical systems was not complete or unavailable at the
time the analyses were performed.

The present situation is very different. Not only do we have a much
longer Doppler data segment for both spacecraft, we also have the
actual telemetry data on the thermal and electric power subsystems for
both Pioneers for the entire lengths of their missions. The electrical
power profile of the spacecraft can be reconstructed with good
accuracy using electrical telemetry measurements (see
Section~\ref{sec:pio-project}, and also~\cite{MDR2005}). The telemetry
also contains measurements from a large number of on-board temperature
sensors.

This information made it possible to construct a detailed thermal
model of the Pioneer spacecraft (see Figures~\ref{fig:thermal_model}
and \ref{fig:tempmap}). As of early 2010, this work is near
completion, and its results are being readied for publication.

\epubtkImage{}{%
  \begin{figure}[ht!]
    \centerline{\includegraphics[width=1.0\linewidth]{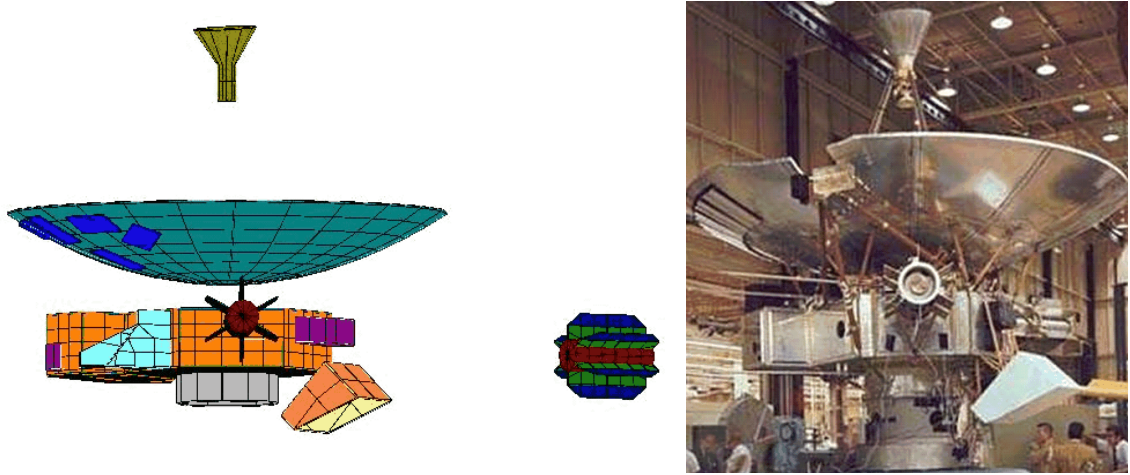}}
    \caption{A geometric model (left) of the Pioneer spacecraft, used
    for finite element analysis, and a photograph (right) of
    Pioneer~10 prior to launch. The geometric model accurately
    incorporates details such as the Medium Gain Antenna (MGA), the
    Asteroid-Meteoroid Detector, and the star sensor shade. Note that
    in the geometric model, the RTGs are shown in the extended
    position; in the photograph, the RTGs are
    stowed. From~\cite{Kinsella-etal:2007, Kinsella-etal:2008}.}
    \label{fig:thermal_model}
\end{figure}}

\epubtkImage{}{%
  \begin{figure}[ht!]
    \centerline{\includegraphics[width=0.8\linewidth]{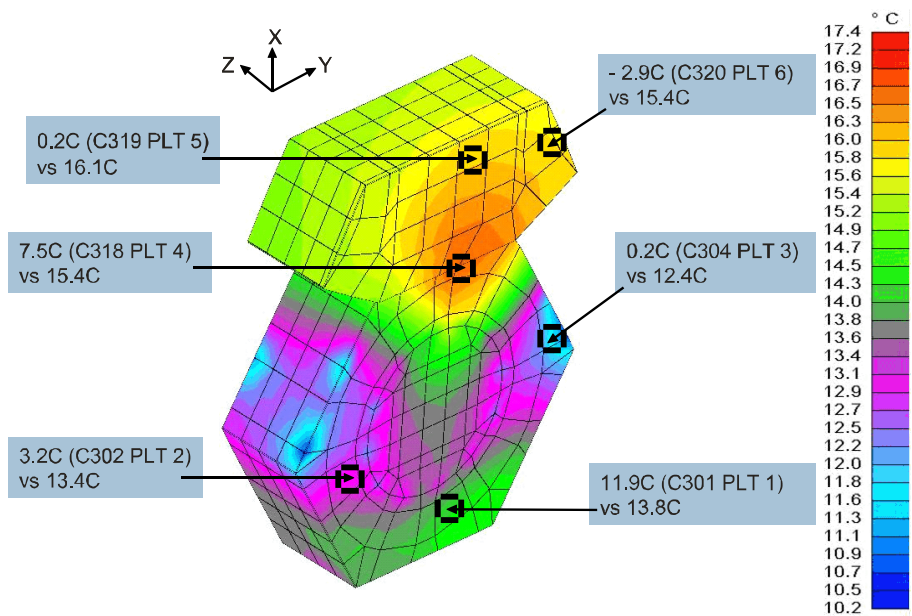}}
    \caption{A ``work-in-progress'' temperature map of the outer
    surface of the Pioneer~10 spacecraft body, comparing temperatures
    calculated via a numerical finite element method vs. temperatures
    measured by platform temperature (PLT) sensors and
    telemetered. While agreement between calculated and telemetered
    temperatures is expected to improve as the model is being
    developed, discrepancies between these values illustrate  the
    difficulties of creating a reliable temperature map using
    numerical methods. (From~\cite{Kinsella-etal:2007,
    Kinsella-etal:2008}).}
    \label{fig:tempmap}
\end{figure}}

\subsection{The thermal recoil force}
\label{sec:therml-recoil-force}

It has been long recognized~\cite{pioprl,pioprd, MDR2005} that
anisotropically emitted thermal radiation can contribute to the
acceleration of the Pioneer~10 and 11 spacecraft, and that the
available thermal inventory on board (in excess of 2~kW), if directed,
is more than sufficient to provide the necessary acceleration
(requiring only $\sim$~65~W of collimated electromagnetic radiation.)

Nonetheless, in 2002, having only limited thermal data and relevant
spacecraft information at hand, the authors of~\cite{pioprd} estimated
the contribution of the thermal recoil force as $(-0.55\pm 0.73)\times
10^{-10}\mathrm{\ m/s}^2$ (the estimate is based on the values reported for
the RTG heat reflected off the craft and nonisotropic radiative
cooling of the spacecraft, items 2b) and 2d) in
Table~\ref{tab:error_budget} correspondingly), i.e., only about 6\% of
the anomalous acceleration.

Since 2002, it has become clear that this figure is in need of a
revision. A quantitative estimate was first offered by Scheffer in
2003~\cite{2003PhRvD..67h4021S}, who calculated a total of 52~W of
directed thermal radiation (not including his estimate on asymmetrical
radiation from the RTGs), or about 80\% of the thrust required to
account fully for the anomalous acceleration. In 2007,
Toth~\cite{TothISSI2007} presented an argument for expressing the
combined recoil force due to electrical and RTG heat as a linear
combination of the RTG thermal power and electrical power on board;
his coefficients (0.012 for the RTGs, 0.36 for electrical heat) yield
a figure of $\sim$~55~W of directed thermal radiation in the
mid-1980s, which, after accounting for the heat from the RHUs (4~W)
and the antenna beam (--7~W) as Scheffer did, translates into a result
that is similar to Scheffer's.

Benefiting from the extensive discussions of the topic of thermal
recoil force during the meetings of the Pioneer Explorer Collaboration
at ISSI\epubtkFootnote{The Pioneer Explorer Collaboration at the
  International Space Science Institute (ISSI), Bern, Switzerland, see
  details
  \url{http://www.issibern.ch/teams/Pioneer/}}~\cite{Kinsella-etal:2007,
Kinsella-etal:2008, ISSI2005, Scheffer:2005, Scheffer:2008,
TothISSI2007}, several researchers tried to model effects of this
force on a Pioneer-like spacecraft using various computer tools. In
2008, Bertolami et al.~\cite{2008PhRvD..78j3001B} made an attempt to
develop a methodology using point-like Lambertian sources to estimate
the thermal recoil force on the Pioneer spacecraft, and obtained an
acceleration estimate that corresponds to $\sim$~67\% of the Pioneer
anomaly, or $\sim$~45~W of directed thermal radiation. In addition, in
2009, Rievers et al.~\cite{Rievers2010, Rievers2009} used finite
element modeling and ray tracing algorithms to compute an acceleration
that corresponds to $\sim$~48~W of directed heat in the mid-1980s.

These recent results support claims that the contribution of the
thermal recoil forces to the Pioneer anomaly was previously
(e.g.,~\cite{pioprd}) underestimated. However, the estimates above
used only rough values for on-board thermal and electrical power and
are valid only for a particular point in Pioneers' missions.  As such,
these estimates can provide only an overall magnitude of the effect
and say little on its temporal behavior. On the other hand, the now
available telemetry and recovered spacecraft design
documentation~\cite{MDR2005} makes it possible to develop a
comprehensive thermal model capable of estimating thermal recoil force
throughout the entire mission of both Pioneers.  Interim results show
that this model is in good agreement with redundant telemetry
observations~\cite{Kinsella-etal, Kinsella-etal:2007,
  Kinsella-etal:2008}. Therefore, the development of a comprehensive,
reliable estimate of the thermal recoil force is now within reach.

Below, we discuss the basic principles of modeling the thermal recoil
force, as well as the application of these principles to the case of the
Pioneer~10 and 11 spacecraft.

\subsubsection{General formalism}
\label{sec:thermal-formalism}

While heat transfer textbooks provide all necessary details about
thermal radiation as a mechanism for energy transfer, momentum
transfer is rarely covered in any detail. This is perhaps not
surprising: the momentum of a photon with energy $E$ is $p=E/c$, and
thus, the recoil force associated with a collimated beam of
electromagnetic radiation with power $P$ is $F=P/c$. For each watt of
radiated power, the corresponding recoil force is only $\sim$~3.33~nN.
Such tiny forces rarely, if ever, need to be taken into
account in terrestrial applications. This is not so in the case of
space applications~\cite{AR1992, DAF2006, SL1994, RA1990, VSA1994,
  ZASEC2005}, in particular in the case of Pioneer~10 and 11: the
observed anomalous acceleration corresponds to a force of less than
220~nN, which can be produced easily by a modest amount ($\sim$~65~W)
of electromagnetic radiation.

A formal treatment of the thermal recoil force must establish a
relationship between heat sources within the radiating object and the
electromagnetic field outside the object~\cite{Toth2009}. This can be
accomplished in stages, first by describing heat conduction inside the
object using Fourier's law~\cite{Kaviany2001, LL2002, MM1999}:
\begin{equation}
\vec{q}=-k\nabla T,
\end{equation}
where $\vec{q}$ is the heat flux (measured in units of power over
area), $T$ is the temperature, and $k$ is the heat conduction
coefficient, a tensorial quantity in the general case, but just a
number for homogeneous and isotropic materials. Heat flux also obeys
the energy conservation equation
\begin{equation}
\nabla\cdot\vec{q}=b-C_h\rho\frac{\partial T}{\partial t},
\end{equation}
where $b$ is the volumetric heat release (measured in units of power
density), $C_h$ is the material's specific heat, and $\rho$ is its
density. For discrete sources,
$b(\vec{x},t)=\sum_{i=1}^nB_i(t)\delta^3(\vec{x}-\vec{x}_i)$, where
$B_i$ is the thermal power of the $i$-th source, $\vec{x}_i$ is its
location, and $\delta$ denotes Dirac's delta function.

At a radiating surface,
\begin{equation}
q=\vec{q}\cdot\vec{a},\label{eq:q}
\end{equation}
where $\vec{a}$ is the unit normal of the radiating surface element,
and $q$ is the surface element's radiant intensity. The radiant
intensity, or energy flux, of a radiating surface is related to its
temperature by the Stefan--Boltzmann law:
\begin{equation}
q(\vec{x},t)=\sigma\epsilon(\vec{x},t, T)T^4(\vec{x},t),
\label{eq:stefan-boltzmann}
\end{equation}
where $\sigma\simeq 5.67\times 10^{-8}\mathrm{\ Wm}^{-2}\mathrm{\ K}^{-4}$ is the
Stefan--Boltzmann constant, while the dimensionless coefficient
$0\le\epsilon\le 1$ is a physical characteristic of the emitting
surface. This coefficient can vary not only as a function of location
and time, but also as a function of
temperature. Equation~(\ref{eq:stefan-boltzmann}) can be used to
calculate the radiative coupling between facing surfaces of the
object, and between the object and its environment (e.g., the deep
sky, which can be modeled as a blackbody with temperature
$T_{\mathrm{sky}}\simeq 2.7\mathrm{\ K}$). Together with internal boundary
conditions (i.e., the power and distribution of internal heat sources,
represented by $b$ above) the problem becomes fully solvable: values
of $\vec{q}$ inside the object, and $q$ on its boundary surface can be
computed~\cite{Toth2009}.

Outside the radiating object, the electromagnetic field is described
by the stress-energy-momentum tensor
\begin{equation}
T^{\mu\nu}=\begin{pmatrix}c^{-2}u&\vec{\mathfrak{p}}\cr
\vec{\mathfrak{p}}&\mathbb{P}\end{pmatrix},
\end{equation}
where $u$ is the energy density of the radiation field,
$\vec{\mathfrak{p}}$ is its momentum density, and $\mathbb{P}$ is the
radiation pressure tensor. The stress-energy-momentum tensor obeys the
conservation equation $\nabla_\mu T^{\mu\nu}=0$, where $\nabla_\mu$ is
the covariant derivative with respect to the coordinate $x^\mu$. This allows one to
develop an expression for the recoil force $\vec{F}$ acting on a
radiating surface $A$ in the form
\begin{equation}
\vec{F}(t)=-\int\mathbb{P}(\vec{x},t)\cdot\vec{dA},
\end{equation}
which, for an isotropic (Lambertian) emitter yields the well-known law
\begin{equation}
\vec{F}(t)=-\frac{2}{3}\frac{1}{c}\int q(\vec{x},t)~\vec{dA}.
\end{equation}
If the geometry of the emitter's exterior (represented by $A$) and the
radiant intensity $q$ along the exterior surface are known, the recoil
force can be computed~\cite{Toth2009}.

The Pioneer~10 and 11 spacecraft have two heat sources that
contribute significant amounts to the thermal recoil force - the
RTGs and the on-board electrical equipment. The case is further
simplified by the fact that the Pioneer~10 and 11 spacecraft are
spinning, as it allows a force computation in only one dimension (see
Section~\ref{sec:nongrav}).

\subsubsection{Contribution of the heat from the RTGs}

\epubtkImage{}{%
  \begin{figure}[!t]
    \centerline{\includegraphics[width=0.65\textwidth]{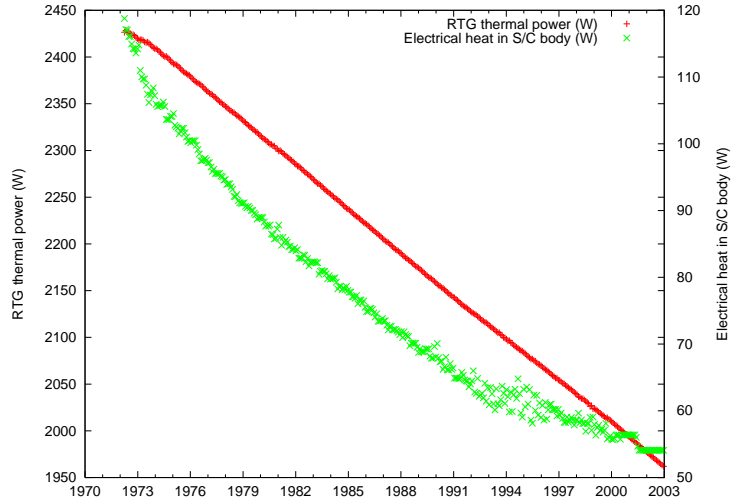}}
    \caption{Heat generated by RTGs (red, approximately straight line,
    scale on left) and electrical equipment (green, scale on right) in
    Pioneer~10 over the lifetime of the spacecraft.}
    \label{fig:power}
\end{figure}}

Although each of the spacecraft has four RTGs, their temporal behavior
is identical (characterized by the half-life of the \super{238}Pu fuel
and the decaying efficiency of the thermocouples). The placement of
the RTGs is symmetrical. Consequently, the set of four RTGs can be
treated as a single heat source, the power of which we denote as
$B_{\mathrm{rtg}}(t)$. The amount of \super{238}Pu fuel on board is well
known from pre-launch test data, and the physics of the fuel's
radioactive decay is well understood~\cite{2006CaJPh..84.1063T,
  2007arXiv0710.2656T, MDR2005}; therefore, the total power of the
RTGs, $P_{\mathrm{rtg}}(t)$ is known:
\begin{equation}
P_{\mathrm{rtg}}(t)=2^{-(t-t_0)/T}P_{\mathrm{rtg}}(t_0),
\end{equation}
where $t_0$ is the time when the power $P_{\mathrm{rtg}}(t_0)$ was
measured, and $T\simeq 87.74$~years is the half-life of the
radioactive fuel.

The amount of power $P_{\mathrm{elec}}(t)$ removed from the RTGs in the
form of electrical power is measured directly by telemetry and is
available for the entire mission durations. Therefore,
$B_{\mathrm{rtg}}(t)$ is given as:
\begin{equation}
B_{\mathrm{rtg}}(t)=P_\mathrm{rtg}(t)-P_{\mathrm{elec}}(t).
\label{eq:sigma-RTG}
\end{equation}

The value of $B_{\mathrm{rtg}}(t)$ can be computed with good accuracy
(Figure~\ref{fig:power}). The initial power $P_{\mathrm{rtg}}(t_0)\simeq
650\mathrm{\ W}$ per RTG was reported with a measurement accuracy of 1~W for
each RTG; the month, though not the exact date, of $t_0$ is
known. Therefore, $P_{\mathrm{rtg}}(t)$ can be calculated with an
accuracy of 0.2\% or better. Currents and voltages from each RTG are
found in the flight telemetry data stream, represented by 6-bit
values. The combined error due mainly to the limited resolution of
this data set amounts to an uncertainty of $\sim$~1~W per RTG. When
all these independent error sources are combined, the resulting figure
is an uncertainty of $\sigma_{\tt rtg} = 2.1\mathrm{\ W}$; the total power of
the four RTGs combined is $\sim$~2600~W at the beginning of the
Pioneer~10 and 11 missions~\cite{Toth2009}.

\subsubsection{Effects of the electrical heat}

Electrical power on board the Pioneer spacecraft was $\sim$~160~W at
the beginning of mission~\cite{2007arXiv0710.2656T}, slowly decreasing
to $\sim$~60~W at the time when the last transmission was received
from Pioneer~10. Some of this power was radiated into space directly
by a shunt radiator plate, some power was consumed by instruments
mounted external to the spacecraft body, and some power was radiated
away in the form of radio waves; however, most electrical power was
converted into heat inside the spacecraft body. This heat escaped the
spacecraft body through three possible routes: a passive thermal
louver system, other leaks and openings, and the spacecraft walls that
were covered by multilayer thermal insulation blankets. While the
distribution of heat sources inside the spacecraft body was highly
inhomogeneous, there is little temporal variation in the distribution
of heat inside the spacecraft body, and the exterior temperatures of
the multilayer insulation remain linear functions of the total
electrical heat. This allows us to treat all electrical heat generated
inside the spacecraft body as another single heat source:
\begin{equation}
B_\mathrm{elec}(t)=\mathrm{known~from~telemetry.}
\label{eq:sigma-elec}
\end{equation}

Therefore, the value of $B_{\mathrm{elec}}(t)$ is available from
telemetry (Figure~\ref{fig:power}). Uncertainties in the estimate of
$B_{\mathrm{elec}}$ are due to several factors. First, telemetry is
again limited in resolution to 6-bit data words. Second, the power
consumption of specific instruments is not known from telemetry, only
their nominal power consumption values are known from
documentation. Third, there are uncertainties due to insufficient
documentation. When these sources of error are combined, the result is
an uncertainty of $\sigma_{\tt elec}= 1.8\mathrm{\ W}$ in the electrical heat
output of the spacecraft body~\cite{Toth2009}.

\subsubsection{The thermal recoil force}
\label{sec:thermal_force}

In addition to heat from the RTGs and electrically generated heat,
there are other mechanisms producing heat on board the Pioneer
spacecraft. First, there are 11 radioisotope heater units (RHUs) on
board, each of which generated 1~W of heat at launch, using \super{238}Pu
fuel with a half-life of 87.74~years. Second, the propulsion system,
when used, can also generate substantial amounts of heat.

Nonetheless, these heat sources can be ignored. The total amount of
heat generated by the RHUs is not only small, most of the RHUs
themselves are mounted near the edge of the high-gain antenna (see
Figure~\ref{fig:propulsion}), and a significant proportion of their
heat is expected to be emitted in a direction perpendicular to the
spin axis. As to the propulsion system, while it can generate
substantial quantities of heat, these events are transient in nature
and are completely masked by uncertainties in the maneuvers themselves,
which are responsible for this heat generation. These arguments can
lead to the conclusion that insofar as the anomalous acceleration is
concerned, only the heat from the RTGs and electrical equipment
contribute noticeably.

As discussed in Subsection~\ref{sec:thermal-formalism} above, knowledge of the physical
properties (thermal properties and geometry) of the spacecraft and its
internal heat sources is sufficient to compute heat, and thus
momentum, transfer between the spacecraft and the sky. This can be
accomplished using direct calculational methods, such as industry
standard finite element thermal-mechanical modeling software. The
availability of redundant telemetry (in particular, the simultaneous
availability of electrical and thermal measurements) makes it possible
to develop a more robust thermal model and also establish reasonable
limits on its accuracy. Such a model has recently been developed at
JPL~\cite{KIN2007} and is yielding valuable results. The results of
this analysis will be published when available (see also
Figures~\ref{fig:thermal_model} and \ref{fig:tempmap}).

It is also possible to conduct a simplified analysis of the Pioneer
spacecraft. First, taking into consideration the spacecraft's spin
means that the thermal recoil force only needs to be calculated in the
spin axis direction (see Section~\ref{sec:nongrav}). Second, it has
been argued in~\cite{Toth2009} that for these spacecraft, the total
recoil force can be accurately modeled as a quantity that is
proportional to some linear combination of the thermal power of the
two dominant heat sources, the RTGs and electrical equipment. Thus, one
may write
\begin{equation}
\vec{F}=\frac{1}{c}(\xi_{\mathrm{rtg}}B_{\mathrm{rtg}}+\xi_{\mathrm{elec}}B_{\mathrm{elec}})\vec{s},
\label{eq:Fer}
\end{equation}
where $\xi_{\mathrm{rtg}}$ and $\xi_{\mathrm{elec}}$ are efficiency factors
associated with RTG thermal power $B_{\mathrm{rtg}}$ and electrical power
$B_{\mathrm{elec}}$, while $\vec{s}$ is a unit vector in the spin axis
direction.

The factors $\xi_{\mathrm{rtg}}$ and $\xi_{\mathrm{elec}}$ can be computed, in
principle, from the geometry and thermal properties of the
spacecraft. However, there also exists another possible
approach~\cite{Toth2009}: after incorporating the force model
Equation~(\ref{eq:Fer}) into the orbital equations of motion, orbit
determination software can solve for these parameters, fitting their
values, along with the spacecraft's initial state vector, maneuvers,
and other parameters, to radiometric Doppler
observations~\cite{Toth2009}. While this approach seems promising, its
success depends on the extent to which orbit determination code can
disentangle the thermal recoil force, solar pressure, and a possible
anomalous contribution from one another based on radiometric Doppler
data alone.

A further complication arises from the fact that although late in
their mission, the physical and thermal configuration of the Pioneer
spacecraft were constant in time, this was not always the
case. Earlier in their mission, when the spacecraft were closer to the
Sun, their internal temperatures were regulated by a thermal louver
system located on the aft side (i.e., the side opposite the high-gain
antenna; see Figure~\ref{fig:louvers}). When these louvers were
partially open (see Figure~\ref{fig:langle}), the effective thermal
emissivity of the aft side was significantly higher, and varied as a
function of internal temperatures (see Figures~\ref{fig:lloss} and
\ref{fig:lperf}). While this effect is difficult to model
analytically, it can be incorporated into a finite element thermal
model accurately.

These recent studies and on-going investigations made it clear that
the figure published in 2002~\cite{pioprd} is likely an
underestimation of the thermal recoil force; far from being
insignificant, the thermal recoil force represents a substantial
fraction of the force required to generate the anomalous acceleration
seen in Pioneer data, and may, in fact, account for all of it. This
possibility clearly demands a thorough, in-depth analysis of the
thermal environment on-board the Pioneers. Meanwhile, all the
needed thermal and power data exist in the form of on-board
telemetry~\cite{2006CaJPh..84.1063T, 2007arXiv0710.2656T,
  MDR2005}. All the tools needed to analyze resulted thermal recoil
forces are now built and tested~\cite{KIN2007, Kinsella-etal:2007,
  Kinsella-etal:2008, Rievers2010, Rievers2009, Toth2008, Toth2009,
  2007arXiv0710.0191T}. The analysis now approaches its most exciting part.

If the anomaly, even in part, is of thermal origin, its magnitude must
decrease with time as the on-board fuel inventory decreases (see
Figure~\ref{fig:power}, for example). Therefore, a thermal model will
necessarily predict a decreasing trend in the anomaly. To what extent
will this trend contradict the previously reported ``constancy'' of
the effect? Or will it support trends already seen as the jerk terms
reported by independent verifications
(Section~\ref{sec:indep-verify})? To that extent we emphasize that the
primary data set for the new investigation of the Pioneer anomalous
acceleration is the much extended set of radiometric Doppler tracking
data available for both spacecraft (Section~\ref{sec:new-data}). It is
clear that if the anomaly was found in the navigational data, it must
be re-evaluated using data of the same nature. This is why the new set
of Doppler data that recently became available, in conjunction with
the newly built tools to evaluate thermal recoil forces discussed
above, is now being used to evaluate the long-term temporal behavior,
direction and other important properties of the Pioneer anomaly
(Section~\ref{sec:unresolved-questions}).

Finally, after a period of tedious preparatory work conducted during
2002\,--\,2009, the study of the Pioneer anomalous acceleration enters
its final stages; the results of this work will be reported.

\newpage
\section{Conclusions}
\label{sec:conclusions}

The Pioneer anomaly entered the phenomenology of modern physics at a
time when researchers have mounted significant efforts to investigate
two other observational anomalies: the flat rotation curves of spiral
galaxies and the accelerated expansion of the universe. Although the
true origin of the anomalous acceleration of the Pioneer~10 and 11
spacecraft is yet to be reported, its existence has already helped us
deepen our understanding of gravity. Unlike the other two anomalies,
the Pioneer anomaly is ``local'' in character, taking place within our
own solar system, involving a spacecraft of our own making. As a
result, the Pioneer anomaly has led to increasing interest in space-based
fundamental physics research focusing on solar system tests of
gravity. This renewed interest is, perhaps, the most significant
contribution of the Pioneer effect to modern physics to date.

The presence of the unexpected acceleration signal in the Pioneer
radiometric Doppler data has led to a re-evaluation of theoretical
frameworks used to test gravity in the solar system. It has also motivated
the development of new physical mechanisms to explain the effect. The
anomaly has necessitated a re-analysis of many physical concepts,
resulting in studies of the behavior of gravitationally bound systems
in an expanding universe, investigations of gravity modification
mechanisms, and efforts to detect the anomaly with other spacecraft,
planets, and other bodies in the solar system. The anomaly has also
motivated the development of new methods to improve the accuracy of
spacecraft navigation, including methods of accounting for the effect
of thermal recoil forces.

In this review we describe the Pioneer~10 and 11 spacecraft. We
provide a significant amount of information on the design, operations
and behavior of the two Pioneers during their entire missions. This
includes information from original project documentation and
descriptions of various data formats and techniques that were used for
acquisition of the Pioneer data. We describe the radiometric Doppler
data and techniques for data preparation and analysis. We also
discussed the Pioneer telemetry data and its value for the anomaly
investigation. We review the observational techniques and physical
models that were used for precision tracking of the Pioneer
spacecraft. We summarize the current knowledge of the physical
properties of the Pioneer anomaly and review various mechanisms
proposed for its explanation.

October 5, 2009 marked the eleventh anniversary of the first
announcement of the Pioneer anomaly~\cite{pioprl} (see
also~\cite{pioprd, moriond}). In the decade that followed, the
existence of the anomalous acceleration in the radiometric Doppler
data received from the Pioneer~10 and 11 spacecraft was confirmed by
several independent researchers \cite{levy-2008, 2002gr.qc.....8046M,
  2007AA...463..393O, Toth2008}. Thus, the existence and approximate
magnitude of the anomaly can be considered established fact. However,
the direction of the acceleration remains unclear: the four principal
directions (sunward, earthward, along the spin axis, or along the
velocity vector) fall within a few degrees of each other and based on
available data, cannot be distinguished easily. The temporal behavior
of the anomaly has been put into question by
studies~\cite{2002gr.qc.....8046M, Toth2008} that demonstrate the
presence of a jerk term. Meanwhile, other studies
(see~\cite{Kinsella-etal:2007, Kinsella-etal:2008, ISSI2005,
  Scheffer:2005, Scheffer:2008, TothISSI2007}, followed
by~\cite{2008PhRvD..78j3001B, Rievers2010, Rievers2009}), indicate
that the magnitude of acceleration due to thermal recoil forces of
on-board origin may be significantly larger than previously
estimated. As a result, the question of the origin of the Pioneer
anomaly remains open, but hopefully not for long.

A comprehensive investigation of the anomaly has recently begun. The
new study relies on the much-extended set of radiometric Doppler data
for both spacecraft in conjunction with the entire record of the
Pioneer~10 and 11 spacecraft telemetry files and large archive of
project documentation. This unique information has already led to the
development of new software tools capturing all aspects of the
in-flight behavior of the Pioneer vehicles throughout their
missions. These efforts may soon reveal the origin of the
anomaly. Our review provides the necessary background for the new
results to appear in the near future. Such an anticipated development
makes the study of the Pioneer anomaly a good subject for
\textit{Living Reviews}.

\section{Acknowledgements}
\label{sec:acknowledgements}

We would like to express our gratitude to our many colleagues who have
either collaborated with us on this manuscript or given us their
wisdom. First among the many people who have helped us with
suggestions, comments, and constructive criticisms, we must thank
Craig B.\ Markwardt of GSFC, whose enthusiastic contribution to the
Pioneer Doppler data recovery efforts was matched by the thoroughness
with which he reviewed and commented on this manuscript. We also thank
John D.\ Anderson, Hansj\"org Dittus, Claus L\"ammerzahl, Agn\`es
Levy, John Moffat, and Serge Reynaud, for their encouraging comments
and suggestions on this manuscript.

We specifically thank Sami Asmar, Curt J.\ Cutler, William M.\
Folkner, Timothy P.\ McElrath, Robert A.\ Jacobson, Michael M.\
Watkins, and James G.\ Williams of JPL, who provided us with
encouragement and valuable comments while this manuscript was in
preparation. We also thank Jordan Ellis, Gene L.\ Goltz, and Neil A.\
Mottinger of JPL for their help in obtaining, understanding and
conditioning of the Pioneer Doppler data.

We are grateful to Gary Kinsella of JPL and Siu-Chun Lee and Daniel
S.\ Lok of Applied Sciences Laboratory who benefited us with their
insightful comments and suggestions regarding the thermal modeling of
Pioneer spacecraft. Louis K. Scheffer of Cadence Design Systems
contributed with useful observations on the thermal modeling, design
and analysis.

Invaluable information on the history, spacecraft design and mission
operations of the Pioneers~10 and 11, as well as the structure of
their telemetry data, came from Lawrence Lasher, Larry Kellogg and
David Lozier formerly of the NASA Ames Research Center.

It is our special pleasure to thank the members of the Pioneer
Explorer Collaboration (POC) that conducted an extensive Pioneer
anomaly investigation at the ISSI for many fruitful discussions of
various aspects of the study of the anomaly. In particular, in
addition to our colleagues on the POC already mentioned above, our
wholehearted thanks are going to Orfeu Bertolami, Francois Bondu,
Bruno Christophe, Jean-Michel Courty, Denis Defrere, Jonathan Fitt,
Frederico Francisco, Bernard Foulon, Paolo Gill, Stefanie Grotjan, Eva
Hackmann, Ulrich Johann, Meike List, Clovis J.\ De Matos, Gilles
M\`etris, Laura Mulin, {\O}ystein Olsen, Jorge Paramos, Sergei
Podgrebenko, Andreas Rathke, Benny Rievers, Wolfgang Seboldt, Stephan
Theil, Paolo Tortora, Pierre Touboul, Patrick Vrancken, Peter Wolf and
many others.

Our gratitude goes to Michael H.\ Salamon of NASA and Hans Mark of
University of Texas in Austin who have kindly provided us with
encouragement, stimulating conversations, and support during various
phases of our work.

We also thank The Planetary Society for support and, in particular,
Louis D.\ Freidman, Charlene M.\ Anderson, and Bruce Betts for their
interest, stimulating conversations and encouragement.

This work was partially performed at the International Space Science
Institute (ISSI), Bern, Switzerland, when both of us visited ISSI as
part of an International Team program. In this respect we thank Roger
M.\ Bonnet, Vittorio Manno, Brigitte Schutte, and Saliba F.\ Saliba of
ISSI for their hospitality and support.

Finally, we would like to thank \textit{Living Reviews in Relativity}
and especially Bernard F.\ Schutz and Clifford M.\ Will for the
rewarding opportunity to prepare this manuscript.

The work described here, in part, was carried out at the Jet
Propulsion Laboratory, California Institute of Technology, under a
contract with the National Aeronautics and Space Administration.

\newpage
\newpage

\part*{Appendices}

\appendix
\renewcommand\theappendix{Appendix~}
\renewcommand\thesection{{\protect\theappendix}\Alph{section}}

\section{Pioneer Spacecraft Geometry}
\label{app:geometry}

Only scarce documentation is available about the exact geometry of the
Pioneer~10 and 11 spacecraft. A sketch of the major components of the
spacecraft (our reconstruction) with approximate dimensions is shown
below.

\epubtkImage{}{%
  \begin{figure}[htbp]
  \centerline{\includegraphics[width=0.75\linewidth]{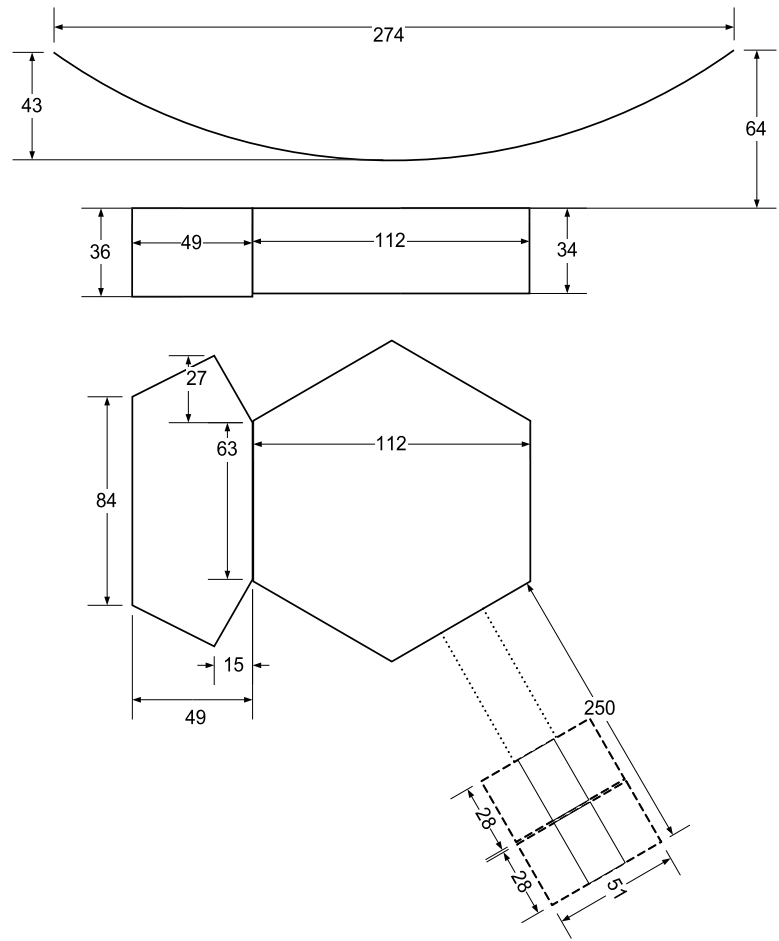}}
  \caption{Side view (above) and top view (below) of the high gain
  antenna, spacecraft body, and one RTG, with the boom length not to
  scale. Approximate measurements are in centimeters. For more
  accurate dimensions of the RTGs, consult Figure~\ref{fig:RTGDIM}.}
\end{figure}}

\newpage

\section{Format of the Orbit Data Files (ODF)}
\label{app:ODF}

The following table contains a concise but complete description of the
structure of Orbit Data Files (ODFs). Records appear in ODFs in the
order indicated. We offer this table to illustrate what information is
available in the form of radio-metric Doppler
measurements~\cite{TRK-2-18}.

\vspace{1em}

{
\newcommand{\odfline}{\vspace{-2pt}\\\hline}
\linespread{1.25}
\fontsize{7}{7}\selectfont
\hskip -3em
\begin{tabular}{|r|c|c|c|}\hline
\multicolumn{4}{|l|}{\bf ~ODF File Label Group Header}\odfline
0 & Primary Key $=101$ & Secondary Key $=0$ & Logical Record Length $=1$\odfline
96 & Group Start Packet Number $=0$ & \multicolumn{2}{c|}{0}\odfline
192 & \multicolumn{3}{c|}{0}\odfline
\multicolumn{4}{|l|}{\bf ~ODF File Label Group Data}\odfline
0 & \multicolumn{2}{c|}{System ID in ASCII} & Program ID in ASCII\odfline
96 & Program ID (cont.) & Spacecraft ID & File Creation Date (YYMMDD)\odfline
192 & File Creation Time (hhmmss) & File Reference Date (19500101) & File Reference Time (000000)\odfline
\multicolumn{4}{|l|}{\bf ~ODF Identifier Group Header}\odfline
0 & Primary Key $=107$ & Secondary Key $=0$ & Logical Record Length $=1$\odfline
96 & Group Start Packet Number $=2$ & \multicolumn{2}{c|}{0}\odfline
192 & \multicolumn{3}{c|}{0}\odfline
\multicolumn{4}{|l|}{\bf ~ODF Identifier Group Data}\odfline
0 & \multicolumn{2}{c|}{\tt "TIMETAG"} & {\tt "OBSR}\odfline
96 & {\tt VBL"} & \multicolumn{2}{c|}{\tt "FREQ. AN}\odfline
192 & \multicolumn{3}{c|}{\tt CILLARY-DATA"}\odfline
\multicolumn{4}{|l|}{\bf ~ODF Orbit Data Group Header}\odfline
0 & Primary Key $=109$ & Secondary Key $=0$ & Logical Record Length $=1$\odfline
96 & Group Start Packet Number $=4$ & \multicolumn{2}{c|}{0}\odfline
192 & \multicolumn{3}{c|}{0}\odfline
\multicolumn{4}{|l|}{\bf ~ODF Orbit Data Group Data}\odfline
\begin{tabular}{r@{}}0\vspace{20pt}\end{tabular} & \multicolumn{3}{@{}l@{}|}{
\begin{tabular*}{\linewidth}{r|l}
~~~0 & Record Time Tag Integer Part (s)\odfline
32 & Record Time Tag Fractional Part (ns)\odfline
42 & Receiving Station Downlink Delay (ns)\odfline
64 & Observable Integer Part (signed)\\
\end{tabular*}}\odfline
\multicolumn{1}{|@{}r@{}|}{\begin{tabular}{r}~~96\vspace{90pt}\odfline\end{tabular}} & \multicolumn{3}{@{}l@{}|}{
\begin{tabular*}{\linewidth}{r|l}
96 & Observable Fractional Part ($10^{-9}$, signed)\odfline
128 & Format ID $=2$\odfline
131 & Receiving Station ID\odfline
138 & Transmitting Station ID\odfline
145 & Network ID ($0=$ DSN Block V, $1=$ Other, $2=$ OTS, $3=$ NSP)\odfline
147 & Data Type ($1x=x$-way Doppler)\odfline
153 & Downlink Band ID ($1=$ S-band)\odfline
155 & Uplink Band ID ($1=$ S-band)\odfline
157 & Exciter Band ID ($1=$ S-band)\odfline
159 & Data Validity Indicator ($0=$ good)\odfline
160 & Second Receiving Station ID\odfline
167 & Spacecraft ID\odfline
177 & Receiver/Exciter Independent Flag\odfline
178 & Reference Frequency (mHz)\\
\end{tabular*}}\\
\multicolumn{1}{|@{}r@{}|}{\begin{tabular}{r}192\vspace{20pt}\end{tabular}} & \multicolumn{3}{@{}l@{}|}{\begin{tabular*}{\linewidth}{r|l}
~ & ~\odfline
224 & Train Axis Angle (mdeg, OTS Doppler)\odfline
244 & Compression Time (0.01 s)\odfline
266 & Transmitting Station Uplink Delay (ns)\\
\end{tabular*}}\odfline
\multicolumn{4}{|l|}{\bf ~ODF Ramp Groups Header}\odfline
0 & Primary Key $=2030$ & Secondary Key $=0$ & Logical Record Length $=1$\odfline
96 & Group Start Packet Number & \multicolumn{2}{c|}{0}\odfline
192 & \multicolumn{3}{c|}{0}\odfline
\multicolumn{4}{|l|}{\bf ~ODF Ramp Groups Data}\odfline
0 & Start Time Integer Part (s) & Start Time Fract. Part (ns) & Rate Integer Part (signed, Hz/s)\odfline
96 & Rate Fract. Part (nHz/s) & \begin{tabular}{c|c}Start Frq. GHz Part & Station ID\end{tabular} & Start Frequency Hz Part)\odfline
192 & Start Frequency Fract. Part (nHz) & End Time Integer Part (s) & End Time Fract. Part (ns)\odfline
\multicolumn{4}{|l|}{\bf ~ODF Clock Offsets Group Header}\odfline
0 & Primary Key $=2040$ & Secondary Key $=0$ & Logical Record Length $=1$\odfline
96 & Group Start Packet Number & \multicolumn{2}{c|}{0}\odfline
192 & \multicolumn{3}{c|}{0}\odfline
\multicolumn{4}{|l|}{\bf ~ODF Clock Offsets Group Data}\odfline
0 & Start Time Integer Part (s) & Start Time Fract. Part (ns) & Offset Integer Part (s, signed)\odfline
96 & Offset Fract. Part (ns) & Primary Station ID & Secondary Station ID\odfline
192 & 0 (spare) & \multicolumn{2}{c|}{0 (reserved for End Time)}\odfline
\multicolumn{4}{|l|}{\bf ~ODF Data Summary Group Header}\odfline
0 & Primary Key $=105$ & Secondary Key $=0$ & Logical Record Length $=1$\odfline
96 & Group Start Packet Number & \multicolumn{2}{c|}{0}\odfline
192 & \multicolumn{3}{c|}{0}\odfline
\multicolumn{4}{|l|}{\bf ~ODF Data Summary Group Data}\odfline
0 & 1st Sample Time Integer Part (s) & 1st Sample Time Fract. Part (ns) & Receiving Station ID\odfline
96 & Doppler Channel Number & Downlink Band ID & Data Type ID\odfline
192 & Number of Samples & Last Sample Time Integer Part (s) & Last Sample Time Fract. Part (ns)\odfline
\multicolumn{4}{|l|}{\bf ~ODF End-Of-File Group Header}\odfline
0 & Primary Key $=-1$ & Secondary Key $=0$ & Logical Record Length $=0$\odfline
96 & Group Start Packet Number & \multicolumn{2}{c|}{0}\odfline
192 & \multicolumn{3}{c|}{0}\odfline
\end{tabular}
}

\newpage

\section{Master Data Records}
\label{app:mdr}


All information received by the DSN from the Pioneer~10 and 11
spacecraft was stored in the form of Master Data Records, which were
structured as shown in the table below. As the structure demonstrates,
in addition to the 6-bit data words (see \ref{app:formatCE})
encoded in the data frames, the MDRs also contain information about
the receiving DSN station~\cite{ARC221, ARC166A, ARC061, ICA84,
ICA155, RCD020}.

\vspace{1em}

{
\centering
\begin{footnotesize}
\begin{picture}(416,284)
\drawline(32,276)(416,276)
\drawline(32,264)(416,264)
\drawline(32,252)(416,252)
\drawline(32,240)(416,240)
\drawline(32,228)(416,228)
\drawline(32,216)(416,216)
\drawline(32,204)(416,204)
\drawline(32,192)(416,192)
\drawline(32,180)(416,180)
\drawline(32,168)(416,168)
\drawline(32,156)(416,156)
\drawline(32,96)(416,96)
\drawline(32,84)(416,84)
\drawline(32,72)(416,72)
\drawline(32,60)(416,60)
\drawline(32,48)(416,48)
\drawline(32,36)(416,36)
\drawline(32,24)(416,24)
\drawline(32,12)(416,12)
\drawline(32,0)(416,0)
\drawline(32,0)(32,276)
\drawline(416,0)(416,276)

\drawline(128,252)(128,264)
\drawline(224,252)(224,264)

\drawline(128,240)(128,252)
\drawline(224,240)(224,252)
\drawline(320,240)(320,252)

\drawline(128,228)(128,240)
\drawline(224,228)(224,240)

\drawline(128,216)(128,228)
\drawline(224,216)(224,228)

\drawline(128,204)(128,216)
\drawline(224,204)(224,216)

\drawline(128,192)(128,204)
\drawline(224,192)(224,204)

\drawline(128,168)(128,180)
\drawline(224,168)(224,180)

\drawline(224,84)(224,96)

\drawline(224,72)(224,84)
\drawline(332,72)(332,84)

\drawline(128,60)(128,72)
\drawline(224,60)(224,72)
\drawline(272,60)(272,72)

\drawline(140,48)(140,60)
\drawline(320,48)(320,60)

\drawline(224,36)(224,48)
\drawline(332,36)(332,48)

\drawline(128,24)(128,36)
\drawline(224,24)(224,36)

\drawline(140,12)(140,24)
\drawline(320,12)(320,24)

\drawline(128,0)(128,12)
\drawline(224,0)(224,12)
\drawline(320,0)(320,12)

\put(12,278){~bit}
\put(12,266){~~~0}
\put(12,254){~~32}
\put(12,242){~~64}
\put(12,230){~~96}
\put(12,218){~128}
\put(12,206){~160}
\put(12,194){~192}
\put(12,182){~224}
\put(12,170){~256}
\put(12,158){~288}
\put(12,146){~320}
\put(12,134){~~~.}
\put(12,122){~~~.}
\put(12,110){~~~.}
\put(12,98){1087}
\put(12,86){1088}
\put(12,74){1120}
\put(12,62){1152}
\put(12,50){1184}
\put(12,38){1216}
\put(12,26){1248}
\put(12,14){1280}
\put(12,2){1312}

\put(200,266){TIME TAG}

\put(68,255){SC/ID}
\put(140,255){TIME COR FLAG}
\put(288,255){DAY OF YEAR}

\put(72,243){UDT}
\put(168,243){DDT}
\put(228,243){SYNC COND CODE}
\put(360,243){DQI}

\put(40,231){\# BIT ERRORS PN}
\put(148,231){YEAR DIGIT}
\put(312,231){SNR}

\put(72,219){DSS}
\put(132,219){LOCK STATUS BITS}
\put(256,219){CONFIGURATION INDICATORS}

\put(40,207){SPCL DATA TYPE}
\put(164,207){GDD}
\put(260,207){\# OF DATA BITS IN RECORD}

\put(40,195){\# AGC SAMP AVER}
\put(132,195){HSD ERR CON BITS}
\put(256,195){RATE OF DATA TRANSMISSION}

\put(136,183){AVERAGE AGC OVER DATA IN RECORD}

\put(64,171){FORMAT}
\put(164,171){SPARE}
\put(272,171){NUMBER OF FRAMES}

\put(223,159){0}

\put(196,123){DATA FRAMES}

\put(64,87){MS CLOCK LSB (FRAME 2 OF 4)}
\put(256,87){MS CLOCK LSB (FRAME 3 OF 4)}

\put(64,75){MS CLOCK LSB (FRAME 4 OF 4)}
\put(236,75){DDA I/P ERRORS (1)}
\put(364,75){DDA}

\put(40,63){COMPUTATIONS (1)}
\put(144,63){DDA STATUS (1)}
\put(232,63){SPARE}
\put(288,63){GROUND RECEIVER AGC}

\put(40,51){DDA I/P ERRORS (2)}
\put(172,51){DDA COMPUTATIONS (2)}
\put(336,51){DDA STATUS (2)}

\put(112,39){SPARE}
\put(236,39){DATA I/P ERRORS (3)}
\put(364,39){DDA}

\put(40,27){COMPUTATIONS (3)}
\put(140,27){DDA STATUS (2)}
\put(304,27){SPARE}

\put(40,15){DDA I/P ERRORS (4)}
\put(172,15){DDA COMPUTATIONS (4)}
\put(336,15){DDA STATUS (4)}

\put(72,3){SCF 1}
\put(168,3){SCF 2}
\put(264,3){SCF 3}
\put(358,3){SCF 4}

\end{picture}
\end{footnotesize}
}

\newpage

\section{Selected Engineering and Science Instrument Telemetry Words}
\label{app:formatCE}

In the following table, we list selected engineering and science
instrument telemetry words (Formats C and E). Analog sensor readings
have word type 'A'; word type 'B' is for bit fields, while 'D'
represents digital values (e.g., counters). This list illustrates the
redundant information that is available from telemetry, describing the
thermal and electrical state of the spacecraft.

\begin{small}
\begin{longtable}{|p{0.05\linewidth}|c|p{0.51\linewidth}|c|}
\hline
Word & Bit & Definition & Type\\\hline\hline
\endfirsthead
\hline
Word & Bit & Definition & Type\\\hline\hline
\endhead
\hline
\endfoot
\hline
\endlastfoot
C\sub{104} && Extended subcommutator ID.&B\\
C\sub{105} && RTG 2 current (0--11~A)&A\\
C\sub{106} && Battery voltage (0--15~V)&A\\
C\sub{107} && DC bus voltage (26--30~V)&A\\
C\sub{108}&1&JPL/HVM power&B\\
&2&ARC/PA power&\\
&3&UC/CPI power&\\
&4&UI/GTT power&\\
&5&GSFC/CRT power&\\
C\sub{109} && Battery charge current (0--0.3~A)&A\\
C\sub{110} && RTG 1 voltage (0--6~V)&A\\
C\sub{111} && Receiver A AGC CONSCAN --4 to +4~dB&A\\
C\sub{113} && RTG 4 voltage (0--6~V)&A\\
C\sub{114} && RTG 3 current (0--11~A)&A\\
C\sub{115} && Battery temperature (--20 to 120\textdegree)F&A\\
C\sub{117} && TRF +5~V output CDU Bus A (0--6~V)&A\\
C\sub{118} && TRF +5~V output CDU Bus B (0--6~V)&A\\
C\sub{119} && DC bus voltage (0--30~V)&A\\
C\sub{121} && Receiver B AGC CONSCAN --4 to +4~dB&A\\
C\sub{122} && Shunt bus current (0--3~A)&A\\
C\sub{123} && RTG 4 current (0--11~A)&A\\
C\sub{124}&1&UCSD/TRD power&B\\
&2&USC/UV power&\\
&3&UA/IPP power&\\
&4&CIT/IR power&\\
&5&GE/AMD power&\\
&6&LaRC/MD power&\\
C\sub{125} && RTG 2 voltage (0--6~V)&A\\
C\sub{126} && Battery discharge current (0--10~A)&A\\
C\sub{127} && RTG 1 current (0--11~A)&A\\
C\sub{128}&1&Battery charge (0=auto, 1=float)&B\\
&2&Battery discharge (0=enabled)&\\
C\sub{129} && DC bus current (0--6~A)&A\\
C\sub{130} && Nitrogen tank temperature (\textdegree\,F)&A\\
C\sub{131} && RTG 3 voltage (0--6~V)&A\\
C\sub{201} && RTG 1 fin root temperature (160--360\textdegree\,F)&A\\
C\sub{202} && RTG 2 fin root temperature (160--360\textdegree\,F)&A\\
C\sub{203} && RTG 3 fin root temperature (160--360\textdegree\,F)&A\\
C\sub{204} && RTG 4 fin root temperature (160--360\textdegree\,F)&A\\
C\sub{205} && TWT A temperature (40--125\textdegree\,F)&A\\
C\sub{206} && Driver a temperature (20--110\textdegree\,F)&A\\
C\sub{207} && TWT A converter temperature (40--125\textdegree\,F)&A\\
C\sub{208} && TWT A cathode current (24--30~mA)&A\\
C\sub{209} && Shunt bus current (0--3~A)&A\\
C\sub{210} && Propellant supply pressure (0--600~PSIA)&A\\
C\sub{211} && TWT A helix current (0--10~mA)&A\\
C\sub{212} && Receiver A loop stress (--100 to 100~kHz)&A\\
C\sub{213} && Receiver B signal strength (--149 to --63~dBm)&A\\
C\sub{214} && TWT B RF output power (26--40.4~dBm)&A\\
C\sub{215} && TWT B cathode current (24--30~mA)&A\\
C\sub{216} && TWT B helix current (0--10~mA)&A\\
C\sub{217} && RTG 4 hot junction temperature (880--1200\textdegree\,F)&A\\
C\sub{218} && RTG 3 hot junction temperature (880--1200\textdegree\,F)&A\\
C\sub{219} && RTG 2 hot junction temperature (880--1200\textdegree\,F)&A\\
C\sub{220} && RTG 1 hot junction temperature (880--1200\textdegree\,F)&A\\
C\sub{221} && TWT B converter temperature (40--125\textdegree\,F)&A\\
C\sub{222} && Receiver A VCO temperature (20--110\textdegree\,F)&A\\
C\sub{223} && Driver B temperature (20--110\textdegree\,F)&A\\
C\sub{224} && TWT A reference voltage (0--28~V)&A\\
C\sub{225} && +Y PSA line temperature (\textdegree\,F)&A\\
C\sub{226} && --Y PSA line temperature (\textdegree\,F)&A\\
C\sub{227} && Receiver B VCO temperature (20--110\textdegree\,F)&A\\
C\sub{228} && TWT B temperature (40--125\textdegree\,F)&A\\
C\sub{229} && Receiver B loop stress (--100 to 100~kHz)&A\\
C\sub{230} && TWT B reference voltage (0--28~V)&A\\
C\sub{231} && TWT A RF output power (26--40.4~dBm)&A\\
C\sub{232} && Receiver A signal strength (--149 to --63~dBm)&A\\
C\sub{301} && Platform temperature 1 (0--140\textdegree\,F)&A\\
C\sub{302} && Platform temperature 2 (0--140\textdegree\,F)&A\\
C\sub{303} && SRA temperature (--10 to 95\textdegree\,F)&A\\
C\sub{304} && Platform temperature 3 (0--140\textdegree\,F)&A\\
C\sub{308}&1&Receiver A signal present&B\\
&2&Receiver B signal present&\\
&3&Receiver A oscillator enabled&\\
&4&Receiver B oscillator enabled&\\
&5&Spin thruster B pulse count&\\
&6&Spin thruster A pulse count&\\
C\sub{309} && Velocity thruster cluster 1 temperature (40--200\textdegree\,F)&A\\
C\sub{310} && Spin thruster cluster temperature (40--200\textdegree\,F)&A\\
C\sub{311} && VPT 1 temperature (400--1800\textdegree\,F)&A\\
C\sub{312} && VPT 2 temperature (400--1800\textdegree\,F)&A\\
C\sub{316}&1&CONSCAN power&D\\
&2&CONSCAN threshold (1=above)&\\
&4&Receiver switch status (1=A/B=Hi/Med)&\\
&5&Transmitter switch status (1=A/B=Med/Hi)&\\
&6&Antenna feed switch status (1=offset)&\\
C\sub{317} && SSA temperature (--30 to 194\textdegree\,F)&A\\
C\sub{318} && Platform temperature 4 (0--140\textdegree\,F)&A\\
C\sub{319} && Platform temperature 5 (0--140\textdegree\,F)&A\\
C\sub{320} && Platform temperature 6 (0--140\textdegree\,F)&A\\
C\sub{321} && Velocity thruster 2 (1B) pulse count&D\\
C\sub{322} && Velocity thruster 4 (2A) pulse count&D\\
C\sub{324}&1&Command memory status (0=processing, 1=standby)&B\\
C\sub{325} && VPT 4 temperature (400--1800\textdegree\,F)&A\\
C\sub{326} && Velocity thruster cluster 2 temperature (40--200\textdegree\,F)&A\\
C\sub{327} && Propellant supply temperature (40--160\textdegree\,)&A\\
C\sub{328} && VPT 3 temperature (400--1800\textdegree\,F)&A\\
C\sub{329} && Velocity thruster 1 (1A) pulse count&D\\
C\sub{330} && Velocity thruster 3 (2B) pulse count&D\\
C\sub{403}&1&Precession pair (1=VPT 1\&4)&D\\
&2--4&Pulse length bits 1--3&\\
&5&$\Delta v$ pair (1=VPT 1\&3)&\\
&6&Spin control direction (0=up)&\\
C\sub{404}&1--5&Star time gate bits 0--4&D\\
&6&$\Delta v$/SCT mode enabled&\\
C\sub{405} && Spin period MSB&D\\
C\sub{406} && Spin period&D\\
C\sub{407} && Spin period LSB&D\\
C\sub{408}&1--5&Roll pulse/index pulse phase error bits 4--0&D\\
&6&Phase error sign (1=roll pulse before index pulse)&\\
C\sub{409}&1&VPT 1 firing status&B\\
&2&VPT 2 firing status&\\
&3&VPT 4 firing status&\\
&4&VPT 3 firing status&\\
&5&SCT 1 firing status&\\
&6&SCT 2 firing status&\\
C\sub{410}&1&Despin on/off&D\\
&2&CONSCAN enabled&\\
C\sub{421}&1--2&Star count bits 0--2&D\\
&4&CEA power status DSL A&\\
&5&CEA power status DSL B&\\
&6&CEA power status PSE&\\
%
E\sub{101} && ARC/PA detectors temperature&A\\
E\sub{102} && ARC/PA electronics temperature&A\\
E\sub{109} && USC/UV electronics temperature&A\\
E\sub{110} && UC/CPI electronics temperature&A\\
E\sub{117} && CIT/IR low range temperature&A\\
E\sub{118} && GE/AMD preamp temperature&A\\
E\sub{125} && GSFC/CRT electronics temperature&A\\
E\sub{128} && GSFC/CRT detector temperature&A\\
E\sub{201} && CIT/IR high range temperature&A\\
E\sub{209} && UCSD/TRD electronics temperature&A\\
E\sub{221} && UI/GTT electronics temperature&A\\
\end{longtable}
\end{small}


\newpage


\bibliography{refs}

\end{document}